\documentclass[times]{elsarticle}
\usepackage{color}
\usepackage{moreverb} 
\usepackage[colorlinks,bookmarksopen,bookmarksnumbered,citecolor=red,urlcolor=red,pdfauthor=author]{hyperref}
\usepackage{amsfonts,amsmath,amssymb}
\usepackage{psfrag,pstricks,pst-node}

\usepackage[section]{placeins}
\usepackage{float}
\usepackage{subfigure} 
\usepackage{setspace}
\usepackage{mathrsfs}
\usepackage{graphicx}
\usepackage[ruled,vlined]{algorithm2e}
\usepackage{fancyhdr}  
\usepackage{psfrag} 
\usepackage[export]{adjustbox} 
\usepackage{enumerate} 
\usepackage{comment}
\usepackage{lscape}
\usepackage{appendix}
\usepackage{algorithmic}
\usepackage{bm}
\usepackage{booktabs}
\usepackage{threeparttable}
\graphicspath{{./figs/}{./model Figure/}}
\DeclareGraphicsExtensions{.pdf,.eps,.PNG,.png}

\usepackage{pgfplots}
\pgfplotsset{compat=1.13}

\usepackage{geometry}
\usepackage{graphicx,array}
\usepackage{subcaption} 
\usepackage{amsmath}
\usepackage{natbib}
\usepackage{amsmath}
\usepackage{threeparttable} 
\usepackage{multirow}
\usepackage{changepage}
\usepackage{booktabs}
\usepackage{fullpage}
\usepackage{times}
\usepackage{fancyhdr,graphicx,amsmath,amssymb}
\usepackage[ruled,vlined]{algorithm2e}
\usepackage{float}
\usepackage[T1]{fontenc}
\usepackage[utf8]{inputenc}
\usepackage{bm}
\usepackage{booktabs}
\usepackage{nomencl}
\usepackage{amsfonts,amssymb}
\usepackage{enumitem}
\usepackage{makecell}
\usepackage{arydshln}
\usepackage{subfigure}
\usepackage{graphicx}
\usepackage{makecell}

\begin{document}
 
\begin{frontmatter}
\renewcommand{\thefootnote}{\fnsymbol{footnotemark}}
 
\fancypagestyle{plain}{%
\fancyhf{} 
\fancyhead[RO,RE]{\thepage} 
}

\title{Stochastic reconstruction of multiphase composite microstructures using statistics-encoded neural network for poro/micro-mechanical modelling}   
    \author[lab1]{Jinlong Fu}
    \author[lab1]{Wei Tan\corref{cor1}} 
    \cortext[cor1]{Corresponding author}
    \ead{wei.tan@qmul.ac.uk}
    \address[lab1]{School of Engineering and Materials Science, Queen Mary University of London, London, E1 4NS, UK}

\begin{abstract}
Fundamental understanding the microstructure-property relationships (MPRs) is crucial for optimising the performances and functionality of multiphase composites. Image-based poro/micro-mechanical modelling offers a powerful non-invasive method to explore MPRs, but the inherent randomness in multiphase composites often necessitates extensive datasets of 3D digital microstructures for reliable statistical analysis. This paper presents a cost-effective machine learning-based framework to efficiently reconstruct numerous virtual 3D microstructures from a limited number of 2D real exemplars, bypassing the prohibitive costs associated with volumetric microscopy for opaque composites. This innovative framework leverages feedforward neural networks to encode morphological statistics in 2D exemplars, referred to as the statistics-encoded neural network (SENN), providing an accurate statistical characterisation of complex multiphase microstructures. Utilising the SENN-based characterisation, 3D morphological statistics can be inferred from 2D measurements through a 2D-to-3D morphology integration scheme, and then statistically equivalent 3D microstructures are synthesised via Gibbs sampling. This framework further incorporates hierarchical characterisation and multi-level reconstruction approaches, allowing for the seamless capture of local, regional, and global microstructural features across multiple length scales. Validation studies are conducted on three representative multiphase composites, and morphological similarity between the reconstructed and reference 3D microstructures is evaluated by comparing a series of morphological descriptors. Additionally, image-based meshing and pore/micro-scale simulations are performed on these digital microstructures to compute effective macroscopic properties, including stiffness, permeability, diffusivity, and thermal conductivity tensors. Results reveal strong statistical equivalence between the reconstructed and reference 3D microstructures in both morphology and physical properties, confirming the SENN-based framework is a high-fidelity tool to reconstruct multiphase microstructures for image-based poro/micro-mechanical analysis.
\end{abstract}

\begin{keyword} 
Multiphase composites; Stochastic reconstruction; Statistics-encoded neural network; Microstructural characterisation; Poro/micro-mechanical modelling; Statistical equivalence
\end{keyword}
\end{frontmatter}

\section{Introduction}
\label{Section1:Introduction}
\vspace{-2pt}
The internal microstructure of composite materials plays a pivotal role in determining their macroscopic properties \cite{torquato2000modeling, yin2008statistical}, including mechanical strength, permeability, effective diffusivity, thermal conductivity, and other critical functionalities. Multiphase composites, in particular, exhibit complex microstructural patterns characterised by heterogeneous phase distributions and intricate interfacial geometries \cite{li2019tunable}.
A fundamental understanding of the microstructure-property relationships (MPRs) in the multiphase composites materials is crucial for optimising their performance in various applications \cite{cecen2018material, kwok2023review, fu2023data}, such as lightweight and durable materials for aerospace and automotive industries, electrodes or solid-state electrolyte for energy storage, biocompatible implants and scaffold in medical devices, corrosion-resistant structures for marine and constructions applications, and impact-resistant materials for offshore wind turbine blades. However, the inherent randomness and stochastic nature of multiphase composites pose significant challenges to precise microstructural analysis and accurate MPR modelling \cite{xu2021method, nan2024transfer}.

One of the most reliable approaches to studying MPRs involves the use of 3D digital microstructures. These 3D datasets allow researchers to analyse the internal organisation of materials and perform image-based numerical simulations to link microstructural features to macroscopic properties \cite{fu2023data, latypov2019materials, bhuiyan2020predicting}. High-resolution 3D images of composite materials are typically obtained using advanced microscopy techniques such as micro-computed tomography (micro-CT) \cite{madra2017clustering} and scanning electron microscopy (SEM) \cite{urushihara2009situ}. While these imaging techniques have provided unprecedented insights into microstructural features, they are often associated with high costs, limited accessibility, and technical constraints, especially when dealing with opaque or fragile materials. Furthermore, the demand for large datasets of 3D digital microstructures to conduct reliable statistical analyses \cite{ostoja2006material, sriramula2009quantification, bessa2017framework} further amplifies these limitations, especially for Monte Carlo analysis.

To address this issue, there has been growing interest in using computational methods to synthesise 3D microstructures from limited data \cite{bostanabad2018computational, fu2023hierarchical}. Stochastic reconstruction techniques \cite{fu2022stochastic, hamza2024multi} leverage statistical descriptions of microstructures to generate virtual 3D samples that are statistically equivalent to real materials. These techniques can broadly be classified into two categories: conventional algorithm-based approaches and computational intelligence-based approaches. Typical conventional methods include stochastic optimisation-based reconstruction \cite{torquato1998reconstructing, ju20143d}, Gaussian random field transformation \cite{liang1998reconstruction}, process-based reconstruction \cite{oren2002process}, multiple-point statistics \cite{hajizadeh2011multiple}, patch-based reconstruction \cite{tahmasebi2012reconstruction}, texture synthesis \cite{liu2015random}, descriptor-based methodologies \cite{xu2014descriptor, mollon20143d}, and packing algorithms \cite{yang2018new}. These methods are often mathematically rigorous, interpretable, and allow for systematic variation of preserved microstructural characteristics by adjusting reconstruction targets or parameters. However, conventional methods face significant limitations when handling complex, multiphase microstructures, as they struggle to accurately capture the intricate geometries, correlations, and dependencies among constituent phases. While capable of generating mathematically equivalent microstructures, these methods fail to produce physically consistent ones, which are critical for preserving fine details required in MPR analysis. In addition, these methods are predominantly developed for bi-phase materials and exhibit limited capability in reconstructing multiphase composite microstructures.

Computational intelligence \citep{engelbrecht2007computational} has unlocked exciting opportunities for stochastic reconstruction of random microstructures, driven by rapid advancements in AI algorithms, processors, data storage, and computing power. A key advantage of deep learning algorithms \citep{lecun2015deep} is their ability to operate without manual feature design, extraction, or selection, streamlining the reconstruction process.
As a result, a wide range of computational intelligence-based approaches have been developed for reconstructing random materials. Simple machine learning algorithms, such as support vector machines \cite{sundararaghavan2005classification}, decision trees \citep{bostanabad2016stochastic, bostanabad2016characterization}, and feedforward neural networks \citep{fu2023hierarchical, fu2021statistical}, require manually designed input features to capture specific morphological characteristics. In contrast, deep learning algorithms automatically extract hierarchical features from raw input images using convolutional or other imag processing operations. Notable methods include generative adversarial networks \citep{mosser2017reconstruction, feng2020end, henkes2022three}, convolutional neural networks \citep{wang2018porous, noguchi2021stochastic}, deep belief networks \cite{cang2017microstructure}, variational autoencoders \cite{laloy2017inversion}, transfer learning \cite{li2018transfer}, generative flow networks \cite{guan2021reconstructing}, recurrent neural networks \cite{zhang2023pm}, U-Nets \cite{zheng2024generative}, and diffusion-based generative models \cite{lee2024denoising}.
Despite their advantages, deep learning models suffer from poor interpretability due to their complex architectures, making it difficult to determine what information learned from the real microstructure is embedded into the reconstructed samples. Most existing methods target bi-phase materials, such as porous media. While some techniques \cite{zheng2024generative, lee2024denoising, gao2021ultra, kench2021generating} have been applied to reconstruct multiphase microstructures by maintaining morphological similarity, they have not demonstrated the ability to preserve various physical properties. Additionally, computational intelligence-based methods often require oversized datasets, extensive computational resources, and face challenges with inefficient convergence, limiting their scalability and applicability for large-scale analyses.

This study presents an innovative Statistics-Encoded Neural Network (SENN)-based framework for stochastic characterisation and reconstruction of 3D multiphase composite microstructures. Unlike conventional methods that rely solely on statistical modelling, the proposed framework leverages machine learning to encode 2D morphological statistics into feedforward neural networks, facilitating the efficient and accurate reconstruction of statistically equivalent 3D microstructures. Once properly trained, SENN models serve as implicit representations of morphological statistics, offering a high level of interpretability.
The SENN-based framework addresses key challenges in stochastic reconstruction of multiphase microstructures by:
\begin{itemize}[topsep=0pt]
\setlength{\itemsep}{0pt}
\setlength{\parsep}{0pt}
\setlength{\parskip}{0pt}
    \item Encoding complex morphological statistics of multiphase microstructures using feedforward neural networks, where only a limited number of 2D exemplars is required;
    \item Employing a 2D-to-3D morphology integration scheme to infer 3D morphological statistics from the 2D measurements embedded in SENN models;
    \item Incorporating hierarchical characterisation to capture local, regional, and global microstructural features across multiple length scales;
    \item Synthesising statistically equivalent 3D microstructures via Gibbs sampling in a multi-level manner, ensuring fidelity to the real material's statistical properties.
\end{itemize}

The versatility and robustness of the novel SENN-based framework are demonstrated through its application to three representative multiphase composites: porous silver-based electrodes (3-phase), porous solid oxide fuel cell anodes (3-phase), and composite cement paste (4-phase). These materials are selected due to their distinct microstructural characteristics and relevance to diverse engineering applications. The reconstructed 3D microstructures are evaluated against reference datasets using various morphological descriptors, including volume fraction, surface area density, triple-phase boundary density, and geometrical tortuosity. Furthermore, macroscopic properties such as stiffness, permeability, diffusivity, and thermal conductivity tensors are computed through image-based numerical simulations to validate the practical applicability of the reconstructed 3D microstructures.

The remainder of this paper is organised as follows: Section \ref{Section2:Statistical_microstructure_characterization} explains the theoretical foundation and statistical characterisation of multiphase microstructures in detail, including the Markov random field assumption and the SENN modelling; Section \ref{Section3:Stochastic_microstructure_reconstruction} outlines the methodology for 2D-to-3D stochastic reconstruction, focusing on hierarchical characterisation, multi-level reconstruction, and Gibbs sampling techniques; Section \ref{Section4:Examples_and_results} presents case studies on the three multiphase composites and evaluates the quality of microstructure reconstruction; Section \ref{Section5:Micro-mechanical_modeling} highlights the results of image-based poro/micro-mechanical modelling on the reconstructed 3D microstructures; and Section \ref{Section6:Discussion_and_conclusions} discusses the limitations and potential future directions and concludes with a summary of the SENN-based framework’s contributions.

\section{Statistical microstructure characterisation} 
\label{Section2:Statistical_microstructure_characterization}
\begin{figure}[h]
\centering
\includegraphics[width = 1.0\linewidth,angle=0,clip=true]{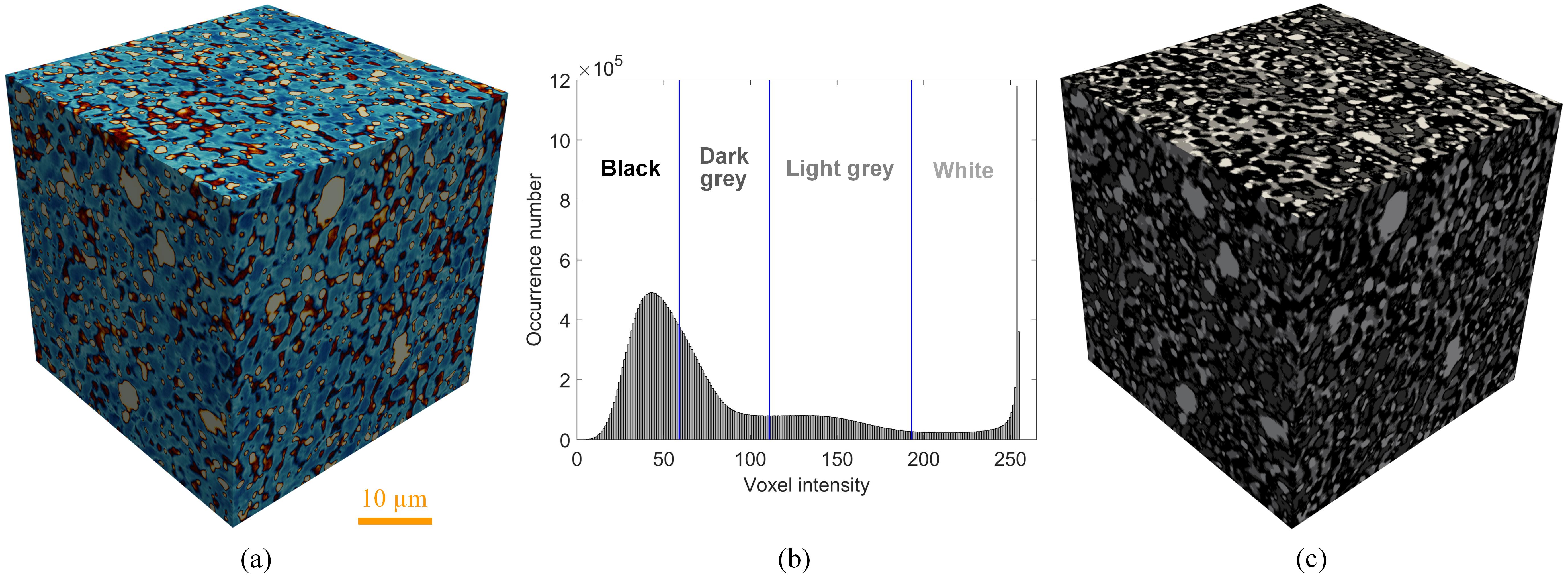} \\
\caption{Graphical illustration of image segmentation: (a) A 3D microscopic image of Titanate ceramic acquired by scanning electron microscopy (SEM) in backscattered electron imaging (BEI) mode; (b) The histogram of voxel intensity values, based on which the segmentation thresholds can be determined; and (c) The digital microstructure with four distinct constituent phases, where the white phase represents metals, the light grey phase is zirconolite, the dark grey phase is hollandite, and the black phase denotes perovskite loveringite or spinel.}
\label{fig:Image_segmentation}
\end{figure}

\vspace{-2pt}
The internal microstructures of opaque composite materials can be digitised via advanced microscopic imaging techniques, such as SEM and micro-CT, as shown in Figure \ref{fig:Image_segmentation}a. Image segmentation is then applied to the digital images to distinguish and separate different constituent phases, as illustrated in Figures \ref{fig:Image_segmentation}b and c. This segmentation enables subsequent analyses, including quantitative characterisation and computational modelling. Statistically quantitative characterisation of multiphase microstructures is a crucial preliminary step in generating equivalent virtual samples, given the significant disorder and randomness existing in composite microstructures.

Consider a 3D regular grid of size $L \times W \times H$, where each site $(i, j, k)$ is associated with a random variable $X_{ijk}\in\mathbb{Q}$, forming a random field $\textsl{\textbf{X}}$, defined as:
\begin{equation}
\textsl{\textbf{X}}=\Big\{X_{ijk}:1\leq i \leq L,\ 1\leq j \leq W,\ 1\leq k \leq H\Big\}\,,
\label{Eq:Regular_lattice}
\end{equation}
where $i\in \mathbb{Z}^+$, $j\in \mathbb{Z}^+$ and $k\in \mathbb{Z}^+$ are the indices of rows, columns and layers in the 3D random field. Assigning a voxel value $X_{ijk}$ into each site $(i, j, k)$ of this 3D grid generates a 3D digital image $\textsl{\textbf{X}}$.
This 3D digital microstructure $\textsl{\textbf{X}}$, in its segmented format (see Figure \ref{fig:Image_segmentation}c) can be represented by an indicator function as follows:
\begin{equation}
X_{ijk}=\left\{
\begin{array}{ll}
\vspace{1ex}
0, &\textup{if} \ (i,j, k)\ \textup{is located at the black phase};\\
1, &\textup{if} \ (i,j, k)\ \textup{is located at the dark grey phase};\\
\, \vdots & \\
n-1, &\textup{if} \ (i,j, k)\ \textup{is located at the white phase};
\end{array}
\right.
\label{Eq:Indication_function}
\end{equation}
where $n$ denotes the number of constituent phases in the multiphase composite.

\subsection{Markov random field assumption}
\label{Subsec:Markov_assumption} 
The full joint probability distribution function (PDF) of voxel variables provides an ideal statistical model for quantitatively characterising the digital microstructure $\textsl{\textbf{X}}$, because the voxel variables $X_{ijk}$ at different locations exhibit statistical correlations. This full joint PDF, denoted by $P(\textsl{\textbf{X}})$ or $P(X_{111}, X_{112},...\,, X_{ijk},...\,,X_{LWH})$, represents the likelihood of a digital microstructure $\textsl{\textbf{X}}$ being assigned a specific configuration of all voxel values. However, deriving the full joint PDF $P(\textsl{\textbf{X}})$ without any prior knowledge necessitates enumerating all possible configurations of voxel variables. This task becomes computationally prohibitive due to the extremely high dimensionality $(L \times W\times H)$ of voxel space. 

The Markov assumption \citep{li2009markov} is employed in this study to simplify the statistical characterisation of composite microstructures by reducing computational complexity. Under this assumption, a digital microstructure is represented as a Markov random field (MRF), where the voxel value at any site depends only on its neighbouring voxels within a specified range. This dependency can be expressed as:
\begin{equation}
P\left(X_{ijk}\,\big|\,\textsl{\textbf{X}}^{(-ijk)}\right)=P\left(X_{ijk}\,\big|\,\textsl{\textbf{N}}_{ijk}\right)\,, 
\label{Eq:MRF_assumption}
\end{equation}
where $\textsl{\textbf{N}}_{ijk}$ represents the set of adjacent voxels around a central voxel $X_{ijk}$, and $\textsl{\textbf{X}}^{(-ijk)}$ denotes all other voxels in the microstructure $\textsl{\textbf{X}}$, excluding $X_{ijk}$. This conditional probability distribution function (CPDF) $P\left(X_{ijk}\big|\textsl{\textbf{N}}_{ijk}\right)$ has significantly lower dimensionality compared to the full joint PDF $P(\textsl{\textbf{X}})$. The reduced dimensionality makes it feasible to approximate $P\left(X_{ijk}\big|\textsl{\textbf{N}}_{ijk}\right)$ by evaluating all possible configurations of the neighbouring voxels. Consequently, the full joint PDF $P(\textsl{\textbf{X}})$ can be efficiently derived from these local statistical properties. This approach enables effective modelling of the microstructure's stochastic behaviour without requiring exhaustive computation.

To model the local statistical characteristics of a digital microstructure, the initial step is to define the neighbourhood geometry surrounding a central voxel $X_{ijk}$. This neighbourhood, denoted as $\textsl{\textbf{N}}_{ijk}$, is confined within a spatial local region characterised by its radii ($R_{xy}$, $R_{yz}$ and $R_{zx}$), as illustrated in Figure \ref{fig:3D_data_template} (a 2D counterpart is illustrated in Figure \ref{fig:Data_collection}b). 
The data template scans the digital microstructure \(\textsl{\textbf{X}}\) to gather local morphology patterns, enabling the approximation of the CPDF $P\left(X_{ijk}\big|\textsl{\textbf{N}}_{ijk}\right)$. 
Each observation of a local morphology pattern, denoted by $\big(X_{ijk}, \textsl{\textbf{N}}_{ijk}\big)$, is refereed as a data event. The CPDF $P\left(X_{ijk}\big|\textsl{\textbf{N}}_{ijk}\right)$ is estimated from occurrence frequencies of such events, as expressed in the following equations:
\begin{equation}
P\left({X_{ijk}}\,\big|\,\textsl{\textbf{N}}_{ijk}\right)=\frac{P\left(X_{ijk}, \textsl{\textbf{N}}_{ijk}\right)}{P\left(\textsl{\textbf{N}}_{ijk}\right)}=\frac{O\left(X_{ijk},\textsl{\textbf{N}}_{ijk}\right)}{O\left(\textsl{\textbf{N}}_{ijk}\right)},
\label{Eq:CPDF}
\end{equation}
where $O(X_{ijk}, \textsl{\textbf{N}}_{ijk})$ and $O(\textsl{\textbf{N}}_{ijk})$ denote the occurrence counts of data events associated with $(X_{ijk}, \textsl{\textbf{N}}_{ijk})$ and $(\textsl{\textbf{N}}_{ijk})$ respectively.
Consequentially, the CPDF $P\left({X_{ijk}}\big|\textsl{\textbf{N}}_{ijk}\right)$ quantifies the probability of a central voxel $X_{ijk}$ taking a specific value, given its neighbouring voxels $\textsl{\textbf{N}}_{ijk}$.
By leveraging the localised CPDF $P\left(X_{ijk}\big|\textsl{\textbf{N}}_{ijk}\right)$, the full joint PDF $P(\textsl{\textbf{X}})$ that statistically characterises the random microstructures can be achieved through Gibbs sampling, which will be detailed in Section \ref{Section3:Stochastic_microstructure_reconstruction}.

\subsection{Statistics-encoded neural network}
\label{Subsec:Statistics-encoded_neural_network} 
Confronting with the common scenario where 3D digital microstructures are scarce while 2D examples are plentiful, this work aims to develop a machine learning-based framework to infer statistically equivalent 3D microstructures from the available 2D exemplars for multiphase composites. 
Achieving this objective necessitates the statistical characterisation of 3D morphological patterns derived from 2D examples as a foundational step.
As discussed earlier, the CPDF $P\left(X_{ijk}\big|\textsl{\textbf{N}}_{ijk}\right)$ encapsulates the statistics of local morphology patterns within random composite microstructures.
Considering the inherent complexity of microstructural features, this work employs machine learning to approximate the CPDF $P\left(X_{ijk}\big|\textsl{\textbf{N}}_{ijk}\right)$, where morphological statistics are integrated into feedforward neural networks, refereed to as $\textit{Statistics-Encoded Neural Network}$ (SENN). 
These models efficiently integrate statistical data, enabling the accurate inference of CPDFs and facilitating the synthesis of statistically consistent 3D microstructures from 2D exemplars.

\begin{figure}[h]
\centering
\includegraphics[width = 1.0\linewidth,angle=0,clip=true]{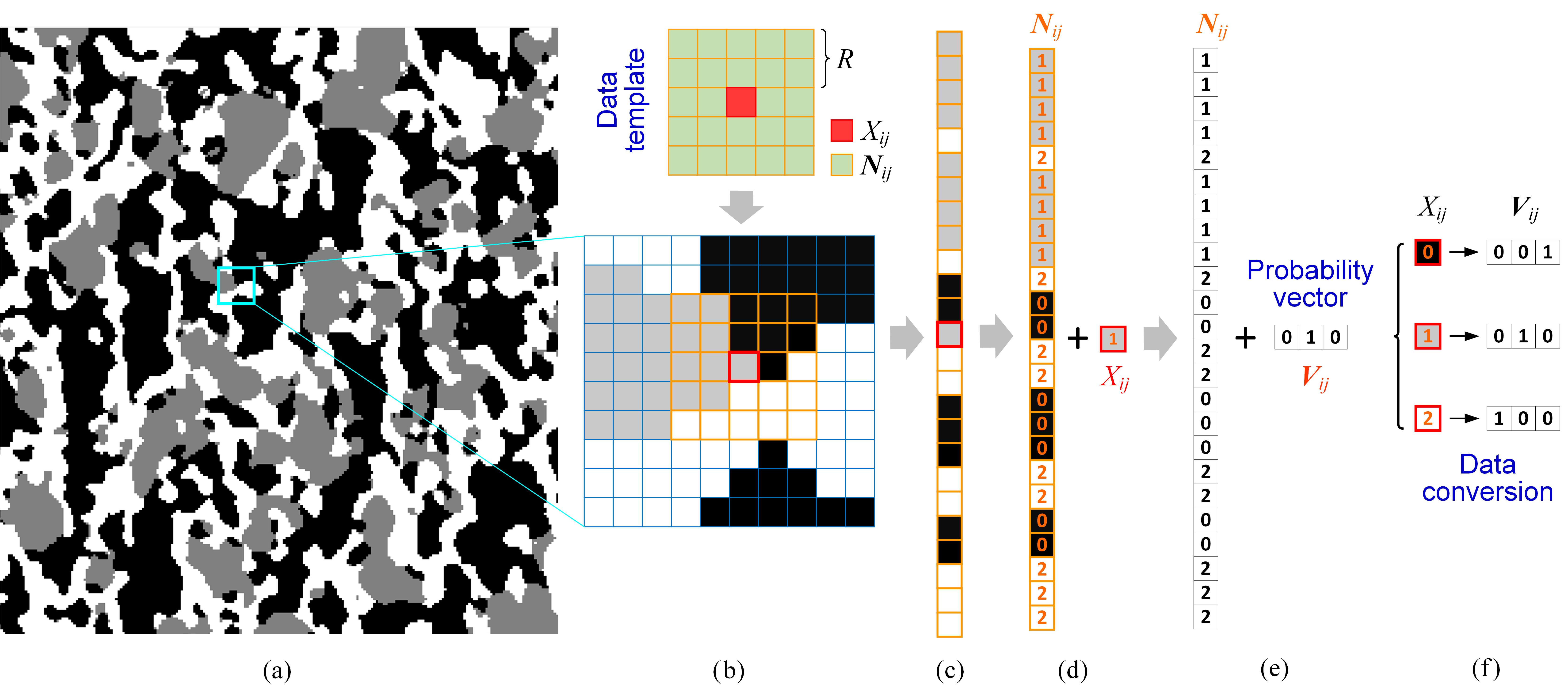} \\
\caption{Graphical illustration of training data preparation: (a) A 2D exemplar of a composite microstructure with three constituent phases ($n=3$); (b) Collection of data events using a square data template defined by radius $R$ to scan the 2D exemplar; (c) Example illustrating a data event $\big(X_{ij}, \textsl{\textbf{N}}_{ij}\big)$, where $X_{ij}$ is the central pixel, and $\textsl{\textbf{N}}_{ij}$ represents its neighbouring pixels; (d) Separation of the central pixel $X_{ij}$ from its neighbouring pixels $\textsl{\textbf{N}}_{ij}$; (e) Conversion of the central pixel value $X_{ij}$ to a probability vector $\textsl{\textbf{V}}_{ij}$; and (f) Data conversion process for training.}
\label{fig:Data_collection}
\end{figure}

As illustrated in Figures \ref{fig:Data_collection}a-d, data events $\big(X_{ij}, \textsl{\textbf{N}}_{ij}\big)$ are efficiently collected by scanning the 2D exemplar in a raster scanning order using the predefined data template. These paired observations $(X_{ij}, \textsl{\textbf{N}}_{ij})$ form an ideal training dataset for fitting a classifier in the supervised learning paradigm. By minimising misclassification likelihood, the class probabilities encoded in the classifier provide an accurate approximation of the CPDF $P\left(X_{ij}\big|\textsl{\textbf{N}}_{ij}\right)$. Among various classifiers, the feedforward neural network (FNN) is particularly effective in handling non-linear classification challenges inherent to image modelling. As depicted in Figures \ref{fig:Data_collection}e and f, the central pixel value $X_{ij}$ is converted into a probability vector $\textsl{\textbf{V}}_{ij}$, ensuring that the trained FNN model's output corresponds to class probabilities equivalent to the CPDF $P\left(X_{ij}\big|\textsl{\textbf{N}}_{ij}\right)$. This data conversion process is reversible, which is essential for stochastic reconstruction of 3D microstructures through probabilistic sampling. Furthermore, this conversion procedure can be readily extended to composite microstructures with more than three constituent phases (i.e., $n\geq 3$).

To fit the FNN model for approximating the CPDF $P\left(X_{ij}\big|\textsl{\textbf{N}}_{ij}\right)$, the neighbourhood pattern $\textsl{\textbf{N}}_{ij}$ is used as the classification feature (input), while the central pixel value $X_{ij}$ serves as the classification category (output). It is crucial that $\textsl{\textbf{N}}_{ij}$ is flatten into a feature vector, and $X_{ij}$ is converted to be a probability vector $\textsl{\textbf{V}}_{ij}$ prior to training the FNN model, resulting in the training dataset $(\textsl{\textbf{V}}_{ij}, \textsl{\textbf{N}}_{ij})$. In a multi-layer FNN model, the input feature vector $\textsl{\textbf{N}}_{ij}$ is propagated through the hidden layers, and the final output, representing the estimated probability vector $\widehat{\textsl{\textbf{V}}}_{ij}$, is computed through a series of forward-propagation equations.
In essence, the fitted FNN model is a vector-valued surrogate that approximates the mapping between  $\textsl{\textbf{N}}_{ij}$ and $\textsl{\textbf{V}}_{ij}$ for the 2D training image $\textsl{\textbf{X}}$, which can be mathematically expressed as follows:
\begin{equation}
\mathcal{FNN}\Big(\textsl{\textbf{N}}_{ij};\textsl{\textbf W}, \textsl{\textbf b}\Big):~ \textsl{\textbf{N}}_{ij}\in\mathbb{R}^{d_N}\longrightarrow \textsl{\textbf{V}}_{ij}\in\mathbb{R}^{n},
\end{equation}
where $\mathcal{FNN}\big(\cdot\big)$ denotes the approximation function of the FNN model, $\textsl{\textbf W}$ is the weight matrices, and $\textsl{\textbf b}$ is the bias vector; $d_N=R^2-1$ denotes the length of the input feature vector $\textsl{\textbf{N}}_{ij}$, and $R$ is the radius of the 2D data template (see Figure \ref{fig:Data_collection}b); and $n$ is the length of the probability vector $\textsl{\textbf{V}}_{ij}$, corresponding to the number of constituent phases within the composite microstructure.

The training process of FNN is to minimise the discrepancy between the target probability vector $\textsl{\textbf{V}}_{ij}$ and the predicted probability vector $\widehat{\textsl{\textbf{V}}}_{ij}$ by iteratively adjusting the weight matrices $\textsl{\textbf W}$ and bias vector $\textsl{\textbf b}$. This optimisation problem can be mathematically formulated as:
\begin{equation}
\begin{aligned}
\mathop{\arg\min}_{\textsl{\textbf W}, \textsl{\textbf b}}~~\mathcal{L}\,\Big(\textsl{\textbf{V}}_{ij},\textsl{\textbf{N}}_{ij};\textsl{\textbf W}, \textsl{\textbf b}\Big)=&~\frac{1}{m}\sum\left(\textsl{\textbf{V}}_{ij}-\widehat{\textsl{\textbf{V}}}_{ij}\right)^2+\lambda\,\Big\|\textsl{\textbf W}\,\Big\|_F\\
=&~\frac{1}{m}\sum\bigg[\textsl{\textbf{V}}_{ij}-\mathcal{FNN}\Big(\textsl{\textbf{N}}_{ij};\textsl{\textbf W}, \textsl{\textbf b}\Big)\bigg]^2+\lambda\,\Big\|\textsl{\textbf W}\,\Big\|_F\,,
\end{aligned}
\label{Eq:loss_function}
\end{equation}
where $\mathcal{L}\,\big(\textsl{\textbf{V}}_{ij},\textsl{\textbf{N}}_{ij};\textsl{\textbf W}, \textsl{\textbf b}\big)$ is the loss function, $m$ is the number of training data points, $\lambda$ is the weight regulation constant, and the Frobenius norm $\big\|\textsl{\textbf W}\,\big\|_F$ the weight matrix, representing weight decay to encourage smoother network responses and reduce overfitting. The parameter settings for training the SENN are given in \ref{appendix}.

Once the FNN model is adequately trained, it can predict the response $\widehat{\textsl{\textbf{V}}}_{ij}$ for any given input feature $\textsl{\textbf{N}}_{ij}$. Since the sigmoid function (i.e., $\sigma(x)=1/(1+{\rm e}^{-x})$) is used as the activation function in the final layer, each element in the output probability vector $\widehat{\textsl{\textbf{V}}}_{ij}$ will range between $0$ and $1$. 
Essentially, the predicted probability vector $\widehat{\textsl{\textbf{V}}}_{ij}$ from the FNN model approximates the CPDF $P\left(X_{ij}\big|\textsl{\textbf{N}}_{ij}\right)$ that characterises the  morphological statistics of the 2D training image $\textsl{\textbf{X}}$. This relationship is expressed as:
\begin{equation}
P\left(X_{ij}\,\big|\,\textsl{\textbf{N}}_{ij}\right)=\left\{
\begin{array}{ll}
\vspace{1.5ex}
P\left(X_{ij}=0\,\big|\,\textsl{\textbf{N}}_{ij}\right)=P\left(\textsl{\textbf{V}}_{ij}=\big[0,0,1\big]\,\big|\,\textsl{\textbf{N}}_{ij}\right)\approx \widehat{\textsl{\textbf{V}}}_{ij}\big(3\big)\\
\vspace{1.5ex}
P\left(X_{ij}=1\,\big|\,\textsl{\textbf{N}}_{ij}\right)=P\left(\textsl{\textbf{V}}_{ij}=\big[0,1,0\big]\,\big|\,\textsl{\textbf{N}}_{ij}\right)\approx\widehat{\textsl{\textbf{V}}}_{ij}\big(2\big)\\
P\left(X_{ij}=2\,\big|\,\textsl{\textbf{N}}_{ij}\right)=P\left(\textsl{\textbf{V}}_{ij}=\big[1,0,0\big]\,\big|\,\textsl{\textbf{N}}_{ij}\right)\approx\widehat{\textsl{\textbf{V}}}_{ij}\big(1\big) \\
\end{array}
\right\}=\widehat{\textsl{\textbf{V}}}_{ij}=\mathcal{FNN}\Big(\textsl{\textbf{N}}_{ij};\textsl{\textbf W}, \textsl{\textbf b}\Big)\ ,
\label{Eq:2D_FNN_CPDF}
\end{equation}
where $\widehat{\textsl{\textbf{V}}}_{ij}\big(t\big)$ denotes the $t$-th element of the output probability vector $\widehat{\textsl{\textbf{V}}}_{ij}$ (here, $1\leq t\leq 3$, as three constituent phases are considered in this illustrative microstructure). Therefore, the fitted FNN model, which  maps data events $\big(X_{ij}, \textsl{\textbf{N}}_{ij}\big)$, served as an implicit representation of the CPDF $P\left(X_{ij}\big|\textsl{\textbf{N}}_{ij}\right)$, and it is referred to as the $\textit{Statistics-Encoded Neural Network}$ (SENN) in this work.

\subsection{Multiphase morphological statistics}
\label{Subsec:Multiphase_morphological_statistics} 
The CPDF $P\left(X_{ij}\big|\textsl{\textbf{N}}_{ij}\right)$ characterises the local morphological statistics associated with the primary microstructural features in the multi-phase composite. To accurately approximate this CPDF, the data template (refer to Figure \ref{fig:Data_collection}b) used for data collection must be sufficiently large to encompass the key morphological features of different constituent phases. As illustrated in Figure \ref{fig:Data_collection}a, the morphological features associated with these constituent phases often differ in size. This variation requires the use of appropriately sized data templates for accurate statistical characterisation. For example, the black phase, with its broader range of features, demands a larger data template than the grey phase to ensure a precise CPDF estimation.

\begin{figure}[h]
\centering
\includegraphics[width = 0.95\linewidth,angle=0,clip=true]{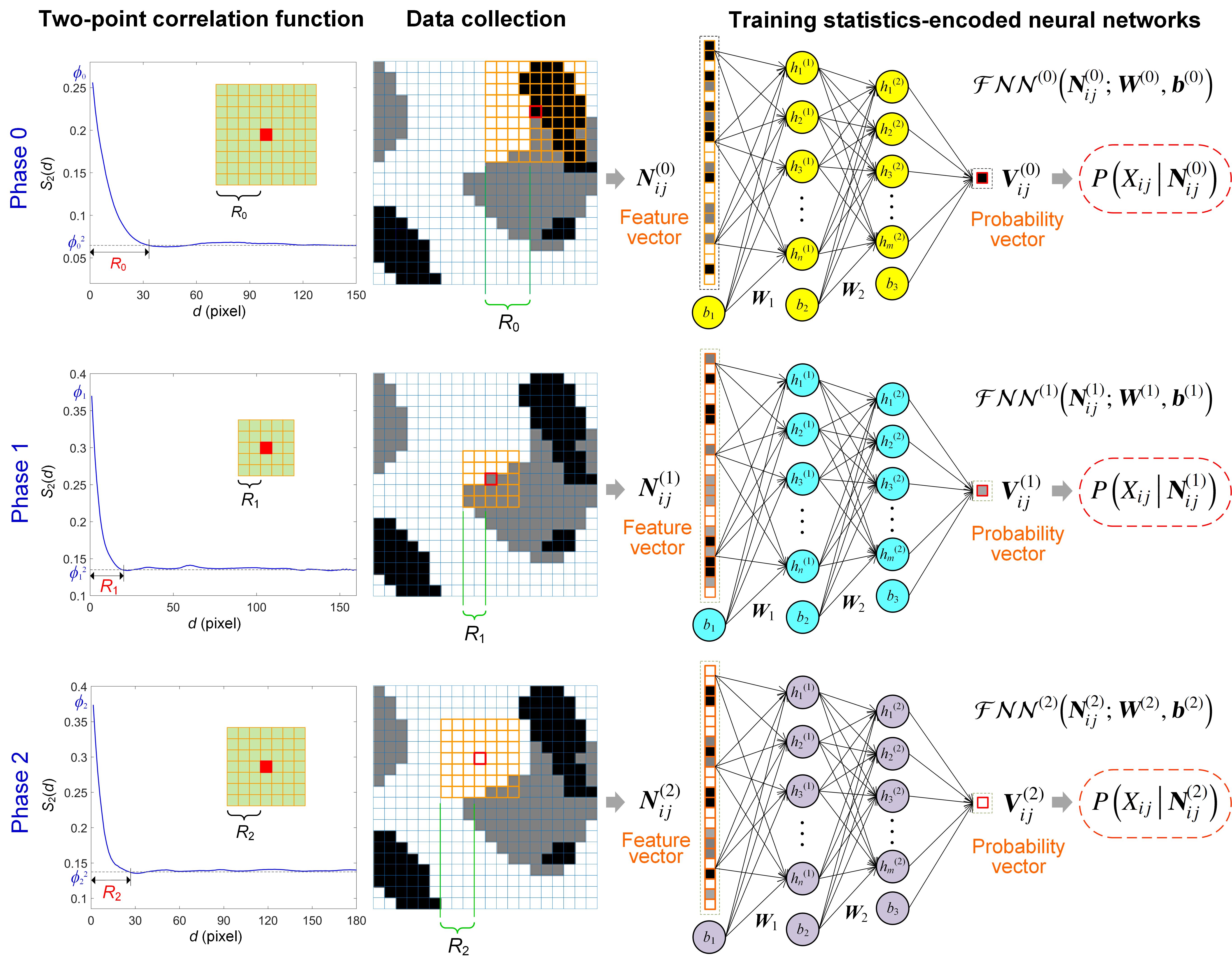} \\
\caption{The procedure of fitting statistics-encoded neural networks (SENNs) to characterise the morphological statistics corresponding to different phases in the composite microstructure (here, a 3-phase composite microstructure is used for illustration).}
\label{fig:Multiphase_SENN}
\end{figure}

In this work, the radius $R$ of a data template is determined using the two-point correlation function (TPCF) $S_2(d)$ \cite{fu2021statistical, cui2021correlation}, which quantifies the probability of finding two points separated by a distance $d$ within a particular constituent phase. 
 As shown in the leftmost column of Figure \ref{fig:Multiphase_SENN}, the TPCF curves stabilise when the distance $d$ between the two points exceeds a certain value $l$ that is called correlation length. This stabilisation indicates that two points can be considered statistically uncorrelated when $d \geq l$. Therefore, the correlation length $l$ extracted from the TPCF curve well defines the correlation range, providing a critical value for the template size $R$.
Additionally, the correlation length $l$ can also be computed by fitting the TPCF $S_2(d)$ to the following exponential function \cite{corson1974correlation}:
\begin{equation}
    S_2(d)=\left(\phi-\phi^2\right){\rm exp}\left(-\frac{3d}{l}\right)+\phi^2,
\label{Eq:Fitting_TPCF}
\end{equation}
where $\phi$ denotes volume fraction of the constituent phase, and the template size $R$ is defined as the smallest integer greater than or equal to the correlation length $l$.

As illustrated in Figure \ref{fig:Multiphase_SENN}, the TPCFs corresponding to three constituent phases ($n=3$) are separately derived from the training image to determine the appropriate sizes of three distinct data templates. Using three sets of data events, each collected with a varying data template, three SENN models can be independently fitted to characterise the morphological statistics pertinent to each constituent phase. The CPDF $P\left(X_{ij}\big|\textsl{\textbf{N}}_{ij}\right)$, which represents multiphase morphological statistics, is computed by averaging the probabilities obtained from each SENN model, as follows:
\begin{equation}
P\left(X_{ij}\,\big|\,\textsl{\textbf{N}}_{ij}\right) = \frac{1}{n}\sum_{t=0}^{n-1}P\left(X_{ij}\,\big|\,\textsl{\textbf{N}}_{ij}^{(t)}\right) \approx \frac{1}{n}\sum_{t=0}^{n-1}\mathcal{FNN}^{(t)}\Big(\textsl{\textbf{N}}_{ij}^{(t)};\textsl{\textbf W}^{(t)}, \textsl{\textbf b}^{(t)}\Big)\,,
\label{Eq:multiphase_SENN}
\end{equation}
where $\textsl{\textbf{N}}_{ij}^{(t)}$ denotes the neighbouring pixels (i.e., the feature vector) defined by the data template for the phase $t$, $\mathcal{FNN}^{(t)}\big(\cdot\big)$ represents the approximation function of the FNN model for the phase $t$, $\textsl{\textbf W}^{(t)}$ is the weight matrix, $\textsl{\textbf b}^{(t)}$ denotes the bias vector, and $n$ is the number of constituent phases (here, $n$ is set at 3, as the example uses a composite microstructure with three phases).

\begin{figure}[h]
\centering
\includegraphics[width = 1.0\linewidth,angle=0,clip=true]{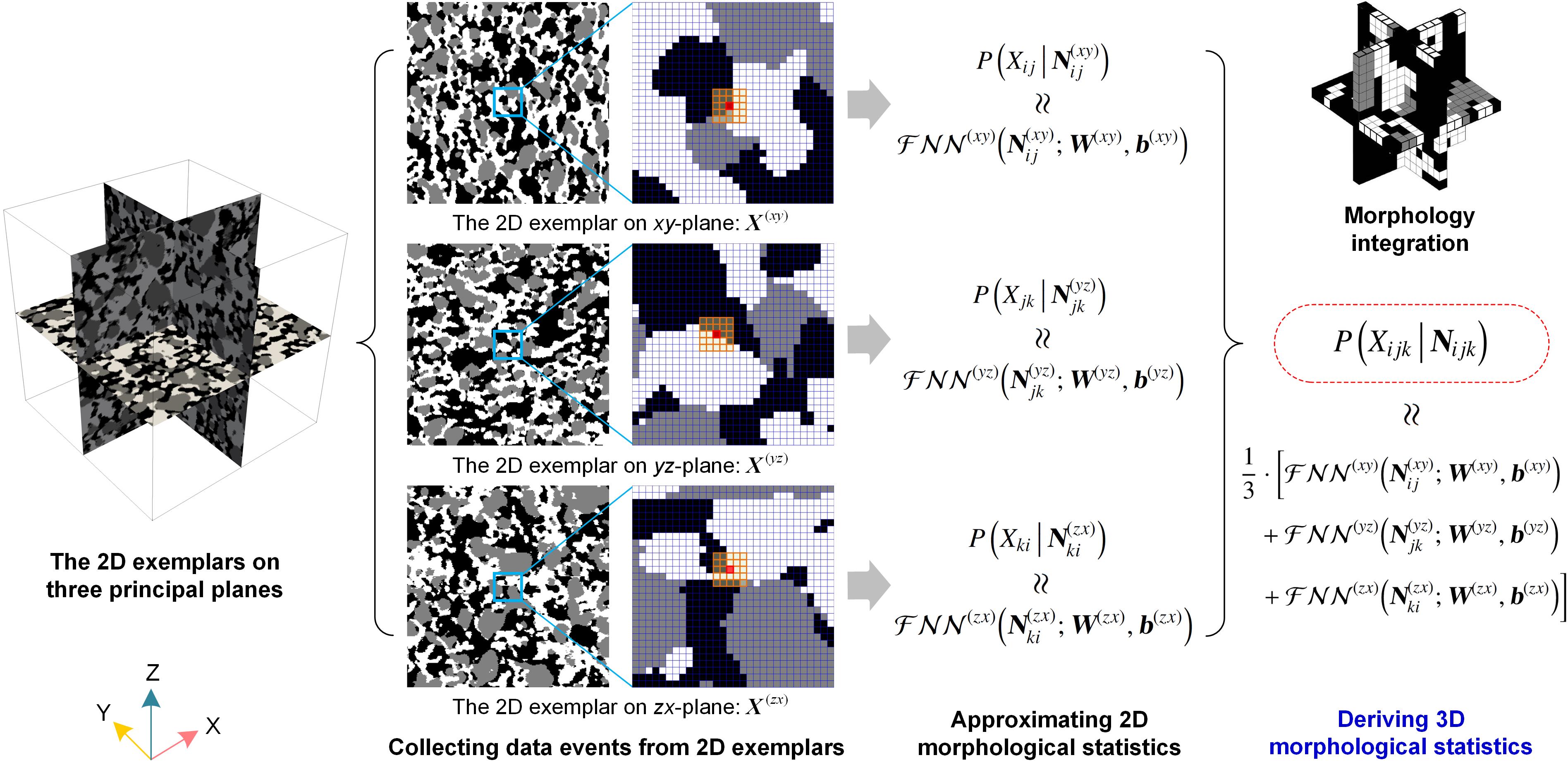} \\
\caption{The computational framework of deriving 3D morphological statistics $P\left(X_{ijk}\big|\textsl{\textbf{N}}_{ijk}\right)$ from 2D morphological statistics $P\left(X_{ij}\big|\textsl{\textbf{N}}_{ij}^{(xy)}\right)$, $P\left(X_{jk}\big|\textsl{\textbf{N}}_{jk}^{(yz)}\right)$ and $P\left(X_{ki}\big|\textsl{\textbf{N}}_{ki}^{(zx)}\right)$ on three orthogonal planes (i.e., $xy$-, $yz$-, and $zx$-planes), where the 2D CPDFs are approximated through fitting the SENN models.}
\label{fig:2D_to_3D_characterization}
\end{figure}

\subsection{2D-to-3D conversion of morphological statistics}
\label{Subsec:2D-to-3D_conversion} 
To address the dimension disparity between 2D morphological statistics $P\left(X_{ij}\,\big|\,\textsl{\textbf{N}}_{ij}\right)$ and 3D microstructure reconstruction, inferring 3D morphological statistics $P\left(X_{ijk}\,\big|\,\textsl{\textbf{N}}_{ijk}\right)$ from 2D measurements is essential. A specialised morphological integration scheme has been developed here for this purpose, enabling the derivation of 3D CPDF $P\left(X_{ijk}\,\big|\,\textsl{\textbf{N}}_{ijk}\right)$ from 2D morphological statistics, as illustrated in Figure \ref{fig:2D_to_3D_characterization}.
The MRF assumption, defined in Eq. (\ref{Eq:MRF_assumption}), is further simplified to model the phase values of a central voxel $X_{ijk}$ as influenced by its neighbouring voxels on three orthogonal planes: $xy$-, $yz$-, and $zx$-planes. This simplification is expressed as:
\begin{equation}
P\left(X_{ijk}\,\big|\,\textsl{\textbf{X}}^{(-ijk)}\right)= P\left(X_{ijk}\,\big|\,\textsl{\textbf{N}}_{ijk}\right)= P\left[X_{ijk}\,\Big|\,\Big(\textsl{\textbf{N}}_{ijk}^{(xy)}\cup \textsl{\textbf{N}}_{ijk}^{(yz)}\cup \textsl{\textbf{N}}_{ijk}^{(zx)}\Big)\right]\,,
\label{Eq:3D_MRF_assumption-simplified}
\end{equation}
where $\textsl{\textbf{N}}_{ijk}^{(xy)}$, $\textsl{\textbf{N}}_{ijk}^{(yz)}$ and $\textsl{\textbf{N}}_{ijk}^{(zx)}$ represent the sets of neighbouring voxels on $xy$-, $yz$- and $zx$-planes, respectively.

For a multiphase composite microstructure, the morphological patterns observed on cross-sectional planes are consistent with those in 3D space. When the range of neighbouring voxels is sufficiently large, a relationship exists between 2D and 3D morphological statistics \citep{fu2023hierarchical, fu2022stochastic}, mathematically expressed as: 
\begin{equation}
P\left(X_{ijk}\,\big|\,\textsl{\textbf{N}}_{ijk}\right) = P\left(X_{ijk}\,\big|\,\textsl{\textbf{N}}_{ijk}^{(xy)}\right) = P\left(X_{ijk}\,\big|\,\textsl{\textbf{N}}_{ijk}^{(yz)}\right) = P\left(X_{ijk}\,\big|\,\textsl{\textbf{N}}_{ijk}^{(zx)}\right)\,,
\label{Eq:morphology-consistency assumption}
\end{equation}
where the CPDFs $P\left(X_{ijk}\big|\textsl{\textbf{N}}_{ijk}\right)$, $P\left(X_{ijk}\big|\textsl{\textbf{N}}_{ijk}^{(xy)}\right)$, $P\left(X_{ijk}\big|\textsl{\textbf{N}}_{ijk}^{(yz)}\right)$ and $P\left(X_{ijk}\big|\textsl{\textbf{N}}_{ijk}^{(zx)}\right)$ statistically describe the primary morphological features in 3D space and on $xy$-, $yz$- and $zx$-planes, respectively. 
It is important to note that while the neighbourhoods $\textsl{\textbf{N}}_{ijk}$, $\textsl{\textbf{N}}_{ijk}^{(xy)}$, $\textsl{\textbf{N}}_{ijk}^{(yz)}$ and $\textsl{\textbf{N}}_{ijk}^{(zx)}$ can differ significantly, the relationships described by this equation remain valid.
Using the morphology-consistency assumption from Eq. (\ref{Eq:morphology-consistency assumption}), the 3D CPDF $P\left(X_{ijk}\big|\textsl{\textbf{N}}_{ijk}\right)$ can be derived from the 2D CPDFs on three orthogonal planes by averaging:
\begin{equation}
P\left(X_{ijk}\,\big|\,\textsl{\textbf{N}}_{ijk}\right) =~\frac{1}{3}\cdot \bigg[P\left(X_{ijk}\,\big|\,\textsl{\textbf{N}}_{ijk}^{(xy)}\right)+P\left(X_{ijk}\,\big|\,\textsl{\textbf{N}}_{ijk}^{(yz)}\right)+P\left(X_{ijk}\,\big|\,\textsl{\textbf{N}}_{ijk}^{(zx)}\right)\bigg]\,,
\label{Eq:3D_conditional_PDF}
\end{equation}
This morphology integration scheme treats the morphological components across the three principal planes equally, so the phase value of a voxel is determined by the information from its neighbouring voxels on all three orthogonal planes.

\begin{figure}[h!]
\centering
\includegraphics[width = 0.575\linewidth,angle=0,clip=true]{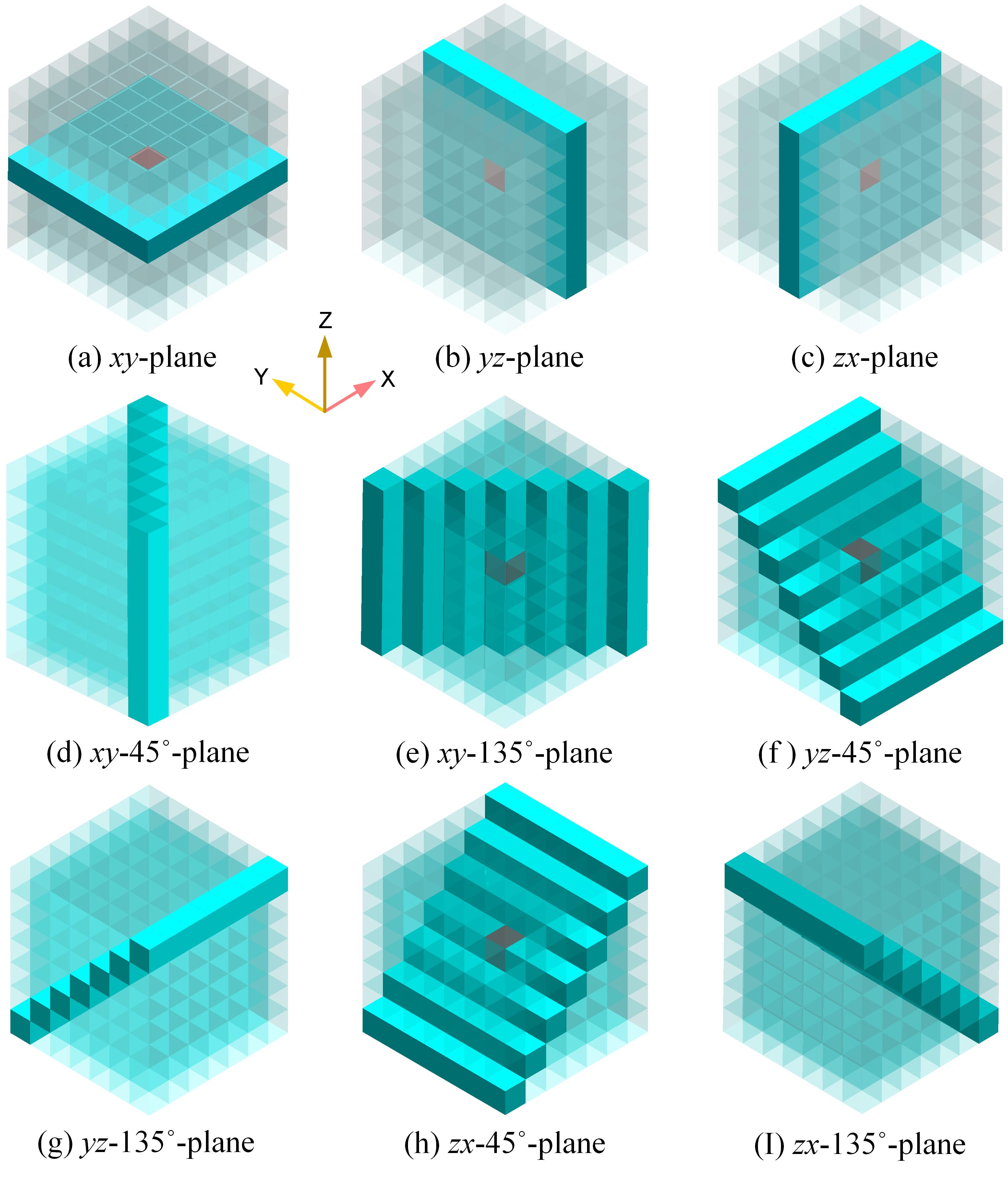} \\
\caption{The 2D measurements on both principal and diagonal planes can be incorporated into the 3D microstructure characterisation.}
\label{Fig:More_2D_examplers}
\end{figure}

By utilising the 2D exemplars on three orthogonal planes, three SENN models can be separately trained to accurately characterise the 2D morphological statistics, as illustrated in Figure \ref{fig:2D_to_3D_characterization}.
As explained by Eq. (\ref{Eq:2D_FNN_CPDF}), these SENN models implicitly represent the 2D CPDFs $P\left(X_{ij}^{(xy)}\Big|\textsl{\textbf{N}}_{ij}^{(xy)}\right)$, $P\left(X_{jk}^{(yz)}\Big|\textsl{\textbf{N}}_{jk}^{(yz)}\right)$ and $P\left(X_{ki}^{(zx)}\Big|\textsl{\textbf{N}}_{ki}^{(zx)}\right)$ respectively. 
These SENN models enable the accurate approximation of the 3D CPDF $P\left(X_{ijk}\big|\textsl{\textbf{N}}_{ijk}\right)$ using Eq. (\ref{Eq:3D_conditional_PDF}):
\begin{equation}
\small
\begin{aligned}
P\left(X_{ijk}\,\big|\,\textsl{\textbf{N}}_{ijk}\right) \approx~\frac{1}{3}\cdot \bigg[\mathcal{FNN}^{(xy)}\Big(\textsl{\textbf{N}}_{ij}^{(xy)};\textsl{\textbf W}^{(xy)}, \textsl{\textbf b}^{(xy)}\Big)+\mathcal{FNN}^{(yz)}\Big(\textsl{\textbf{N}}_{jk}^{(yz)};\textsl{\textbf W}^{(yz)}, \textsl{\textbf b}^{(yz)}\Big)+\mathcal{FNN}^{(zx)}\Big(\textsl{\textbf{N}}_{ki}^{(zx)};\textsl{\textbf W}^{(zx)}, \textsl{\textbf b}^{(zx)}\Big)\bigg]\,,
\end{aligned}
\label{Eq:3D_FNN_CPDF}
\end{equation}
where $\mathcal{FNN}^{(xy)}\big(\cdot\big)$, $\mathcal{FNN}^{(yz)}\big(\cdot\big)$ and $\mathcal{FNN}^{(zx)}\big(\cdot\big)$ represent the approximation functions of the SENN models for statistically characterising the 2D exemplars on the $xy$-, $yz$-, and $zx$-planes, respectively. The main steps of the SENN-based statistical characterisation for multiphase microstructures are summarised in Algorithm \ref{Algorithm:Statistical_characterization}.

\begin{algorithm}[h]
	\SetAlgoLined
	\small
    \setstretch{1.3}
	\textbf{Training images:} The available 2D exemplars on $xy$-, $yz$-, and $zx$-planes: $\textsl{\textbf{X}}^{(xy)}$, $\textsl{\textbf{X}}^{(yz)}$ and $\textsl{\textbf{X}}^{(zx)}$.\\   
    \smallskip
	\textbf{(1) Determine the sizes of 2D data templates:}\\
    Compute two-point correlation functions (TPCFs) for all constituent phases in the selected 2D exemplars;\\ 
    Measure the correlation lengths $l$ from TPCF curves to determine the sizes of 2D data templates for different constituent phases (using the Eq. (\ref{Eq:Fitting_TPCF})): $R_0^{(xy)}$, $R_1^{(xy)}$ and $R_2^{(xy)}$; $R_0^{(yz)}$, $R_1^{(yz)}$ and $R_2^{(yz)}$; $R_0^{(zx)}$, $R_1^{(zx)}$ and $R_2^{(zx)}$.\\
    $\textbf{Return:}$ A set of 2D data templates: $DT_0^{(xy)}$, $DT_1^{(xy)}$ and $DT_2^{(xy)}$; $DT_0^{(yz)}$, $DT_1^{(yz)}$ and $DT_2^{(yz)}$; $DT_0^{(zx)}$, $DT_1^{(zx)}$ and $DT_2^{(zx)}$.\\
	\smallskip
	\textbf{(2) Training data preparation:}\\
	\For{$R_{0}^{(xy)}+1\leq i \leq H^{\rm (xy)}-R_{0}^{(xy)}$} 
	{
		\For{$R_{0}^{(xy)}+1\leq j \leq W^{\rm (xy)}-R_{0}^{(xy)}$}
		{Collect the data event $\left(X_{ij}^{(xy:\ 0)}, \textsl{\textbf{N}}_{ij}^{(xy:\ 0)}\right)$ from $\textsl{\textbf{X}}^{(xy)}$ using the 2D data template $DT_{0}^{(xy)}$;\\
		Convert the central pixel value $X_{ij}^{(xy:\ 0)}$ into a probability vector $\textsl{\textbf{V}}_{ij}^{(xy:\ 0)}$;\\
		Move the 2D data template $DT_{0}^{(xy)}$ to the next position in the raster scan order: $j=j+1$;		
		}
		Move the 2D data template $DT_{0}^{(xy)}$ to the next line: $i=i+1$;
	}
Store the training dataset $\left(\textsl{\textbf{V}}_{ij}^{(xy:\ 0)}, \textsl{\textbf{N}}_{ij}^{(xy:\ 0)}\right)$ for the constituent phase 0 in the 2D exemplar $\textsl{\textbf{X}}^{(xy)}$;\\
Repeat the above procedures to collect training data for different constituent phases in $\textsl{\textbf{X}}^{(xy)}$, $\textsl{\textbf{X}}^{(yz)}$ and $\textsl{\textbf{X}}^{(zx)}$;\\
$\textbf{Return:}$ The training datasets: $\left(\textsl{\textbf{V}}_{ij}^{(xy:\ 0)}, \textsl{\textbf{N}}_{ij}^{(xy:\ 0)}\right)$, $\left(\textsl{\textbf{V}}_{ij}^{(xy:\ 1)}, \textsl{\textbf{N}}_{ij}^{(xy:\ 1)}\right)$ and $\left(\textsl{\textbf{V}}_{ij}^{(xy:\ 2)}, \textsl{\textbf{N}}_{ij}^{(xy:\ 2)}\right)$; 
$\left(\textsl{\textbf{V}}_{jk}^{(yz:\ 0)}, \textsl{\textbf{N}}_{jk}^{(yz:\ 0)}\right)$, $\left(\textsl{\textbf{V}}_{jk}^{(yz:\ 1)}, \textsl{\textbf{N}}_{jk}^{(yz:\ 1)}\right)$ and $\left(\textsl{\textbf{V}}_{jk}^{(yz:\ 2)}, \textsl{\textbf{N}}_{jk}^{(yz:\ 2)}\right)$;
$\left(\textsl{\textbf{V}}_{ki}^{(zx:\ 0)}, \textsl{\textbf{N}}_{ki}^{(zx:\ 0)}\right)$, $\left(\textsl{\textbf{V}}_{ki}^{(zx:\ 1)}, \textsl{\textbf{N}}_{ki}^{(zx:\ 1)}\right)$ and $\left(\textsl{\textbf{V}}_{ki}^{(zx:\ 2)}, \textsl{\textbf{N}}_{ki}^{(zx:\ 2)}\right)$.\\
\smallskip
\smallskip
\textbf{(3) Training SENN models for statistical characterisation:}\\
Use the training dataset $\left(\textsl{\textbf{V}}_{ij}^{(xy:\ 0)}, \textsl{\textbf{N}}_{ij}^{(xy:\ 0)}\right)$ to fit a SENN model $\mathcal{FNN}^{(xy:\ 0)}$ for approximating $P\left(X_{ij}^{(xy:\ 0)}\,\big|\,\textsl{\textbf{N}}_{ij}^{(xy:\ 0)}\right)$;\\
Repeat the above procedure to fit SENN models using the prepared training datasets for approximating the following CPDFs: $P\left(X_{ij}^{(xy:\ 1)}\,\big|\,\textsl{\textbf{N}}_{ij}^{(xy:\ 1)}\right)$ and $P\left(X_{ij}^{(xy:\ 2)}\,\big|\,\textsl{\textbf{N}}_{ij}^{(xy:\ 2)}\right)$; $P\left(X_{jk}^{(yz:\ 0)}\,\big|\,\textsl{\textbf{N}}_{jk}^{(yz:\ 0)}\right)$, $P\left(X_{jk}^{(yz:\ 1)}\,\big|\,\textsl{\textbf{N}}_{jk}^{(yz:\ 1)}\right)$ and $P\left(X_{jk}^{(yz:\ 2)}\,\big|\,\textsl{\textbf{N}}_{jk}^{(yz:\ 2)}\right)$; $P\left(X_{ki}^{(zx:\ 0)}\,\big|\,\textsl{\textbf{N}}_{ki}^{(zx:\ 0)}\right)$, $P\left(X_{ki}^{(zx:\ 1)}\,\big|\,\textsl{\textbf{N}}_{ki}^{(zx:\ 1)}\right)$ and $P\left(X_{ki}^{(zx:\ 2)}\,\big|\,\textsl{\textbf{N}}_{ki}^{(zx:\ 2)}\right)$;\\ 
Derive the following 2D CPDFs using Eq. (\ref{Eq:multiphase_SENN}) for statistical characterisation of the 2D exemplars $\textsl{\textbf{X}}^{(xy)}$, $\textsl{\textbf{X}}^{(yz)}$ and $\textsl{\textbf{X}}^{(zx)}$, respectively: $P\left(X_{ij}^{(xy)}\,\big|\,\textsl{\textbf{N}}_{ij}^{(xy)}\right)$, $P\left(X_{jk}^{(yz)}\,\big|\,\textsl{\textbf{N}}_{jk}^{(yz)}\right)$ and $P\left(X_{ki}^{(zx)}\,\big|\,\textsl{\textbf{N}}_{ki}^{(zx)}\right)$;\\
Infer the 3D CPDF $P\left(X_{ijk}\,\big|\,\textsl{\textbf{N}}_{ijk}\right)$ from the 2D morphological statistics on three orthogonal planes using Eq. (\ref{Eq:3D_FNN_CPDF}).\\ 
$\textbf{Return:}$ A set of SENN models representing the 2D morphological statistics: $\mathcal{FNN}^{(xy:\ 0)}$, $\mathcal{FNN}^{(xy:\ 1)}$ and $\mathcal{FNN}^{(xy:\ 2)}$; $\mathcal{FNN}^{(yz:\ 0)}$, $\mathcal{FNN}^{(yz:\ 1)}$ and $\mathcal{FNN}^{(yz:\ 2)}$; $\mathcal{FNN}^{(zx:\ 0)}$, $\mathcal{FNN}^{(zx:\ 1)}$ and $\mathcal{FNN}^{(zx:\ 2)}$.
\caption{\small{Statistical characterisation of multiphase composite microstructures by fitting SENN models}}
\smallskip
(\footnotesize Note: $H^{(xy)}$ and $W^{(xy)}$ represent the height and width of the 2D exemplar $\textsl{\textbf{X}}^{(xy)}$, respectively.)
\label{Algorithm:Statistical_characterization}
\end{algorithm}

In summary, for isotropic multiphase microstructures, the morphological characteristics are equivalent in all directions. Consequently, cross-sectional images in a single direction are sufficient to serve as 2D exemplars for training the SENN model. In contrast, anisotropic microstructures exhibit significant variation in morphological patterns across different directions. To capture this anisotropy, 2D slices from three orthogonal planes are required, necessitating the training of three separate SENN models: $\mathcal{FNN}^{(xy)}\big(\cdot\big)$, $\mathcal{FNN}^{(yz)}\big(\cdot\big)$ and $\mathcal{FNN}^{(zx)}\big(\cdot\big)$ should be trained separately.
Additionally, incorporating 2D exemplars from diagonal planes into the morphology integration scheme can further enhance the characterisation of 3D microstructures, as illustrated in Figure \ref{Fig:More_2D_examplers}. In this case, the 3D morphological statistics $P\left(X_{ijk}\big|\textsl{\textbf{N}}_{ijk}\right)$ can be estimated as a weighted average of the 2D CPDFs from nine different planes:
\begin{equation}
\begin{aligned}
P\left(X_{ijk}\,\big|\,\textsl{\textbf{N}}_{ijk}\right) &\approx~\frac{1}{9}\cdot \bigg[\mathcal{FNN}^{(xy)}\Big(\textsl{\textbf{N}}_{ij}^{(xy)}\Big)+\mathcal{FNN}^{(yz)}\Big(\textsl{\textbf{N}}_{jk}^{(yz)}\Big)+\mathcal{FNN}^{(zx)}\Big(\textsl{\textbf{N}}_{ki}^{(zx)}\Big)\\
& \ \ \ \ \ + \mathcal{FNN}^{(xy-45^{\circ})}\Big(\textsl{\textbf{N}}_{ij}^{(xy-45^{\circ})}\Big)+\mathcal{FNN}^{(yz-45^{\circ})}\Big(\textsl{\textbf{N}}_{jk}^{(yz-45^{\circ})}\Big)+\mathcal{FNN}^{(zx-45^{\circ})}\Big(\textsl{\textbf{N}}_{ki}^{(zx-45^{\circ})}\Big)\\
& \ \ \ \ \ + \mathcal{FNN}^{(xy-135^{\circ})}\Big(\textsl{\textbf{N}}_{ij}^{(xy-135^{\circ})}\Big)+\mathcal{FNN}^{(yz-135^{\circ})}\Big(\textsl{\textbf{N}}_{jk}^{(yz-135^{\circ})}\Big)+\mathcal{FNN}^{(zx-135^{\circ})}\Big(\textsl{\textbf{N}}_{ki}^{(zx-135^{\circ})}\Big)\bigg]\,,\\
\end{aligned}
\label{Eq:3D_FNN_CPDF_2}
\end{equation}
where $\mathcal{FNN}^{(xy-45^{\circ})}\big(\cdot\big)$, $\mathcal{FNN}^{(yz-45^{\circ})}\big(\cdot\big)$, $\mathcal{FNN}^{(zx-45^{\circ})}\big(\cdot\big)$, $\mathcal{FNN}^{(xy-135^{\circ})}\big(\cdot\big)$, $\mathcal{FNN}^{(yz-135^{\circ})}\big(\cdot\big)$ and $\mathcal{FNN}^{(zx-135^{\circ})}\big(\cdot\big)$ represent the approximation functions of the SENN models for statistically characterizing the 2D exemplars on the $xy$-$45^{\circ}$-, $yz$-$45^{\circ}$-, $zx$-$45^{\circ}$-, $xy$-$135^{\circ}$-, $yz$-$135^{\circ}$-, and $zx$-$135^{\circ}$-planes, respectively.

\section{Stochastic microstructure reconstruction}
\label{Section3:Stochastic_microstructure_reconstruction}
\vspace{-2pt}
In the preceding section, a statistical characterisation approach has been developed to derive 3D morphological statistics $P\left(X_{ijk}\big|\textsl{\textbf{N}}_{ijk}\right)$ from 2D measurements, where 2D morphological statistics are approximated by training SENN models on representative 2D exemplars. The focus now shifts to leveraging this machine learning-based characterisation to synthesise 3D microstructure samples that are statistically equivalent to the 2D exemplars. In this section, a stochastic 2D-to-3D reconstruction approach is introduced for generating 3D multiphase microstructures, which utilises the 2D morphological statistics embedded in the pretrained SENN models. A Gibbs sampler is employed to iteratively generate voxel values, progressively updating the 3D realisation to converge toward the target probability distribution.

\subsection{Voxel generation via probability sampling}
\label{Subsec:Probability_sampling}
Generating observations of multivariate random variables through direct sampling can be challenging without complete knowledge of the full joint PDF. However, Gibbs sampling \cite{casella1992explaining, raftery1992many} provides an effective solution by leveraging conditional PDFs to produce a sequence of observations that approximate the target multivariate probability distribution. The 3D CPDF $P\left(X_{ijk}\big|\textsl{\textbf{N}}_{ijk}\right)$, as defined in Eq. \ref{Eq:3D_FNN_CPDF}, characterises the local morphology of multiphase microstructures. This CPDF, efficiently derived from pretrained SENN models, supplies the conditional statistics necessary to construct a 3D Markov random field (MRF). The Gibbs sampler iteratively updates this 3D MRF using the 3D CPDF, guiding it toward the underlying full joint PDF. After sufficient iterations, the 3D MRF converges to a stationary state that closely approximates the target full joint PDF, regardless of the initial conditions. This equilibrium state of the 3D MRF represents a reconstructed 3D microstructure sample that is statistically equivalent to the 2D exemplars.

\begin{figure}[h]
\centering
\includegraphics[width = 0.6\linewidth,angle=0,clip=true]{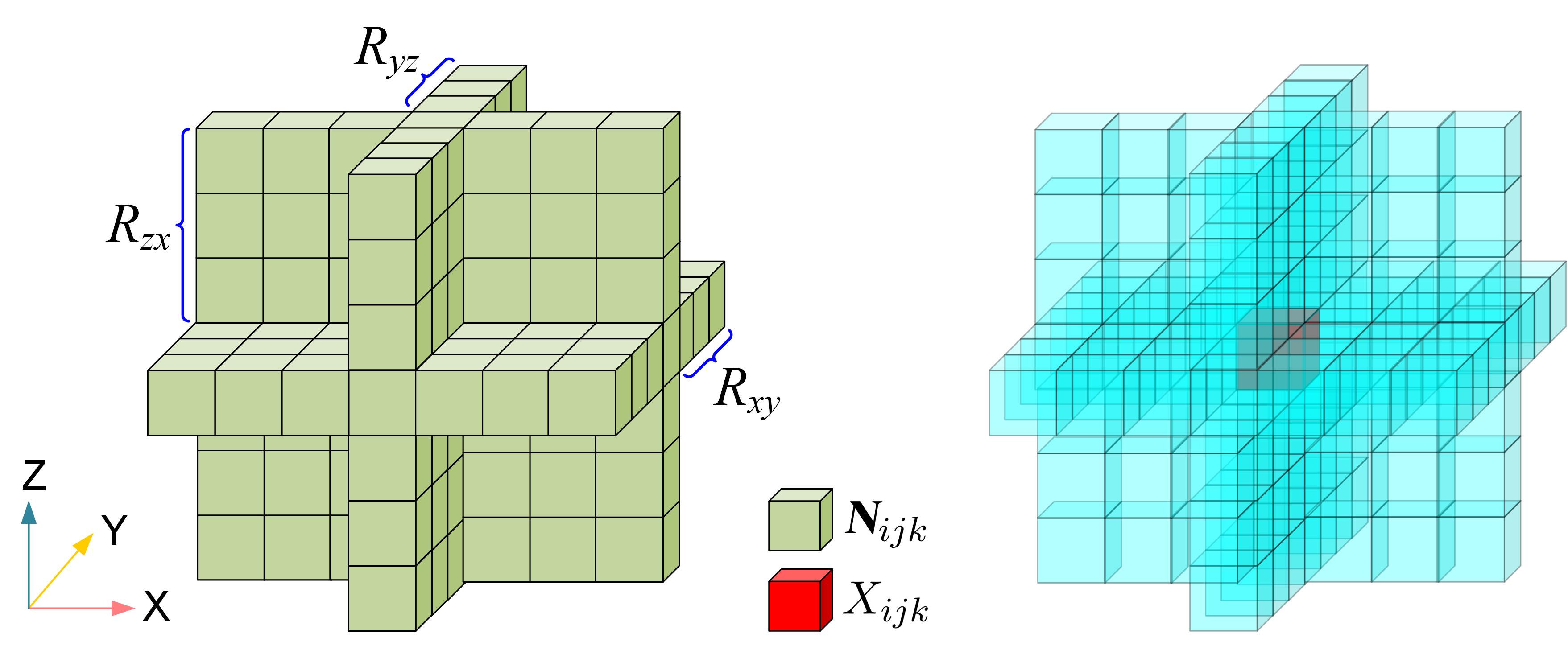} \\
\caption{A 3D data template composed of three 2D data templates (see Figure \ref{fig:Data_collection}b) on three orthogonal planes. The entire neighbourhood data $\textsl{\textbf{N}}_{ijk}$ in 3D space consist of neighbouring voxels $\textsl{\textbf{N}}_{ijk}^{(xy)}$, $\textsl{\textbf{N}}_{ijk}^{(yz)}$ and $\textsl{\textbf{N}}_{ijk}^{(zx)}$ on $xy$-, $yz$- and $zx$-planes, and the neighbourhood ranges are measured by $R_{xy}$, $R_{yz}$ and $R_{zx}$, respectively.}
\label{fig:3D_data_template}
\end{figure}

To achieve the objective of 2D-to-3D microstructure reconstruction, a specially designed 3D data template is employed to generate voxels using Gibbs sampling, as illustrated in Figure \ref{fig:3D_data_template}. This 3D data template integrates three 2D data templates on orthogonal planes, which were previously utilised for statistically characterising the 2D exemplars (as shown in Figure \ref{fig:Data_collection}). The geometric consistency between the 3D and 2D data templates ensures seamless application of the derived 3D CPDF $P\left(X_{ijk}\big|\textsl{\textbf{N}}_{ijk}\right)$, as defined in Eq. (\ref{Eq:3D_FNN_CPDF}), during the microstructure reconstruction process.

Gibbs sampling \cite{casella1992explaining, raftery1992many} iteratively generates and updates voxel values in the initial 3D MRF $\textsl{\textbf{Y}}_0$, using the CPDF $P\left(X_{ijk}\big|\textsl{\textbf{N}}_{ijk}\right)$. This process ultimately converges to the full joint PDF $P(\textsl{\textbf{Y}})$. 
The 3D MRF is characterised by three key factors: the initial probability $P\big(\textsl{\textbf{Y}}_0\big)$, the transition probability $P\left(X_{ijk}\big|\textsl{\textbf{N}}_{ijk}\right)$, and the stationary probability $P(\textsl{\textbf{Y}})$. The iterative process of Gibbs sampling is expressed as:
\begin{equation}
P\big(\textsl{\textbf{Y}}_0\big)\prod P\left(Y_{ijk}\,\big|\,\textsl{\textbf{N}}_{ijk}\right) \longrightarrow P\big(\textsl{\textbf{Y}}\,\big)\,,
\end{equation}
where $\textsl{\textbf{Y}}$ represents the equilibrium state of the 3D MRF, which can be regarded as a reconstructed 3D microstructure sample statistically equivalent to the 2D exemplars. This approach effectively leverages plane-wise neighbourhood information to iteratively reconstruct 3D microstructures, bridging the gap between 2D statistical characterisation and 3D microstructure reconstruction.

\begin{figure}[h]
\centering
\includegraphics[width = 0.90\linewidth,angle=0,clip=true]{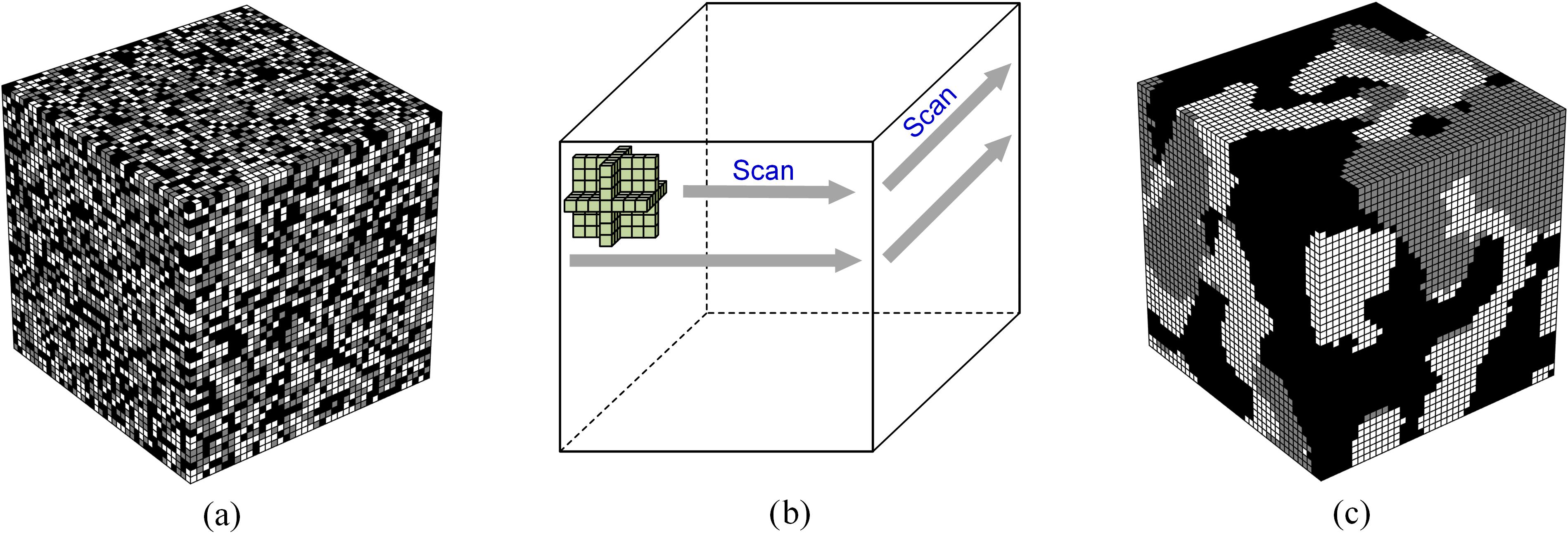} \\
\caption{Graphical illustration of probability sampling of voxel values: (a) A 3D random field is used as the initial assumption $\textsl{\textbf{Y}}_0$; (b) The predefined 3D data template is utilised to collect neighbouring voxels $\textsl{\textbf{N}}_{ijk}$ and generate/update the value of the central voxel $Y_{ijk}$ via Gibbs sampling; and (c) A reconstructed 3D microstructure sample (i.e., a realization of 3D MRF) that is statistically equivalent to the 2D examples.}
\label{fig:voxel_generation}
\end{figure}

In this methodology, a 3D Gaussian random field (white noise) is initially generated and then segmented into a discrete field with $n$ distinct voxel values, as defined by the indicator function in Eq. (\ref{Eq:Indication_function}). The obtained discrete random field serves as the initial assumption $\textsl{\textbf{Y}}_0$ for stochastic microstructure reconstruction. Subsequently, the predefined 3D data template (as illustrated in Figure \ref{fig:3D_data_template}) is then employed to scan $\textsl{\textbf{Y}}_0$ voxel by voxel in the raster scanning order (i.e., left-to-right and top-to-bottom).
The collected neighbouring voxels $\textsl{\textbf{N}}_{ijk}$ in each scanning step are decomposed into feature vectors $\textsl{\textbf{N}}_{ijk}^{(xy)}$, $\textsl{\textbf{N}}_{ijk}^{(yz)}$ and $\textsl{\textbf{N}}_{ijk}^{(zx)}$ on $xy$-, $yz$- and $zx$-planes, respectively. These feature vectors are input into the pretrained SENN models $\mathcal{FNN}^{(xy)}$, $\mathcal{FNN}^{(yz)}$ and $\mathcal{FNN}^{(zx)}$ to rapidly approximate the 3D CPDF $P\left(Y_{ijk}\big|\textsl{\textbf{N}}_{ijk}\right)$, as elucidated by Eq. (\ref{Eq:3D_FNN_CPDF}). Given this probability $P\left(Y_{ijk}\big|\textsl{\textbf{N}}_{ijk}\right)$, Gibbs sampler can assign a new phase value to the current (central) voxel $Y_{ijk}$ through probability sampling. This newly generated voxel is then integrated into the feature vectors to update subsequent adjacent voxels within the predefined neighbourhood ranges. By iterating these steps, the morphological patterns of the 3D random field gradually converge toward those of the 2D exemplars. Once equilibrium is achieved, a 3D microstructure sample statistically equivalent to the 2D exemplars is reconstructed, as illustrated in Figure \ref{fig:voxel_generation}c.

\begin{algorithm}
	\SetAlgoLined
    \small
    \setstretch{1.3}
	\KwData{A 3D lattice grid of size $I\times J\times K$.}
	\textbf{Initialization:} Assign a discrete random field to the 3D lattice grid and use it as the initial assumption $\textsl{\textbf{Y}}_{0}$.\\
    \textbf{SENN models:} The pretrained SENN models statistically characterising the 2D exemplars: $\mathcal{FNN}^{(xy)}$, $\mathcal{FNN}^{(yz)}$ and $\mathcal{FNN}^{(zx)}$.\\
    \textbf{Data templates:} The 3D data template $DT$ with the predefined size.\\
    \smallskip
    Add a periodic or reflective boundary to $\textsl{\textbf{Y}}_0$;\\
	\While{\ \rm the\ convergence\ criterion is not satisfied\ \ }
	{
    \For{{\ \rm each\ voxel\ variable} $Y_{ijk}$ {\rm in} $\textsl{\textbf{Y}}_0$\ \ }
		{Extract the neighbouring data $\textsl{\textbf{N}}_{ijk}$ from $\textsl{\textbf{Y}}_0$ using the 3D data template $DT$;\\
        Rearrange and split $\textsl{\textbf{N}}_{ijk}$ into three feature vectors $\textsl{\textbf{N}}_{ijk}^{(xy)}$, $\textsl{\textbf{N}}_{ijk}^{(yz)}$ and $\textsl{\textbf{N}}_{ijk}^{(zx)}$ on $xy$-, $yz$- and $zx$-planes, respectively;\\
        Input the above feature vectors into corresponding SENN models $\mathcal{FNN}^{(xy)}$, $\mathcal{FNN}^{(yz)}$ and $\mathcal{FNN}^{(zx)}$;\\
        Derive the 3D CPDF $P\left(Y_{ijk}\big|\textsl{\textbf{N}}_{ijk}\right)$ from 2D morphological statistics through Eq. (\ref{Eq:3D_FNN_CPDF}), given by: \\
        $P\left(X_{ijk}\,\big|\,\textsl{\textbf{N}}_{ijk}\right) \approx~\frac{1}{3}\cdot \bigg[\mathcal{FNN}^{(xy)}\Big(\textsl{\textbf{N}}_{ij}^{(xy)};\textsl{\textbf W}^{(xy)}, \textsl{\textbf b}^{(xy)}\Big)+\mathcal{FNN}^{(yz)}\Big(\textsl{\textbf{N}}_{jk}^{(yz)};\textsl{\textbf W}^{(yz)}, \textsl{\textbf b}^{(yz)}\Big)+\mathcal{FNN}^{(zx)}\Big(\textsl{\textbf{N}}_{ki}^{(zx)};\textsl{\textbf W}^{(zx)}, \textsl{\textbf b}^{(zx)}\Big)\bigg]$;\\
        Generate a new voxel value $Y_{ijk}^{(\rm new)}$ through probability sampling from the 3D CPDF    $P\left(Y_{ijk}\big|\textsl{\textbf{N}}_{ijk}\right)$;\\
	    Update the voxel value $Y_{ijk}$ using the newly generated $Y_{ijk}^{(\rm new)}$;	
		}
    Reset the periodic or reflective boundary condition for $\textsl{\textbf{Y}}_{\rm new}$.\\
	}
    Remove the periodic or reflective boundary condition for $\textsl{\textbf{Y}}_{\rm new}$. \\
	$\textbf{Return:}$ A reconstructed 3D microstructure sample $\textsl{\textbf{Y}}_{\rm new}$ that is statistically equivalent to the 2D exemplars.\\
    \qquad \quad \ (A constructed Markov random field $\textsl{\textbf{Y}}_{\rm new}$ that closely approximates the target full joint PDF $P(\textsl{\textbf{Y}})$).
	\caption{\small{Stochastic reconstruction of 3D multiphase microstructures using SENN-based 2D characterisation}}
	\label{Algorithm:microstructure_reconstruction}
\end{algorithm}

The primary workflow of this SENN-based framework for 2D-to-3D reconstruction of multiphase microstructures are summarised in Algorithm \ref{Algorithm:microstructure_reconstruction}. During the reconstruction process, the border voxels lack sufficient neighbourhood information to form feature vectors for the pretrained SENN models, thus preventing their voxel values from being updated. To address this issue, a periodic or reflective boundary condition can be incorporated into the intermediate constructed MRF, transforming the original "border voxels" into inner voxels that can be updated during the construction process. The thickness of the periodic or reflective boundary should align with the dimension of the 3D data template used for microstructure reconstruction. Additionally, it is essential to reset the periodic or reflective boundary after each round of Gibbs sampling until stability is achieved. Further details on periodic or reflective boundaries can be found in relevant references \citep{liu2015random, fu2021statistical}.

\subsection{Parallel voxel generation}
\label{Subsec:Paralle_voxel_generation}
\begin{figure}[h]
	\centering
	\includegraphics[width=0.30\linewidth]{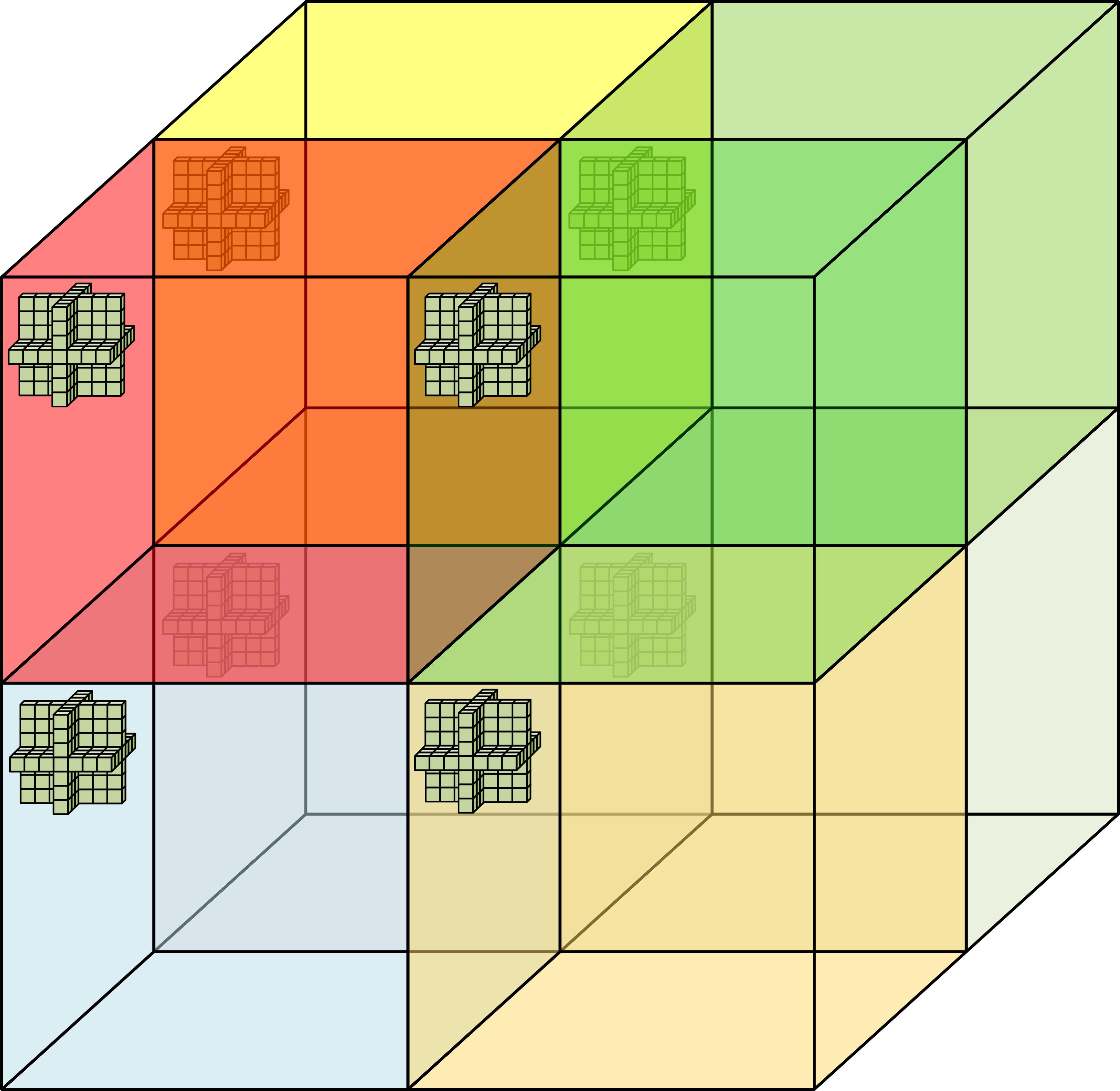}
	\caption{The initial 3D random field is divided into multiple subdomains, where voxels located beyond each other's neighbourhood ranges can be parallelly generated using independent Gibbs samplers.}
	\label{Parallel_reconstruction}
\end{figure}

Generating voxels sequentially can make the microstructure reconstruction process highly time-consuming, as representative elementary volumes (REVs) of digital microstructures often contain millions of voxels. Leveraging the MRF assumption defined in Eq. (\ref{Eq:3D_MRF_assumption-simplified}), distant voxels can be treated as independent. This allows for simultaneous generation of voxels beyond each other's neighbourhood ranges using independent Gibbs samplers. As shown in Figure \ref{Parallel_reconstruction}, the initial 3D random field can be divided into multiple subdomains, enabling parallel voxel generation and updating within these subdomains. This parallel approach significantly can accelerate the microstructure reconstruction process, especially as the number of subdomains increases. However, achieving large-scale parallelism, where thousands of voxels are generated concurrently, requires substantial data storage capacity. Consequently, implementing a parallel voxel generation strategy necessitates balancing computational speed with memory requirements to ensure efficiency.

\subsection{Multi-level reconstruction}
\label{Subsec:Multi-level_reconstruction}
As detailed in Section \ref{Section2:Statistical_microstructure_characterization}, the data template must be sufficiently large to capture the primary microstructural characteristics of multiphase composites. While effective for random microstructures with short-range correlations \citep{fu2022stochastic, fu2021statistical}, capturing long-range features in multiphase microstructures poses challenges. Simply enlarging the data template can drastically increase memory requirements for storing training data, complicating the SENN model training process for CPDF approximation. Furthermore, prioritising global characteristics may inadvertently dilute critical local features.

\begin{figure}[h]
	\centering
	\includegraphics[width=1.0\linewidth]{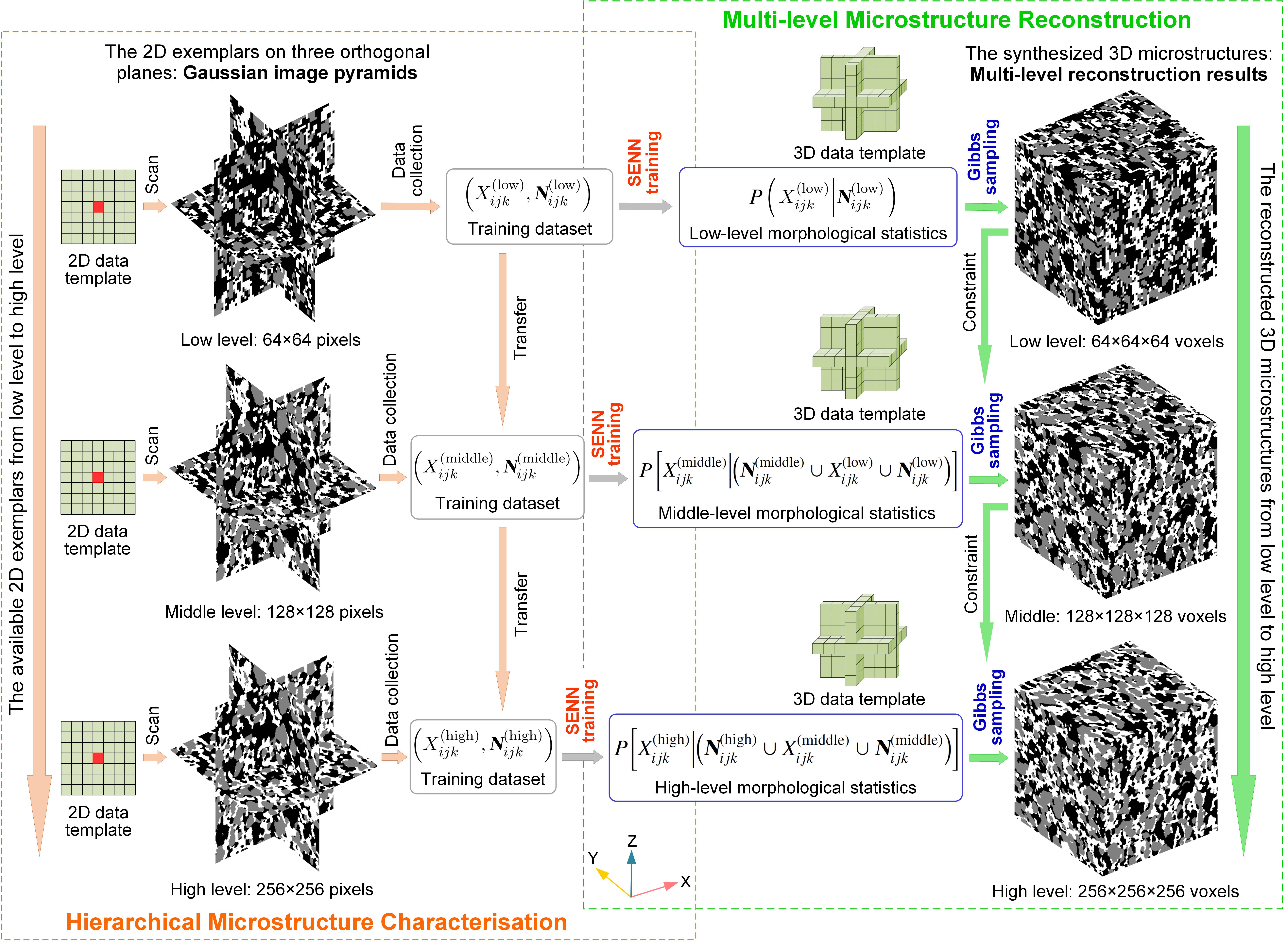}
	\caption{Graphical illustration of hierarchical characterisation and multi-level reconstruction for multiphase composite microstructures.}
	\label{Fig:hierachical_characterization_and_multi-level_reconstruction}
\end{figure}

To address the limitations of single-level microstructure characterisation, a hierarchical characterisation approach has been developed to statistically measure the local, regional, and global features of multiphase microstructures, enabling the derivation of 3D morphological statistics across multiple length scales. The methodology underlying this hierarchical characterisation is graphically illustrated in Figure \ref{Fig:hierachical_characterization_and_multi-level_reconstruction}, with the procedure summarised as follows:
\begin{itemize}
\small
    \item $\textbf{Gaussian image pyramids:}$ The original high-resolution 2D exemplars are transformed into Gaussian image pyramids \cite{adelson1984pyramid}, comprising 2D images at three resolution levels: $\textsl{\textbf{X}}^{({\rm low})}$, $\textsl{\textbf{X}}^{({\rm middle})}$, and $\textsl{\textbf{X}}^{({\rm high})}$. These pyramids enable the stepwise characterisation of global, regional, and local microstructural features, allowing large-scale metrics to be effectively captured using relatively small data templates.
        
    \item $\textbf{Low-level characterisation:}$ The 2D low-level exemplars, denoted by $\textsl{\textbf{X}}^{(\rm low)}$, are statistically characterised by training SENN models to approximate 2D CPDFs on three orthogonal planes, as detailed in Section \ref{Section2:Statistical_microstructure_characterization}. Based on Eqs. (\ref{Eq:3D_conditional_PDF}) and (\ref{Eq:3D_FNN_CPDF}), the 3D morphological statistics $P\left(X_{ijk}^{(\rm low)}\,\big|\,\textsl{\textbf{N}}_{ijk}^{(\rm low)}\right)$ at the low level can be inferred from the 2D CPDFs on three orthogonal planes, given by:
   \begin{equation}
   \begin{aligned} 
   P\left(X_{ijk}^{(\rm low)}\,\Big|\,\textsl{\textbf{N}}_{ijk}^{(\rm low)}\right)& = P\left[X_{ijk}^{(\rm low)}\,\Big|\,\Big(\textsl{\textbf{N}}_{ijk}^{({\rm low}: \ xy)}\cup\textsl{\textbf{N}}_{ijk}^{({\rm low}: \ yz)}\cup\textsl{\textbf{N}}_{ijk}^{({\rm low}: \ zx)}\Big)\right]\\
   &= \frac{1}{3}\cdot\bigg[P\left(X_{ijk}^{(\rm low)}\,\Big|\,\textsl{\textbf{N}}_{ijk}^{({\rm low}: \ xy)}\right)+P\left(X_{ijk}^{(\rm low)}\,\Big|\,\textsl{\textbf{N}}_{ijk}^{({\rm low}:\ yz)}\right)+P\left(X_{ijk}^{(\rm low)}\,\Big|\,\textsl{\textbf{N}}_{ijk}^{({\rm low}:\ zx)}\right)\bigg]\\
   &\approx \frac{1}{3} \cdot\bigg[\mathcal{FNN}_{xy}^{(\rm low)} \Big(\textsl{\textbf{N}}_{ijk}^{({\rm low}: \ xy)}\Big)+\mathcal{FNN}_{yz}^{(\rm low)}\Big(\textsl{\textbf{N}}_{ijk}^{({\rm low}: \ yz)}\Big)+\mathcal{FNN}_{zx}^{(\rm low)}\Big(\textsl{\textbf{N}}_{ijk}^{({\rm low}: \ zx)}\Big)\bigg]\,, \\
   \end{aligned} 
   \label{Eq:3D_conditional_PDF_low}
   \end{equation}
   where $\textsl{\textbf{N}}_{ijk}^{({\rm low}:\ xy)}$, $\textsl{\textbf{N}}_{ijk}^{({\rm low}:\ yz)}$ and $\textsl{\textbf{N}}_{ijk}^{({\rm low}:\ zx)}$ represent the neighbouring voxels on $xy$-, $yz$- and $zx$-planes, respectively; and meanwhile, $\mathcal{FNN}_{xy}^{(\rm low)}\big(\cdot\big)$, $\mathcal{FNN}_{yz}^{(\rm low)}\big(\cdot\big)$ and $\mathcal{FNN}_{zx}^{(\rm low)}\big(\cdot\big)$ denote the approximation functions of the SENN models used to statistically characterise the low-level 2D exemplars on the respective planes.

   \item $\textbf{Middle-level characterisation:}$ The 2D middle-level exemplars $\textsl{\textbf{X}}^{(\rm middle)}$ are utilised alongside the 2D low-level exemplars $\textsl{\textbf{X}}^{(\rm low)}$, to train another set of SENN models for the statistical characterisation at the middle level. 
   The input feature vector for SENN comprises voxel values from $\textsl{\textbf{N}}_{ijk}^{(\rm middle)}$, $X_{ijk}^{(\rm low)}$, and $\textsl{\textbf{N}}_{ijk}^{(\rm low)}$, while the output is the probability vector $\textsl{\textbf{V}}_{ijk}^{(\rm middle)}$ corresponding to $X_{ijk}^{(\rm middle)}$. 
   This operation not only captures statistical information from $\textsl{\textbf{X}}^{(\rm middle)}$ but also learns correlations between $\textsl{\textbf{X}}^{(\rm middle)}$ and $\textsl{\textbf{X}}^{(\rm low)}$, such as feature consistency and morphology continuity across adjacent levels.
   The 3D morphological statistics, $P\Big[X_{ijk}^{(\rm middle)}\,\big|\,\Big(\textsl{\textbf{N}}_{ijk}^{(\rm middle)}\cup X_{ijk}^{(\rm low)}\cup \textsl{\textbf{N}}_{ijk}^{(\rm low)}\Big)\Big]$, at the middle level can be derived from the 2D CPDFs on the three orthogonal planes as follows:
   \begin{equation}
   \begin{aligned} 
   P\left[X_{ijk}^{(\rm middle)}\,\Big|\,\Big(\textsl{\textbf{N}}_{ijk}^{(\rm middle)}\cup X_{ijk}^{(\rm low)}\cup \textsl{\textbf{N}}_{ijk}^{(\rm low)}\Big)\right] = 
   \frac{1}{3} \cdot \Bigg\{&P\left[X_{ijk}^{(\rm middle)}\,\Big|\,\Big(\textsl{\textbf{N}}_{ijk}^{({\rm middle}:\ xy)}\cup X_{ijk}^{(\rm low)}\cup \textsl{\textbf{N}}_{ijk}^{({\rm low}:\ xy)}\Big)\right]\\
   +&P\left[X_{ijk}^{(\rm middle)}\,\Big|\,\Big(\textsl{\textbf{N}}_{ijk}^{({\rm middle}:\ yz)}\cup X_{ijk}^{(\rm low)}\cup \textsl{\textbf{N}}_{ijk}^{({\rm low}:\ yz)}\Big)\right]\\
   +&P\left[X_{ijk}^{(\rm middle)}\,\Big|\,\Big(\textsl{\textbf{N}}_{ijk}^{({\rm middle}:\ zx)}\cup X_{ijk}^{(\rm low)}\cup \textsl{\textbf{N}}_{ijk}^{({\rm low}:\ zx)}\Big)\right]\Bigg\}\\
   \approx \frac{1}{3} \cdot\bigg[&\mathcal{FNN}_{xy}^{(\rm middle)} \Big(\textsl{\textbf{N}}_{ijk}^{({\rm middle}:\ xy)}\cup X_{ijk}^{(\rm low)}\cup \textsl{\textbf{N}}_{ijk}^{({\rm low}:\ xy)}\Big)\\
   +&\mathcal{FNN}_{yz}^{(\rm middle)} \Big(\textsl{\textbf{N}}_{ijk}^{({\rm middle}:\ yz)}\cup X_{ijk}^{(\rm low)}\cup \textsl{\textbf{N}}_{ijk}^{({\rm low}:\ yz)}\Big)\\
   +&\mathcal{FNN}_{zx}^{(\rm middle)} \Big(\textsl{\textbf{N}}_{ijk}^{({\rm middle}:\ zx)}\cup X_{ijk}^{(\rm low)}\cup \textsl{\textbf{N}}_{ijk}^{({\rm low}:\ zx)}\Big)\bigg]\,,\\
   \end{aligned} 
   \label{Eq:3D_conditional_PDF_middle}
   \end{equation}
   where $\textsl{\textbf{N}}_{ijk}^{({\rm middle}:\ xy)}$, $\textsl{\textbf{N}}_{ijk}^{({\rm middle}:\ yz)}$ and $\textsl{\textbf{N}}_{ijk}^{({\rm middle}:\ zx)}$ represent the neighbouring voxels on $xy$-, $yz$- and $zx$-planes of the middle-level image $\textsl{\textbf{X}}^{(\rm middle)}$, respectively; The 2D CPDFs $P\Big[X_{ijk}^{(\rm middle)}\,\big|\,\Big(\textsl{\textbf{N}}_{ijk}^{({\rm middle}:\ xy)}\cup X_{ijk}^{(\rm low)}\cup \textsl{\textbf{N}}_{ijk}^{({\rm low}:\ xy)}\Big)\Big]$, $P\Big[X_{ijk}^{(\rm middle)}\,\big|\,\Big(\textsl{\textbf{N}}_{ijk}^{({\rm middle}:\ yz)}\cup X_{ijk}^{(\rm low)}\cup \textsl{\textbf{N}}_{ijk}^{({\rm low}:\ yz)}\Big)\Big]$ and $P\Big[X_{ijk}^{(\rm middle)}\,\big|\,\Big(\textsl{\textbf{N}}_{ijk}^{({\rm middle}:\ zx)}\cup X_{ijk}^{(\rm low)}\cup \textsl{\textbf{N}}_{ijk}^{({\rm low}:\ zx)}\Big)\Big]$ can be obtained from the pretrained SENN models; $\mathcal{FNN}_{xy}^{(\rm middle)}\big(\cdot\big)$, $\mathcal{FNN}_{yz}^{(\rm middle)}\big(\cdot\big)$ and $\mathcal{FNN}_{zx}^{(\rm middle)}\big(\cdot\big)$ denote the approximation functions of the SENN models to statistically characterise the middle-level 2D exemplars on $xy$-, $yz$- and $zx$-planes, respectively.

   \item $\textbf{High-level characterisation:}$ The 2D high-level exemplars $\textsl{\textbf{X}}^{(\rm high)}$ are combined with the 2D middle-level exemplars $\textsl{\textbf{X}}^{(\rm middle)}$ to train a new set of SENN models. The input feature vector for SENN contain the voxel values of $\textsl{\textbf{N}}_{ijk}^{(\rm high)}$, $X_{ijk}^{(\rm middle)}$ and $\textsl{\textbf{N}}_{ijk}^{(\rm middle)}$.
   This characterisation manner is not only to extract statistical informatics from $\textsl{\textbf{X}}^{(\rm high)}$, but also to learn the correlations between $\textsl{\textbf{X}}^{(\rm high)}$ and $\textsl{\textbf{X}}^{(\rm middle)}$, thereby preserving the feature consistency and the morphology continuity between adjacent levels. The 3D morphological statistics $P\Big[X_{ijk}^{(\rm high)}\,\big|\,\Big(\textsl{\textbf{N}}_{ijk}^{(\rm high)}\cup X_{ijk}^{(\rm middle)}\cup\textsl{\textbf{N}}_{ijk}^{(\rm middle)}\Big)\Big]$ at the high level can be approximated from the 2D CPDFs on three orthogonal planes, as follows:
   \begin{equation}
   \begin{aligned} 
   P\left[X_{ijk}^{(\rm high)}\,\Big|\,\Big(\textsl{\textbf{N}}_{ijk}^{(\rm high)}\cup X_{ijk}^{(\rm middle)}\cup\textsl{\textbf{N}}_{ijk}^{(\rm middle)}\Big)\right] =
   \frac{1}{3} \cdot \Bigg\{&P\left[X_{ijk}^{(\rm high)}\,\Big|\,\Big(\textsl{\textbf{N}}_{ijk}^{({\rm high}:\ xy)}\cup X_{ijk}^{(\rm middle)}\cup \textsl{\textbf{N}}_{ijk}^{({\rm middle}:\ xy)}\Big)\right]\\
   +&P\left[X_{ijk}^{(\rm high)}\,\Big|\,\Big(\textsl{\textbf{N}}_{ijk}^{({\rm high}:\ yz)}\cup X_{ijk}^{(\rm middle)}\cup\textsl{\textbf{N}}_{ijk}^{({\rm middle}:\ yz)}\Big)\right]\\
   +&P\left[X_{ijk}^{(\rm high)}\,\Big|\,\Big(\textsl{\textbf{N}}_{ijk}^{({\rm high}:\ zx)}\cup X_{ijk}^{(\rm middle)}\cup\textsl{\textbf{N}}_{ijk}^{({\rm middle}:\ zx)}\Big)\right]\Bigg\}\\
   \approx \frac{1}{3} \cdot\bigg[&\mathcal{FNN}_{xy}^{(\rm high)} \Big(\textsl{\textbf{N}}_{ijk}^{({\rm high}:\ xy)}\cup X_{ijk}^{(\rm middle)}\cup\textsl{\textbf{N}}_{ijk}^{({\rm middle}:\ xy)}\Big)\\
   +&\mathcal{FNN}_{yz}^{(\rm high)} \Big(\textsl{\textbf{N}}_{ijk}^{({\rm high}:\ yz)}\cup X_{ijk}^{(\rm middle)}\cup\textsl{\textbf{N}}_{ijk}^{({\rm middle}:\ yz)}\Big)\\
   +&\mathcal{FNN}_{zx}^{(\rm high)} \Big(\textsl{\textbf{N}}_{ijk}^{({\rm high}:\ zx)}\cup X_{ijk}^{(\rm middle)}\cup\textsl{\textbf{N}}_{ijk}^{({\rm middle}:\ zx)}\Big)\bigg]\,,\\
   \end{aligned} 
   \label{Eq:3D_conditional_PDF_high}
   \end{equation}
   where $\textsl{\textbf{N}}_{ijk}^{({\rm high}:\ xy)}$, $\textsl{\textbf{N}}_{ijk}^{({\rm high}:\ yz)}$ and $\textsl{\textbf{N}}_{ijk}^{({\rm high}:\ zx)}$ represent the neighbouring voxels on $xy$-, $yz$- and $zx$-planes of the high-level images $\textsl{\textbf{X}}^{(\rm high)}$, respectively; The 2D CPDFs $P\Big[X_{ijk}^{(\rm high)}\,\big|\,\Big(\textsl{\textbf{N}}_{ijk}^{({\rm high}:\ xy)}\cup X_{ijk}^{(\rm middle)}\cup\textsl{\textbf{N}}_{ijk}^{({\rm middle}:\ xy)}\Big)\Big]$, $P\Big[X_{ijk}^{(\rm high)}\,\big|\,\Big(\textsl{\textbf{N}}_{ijk}^{({\rm high}:\ yz)}\cup X_{ijk}^{(\rm middle)}\cup\textsl{\textbf{N}}_{ijk}^{({\rm middle}:\ yz)}\Big)\Big]$ and $P\Big[X_{ijk}^{(\rm high)}\,\big|\,\Big(\textsl{\textbf{N}}_{ijk}^{({\rm high}:\ zx)}\cup X_{ijk}^{(\rm middle)}\cup\textsl{\textbf{N}}_{ijk}^{({\rm middle}:\ zx)}\Big)\Big]$ can be approximated from the pretrained SENN models; and $\mathcal{FNN}_{xy}^{(\rm high)}\big(\cdot\big)$, $\mathcal{FNN}_{yz}^{(\rm high)}\big(\cdot\big)$ and $\mathcal{FNN}_{zx}^{(\rm high)}\big(\cdot\big)$ denote the approximation functions of the SENN models for statistically characterizing the high-level 2D exemplars on $xy$-, $yz$- and $zx$-planes, respectively.  
\end{itemize}

In order to fully utilise the hierarchical SENN-based characterisation results, a multi-level reconstruction approach has thus been specially developed to generate statistically equivalent 3D microstructure samples for multiphase composites, as graphically illustrated in Figure \ref{Fig:hierachical_characterization_and_multi-level_reconstruction}. During microstructure reconstruction process, the pretrained SENN models serve as the dictionaries of conditional probabilities, for generating/updating voxels through Gibbs sampling.
In the low-level reconstruction step, long-distance features present in the 2D exemplars can be efficiently created and preserved in the 3D reconstructed microstructure sample. Conditional on the low-level reconstruction result, the middle-level reconstruction step generates and incorporates regional microstructural characteristics while ensuring morphological consistency and continuity with the low-level reconstruction. 
As for high-level construction, local microstructural details are seamlessly integrated into the 3D reconstructed sample to make it morphologically similar to real microstructures.
This innovative multi-level reconstruction approach reproduces global, regional, and local morphological patterns separately while ensuring seamless fusion across different length scales. By leveraging the interrelation between the 3D CPDFs at each level, it achieves a cohesive, statistically accurate representation of multiphase microstructures through Gibbs sampling.

\subsection{Convergence criterion}
\label{subsubsec:convergence criterion}
As detailed in Algorithm \ref{Algorithm:microstructure_reconstruction}, Gibbs sampling iterates until the reconstructed microstructure sample reaches a steady state. Convergence is achieved when the difference between consecutive iterations falls below a specified tolerance. The convergence criterion is given by:
\begin{equation}
\varepsilon_i = \frac{\big\|\textbf{Y}_i-\textbf{Y}_{i-1}\big\|_2}{\big\|\textbf{Y}_{i-1}\big\|_2} \leq \eta\,,
\end{equation}
Here, \(\varepsilon_i\) is the relative \(L_2\)-norm error, quantifying the difference between the current \(\mathbf{Y}_i\) and previous \(\mathbf{Y}_{i-1}\) samples. The iteration index is represented by \(i\), and \(\eta \ll 1\) is a user-defined tolerance threshold. When \(\varepsilon_i\) falls below \(\eta\), the reconstruction process is considered stable.

\section{Examples and results} 
\label{Section4:Examples_and_results}
\vspace{-2pt}
To evaluate the effectiveness and robustness of the proposed SENN-based framework for multiphase microstructure reconstruction, it is applied to generate virtual microstructure samples for three representative multiphase materials with distinct morphologies. SENN models are trained to approximate morphological statistics required for microstructural characterisation, using training data gathered from available 2D exemplars. Based on the SENN-based characterisation, a group of thirty 3D microstructure samples are stochastically synthesised, preserving the morphological patterns observed in the 2D training data. Finally, morphological descriptors \citep{fu2023data, fu2022stochastic, cui2021correlation} are extracted from both the reconstructed and reference 3D microstructures to quantitatively assess their morphological similarity and statistical equivalence.

\subsection{Example 1: Porous silver-based electrode}
\label{Subsection:Example1}
\begin{figure}[h]
	\centering
	\includegraphics[width=1.0\linewidth]{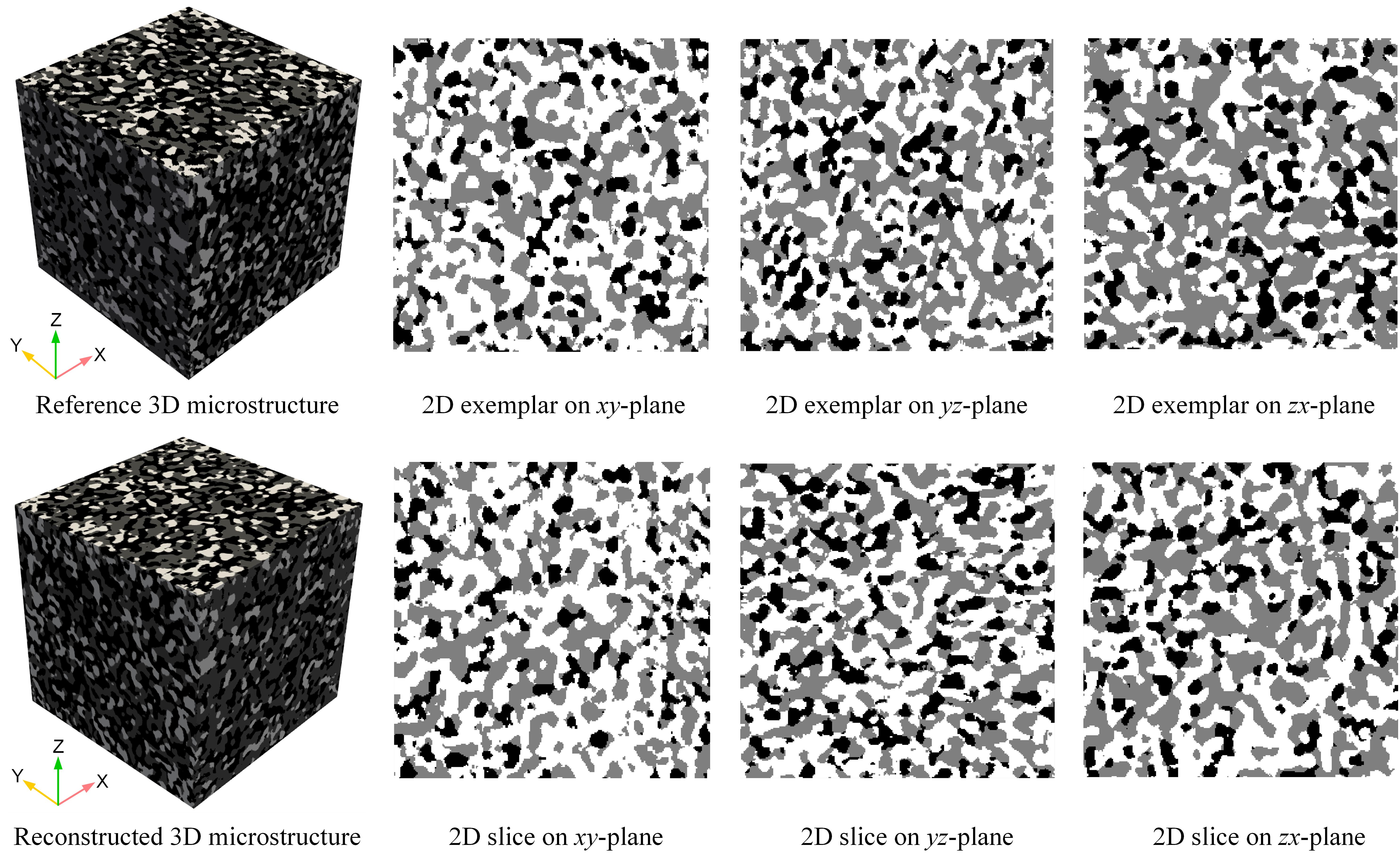}
	\caption{Visual comparison between the reference (top row) and the reconstructed (bottom row) 2D/3D microstructures of a porous silver-based electrode, highlighting its three distinct constituent phases: black regions represent silver, grey regions represent PTFE, and white regions represent pore space.}
	\label{Ex1:Reconstruction_result}
\end{figure}

Porous silver-based electrodes \cite{li2016formation, neumann2019pluri} are widely employed in electrochemical applications such as solid oxide fuel cells, oxygen sensors, and batteries, due to their superior electrical conductivity, catalytic performance, and chemical stability. In this study, a porous silver-based electrode is chosen as a representative multiphase composite material to demonstrate the effectiveness of the proposed SENN-based reconstruction approach. The electrode consists of three constituent phases: silver, polytetrafluoroethylene (PTFE), and pore space, as depicted in Figure \ref{Ex1:Reconstruction_result}.

As illustrated in the upper row of Figure \ref{Ex1:Reconstruction_result}, representative 2D exemplars on three orthogonal planes are selected to statistically reconstruct virtual 3D microstructure samples. From these original 2D exemplars, Gaussian image pyramids with three levels are constructed for hierarchical characterisation of the porous microstructure, as detailed in Section \ref{Subsec:Multi-level_reconstruction}. Three sets of SENN models are then trained to approximate the 2D morphological statistics on the orthogonal planes, enabling the derivation of 3D CPDFs across multiple length scales. These 3D CPDFs include $P\Big(X_{ijk}^{(\rm low)}\,\big|\,\textsl{\textbf{N}}_{ijk}^{(\rm low)}\Big)$, $P\Big[X_{ijk}^{(\rm middle)}\,\big|\,\Big(\textsl{\textbf{N}}_{ijk}^{(\rm middle)}\cup X_{ijk}^{(\rm low)}\cup \textsl{\textbf{N}}_{ijk}^{(\rm low)}\Big)\Big]$ and $P\Big[X_{ijk}^{(\rm high)}\,\big|\,\Big(\textsl{\textbf{N}}_{ijk}^{(\rm high)}\cup X_{ijk}^{(\rm middle)}\cup\textsl{\textbf{N}}_{ijk}^{(\rm middle)}\Big)\Big]$. This hierarchical characterisation aims to capture the long-range connectivity of pore networks. Besides, parameter settings for training data perpetration and SENN modelling are summarised in Table \ref{Tab:SENN_parameters}.

\vspace{-2pt}

\begin{figure}[H]\footnotesize
	\begin{minipage}[t]{0.333\textwidth}
		\centering  
		\includegraphics[width=0.96\textwidth]{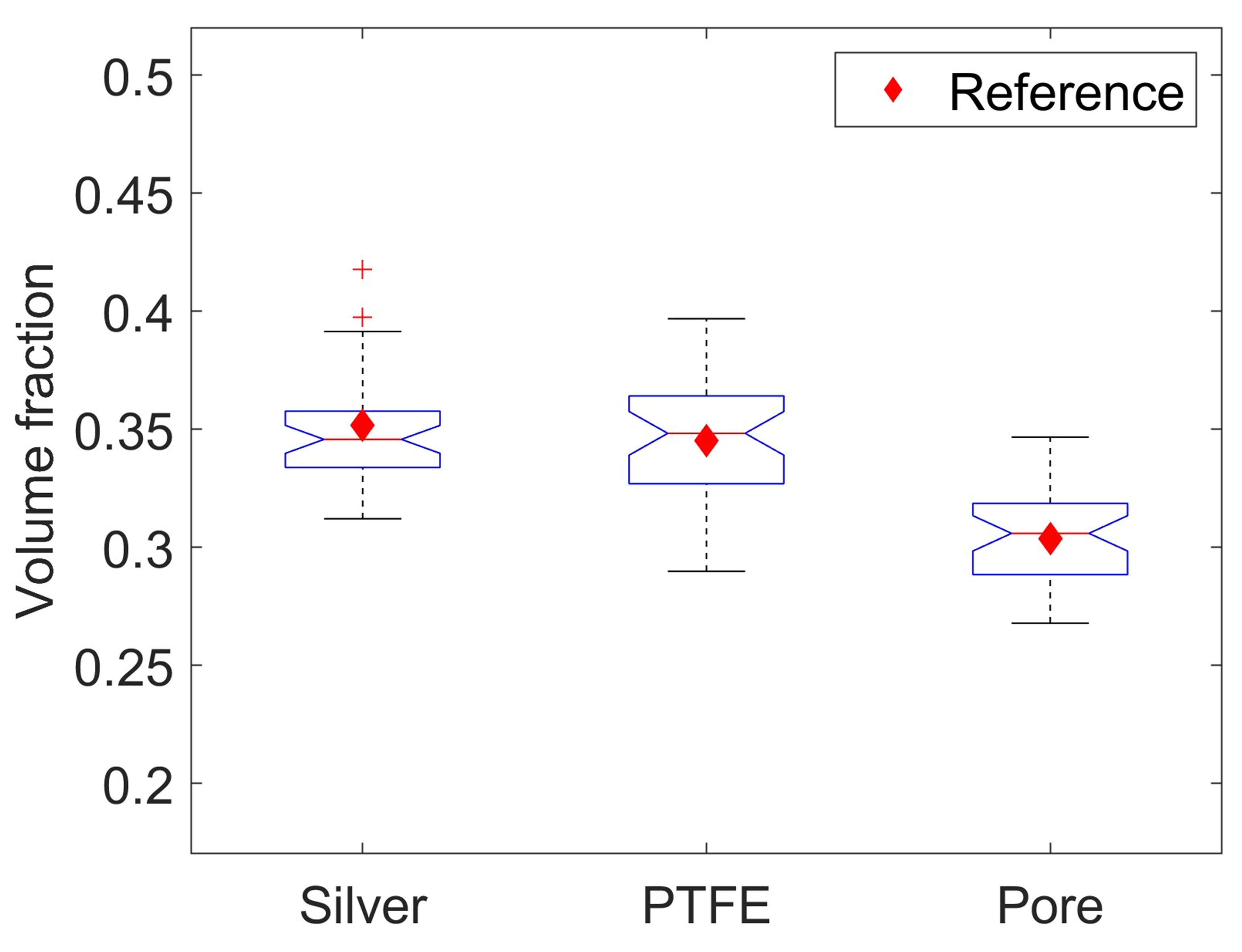}
		\text{(a) Volume fraction}
	\end{minipage}  
    \smallskip
	\begin{minipage}[t]{0.333\textwidth}  
		\centering  
		\includegraphics[width=0.96\textwidth]{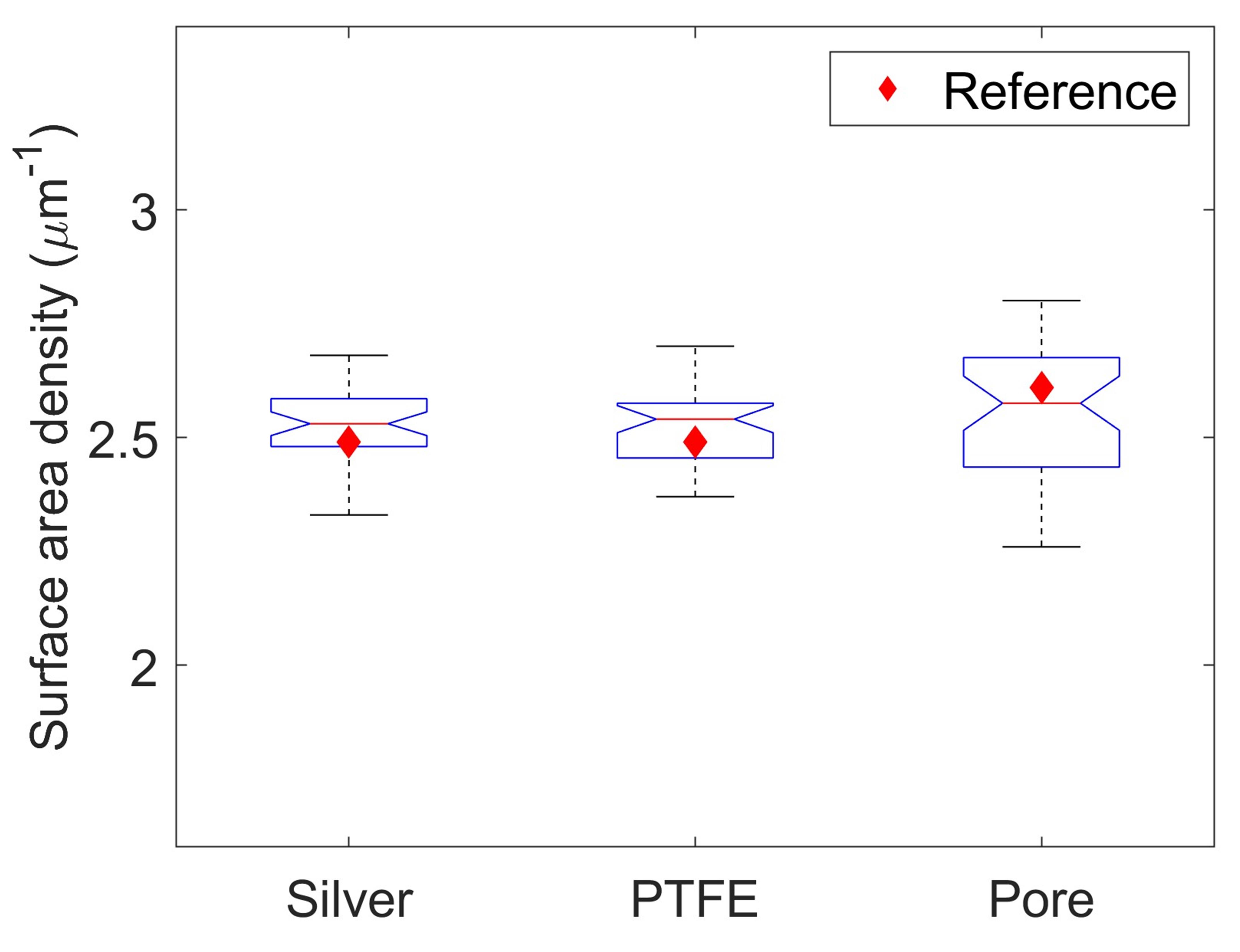}
		\text{(b) Surface area density}
	\end{minipage}  
    \smallskip
	\begin{minipage}[t]{0.333\textwidth}  
		\centering  
		\includegraphics[width=0.97\textwidth]{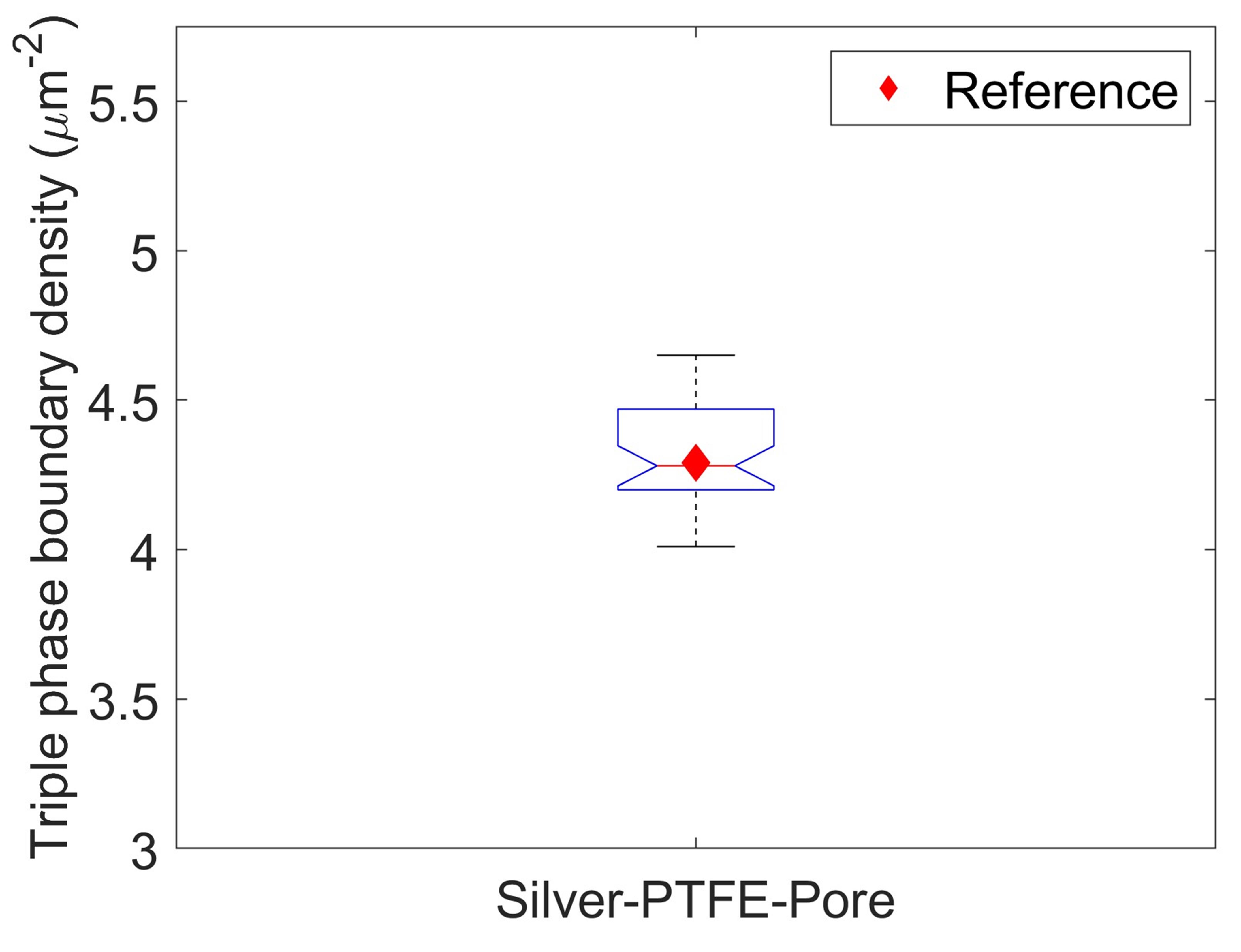}
		\text{(c) Triple phase boundary density}
	\end{minipage}  
    \smallskip
	\begin{minipage}[t]{0.333\textwidth}  
		\centering  
		\includegraphics[width=0.96\textwidth]{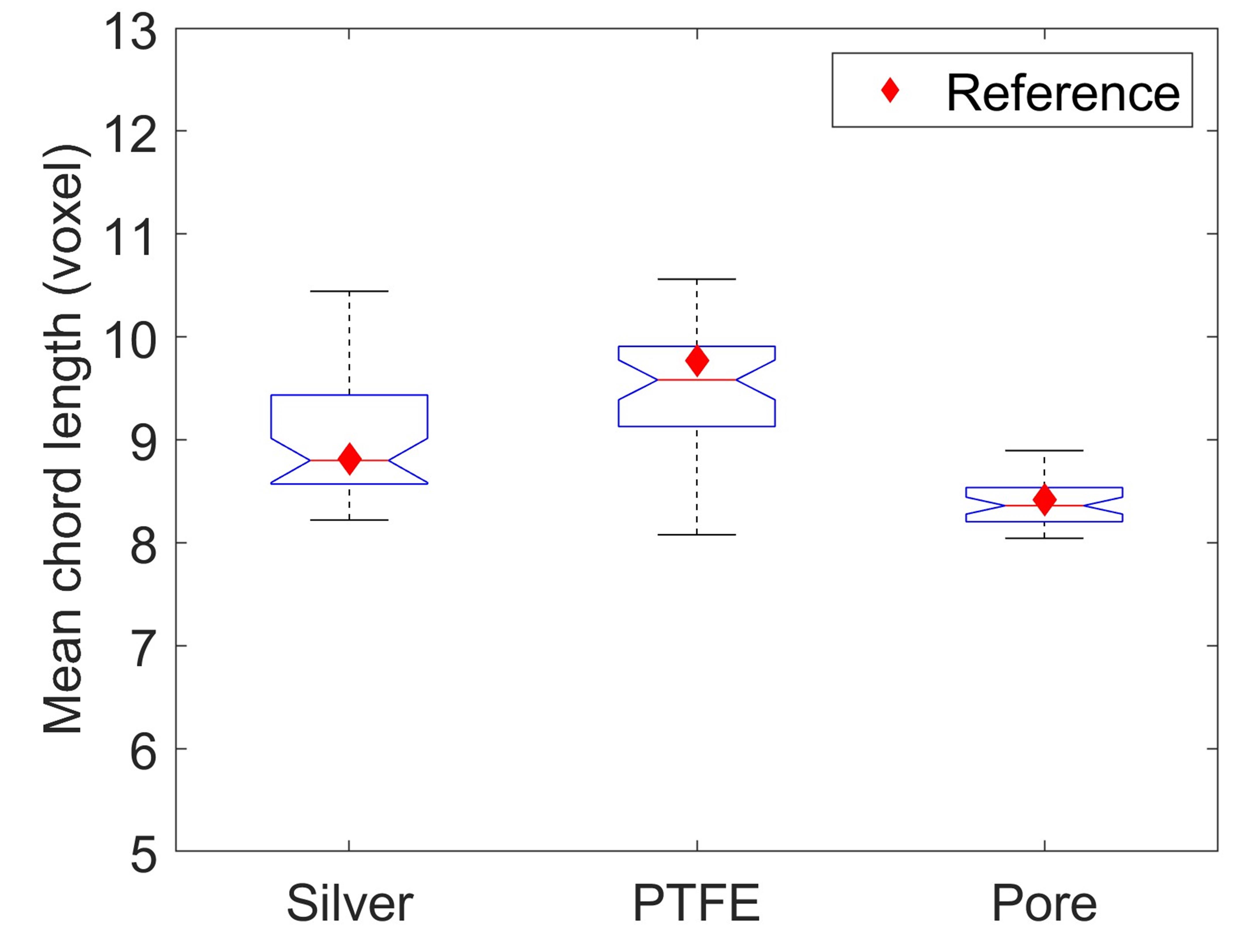} 
		\text{(d) Mean chord length}
	\end{minipage}  
    \begin{minipage}[t]{0.33\textwidth} 
		\centering  
		\includegraphics[width=0.98\textwidth]{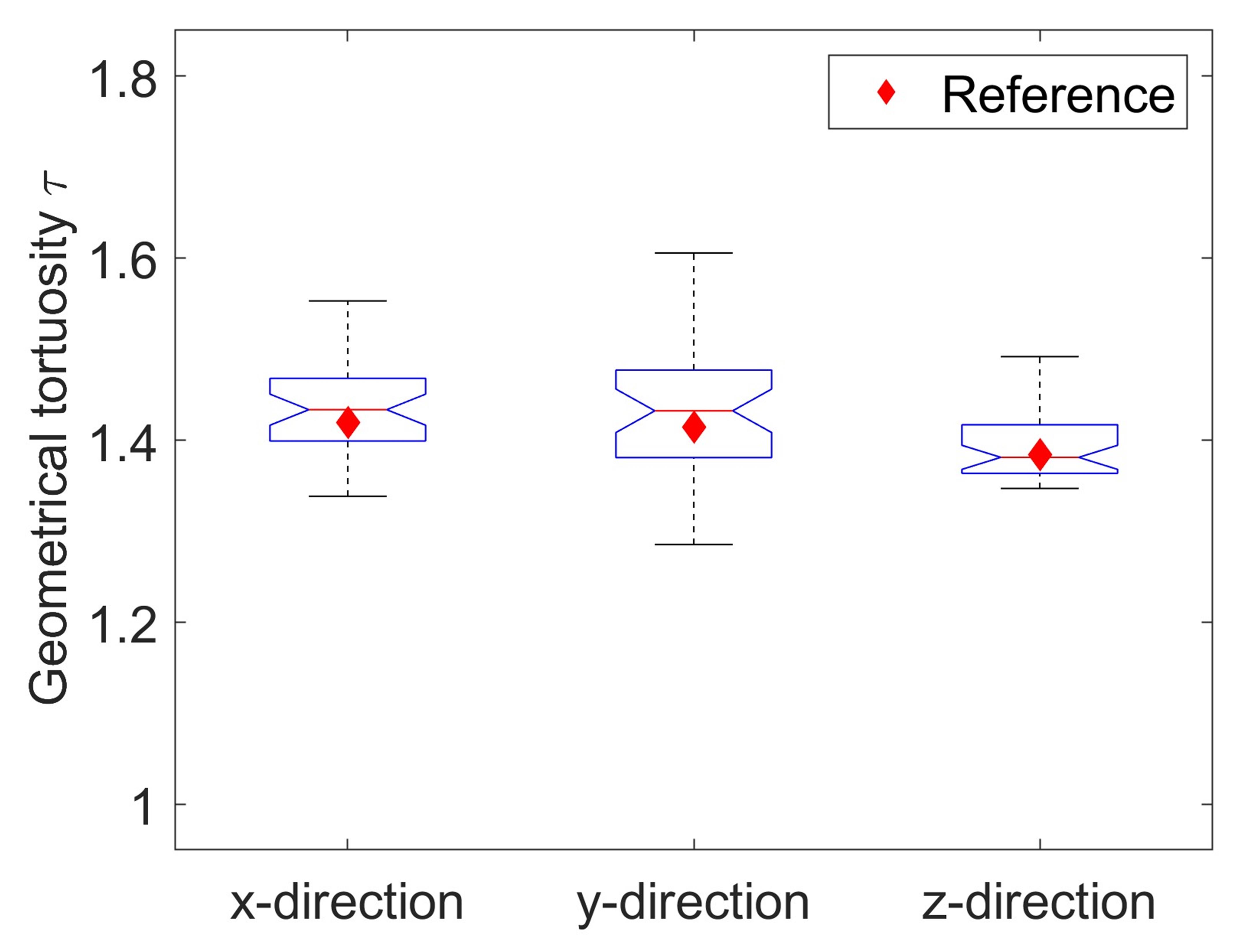}
		\text{(e) Geometrical tortuosity}
	\end{minipage}
	\begin{minipage}[t]{0.333\textwidth}  
		\centering  
		\includegraphics[width=0.98\textwidth]{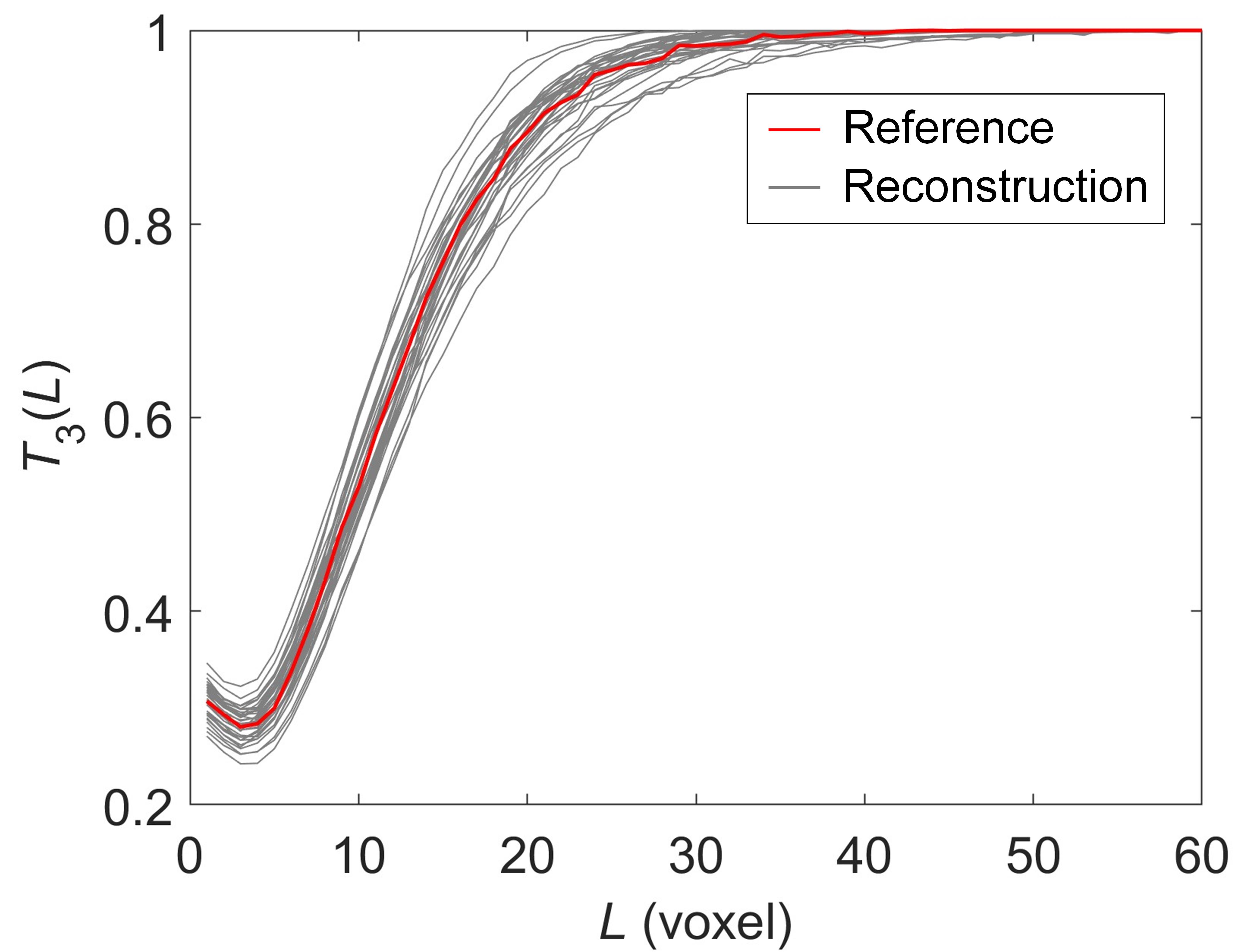}  
		\text{(f) Total fraction of percolation cells}
	\end{minipage}
	\caption{Quantitative comparison between the reference and reconstructed 3D microstructures by evaluating morphological descriptors.}
	\label{Fig:Ex1_descriptors}
\end{figure}

Using the SENN-approximated morphological statistics, a group of thirty 3D microstructure samples are synthesised, with representative results shown in the lower row of Figure \ref{Ex1:Reconstruction_result}. Visual inspection reveals that the reconstructed 3D microstructure is morphologically similar to the 2D exemplars and nearly indistinguishable from the 3D reference microstructure. To quantitatively evaluate the reconstruction quality, a series of morphological descriptors are extracted from both the reconstructed and reference 3D microstructures for comparison. These descriptors include volume fraction, surface area density \cite{fu2022stochastic}, triple-phase boundary density \citep{bertei2017fractal}, mean chord length \cite{cui2021correlation}, geometrical tortuosity \cite{fu2021tortuosity}, and total fraction of percolation cells (TFOPC) \citep{fu2023data}, as presented in Figure \ref{Fig:Ex1_descriptors}.

The first four descriptors (Figure \ref{Fig:Ex1_descriptors}a–d) are calculated for all three phases of the porous electrode, while the final two descriptors (Figure \ref{Fig:Ex1_descriptors}e–f) are specific to the pore space to quantify its long-distance connectivity. Descriptor values or curves derived from the thirty reconstructed microstructures exhibit slight variations around those obtained from the reference 3D microstructure. This variation arises from probabilistic voxel sampling during the statistical reconstruction process. Despite these variations, the mean values or curves of the descriptors extracted from the reconstructed microstructures closely align with the corresponding reference values or curves. This consistency, combined with the slight variance, demonstrates the statistical equivalence between the reference and reconstructed 3D microstructures. In addition, the excellent agreement observed for geometrical tortuosity and TFOPC further confirms that the developed hierarchical reconstruction approach effectively captures the long-range connectivity of multiphase porous media.

\subsection{Example 2: Porous SOFC anode}
\label{Subsection:Example2}
\begin{figure}[H]
	\centering
	\includegraphics[width=1.0\linewidth]{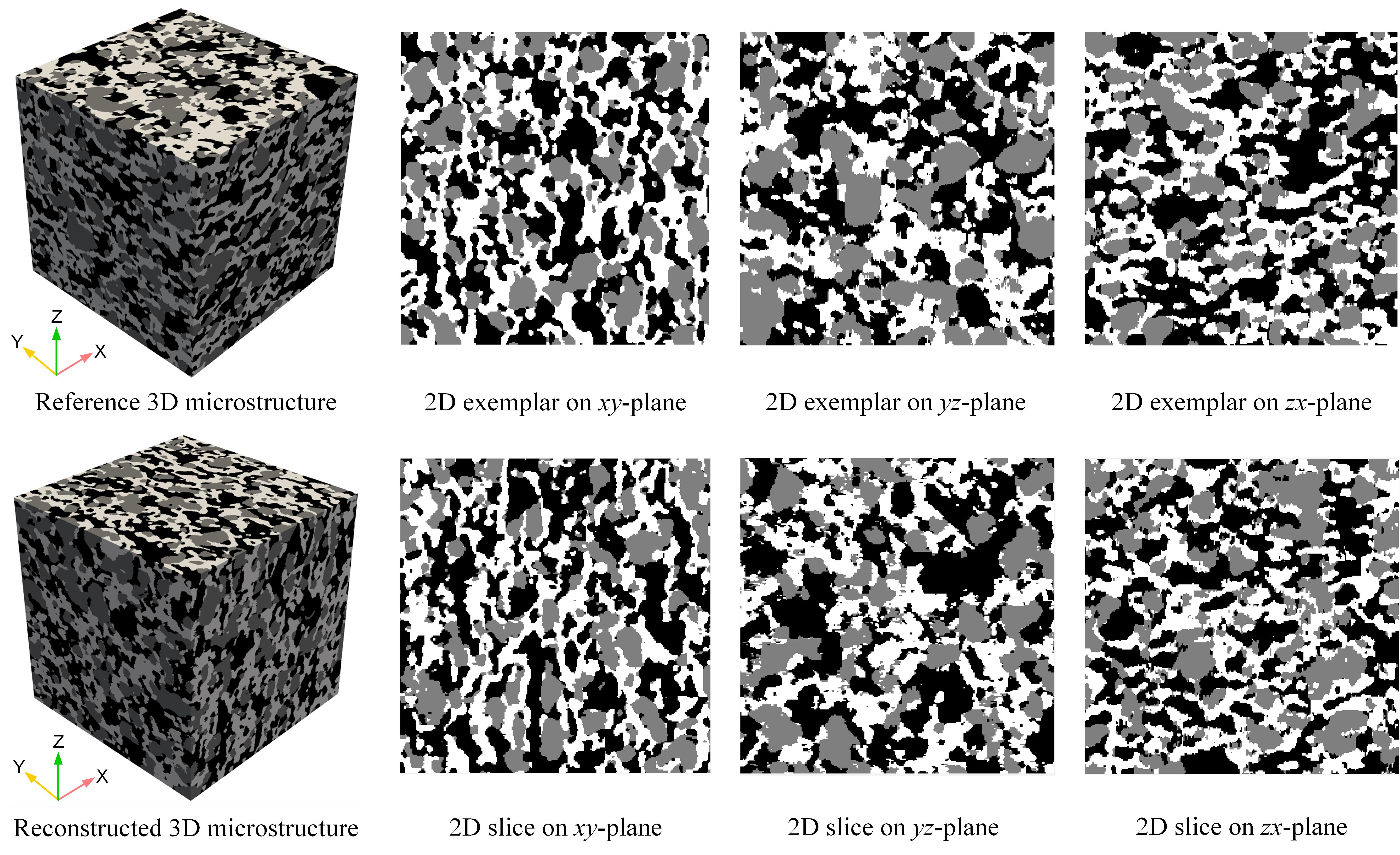}
	\caption{Visual comparison between the reference (top row) and the reconstructed (bottom row) 2D/3D microstructures of a porous SOFC anode with three distinct constituent phases: black regions represent YSZ, grey regions represent Ni, and white regions represent pore space.}
	\label{Fig:Ex2_reconstruction}
\end{figure}

\vspace{-2pt}
Solid oxide fuel cells (SOFCs) \cite{toros2016microstructural, liu2022correlation} are advanced electrochemical devices that efficiently convert the chemical energy of hydrogen into electricity. They consist of a dense electrolyte sandwiched between an anode and a cathode. Porous nickel/yttria-stabilized zirconia (Ni-YSZ) cermet \cite{liu2022correlation, cooper2016taufactor} is widely used as an anode material in SOFCs. In this study, it serves as a representative multiphase composite to evaluate the performance of the proposed SENN-based reconstruction approach. As shown in Figure \ref{Fig:Ex2_reconstruction}, this Ni-YSZ-based SOFC anode comprises three constituent phases: YSZ (black), Ni (grey), and pore space (black).

As shown in the upper row of Figure \ref{Fig:Ex2_reconstruction}, the morphological differences between the 2D exemplars along different directions reveal the anisotropy of the porous microstructure in the SOFC anode. Representative 2D exemplars on nine different planes are selected as training images to construct the SENN-based characterisation.
To capture the long-range connectivity of the pore networks (white phase), the SOFC anode is statistically characterised at three hierarchical length scales, as described in Section \ref{Subsec:Multi-level_reconstruction}. Three-level Gaussian image pyramids are constructed from the original high-resolution 2D exemplars. Using training data derived from these Gaussian pyramids, a set of SENN models is fitted to infer 3D morphological statistics across different length scales, represented as $P\left(X_{ijk}^{(\rm low)}\big|\textsl{\textbf{N}}_{ijk}^{(\rm low)}\right)$, $P\left[X_{ijk}^{(\rm middle)}\big|\Big(\textsl{\textbf{N}}_{ijk}^{(\rm middle)}\cup X_{ijk}^{(\rm low)}\cup\textsl{\textbf{N}}_{ijk}^{(\rm low)}\Big)\right]$ and $P\left[X_{ijk}^{(\rm high)}\big|\Big(\textsl{\textbf{N}}_{ijk}^{(\rm high)}\cup X_{ijk}^{(\rm middle)}\cup \textsl{\textbf{N}}_{ijk}^{ (\rm middle) }\Big)\right]$. Besides, the parameter settings for training data preparation and SENN modelling are summarised in Table \ref{Tab:SENN_parameters}.

Leveraging the three-level morphological statistics embedded in the pretrained SENN models, a group of thirty 3D microstructure samples are stochastically synthesised. A representative sample is shown in the bottom row of Figure \ref{Fig:Ex2_reconstruction}. Visual comparison indicates that the reconstructed 3D microstructure closely resembles the reference sample, particularly in reproducing anisotropic morphology patterns. To quantitatively assess the morphological similarity, four descriptors are extracted from both the reference and reconstructed 3D microstructures. These descriptors include volume fraction, surface area density \cite{cooper2016taufactor}, triple-phase boundary density \cite{liu2022correlation}, and the two-point correlation function (TPCF) \citep{fu2022stochastic}. These metrics are essential for evaluating the macroscopic behaviour of porous SOFC anodes, such as elastic properties and effective diffusivity \cite{fu2023data, liu2022correlation}.

\begin{figure}[h]\footnotesize
	\begin{minipage}[t]{0.333\textwidth}
		\centering  
		\includegraphics[width=0.96\textwidth]{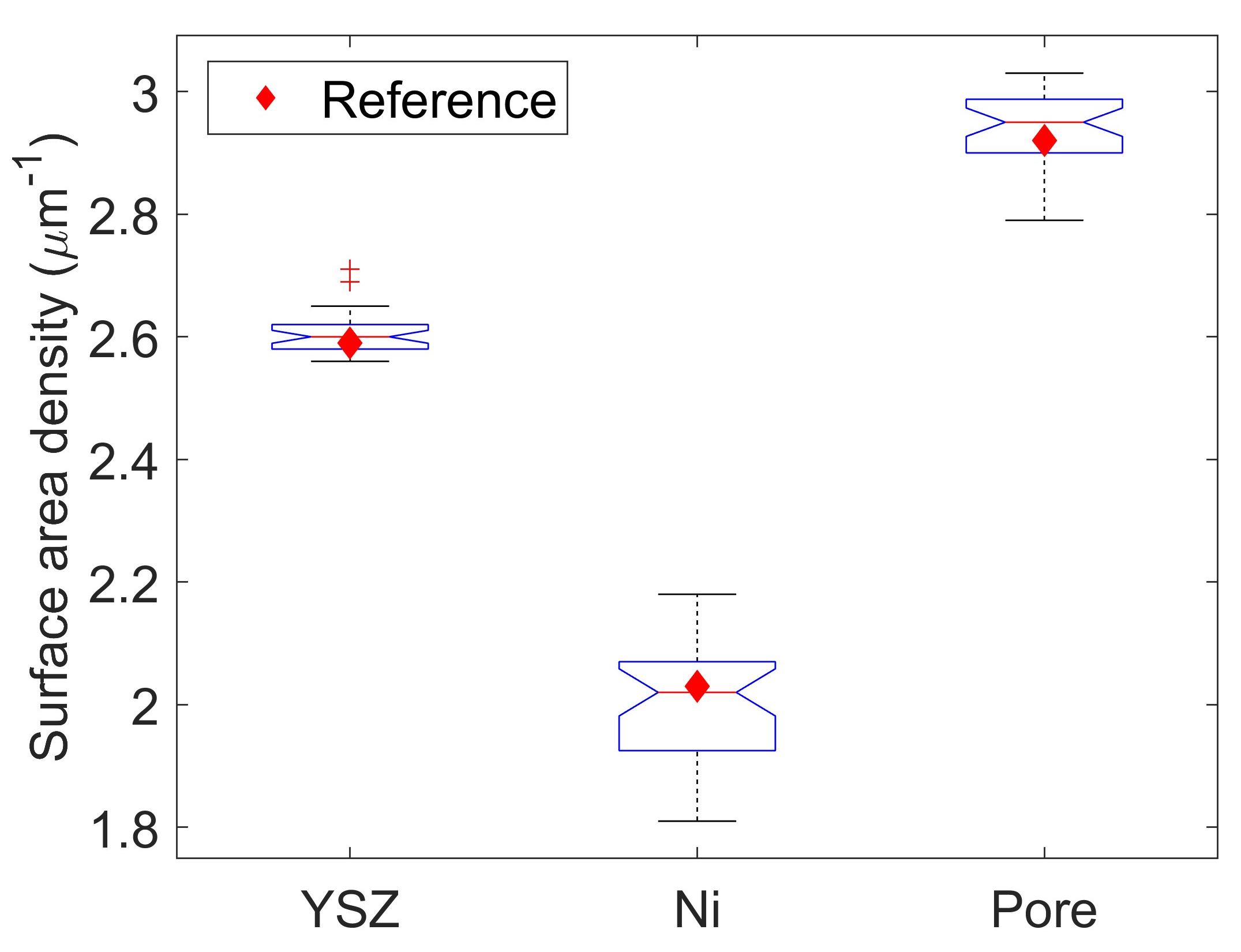}
		\text{(a) Surface area density}
	\end{minipage}  
    \smallskip
	\begin{minipage}[t]{0.333\textwidth}  
		\centering  
		\includegraphics[width=0.96\textwidth]{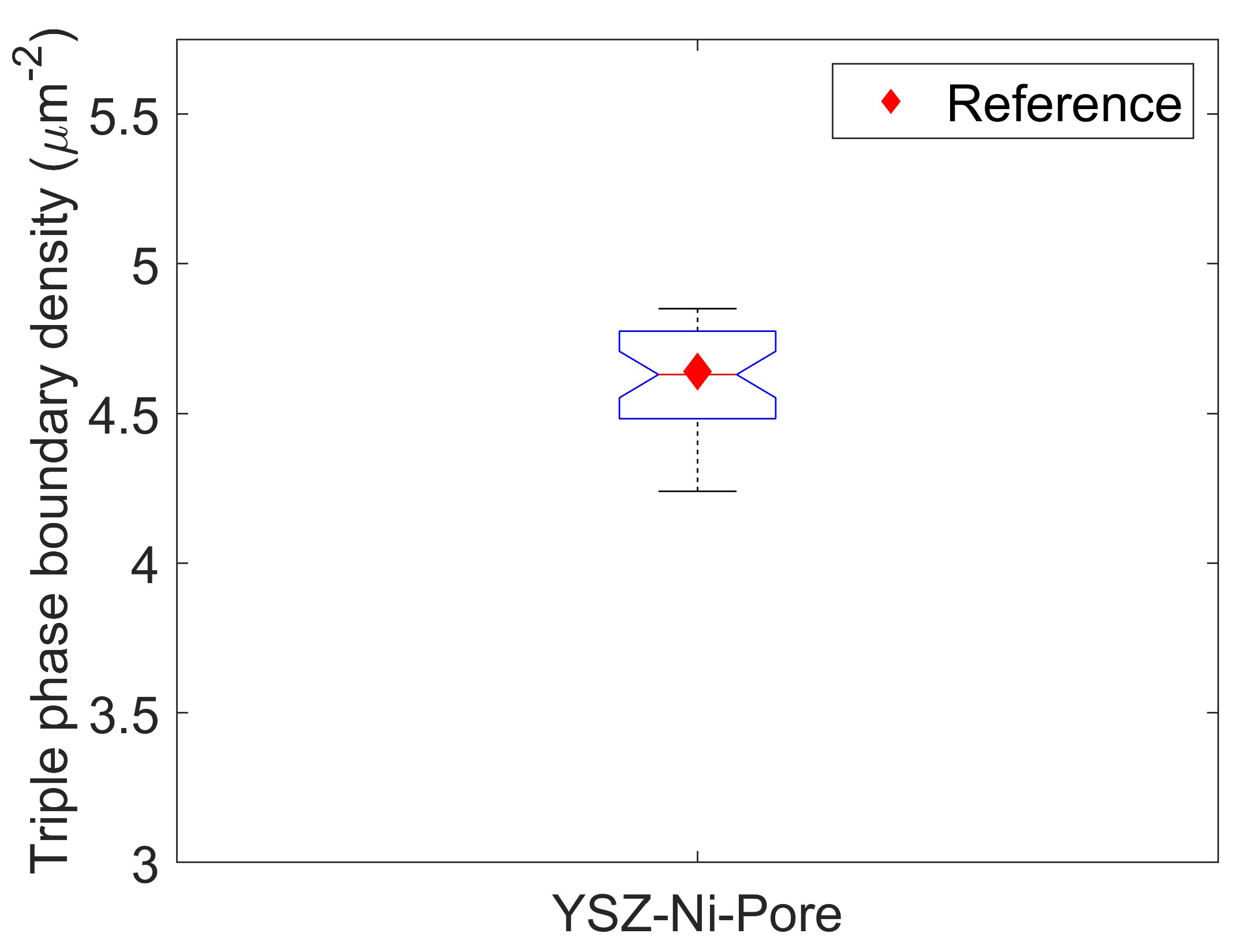}
		\text{(b) Triple phase boundary density}
	\end{minipage}  
    \smallskip
	\begin{minipage}[t]{0.333\textwidth}  
		\centering  
		\includegraphics[width=0.96\textwidth]{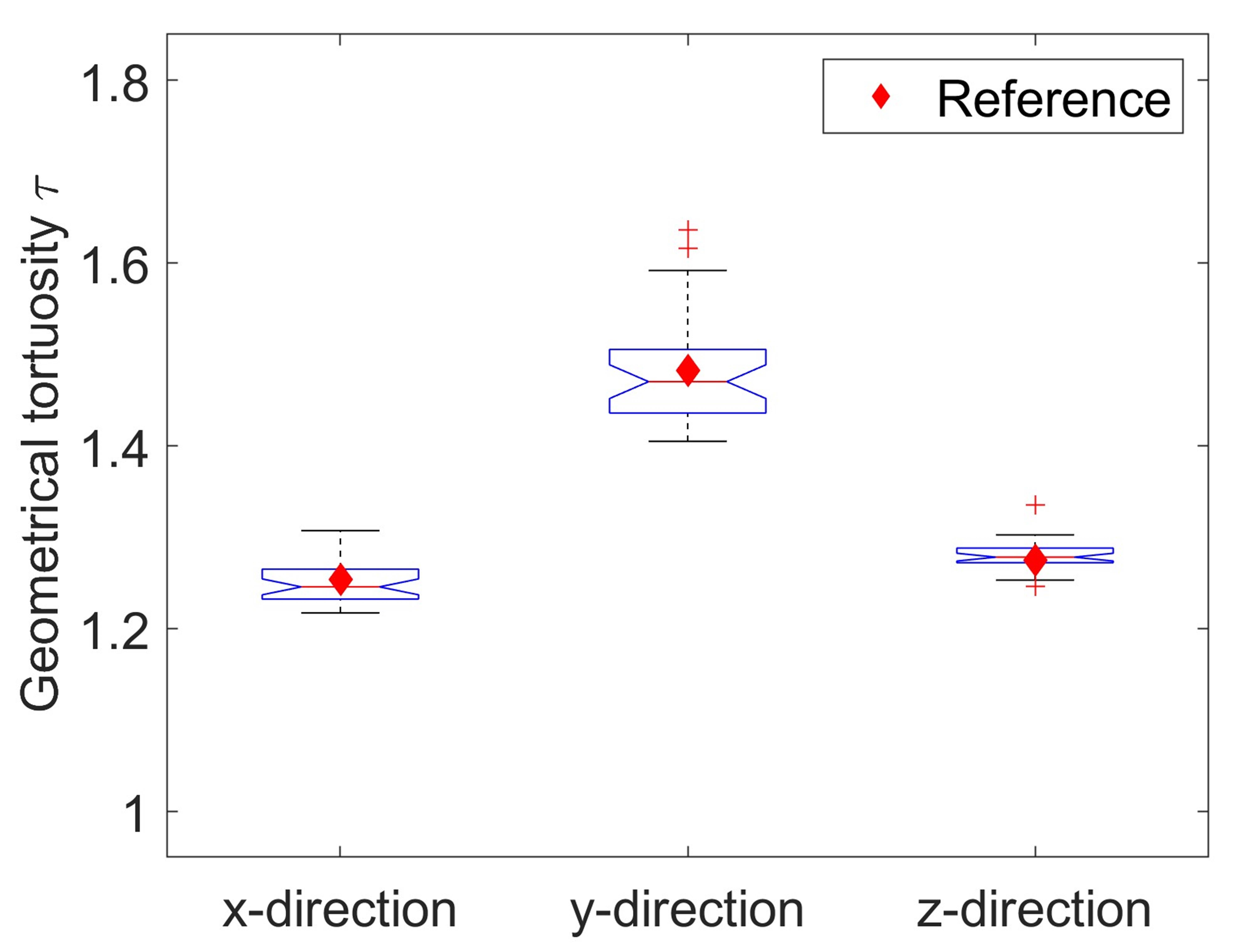}
		\text{(c) Geometrical tortuosity}
	\end{minipage}  
    \smallskip
	\begin{minipage}[t]{0.33\textwidth} 
		\centering  
		\includegraphics[width=0.98\textwidth]{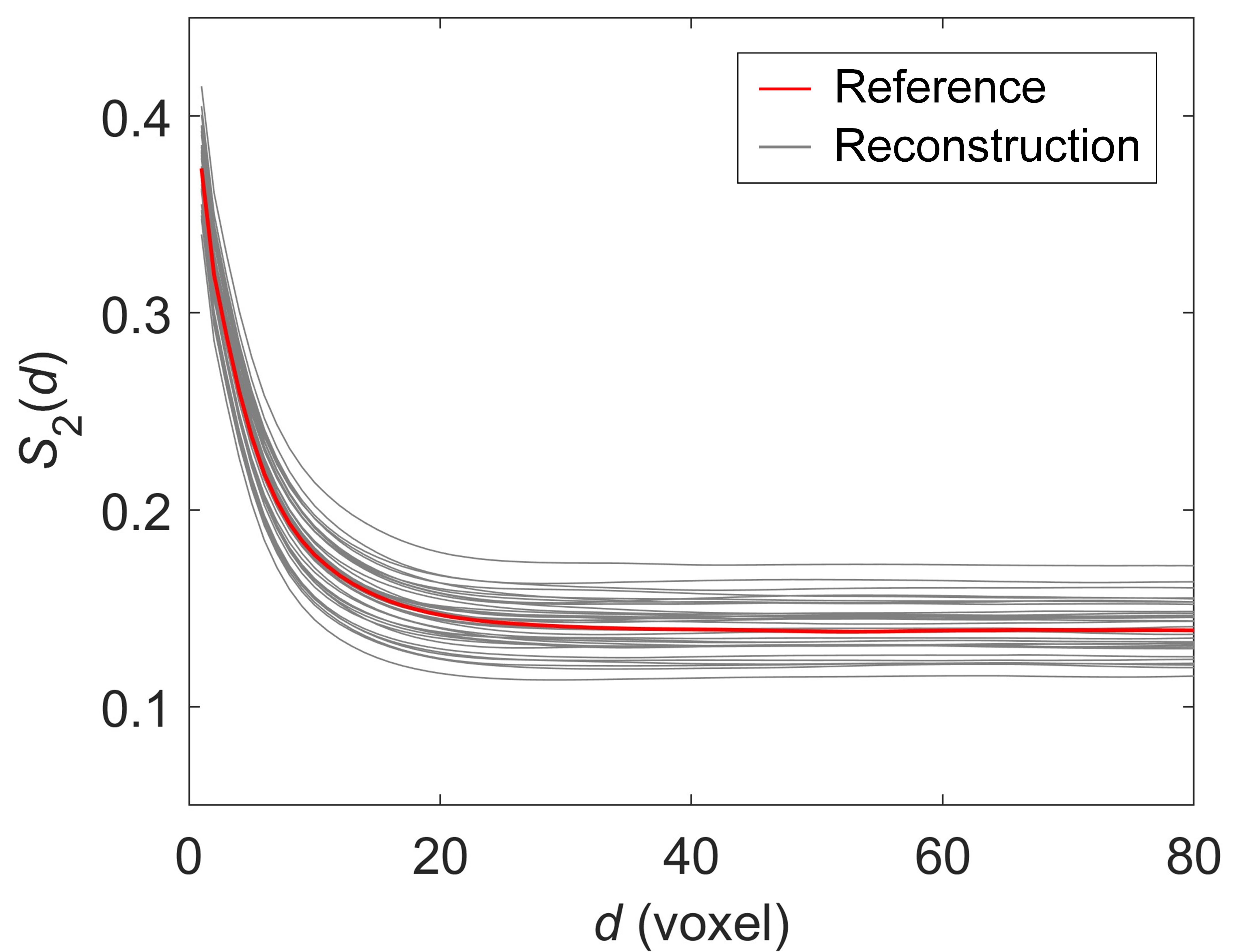}
		\text{(d) TPCF: YSZ phase}
	\end{minipage}  
	\begin{minipage}[t]{0.333\textwidth}  
		\centering  
		\includegraphics[width=0.98\textwidth]{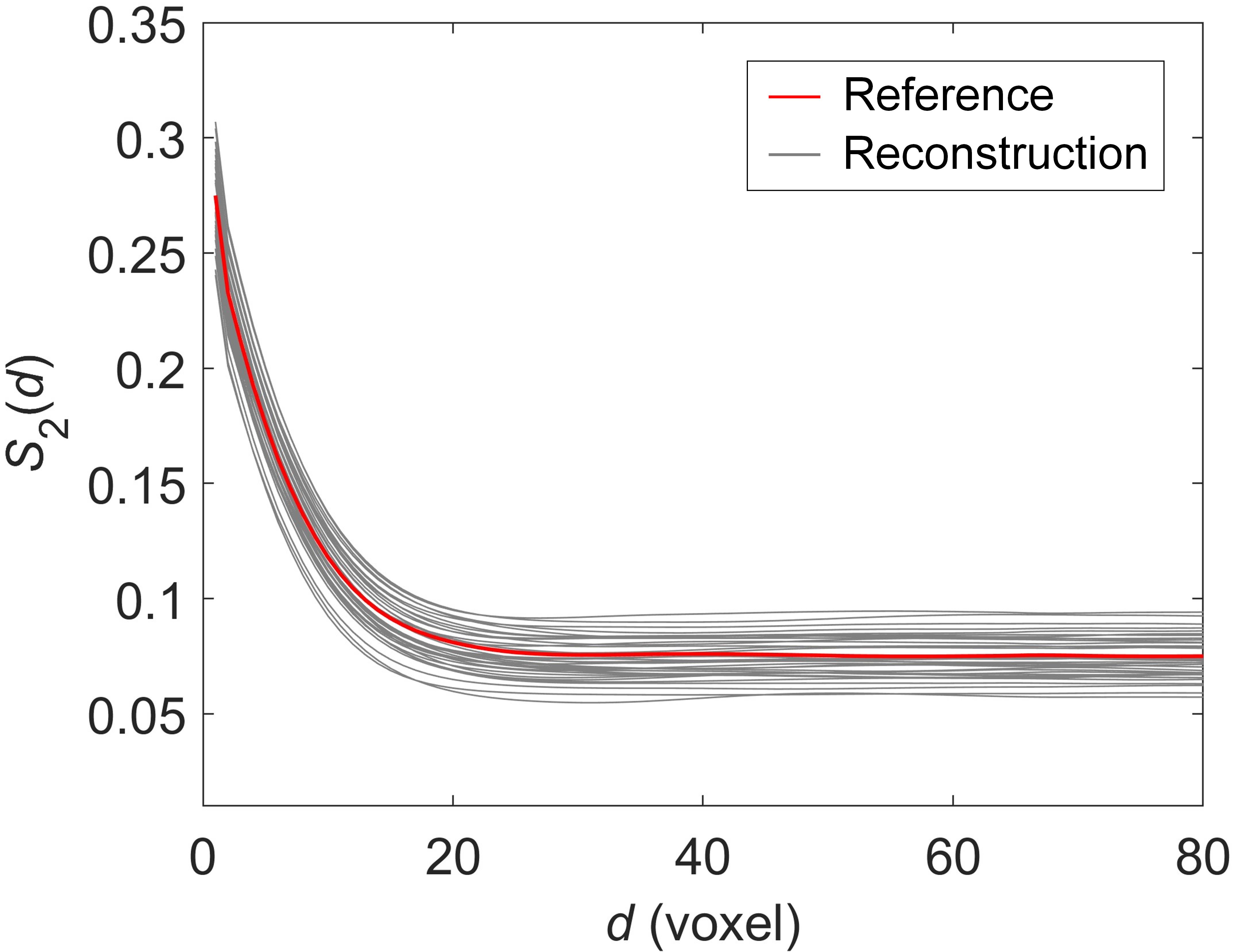} 
		\text{(e) TPCF: Ni phase}
	\end{minipage}  
	\begin{minipage}[t]{0.333\textwidth}  
		\centering  
		\includegraphics[width=0.98\textwidth]{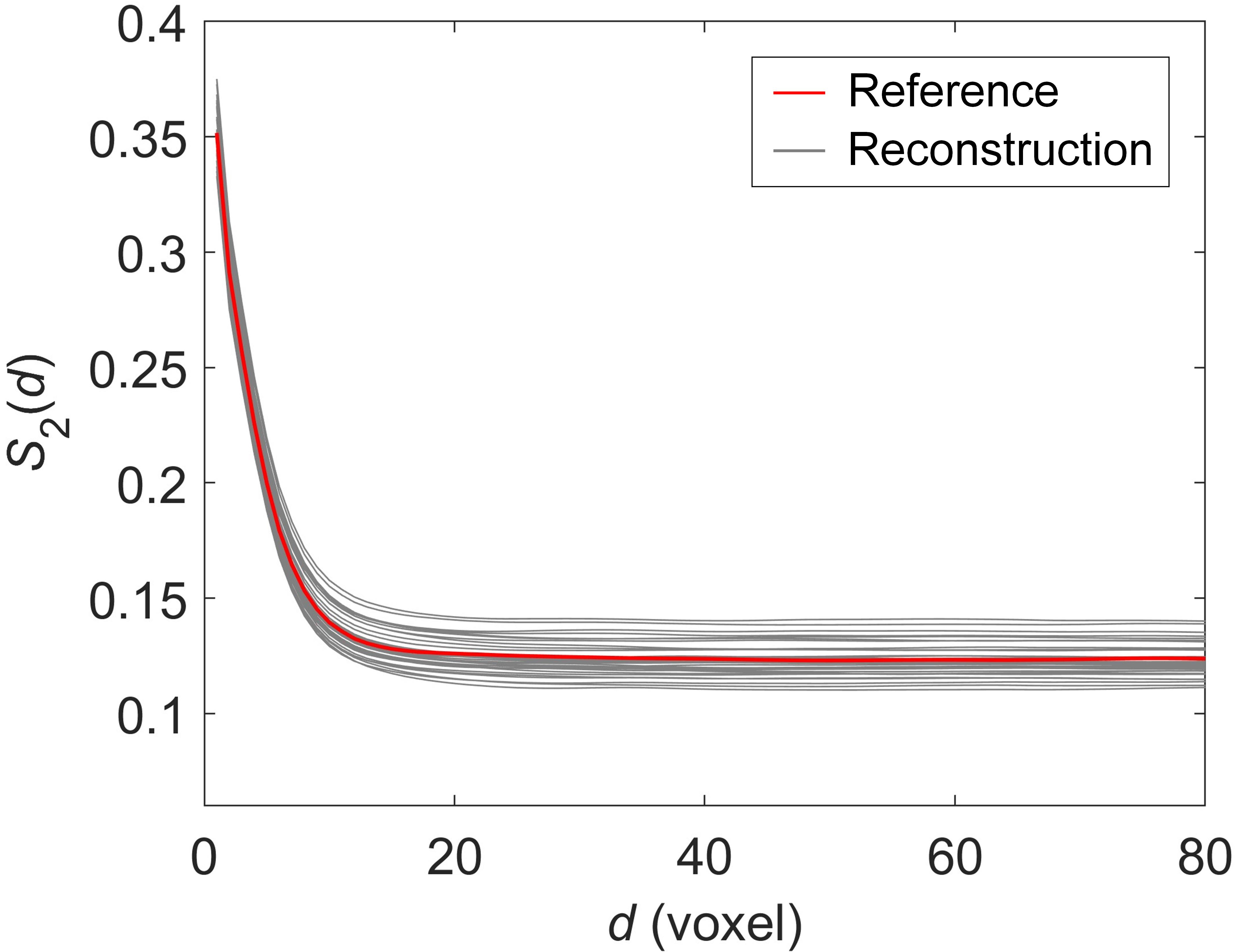}  
		\text{(f) TPCF: Pore space}
	\end{minipage}
	\caption{Quantitative comparison between the reference and reconstructed 3D microstructures by evaluating morphological descriptors.}
	\label{Fig:Ex2_descriptors}
\end{figure}

The four morphological descriptors are computed for all three constituent phases of the porous SOFC anode, with results shown in Figure \ref{Fig:Ex2_descriptors}. Overall, the descriptor values or curves extracted from the reconstructed 3D microstructure samples exhibit minor fluctuations around those derived from the reference 3D microstructure. This variation is expected due to the inherent randomness of the statistical reconstruction process. Despite these fluctuations, the mean values or curves of the descriptors from the reconstructed microstructures closely match their corresponding reference values or curves. This observed consistency, coupled with the minor variations, confirms the statistical equivalence between the reconstructed and reference 3D microstructures. Moreover, the excellent agreements in geometrical tortuosities along different directions (as shown in Figure \ref{Fig:Ex2_descriptors}c) highlight the effectiveness of the proposed multi-level reconstruction method in accurately capturing the anisotropy and long-range connectivity of multiphase porous media.

\subsection{Example 3: Composite cement paste}
\label{Subsection:Example3}
\begin{figure}[h]
	\centering
	\includegraphics[width=1.0\linewidth]{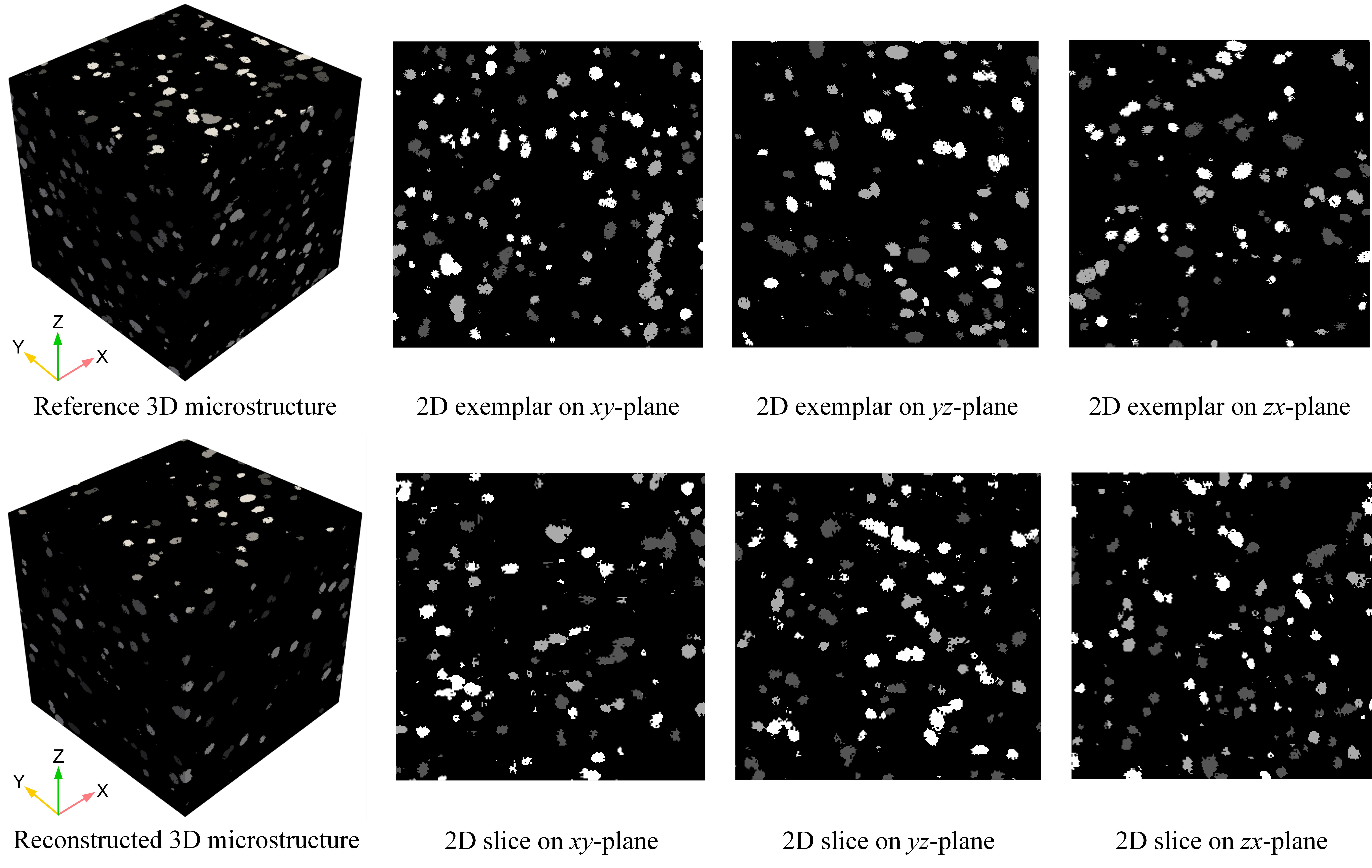}
	\caption{Visual comparison between the reference (top row) and the reconstructed (bottom row) 3D microstructure samples for a composite cement paste with four distinct constituent phases: the black region represents calcium silicate hydrate (C-S-H), the dark grey region represents anhydrous grains, the light grey region represents portlandite (calcium hydroxide) crystals, and the white region represents water-filled pores.}
	\label{Fig:Ex3_reconstruction}
\end{figure}

\vspace{-2pt}
Composite cement paste \cite{sanahuja2007modelling, ricklefs2017thermal}, a critical component in construction materials, consists of a cement matrix interspersed with various additives or inclusions, such as fly ash, slag, and silica fume. The heterogeneous microstructure of composite cement paste significantly influences its macroscopic properties, including strength, permeability, shrinkage resistance, and thermal conductivity. In this work, composite cement paste is used as a representative example to test the capability of the proposed SENN-based reconstruction approach. As shown in Figrue \ref{Fig:Ex3_reconstruction} (upper row), it consists of four constituent phases, including calcium silicate hydrate (C-S-H), anhydrous grains, portlandite crystals (calcium hydroxide) and water-filled pores. This four-phase microstructure provides an excellent platform for evaluating the method’s ability to capture complex morphologies and statistical characteristics.

As illustrated in the upper row of Figure \ref{Fig:Ex3_reconstruction}, representative 2D exemplars from three principal planes are selected for statistical characterisation and subsequent 3D microstructure reconstruction.
Two-level Gaussian image pyramids are constructed from the original 2D exemplars, serving as the basis for collecting training data to fit a set of SENN models. These SENN models accurately approximate 2D morphological statistics on different planes, enabling the inference of CPDFs across two length scales, represented as $P\left(X_{ijk}^{(\rm low)}\big|\textsl{\textbf{N}}_{ijk}^{(\rm low)}\right)$ and $P\left[X_{ijk}^{(\rm high)}\big|\Big(\textsl{\textbf{N}}_{ijk}^{(\rm high)}\cup X_{ijk}^{(\rm middle)}\cup \textsl{\textbf{N}}_{ijk}^{ (\rm middle) }\Big)\right]$, respectively. Further details on hierarchical characterisation are provided in Section \ref{Subsec:Multi-level_reconstruction}, while the parameter settings for SENN model training are summarised in Table \ref{Tab:SENN_parameters}.

Stochastic reconstruction is performed using the 3D CPDFs derived from the pretrained SENN models, producing a group of thirty 3D microstructure samples. A representative reconstruction result is shown in the lower row of Figure \ref{Fig:Ex3_reconstruction}. Visual inspection reveals that the reconstructed 3D microstructure closely resembles both the 2D exemplars and the reference 3D microstructure, demonstrating the effectiveness of the proposed approach in accurately reproducing morphological patterns. To quantitatively assess the reconstruction quality, a set of morphological descriptors is extracted from the reconstructed and reference 3D microstructures. These descriptors include volume fraction, surface area density, mean chord length, and the two-point cluster correlation function (TPCCF) \cite{fu2023data,cui2021correlation}. These metrics are crucial for evaluating the macroscopic behaviour of composite cement paste, such as mechanical strength \cite{sanahuja2007modelling} and thermal conductivity \cite{ricklefs2017thermal}.

\begin{figure}[h]\footnotesize
	\begin{minipage}[t]{0.333\textwidth}
		\centering  
		\includegraphics[width=0.98\textwidth]{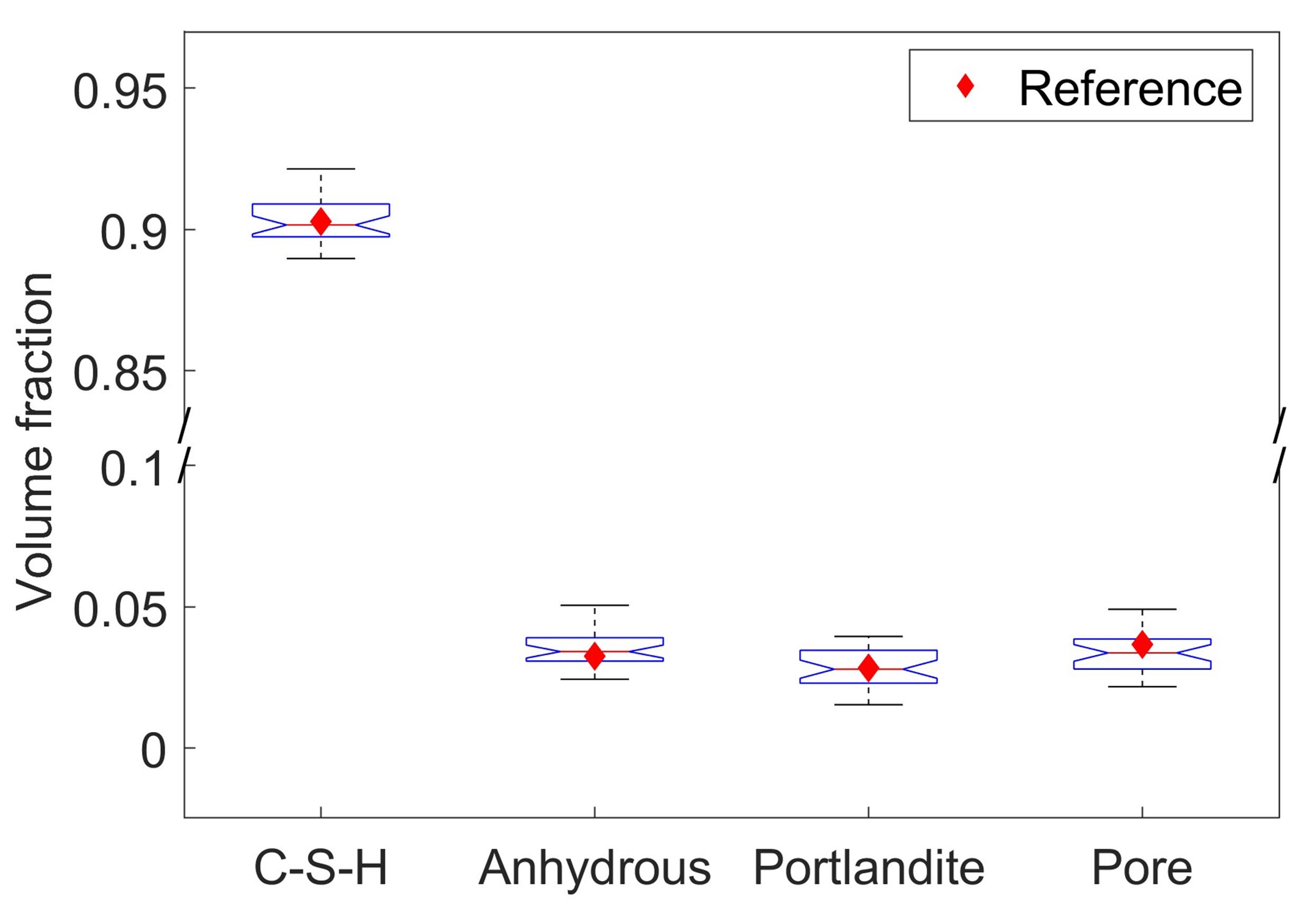}
		\text{(a) Volume fraction}
	\end{minipage}  
    \smallskip
	\begin{minipage}[t]{0.333\textwidth}  
		\centering  
		\includegraphics[width=0.98\textwidth]{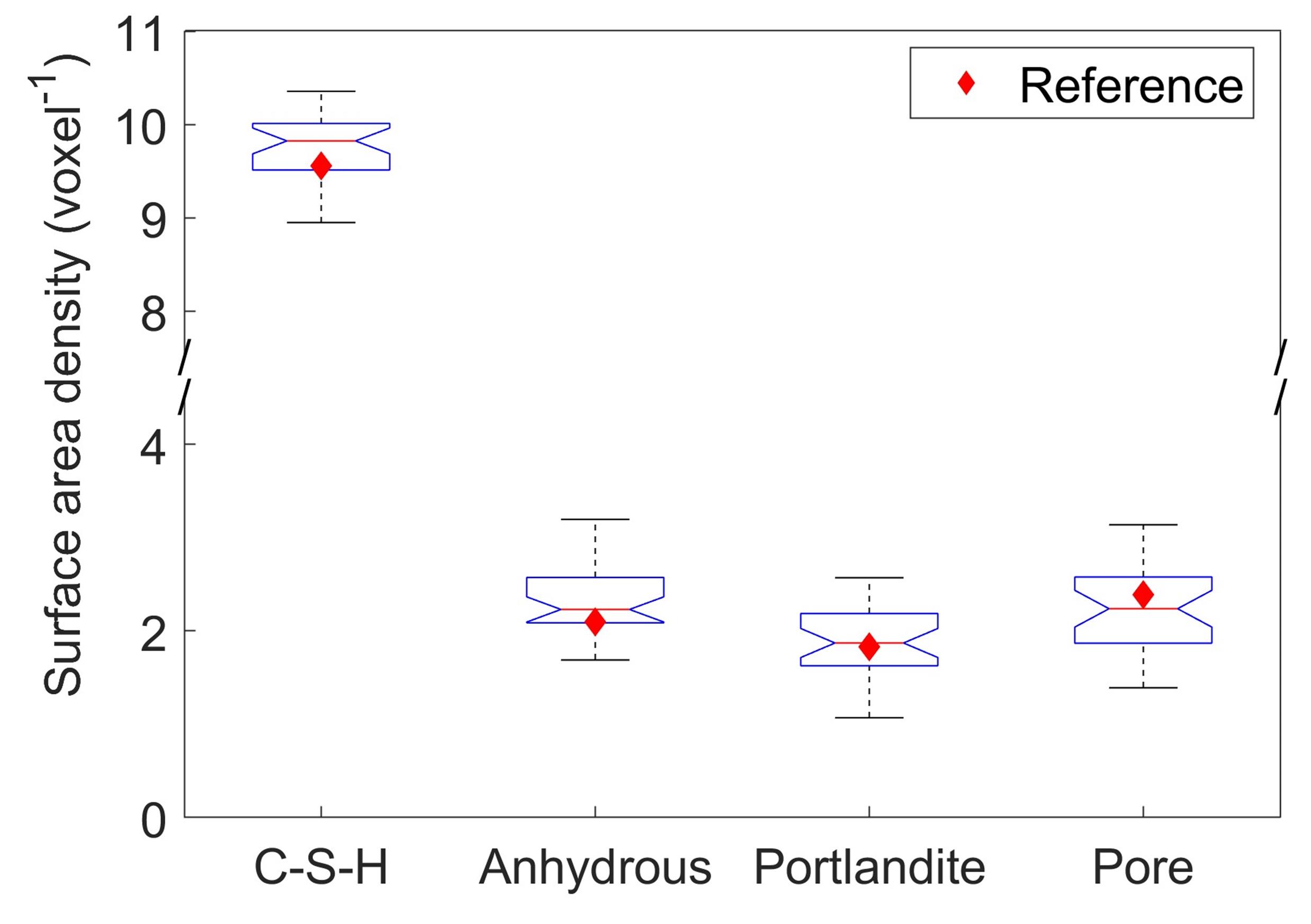}
		\text{(b) Surface area density}
	\end{minipage}  
    \smallskip
	\begin{minipage}[t]{0.333\textwidth}  
		\centering  
		\includegraphics[width=0.98\textwidth]{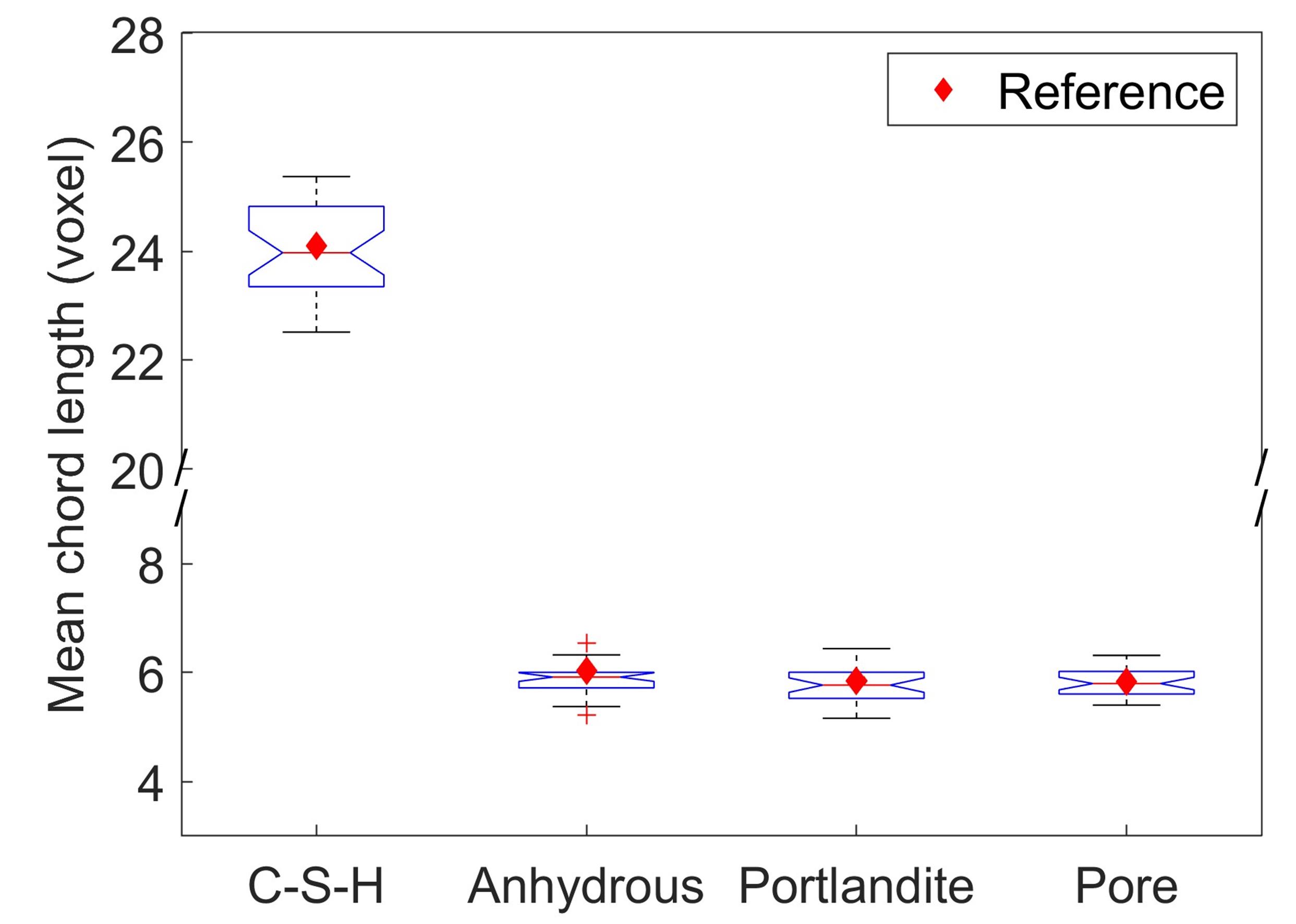}
		\text{(c) Mean chord length}
	\end{minipage}  
    \smallskip
	\begin{minipage}[t]{0.33\textwidth} 
		\centering  
		\includegraphics[width=0.98\textwidth]{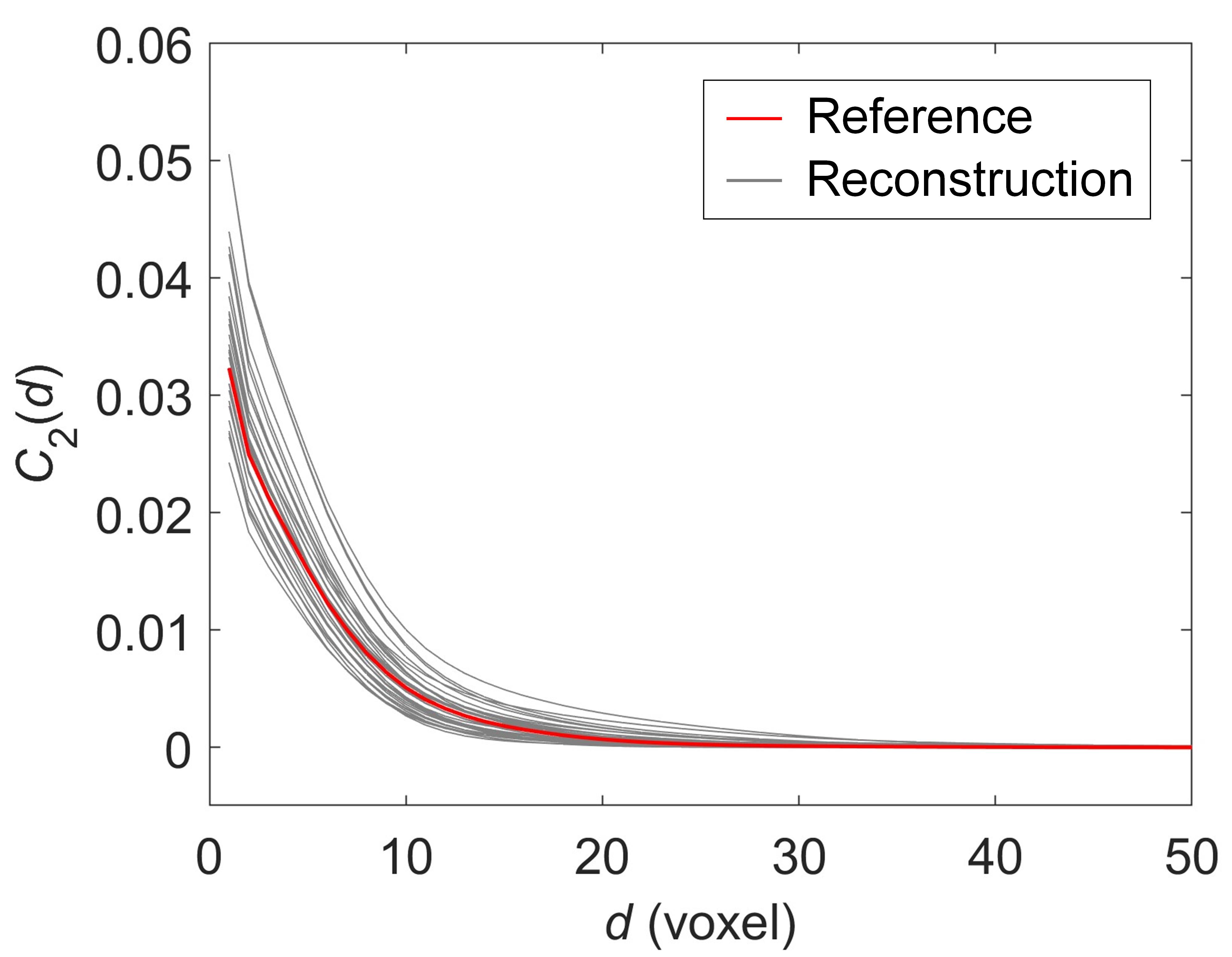}
		\text{(d) TPCCF: Anhydrous phase}
	\end{minipage}  
	\begin{minipage}[t]{0.333\textwidth}  
		\centering  
		\includegraphics[width=0.98\textwidth]{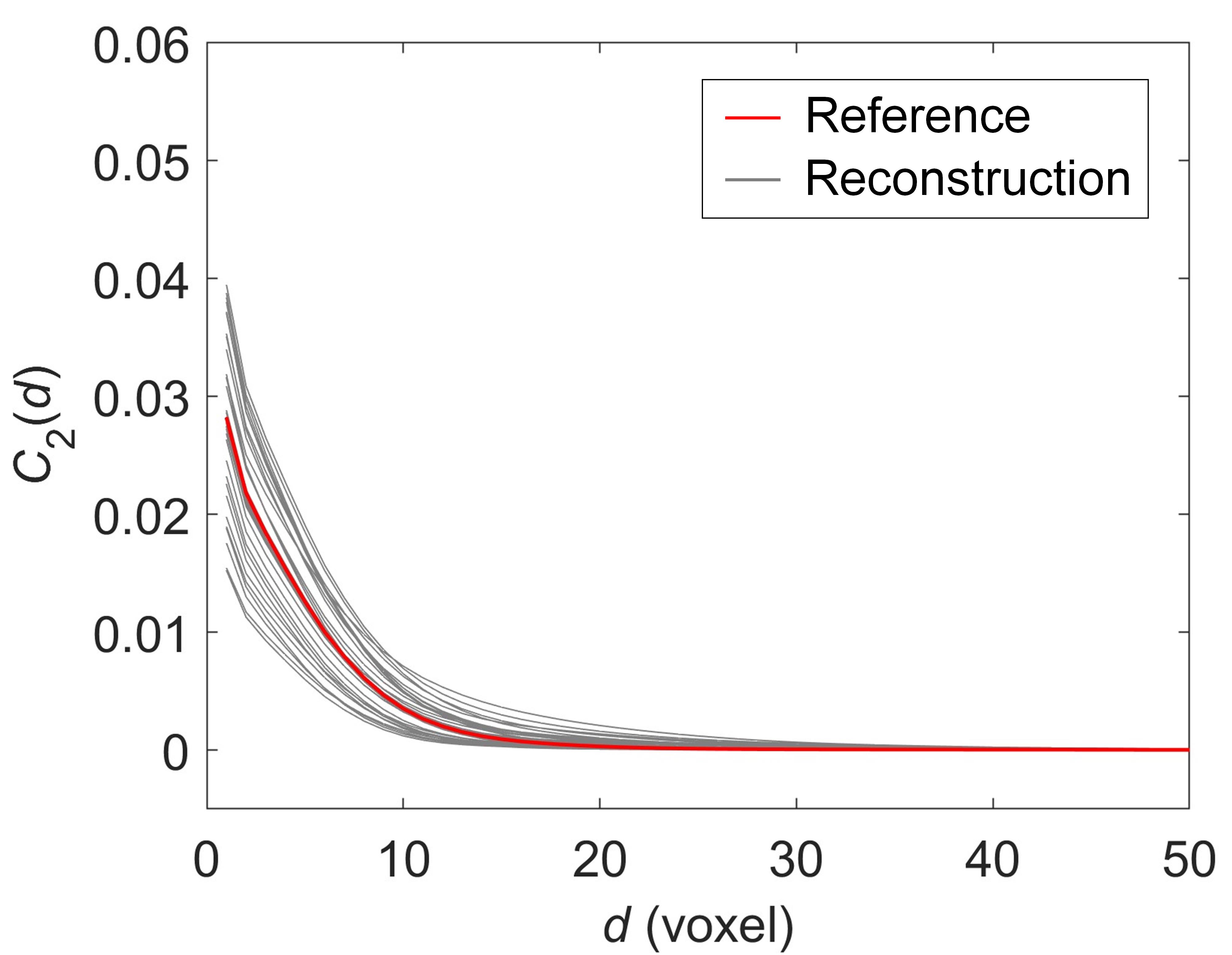} 
		\text{(e) TPCCF: Portlandite phase}
	\end{minipage}  
	\begin{minipage}[t]{0.333\textwidth}  
		\centering  
		\includegraphics[width=0.98\textwidth]{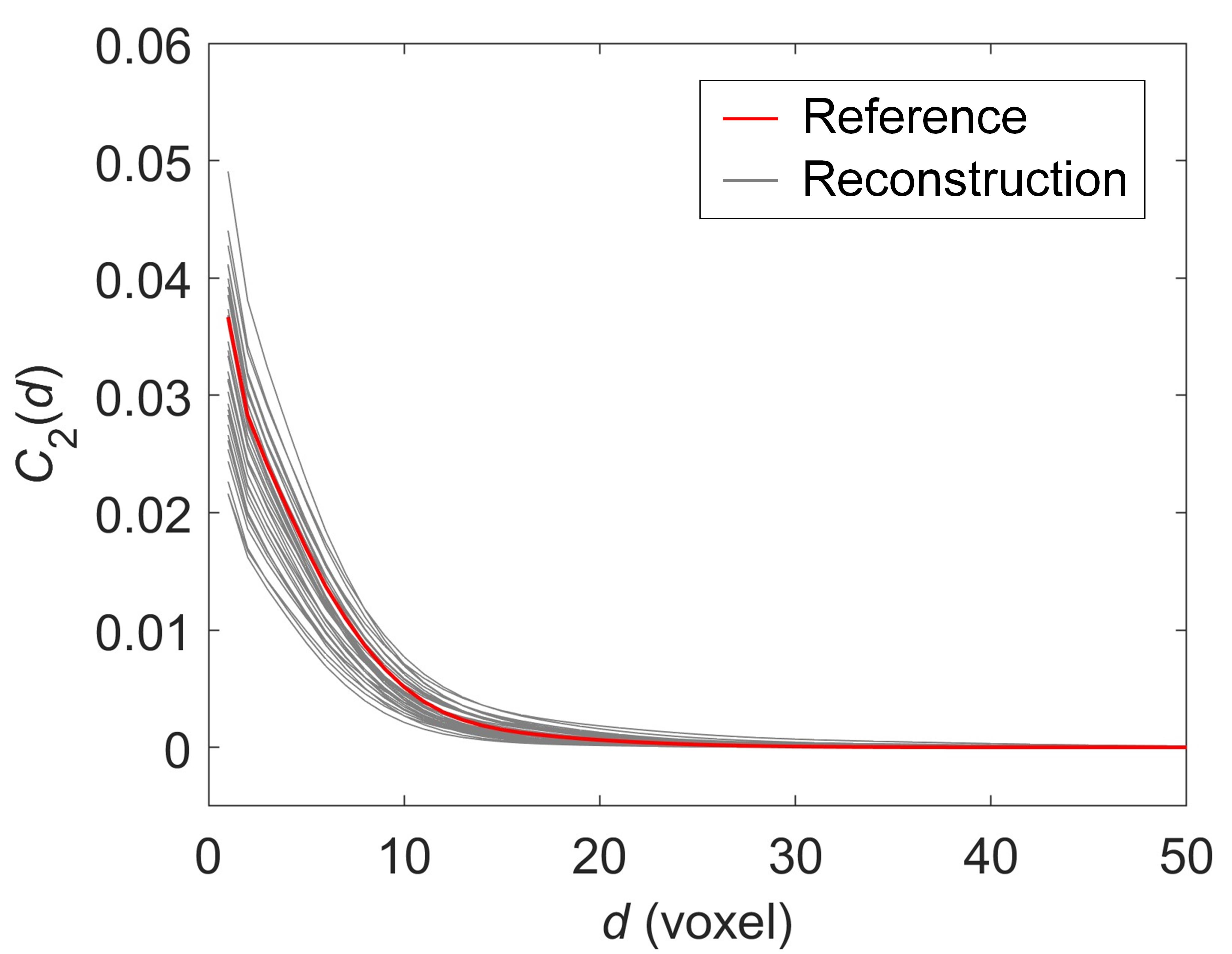}  
		\text{(f) TPCCF: Water-filled pore}
	\end{minipage}
	\caption{Quantitative comparison between the reference and reconstructed 3D microstructures by evaluating morphological descriptors.}
	\label{Fig:Ex3_descriptors}
\end{figure}

The first three descriptors are evaluated for all four constituent phases of the composite cement paste, while the TPCCF is specifically calculated for the three secondary phases to measure cluster sizes. The descriptor values or curves derived from the thirty reconstructed 3D microstructures show slight variations compared to those obtained from the reference 3D microstructure, primarily due to the stochastic nature of voxel generation during the reconstruction process. Nonetheless, the mean values or curves of the reconstructed descriptors closely align with their reference counterparts. This close agreement, alongside the minor variations, validates the statistical equivalence between the reconstructed and reference 3D microstructures. Additionally, the accurate reproduction of mean chord length and TPCCF confirms the capability of the proposed multi-level reconstruction method to effectively capture and preserve the grain and pore size characteristics of multiphase porous media.

\section{Image-based poro/micro-mechanical modelling} 
\label{Section5:Micro-mechanical_modeling}
\vspace{-2pt}
High-resolution 3D digital microstructures of multiphase composites are essential for image-based poro/micro-mechanical modelling, enabling advanced investigations such as materials design and the exploration of MPRs \cite{terada1997digital, fang2016smoothing, evans2023review}. These models facilitate accurate simulations of various physical phenomena, but generating precise meshes from 3D image data poses significant challenges due to the intricate geometries and complex topologies of multiphase composites \cite{ulrich1998finite, perre2005meshpore}.
To overcome these challenges, this study employs advanced image-based meshing techniques \cite{perre2005meshpore} to generate high-quality volumetric meshes from digital microstructures. These meshes serve as a foundation for numerical simulations using homogenization methods, enabling the computation of effective material properties. The evaluated physical properties include stiffness, permeability, effective diffusivity, and effective thermal conductivity tensors. These quantitative metrics provide a robust means of assessing the physical equivalence between the reference and reconstructed 3D microstructures.
Such evaluations play a critical role in verifying the effectiveness and practicality of the proposed SENN-based reconstruction method. Additionally, the physical properties of the constituent phases in the studied multiphase composites are summarised in Table \ref{Tab:Materials_properties}, offering essential inputs for numerical simulations and further analyses.

\begin{table}[h]
\fontsize{8}{10}\selectfont
\caption{The physical properties of constituent phases in multiphase composite materials}
\centering
\label{Tab:Materials_properties}
\begin{tabular}{lllcc}
\toprule
\textbf{Multiphase composite} & \textbf{Constituent phase} & \textbf{Elastic modulus $\bm E$} & \textbf{Possion's ratio $\bm\nu$} & \textbf{Thermal conductivity $\bm\chi$}\\
\midrule
 \multirow{3}{*}{Silver-based electrode} & Silver & 76.00 GPa & 0.380 & -- \\
                                         & PTFE & 0.50 GPa & 0.460 & -- \\
                                         & Pore & 0.00      & 0.000 & -- \\
\midrule
 \multirow{3}{*}{Porous SOFC anode} & YSE & 215.00 GPa & 0.317 & -- \\
                                    & Ni  & 200.00 GPa & 0.300 & -- \\
                                    & Pore& 0.00      & 0.000 & -- \\
\midrule
\multirow{4}{*}{Composite cement past} & Calcium silicate hydrate  & 25.0 GPa   & 0.200 & 0.40 W/m$\cdot$K\\           
                                         & Anhydrous grain           & 138.0 GPa  & 0.300 & 1.40 W/m$\cdot$K \\
                                         & Portlandite crystal       & 35.4 GPa   & 0.300 & 1.15 W/m$\cdot$K \\
                                         & Water-filled pore         & 0.0001 GPa & 0.499 & 0.61 W/m$\cdot$K \\
\bottomrule 
\end{tabular}
\end{table}

\subsection{Example 1: Mechanical and transport properties}
\begin{figure}[H]\footnotesize
	\begin{minipage}[t]{0.36\textwidth}
		\centering  
		\includegraphics[width=1.0\textwidth]{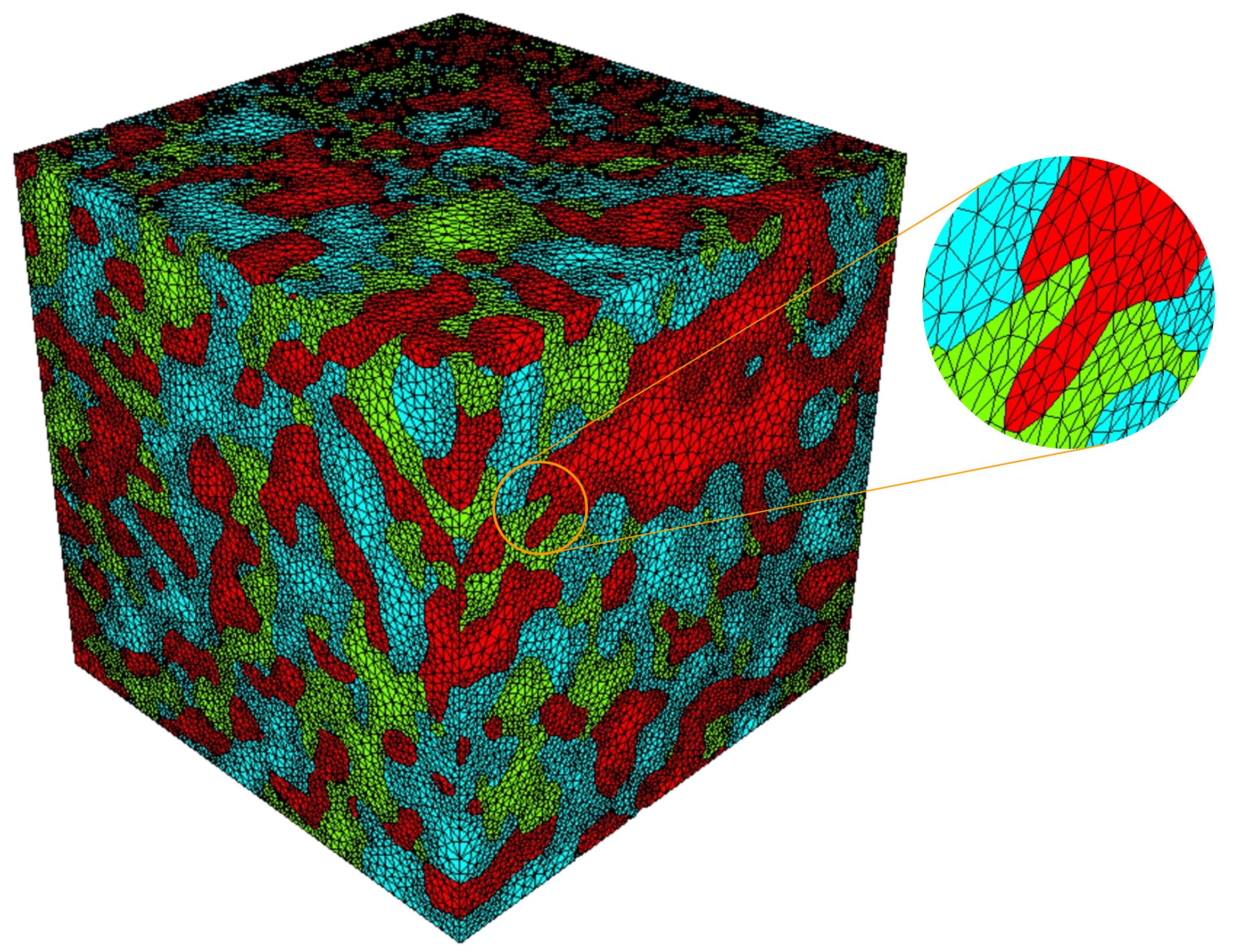}
		\text{(a) Image-based meshing}
	\end{minipage}  
    \smallskip
	\begin{minipage}[t]{0.32\textwidth}  
		\centering  
		\includegraphics[width=0.95\textwidth]{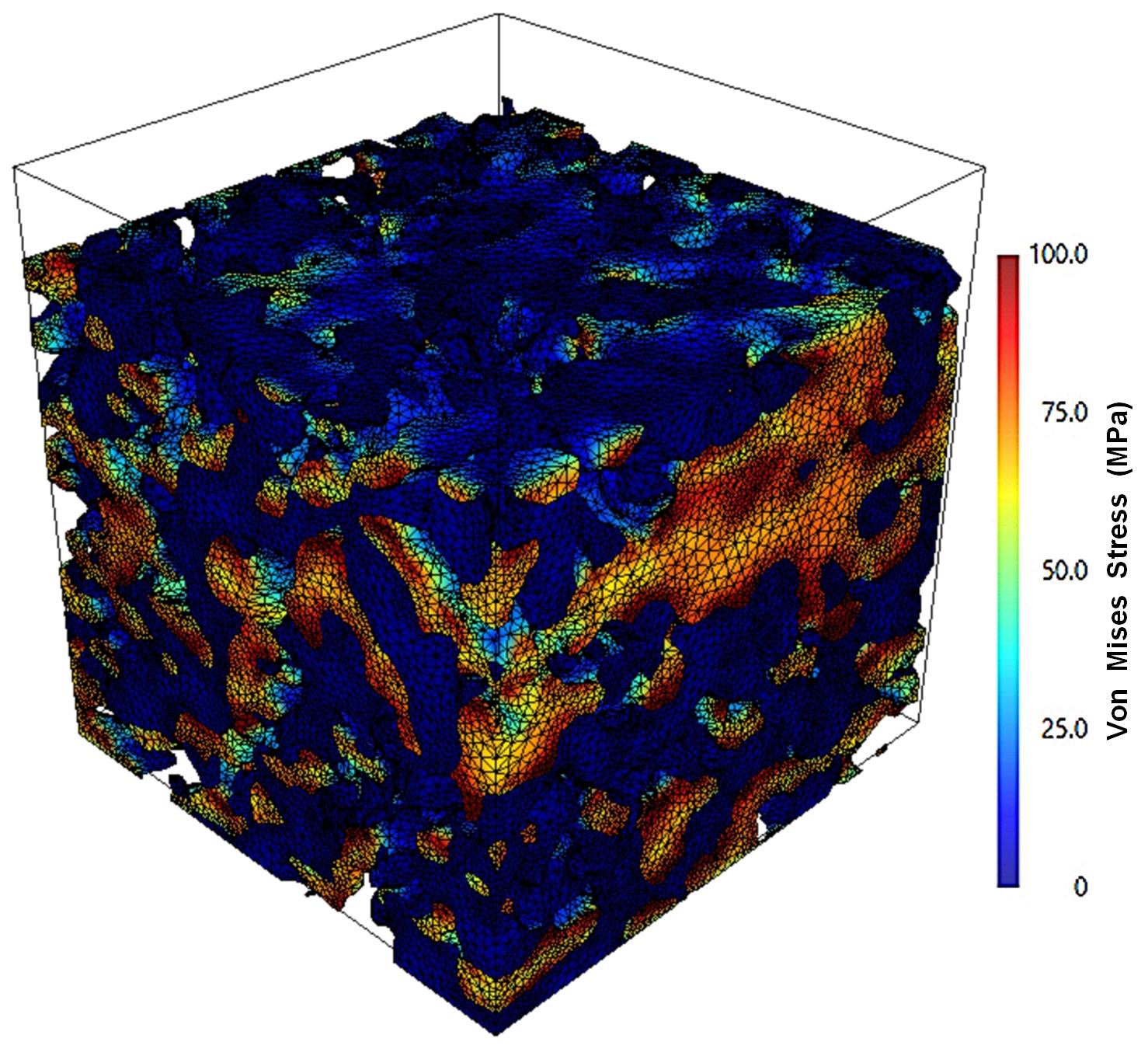}
		\text{(b) Stress field in solid phases}
	\end{minipage}  
    \smallskip
	\begin{minipage}[t]{0.32\textwidth}  
		\centering  
		\includegraphics[width=0.95\textwidth]{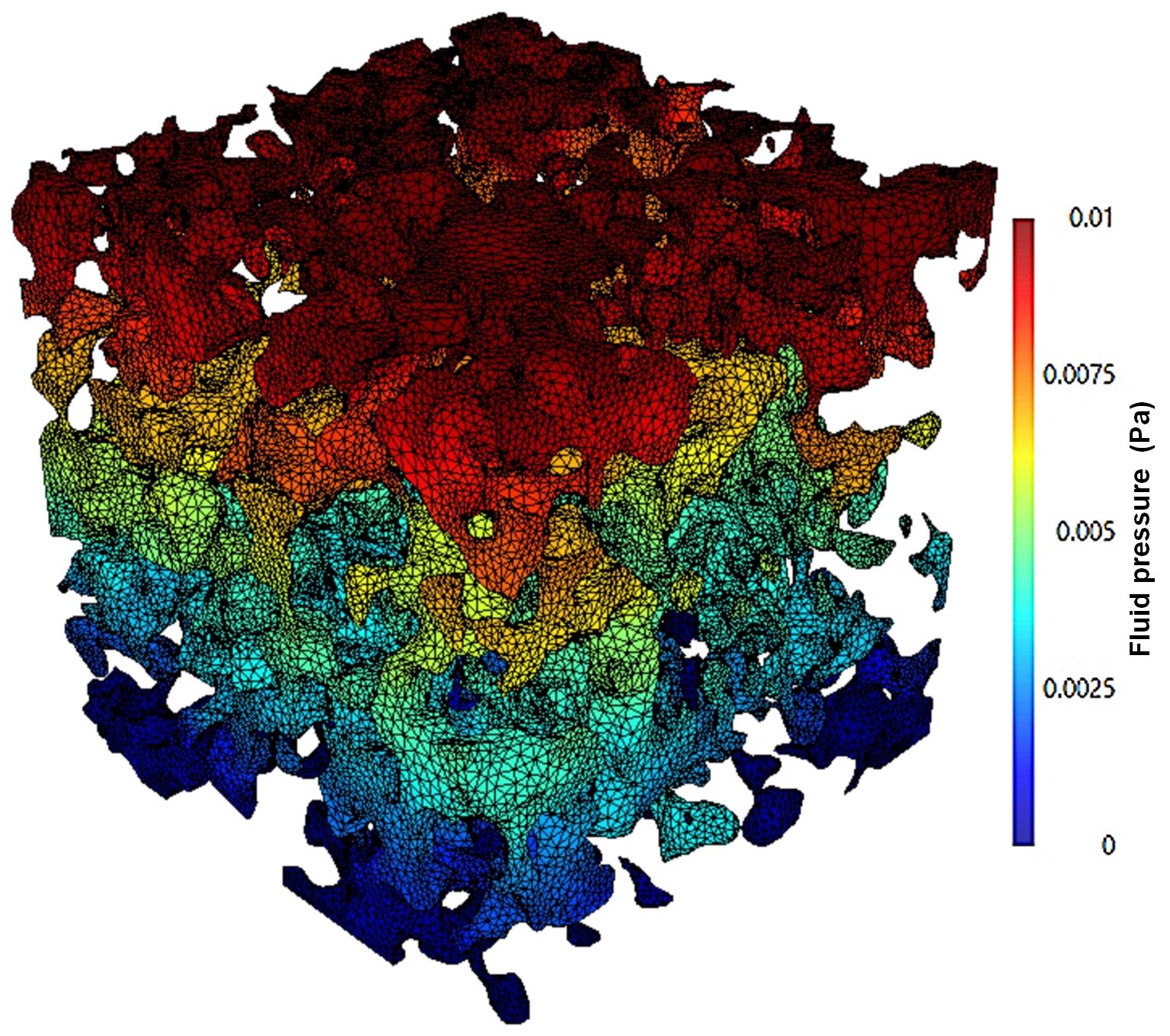}
		\text{(c) Pressure field in pore space}
	\end{minipage}  
    \smallskip
	\begin{minipage}[t]{0.36\textwidth} 
		\centering  
		\includegraphics[width=1.0\textwidth]{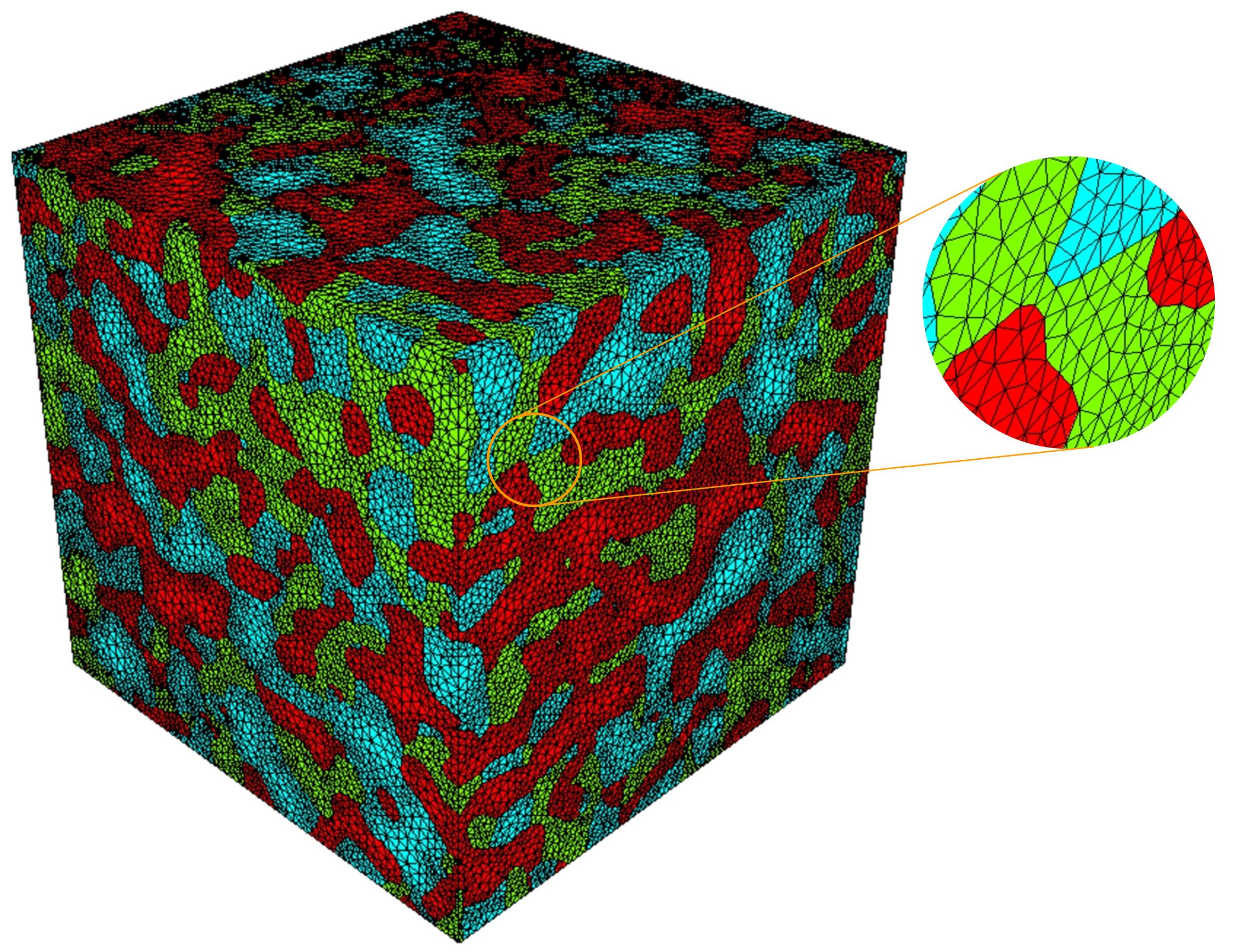}
		\text{(d) Image-based meshing}
	\end{minipage}  
	\begin{minipage}[t]{0.32\textwidth}  
		\centering  
		\includegraphics[width=0.95\textwidth]{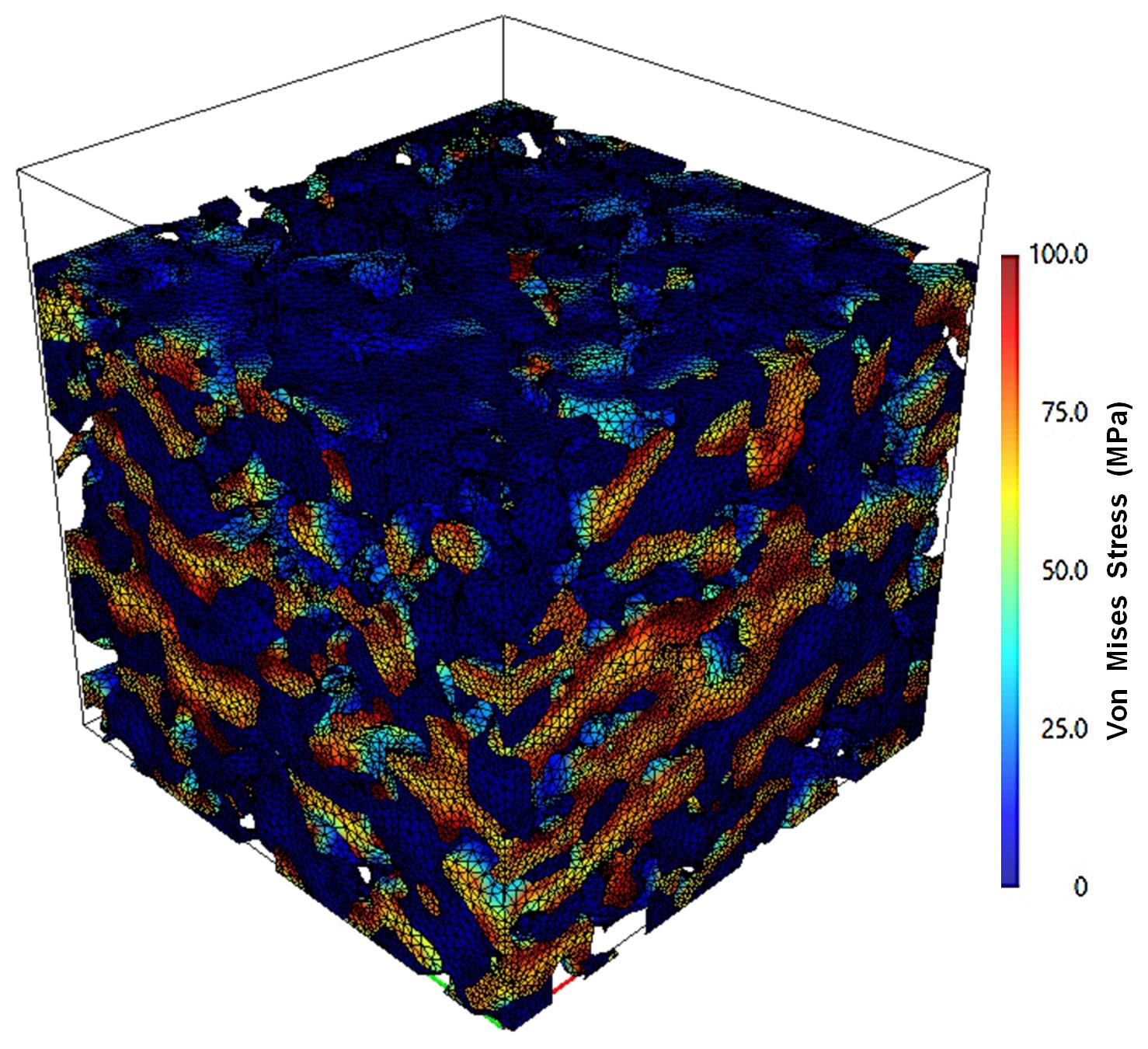} 
		\text{(e) Stress field in solid phases}
	\end{minipage}  
	\begin{minipage}[t]{0.32\textwidth}  
		\centering  
		\includegraphics[width=0.95\textwidth]{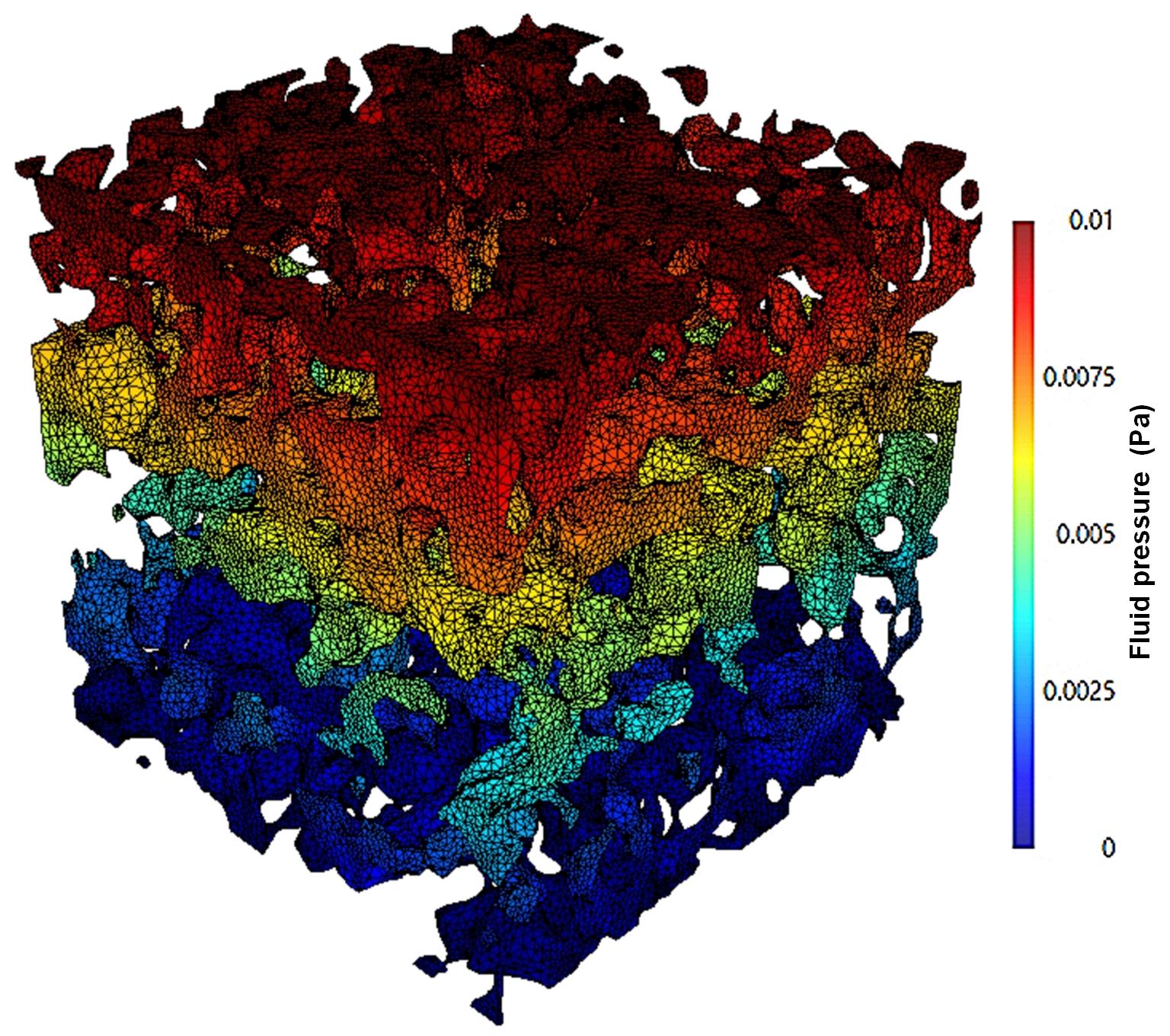}  
		\text{(f) Pressure field in pore space}
	\end{minipage}
	\caption{Image-based poro/micro-mechanical modelling on the reference (top raw) and the reconstructed (bottom raw) 3D microstructures of the porous silver-based electrode: (a) and (d) Mesh generation on multiphase microstructures: red areas denote silver, cyan areas denote PTFE, and green areas represents pore space; (b) and (e) Von Mises stress fields in the solid phases under axial compression; (c) and (f) Pressure fields of viscous fluid flow within pore space.}
	\label{Fig:Ex1_numerical_simulation}
\end{figure}

\vspace{-2pt}
\label{Subsection5.1}
The performance and durability of porous silver-based electrodes in electrochemical applications rely heavily on their mechanical and transport properties, such as stiffness and permeability. These properties are shaped by the internal microstructure, phase distribution, porosity, and interactions among the silver matrix, PTFE, and pore spaces. Gaining a detailed understanding of these factors is essential for optimising devices like SOFCs and batteries. To analyse these characteristics, image-based meshing is conducted on both the reference and reconstructed 3D microstructures for subsequent poro/micro-mechanical modelling. As shown in Figures \ref{Fig:Ex1_numerical_simulation}a and d, unstructured tetrahedral meshes are generated to accurately represent the intricate interfaces and shared nodes among the multiple phases in this heterogeneous material.

To compute stiffness tensors, the finite element method (FEM) is conducted on image-based meshes of multiphase electrode microstructures to simulate axial and shear deformations. Figures \ref{Fig:Ex1_numerical_simulation}b and e depict stress fields in the reference and reconstructed 3D microstructures under axial compression, providing insights into the mechanical behaviour of the electrode and validating the fidelity of the reconstruction process. Using Voigt notation, the stiffness tensor \cite{zheng2021data} connects stress to strain through the generalized Hooke’s law:
\begin{equation}
\begin{bmatrix}
\sigma_{11} \\
\sigma_{22} \\
\sigma_{33} \\
\sigma_{12} \\
\sigma_{23} \\
\sigma_{31}
\end{bmatrix}
=
\begin{bmatrix}
e_{11} & e_{12} & e_{13} & 0      & 0      & 0 \\
e_{21} & e_{22} & e_{23} & 0      & 0      & 0 \\
e_{31} & e_{32} & e_{33} & 0      & 0      & 0 \\
0      &   0    &    0   & e_{44} & 0      & 0 \\
0      &   0    &    0   &    0    & e_{55} & 0 \\
0      &   0    &    0   &    0    &   0     & e_{66}
\end{bmatrix}
\begin{bmatrix}
\varepsilon_{11} \\
\varepsilon_{22} \\
\varepsilon_{33} \\
\varepsilon_{12} \\
\varepsilon_{23} \\
\varepsilon_{31}
\end{bmatrix}
\label{Eq:Hooke's_law}
\end{equation}
where $\sigma_{ij}$ denotes stress components, $\varepsilon_{ij}$ represents strain components, and $e_{ij}$ corresponds to stiffness components. 

\begin{figure}[H]\footnotesize
	\begin{minipage}[t]{0.5\textwidth}
		\centering  
		\includegraphics[width=0.8\textwidth]{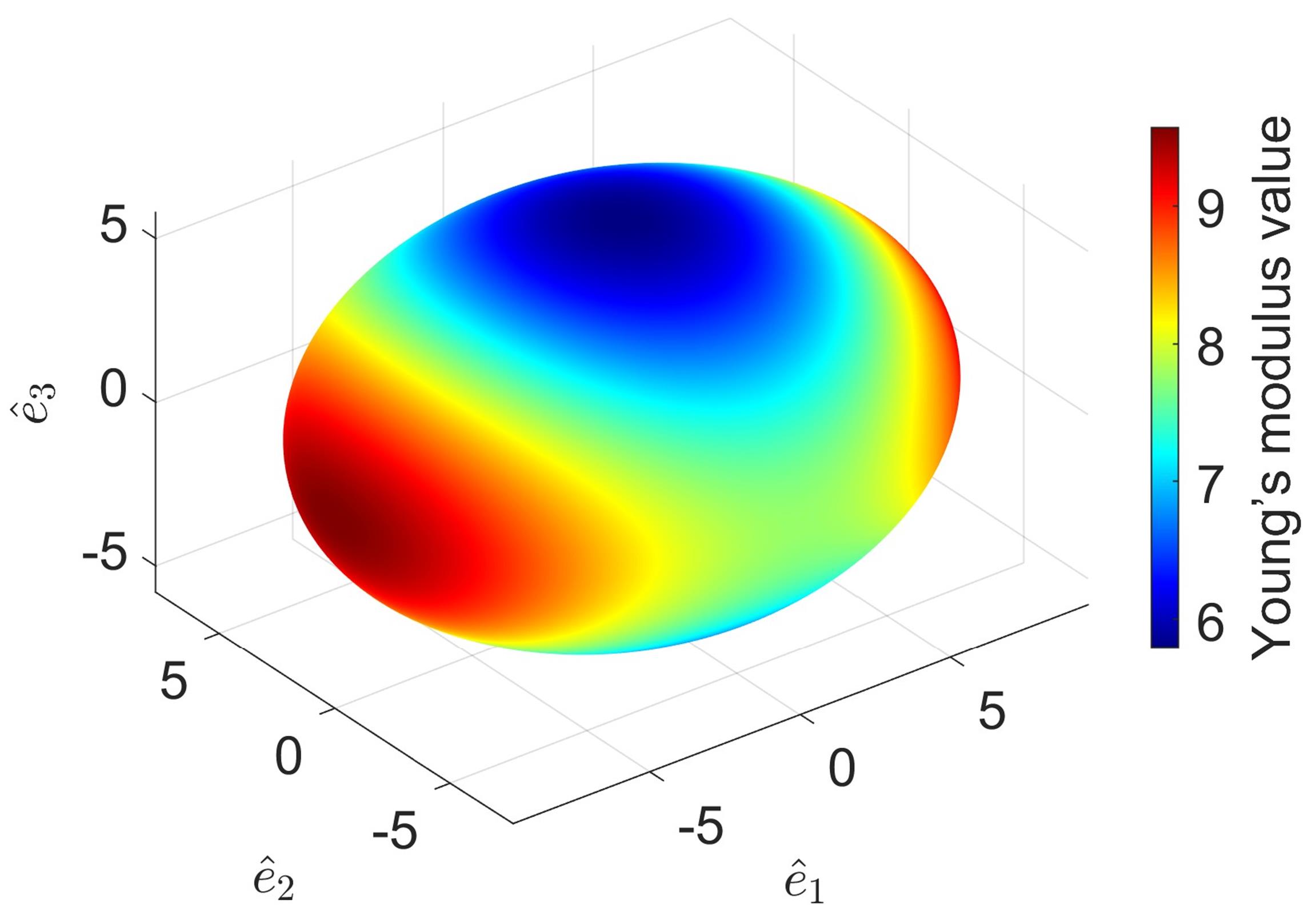}
		\text{(a) Volumetric representation of stiffness tensor (reference)}
	\end{minipage}  
    \smallskip\smallskip 
	\begin{minipage}[t]{0.5\textwidth}  
		\centering  
		\includegraphics[width=0.8\textwidth]{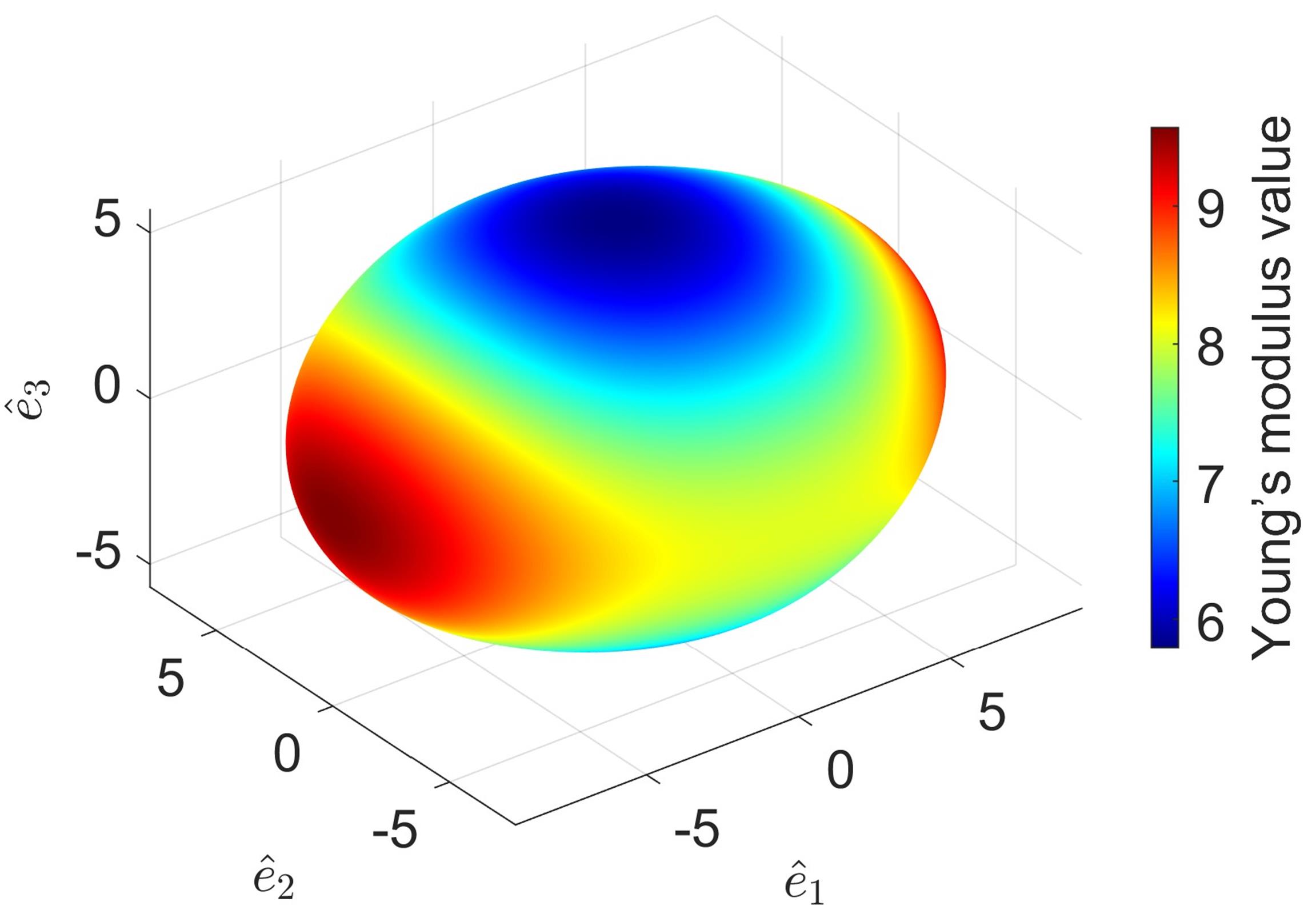}
		\text{(b) Volumetric representation of stiffness tensor (reconstruction)}
	\end{minipage}  
    \smallskip\smallskip  
    \begin{minipage}[t]{0.33\textwidth} 
		\centering  
		\includegraphics[width=0.98\textwidth]{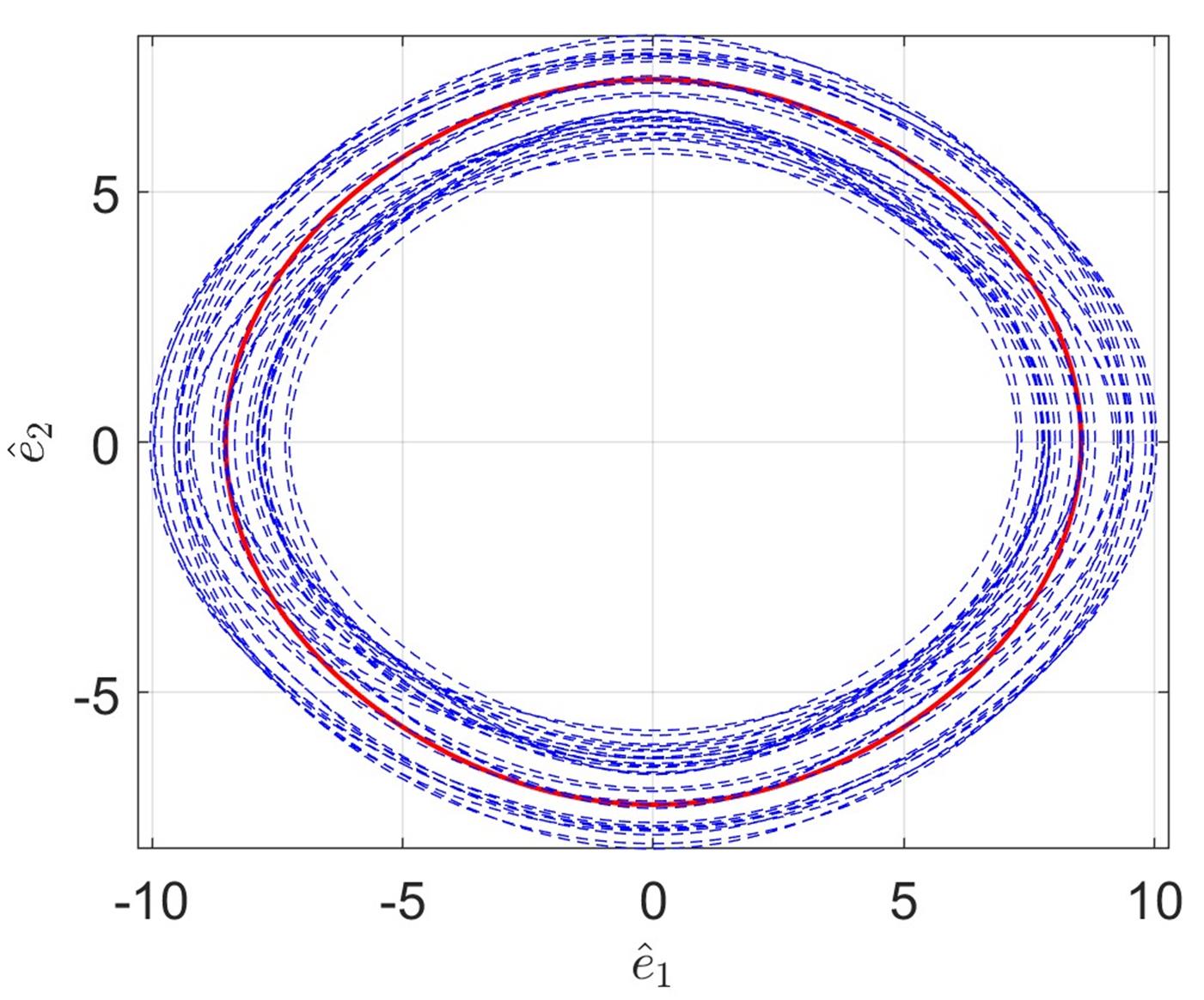}
		\text{(c) Stiffness tensor on $xy$-plane}
	\end{minipage}  
	\begin{minipage}[t]{0.33\textwidth}  
		\centering  
		\includegraphics[width=0.98\textwidth]{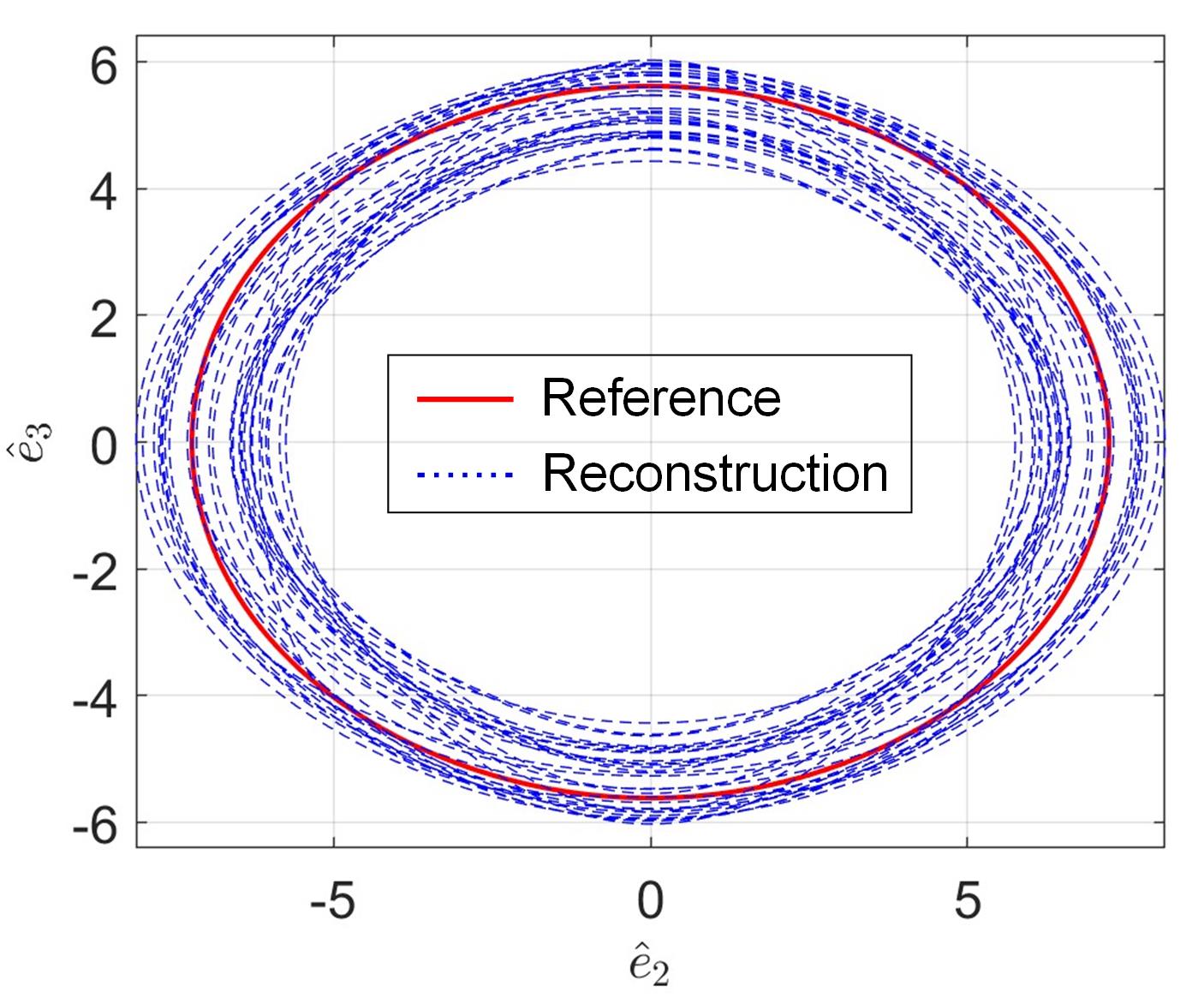} 
		\text{(d) Stiffness tensor on $yz$-plane}
	\end{minipage}  
	\begin{minipage}[t]{0.33\textwidth}  
		\centering  
		\includegraphics[width=0.99\textwidth]{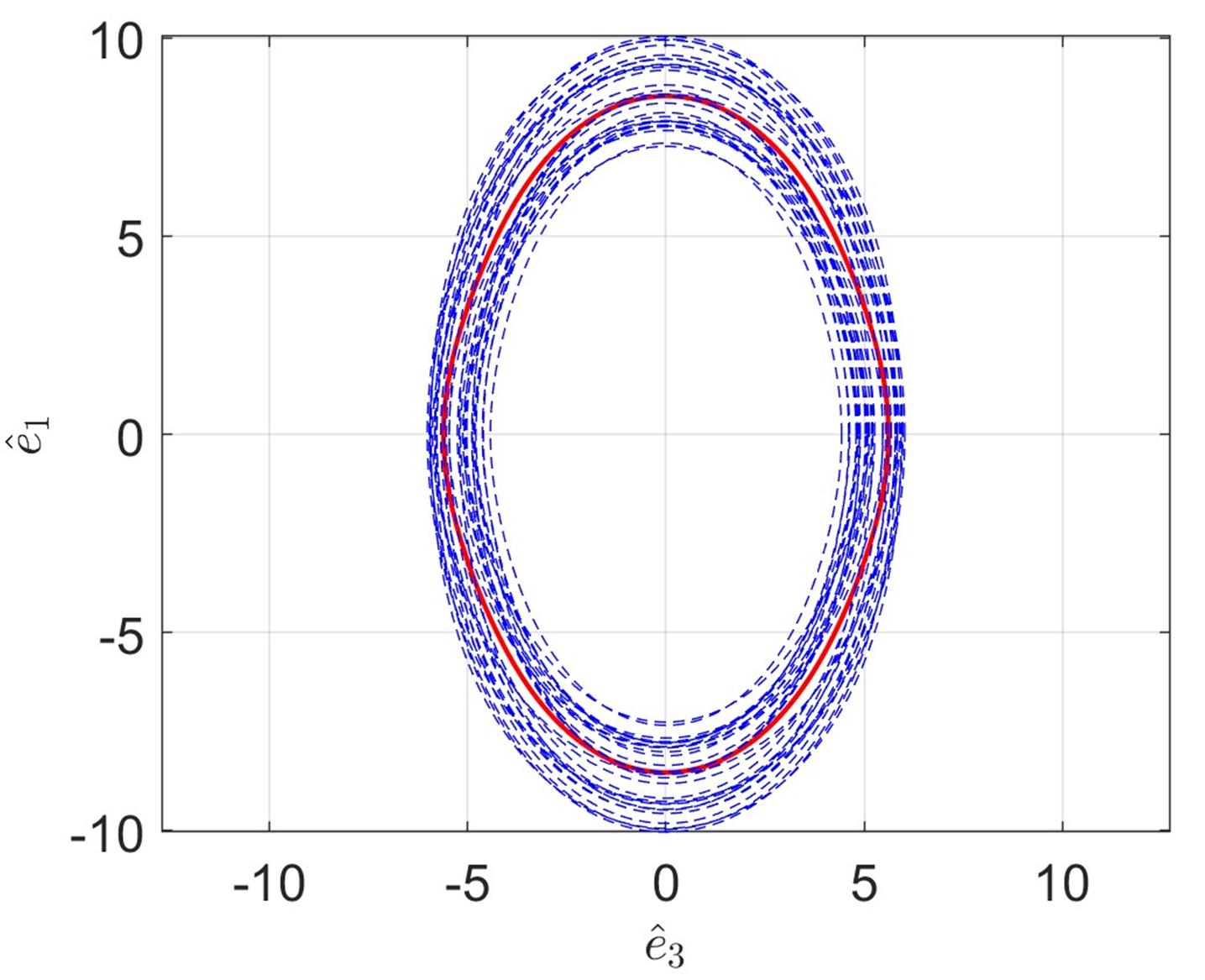}  
		\text{(e) Stiffness tensor on $zx$-plane}
	\end{minipage}
	\caption{Comparison of stiffness tensors between the reference and reconstructed 3D microstructures of the silver-based electrode.}
	\label{Fig:Ex1_stiffness_tensor}
\end{figure}

Volumetric stiffness tensors, as visualized in Figures \ref{Fig:Ex1_stiffness_tensor}a and b, are computed for both reference and reconstructed microstructures. The similarity in elasticity surfaces for these microstructures indicates strong agreement in their mechanical properties. Further, the volumetric stiffness tensors of thirty reconstructed microstructures are projected onto three principal planes, as shown in Figures \ref{Fig:Ex1_stiffness_tensor}c-e. The stiffness tensors of reconstructed samples exhibit minor fluctuations around the reference one, with their mean values closely matching the reference. This observed agreement, combined with minor variability, confirms the statistical equivalence of the reference and reconstructed microstructures in terms of elasticity.

To compute the permeability tensor, the Navier-Stokes equations are numerically solved on the image-based meshes of the porous microstructures via FEM. This involves applying a pressure gradient along specific directions and simulating fluid flow through the pore space. Figures \ref{Fig:Ex1_numerical_simulation}c and f depict the simulated flow pressure fields within the pore spaces of the reference and reconstructed 3D microstructures, revealing the intricate flow paths. The permeability tensor is derived from Darcy’s law \cite{fu2020resolution}, which establishes the relationship between the volumetric flow rate and the pressure gradient as follows:
\begin{equation}
\begin{bmatrix}
q_{x} \\
q_{y} \\
q_{z}
\end{bmatrix}
=-\frac{1}{\mu}
\begin{bmatrix}
\kappa_{11} & \kappa_{12} & \kappa_{13} \\
\kappa_{21} & \kappa_{22} & \kappa_{23} \\
\kappa_{31} & \kappa_{32}& \kappa_{33}
\end{bmatrix}
\begin{bmatrix}
\smallskip
\frac{ \partial P}{ \partial x } \\\smallskip
\frac{ \partial P}{ \partial y } \\
\frac{ \partial P}{ \partial z }
\end{bmatrix}
\label{Eq:Darcy's_law}
\end{equation}
where $q_x$, $q_y$ and $q_z$ denote volumetric flow rates along $x$-, $y$- and $z$-directions, respectively; $\mu$ is the dynamic viscosity; $P$ represents pressure ; and $\kappa_{ij}$ represents the permeability tensor components, quantifying the flow response in the $i$-direction to a pressure gradient in the $j$-direction.

\begin{figure}[h]\footnotesize
	\begin{minipage}[t]{0.5\textwidth}
		\centering  
		\includegraphics[width=0.8\textwidth]{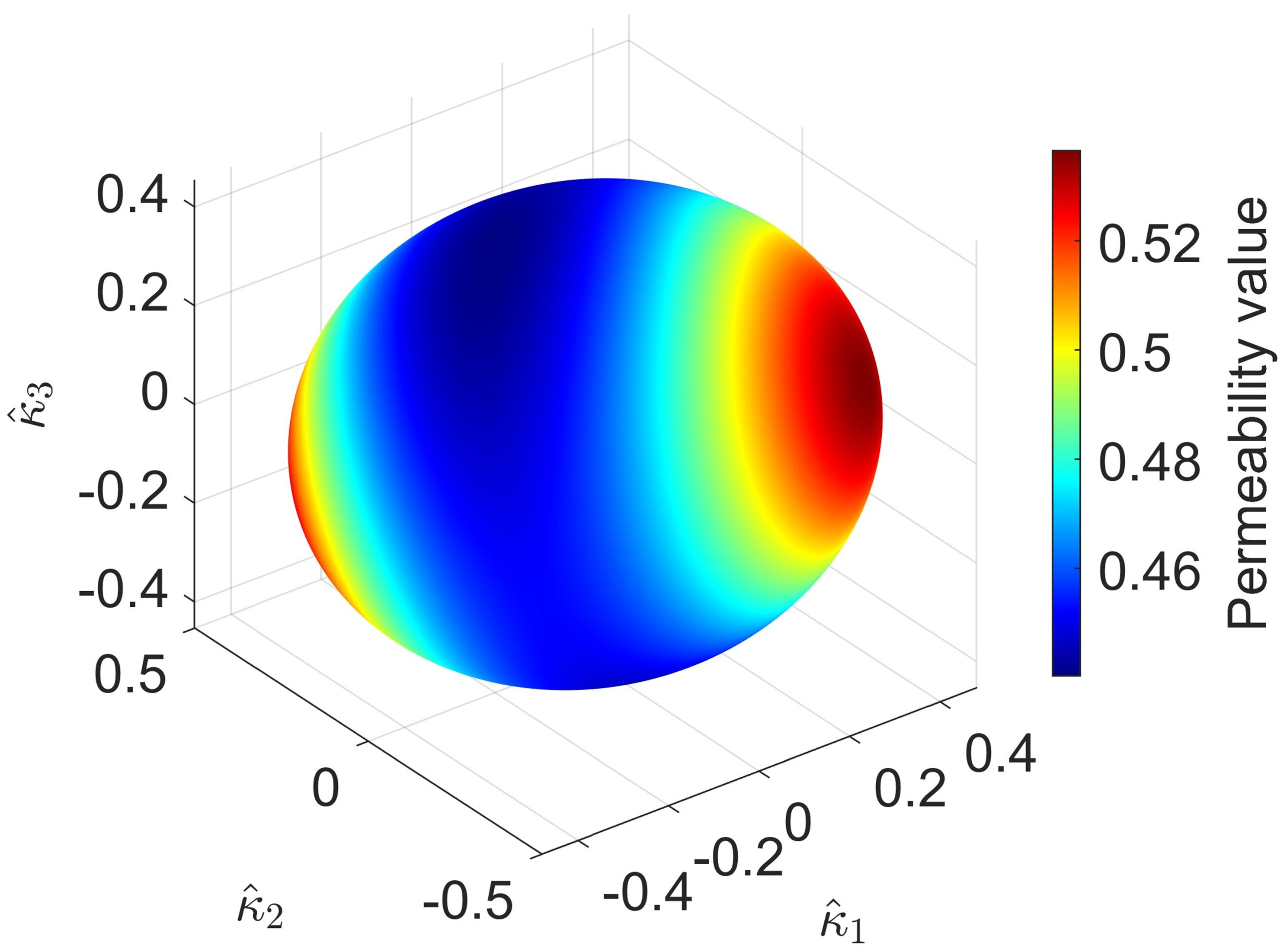}
		\text{(a) Volumetric representation of permeability tensor (reference)}
	\end{minipage}  
    \smallskip\smallskip 
	\begin{minipage}[t]{0.5\textwidth}  
		\centering  
		\includegraphics[width=0.8\textwidth]{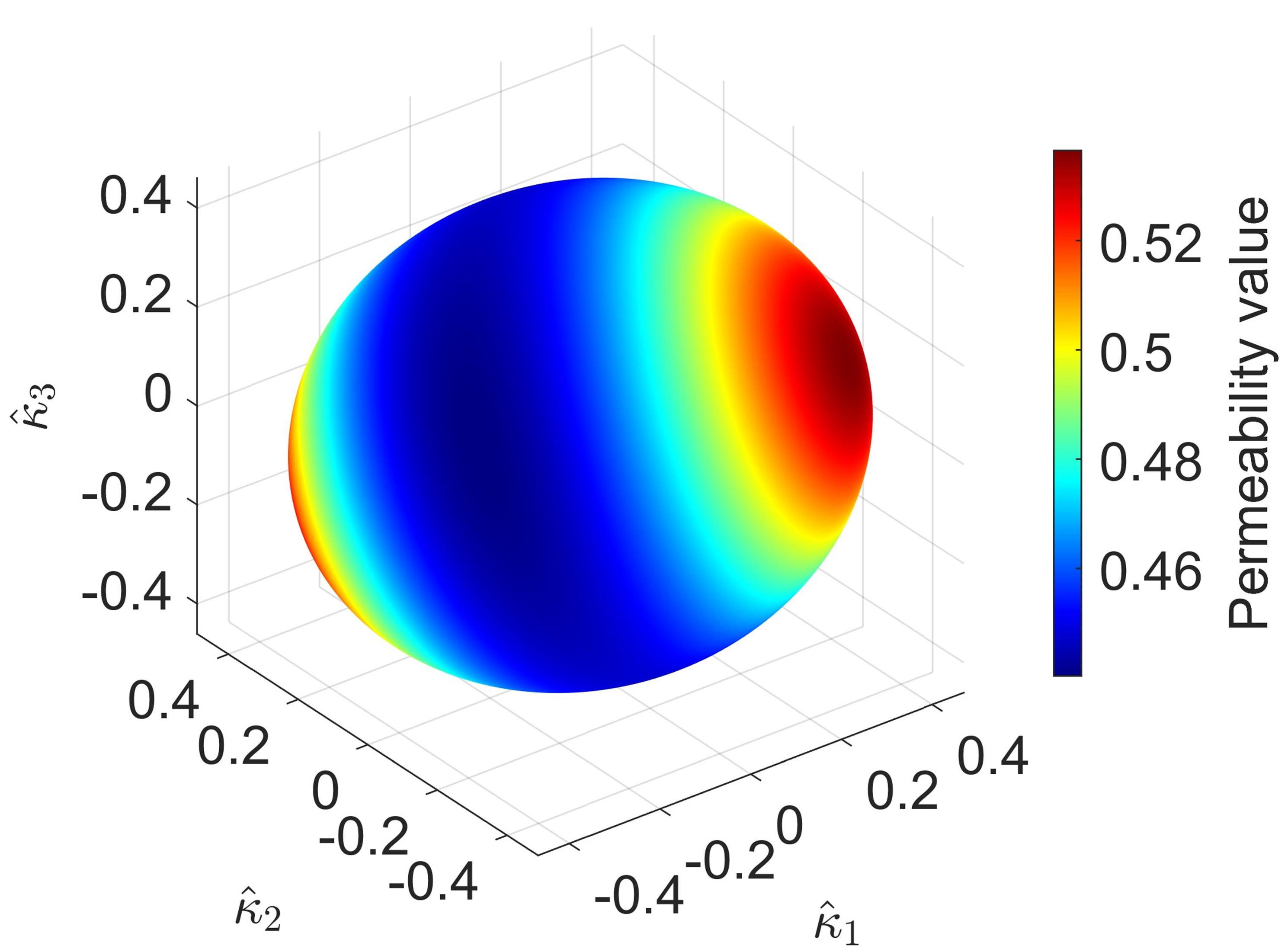}
		\text{(b) Volumetric representation of permeability tensor (reconstruction)}
	\end{minipage}  
    \smallskip\smallskip  
    \begin{minipage}[t]{0.33\textwidth} 
		\centering  
		\includegraphics[width=0.98\textwidth]{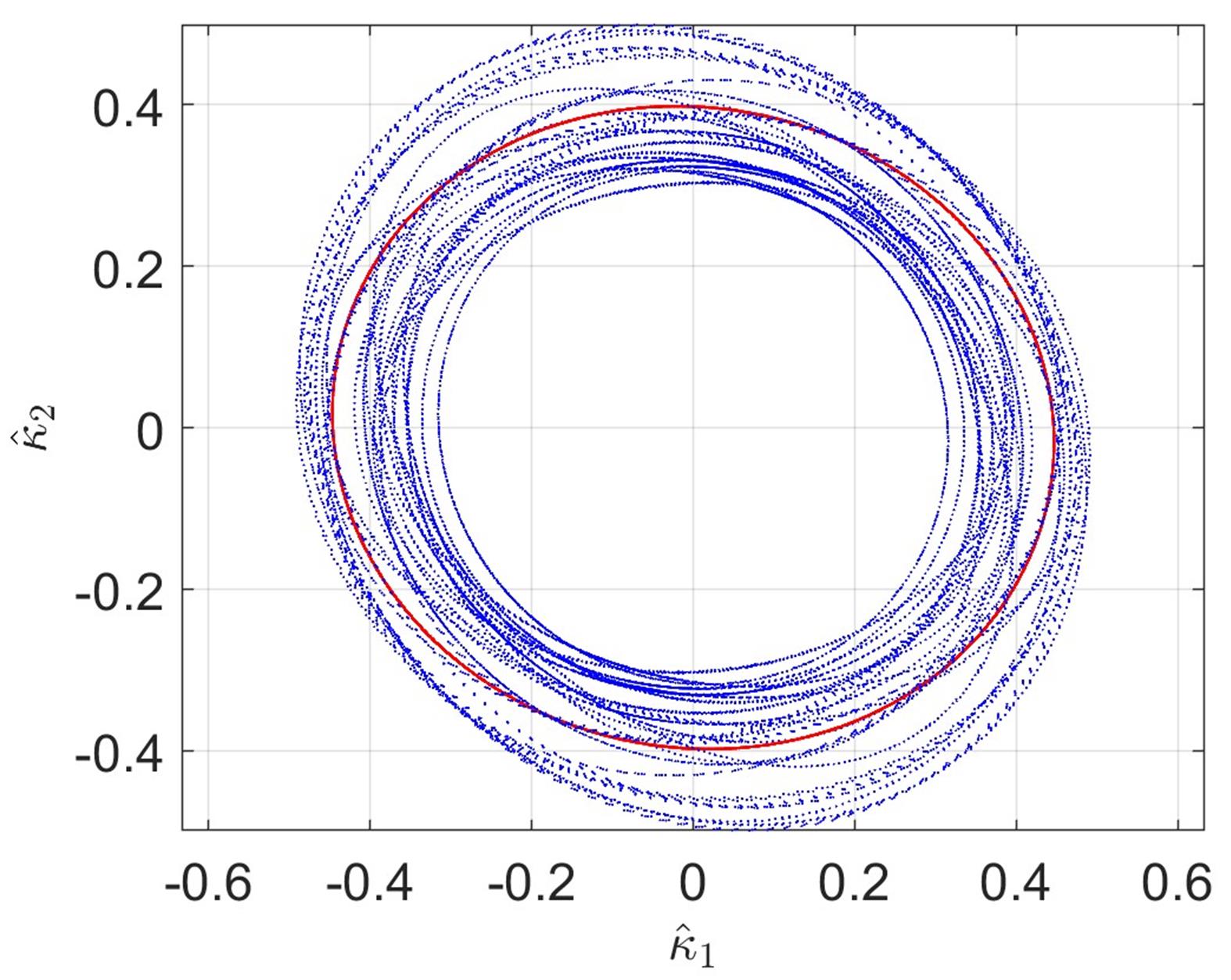}
		\text{(c) Permeability tensor on $xy$-plane}
	\end{minipage}  
	\begin{minipage}[t]{0.33\textwidth}  
		\centering  
		\includegraphics[width=0.98\textwidth]{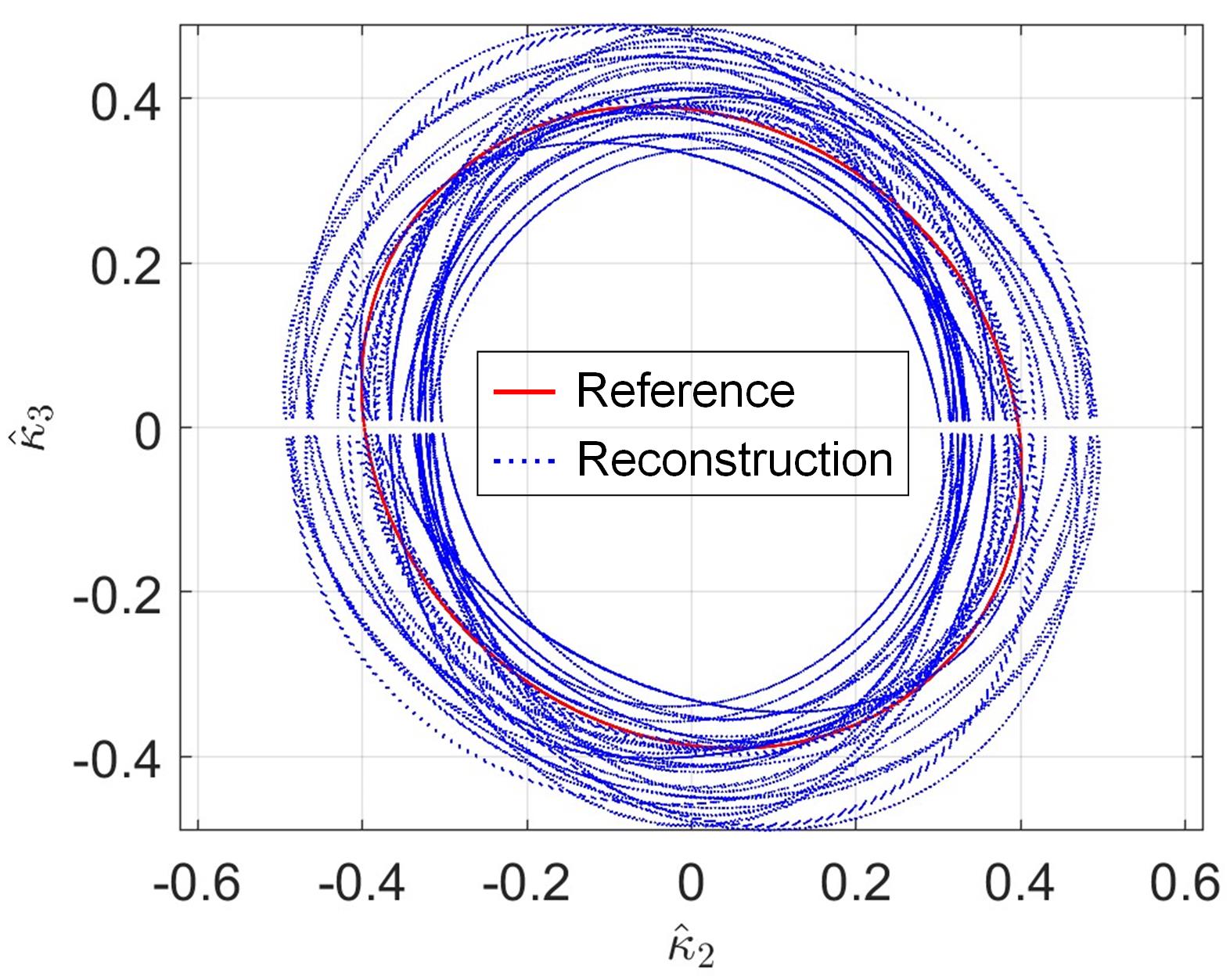} 
		\text{(d) Permeability tensor on $yz$-plane}
	\end{minipage}  
	\begin{minipage}[t]{0.33\textwidth}  
		\centering  
		\includegraphics[width=0.98\textwidth]{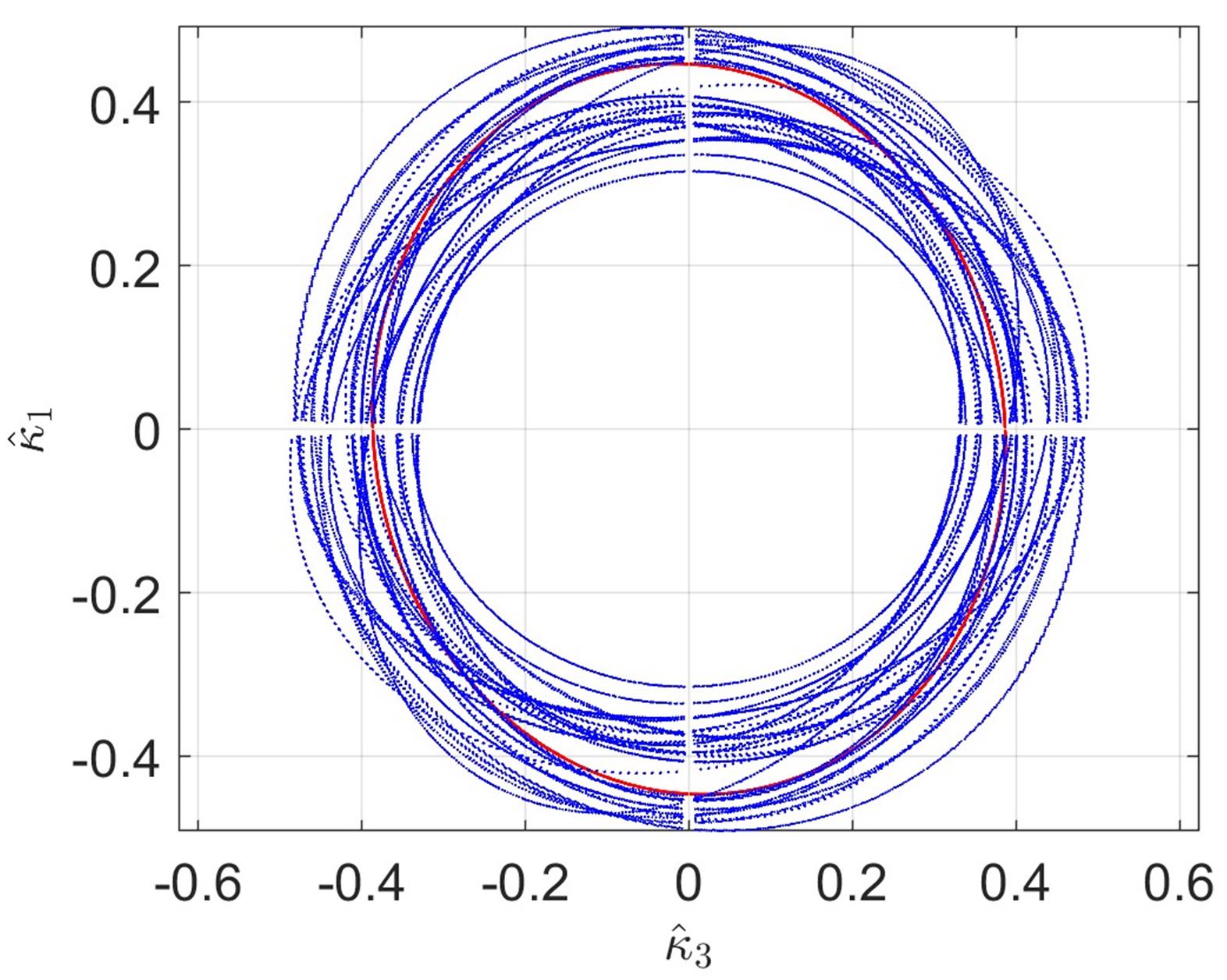}  
		\text{(e) Permeability tensor on $zx$-plane}
	\end{minipage}
	\caption{Comparison of permeability tensors between the reference and reconstructed 3D microstructures of the silver-based electrode.}
	\label{Fig:Ex1_permeability_tensor}
\end{figure}

Volumetric representations of the permeability tensors, as depicted in Figures \ref{Fig:Ex1_permeability_tensor}a and b, exhibit strong alignment between the reference and reconstructed 3D microstructures. Additionally, permeability values derived from thirty reconstructed samples show fluctuations around the reference value, with the mean closely matching the reference permeability tensor. This alignment is further evidenced by the projections of the tensor onto principal planes in Figures \ref{Fig:Ex1_permeability_tensor}c-e. This consistency demonstrates that the reconstructed 3D microstructure samples effectively replicate the transport efficiency of the reference microstructure, validating the capability of the proposed reconstruction method in accurately preserving fluid transport properties.

\subsection{Example 2: Mechanical and diffusional properties}
\label{Subsection5.2}
\begin{figure}[h]\footnotesize
	\begin{minipage}[t]{0.36\textwidth}
		\centering  
		\includegraphics[width=1.0\textwidth]{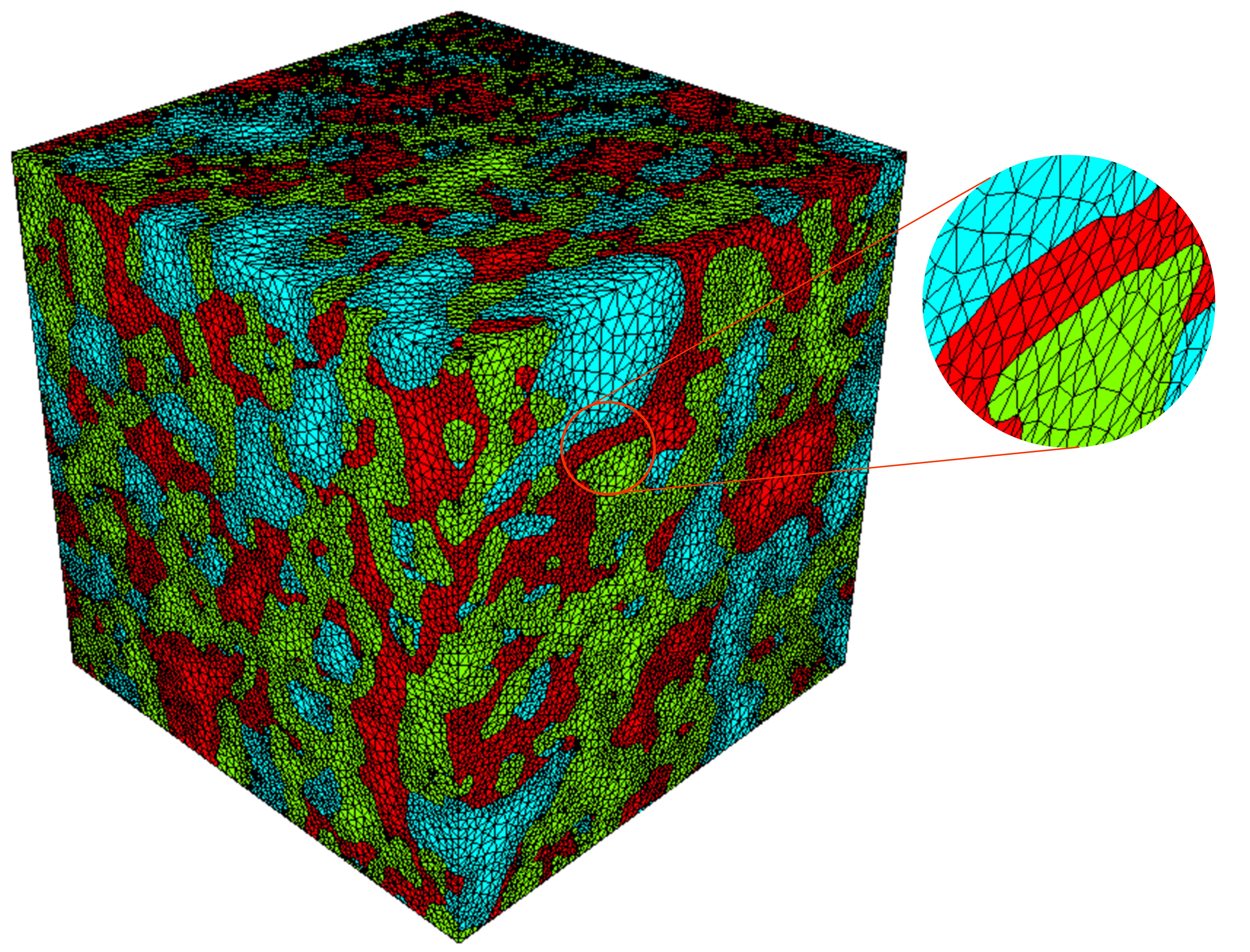}
		\text{(a) Image-based meshing}
	\end{minipage}  
    \smallskip
	\begin{minipage}[t]{0.32\textwidth}  
		\centering  
		\includegraphics[width=0.95\textwidth]{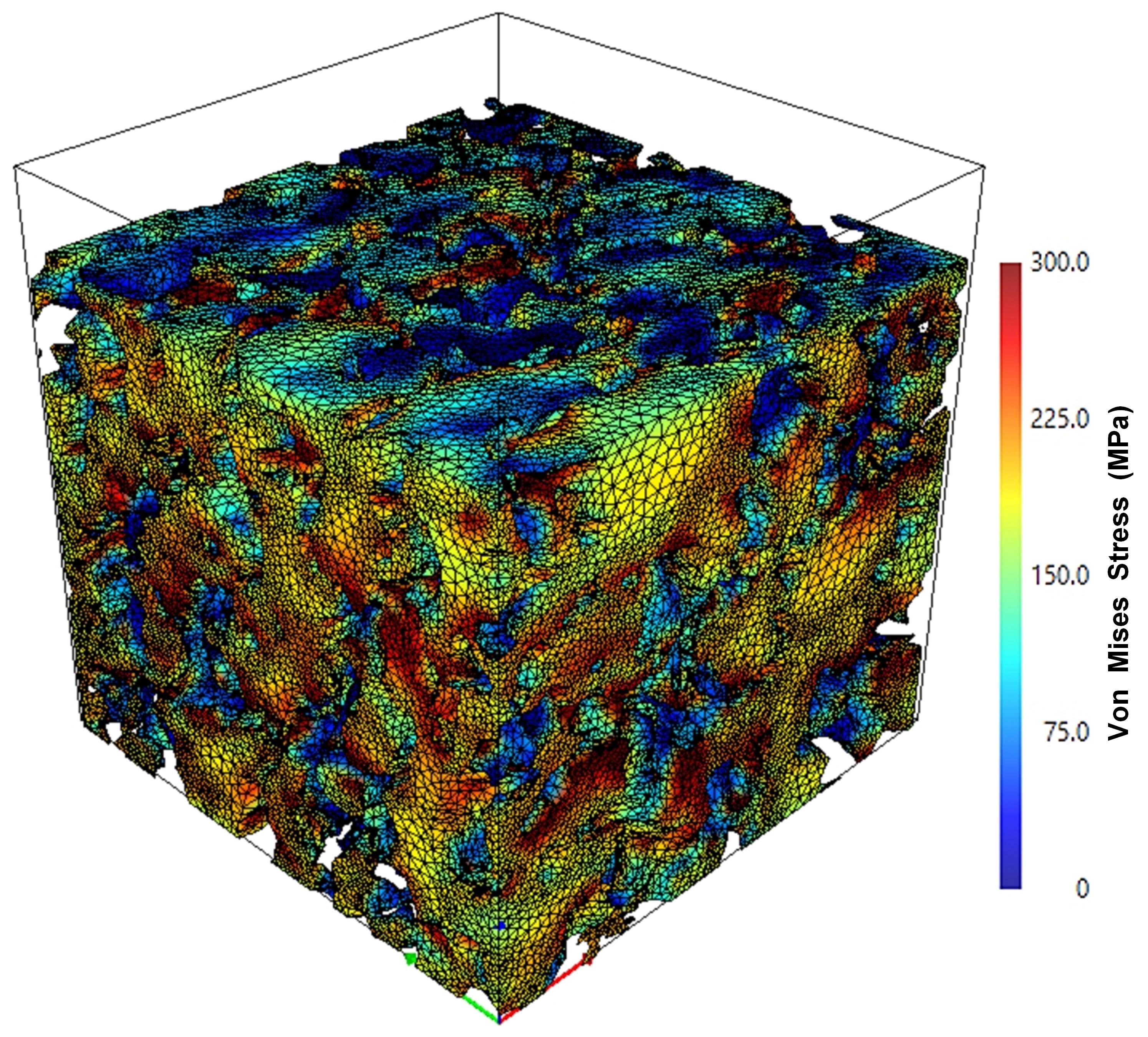}
		\text{(b) Stress field in solid phases}
	\end{minipage}  
    \smallskip
	\begin{minipage}[t]{0.32\textwidth}  
		\centering  
		\includegraphics[width=0.95\textwidth]{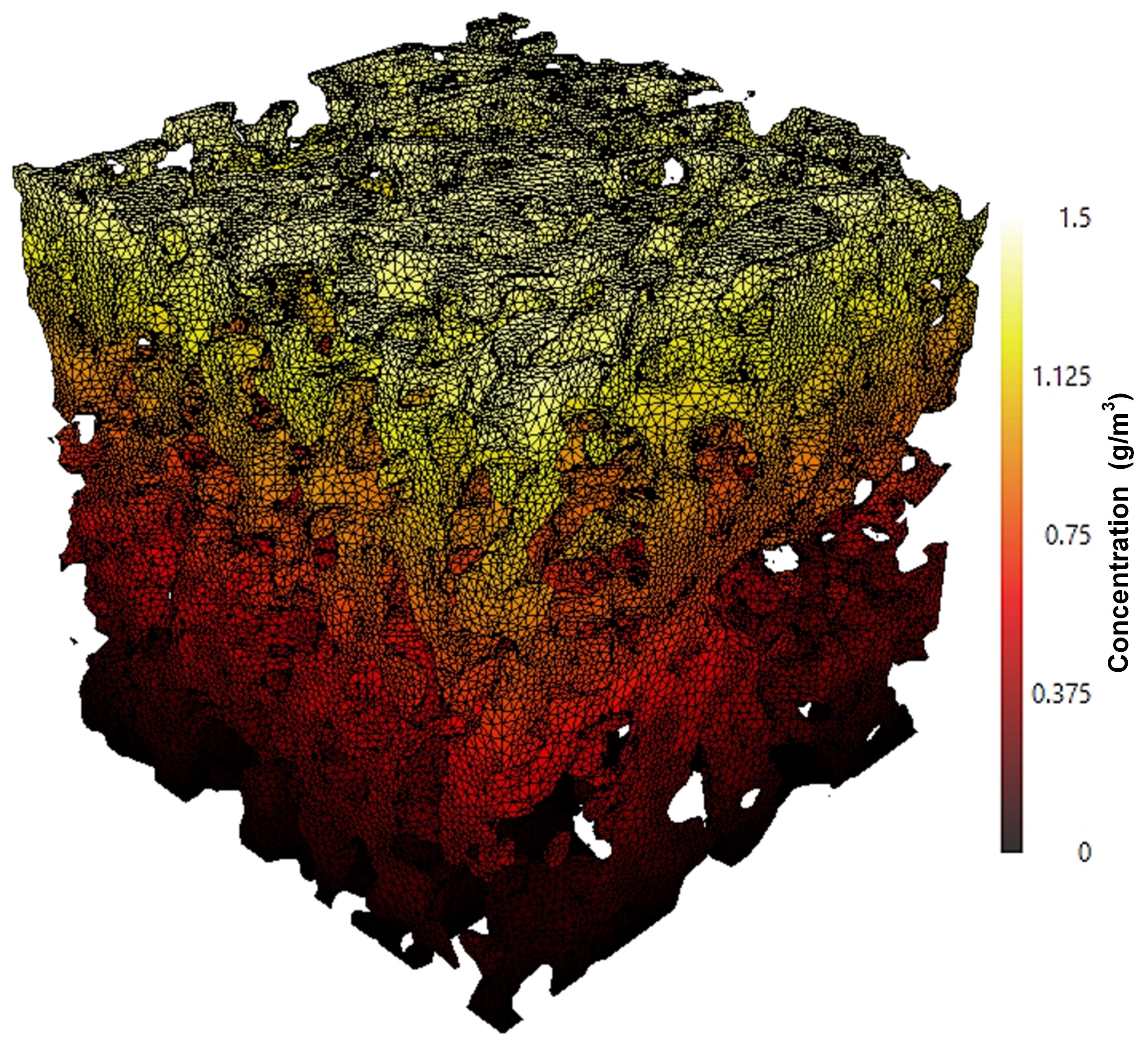}
		\text{(c) Concentration field in pore space}
	\end{minipage}  
    \smallskip
	\begin{minipage}[t]{0.36\textwidth} 
		\centering  
		\includegraphics[width=1.0\textwidth]{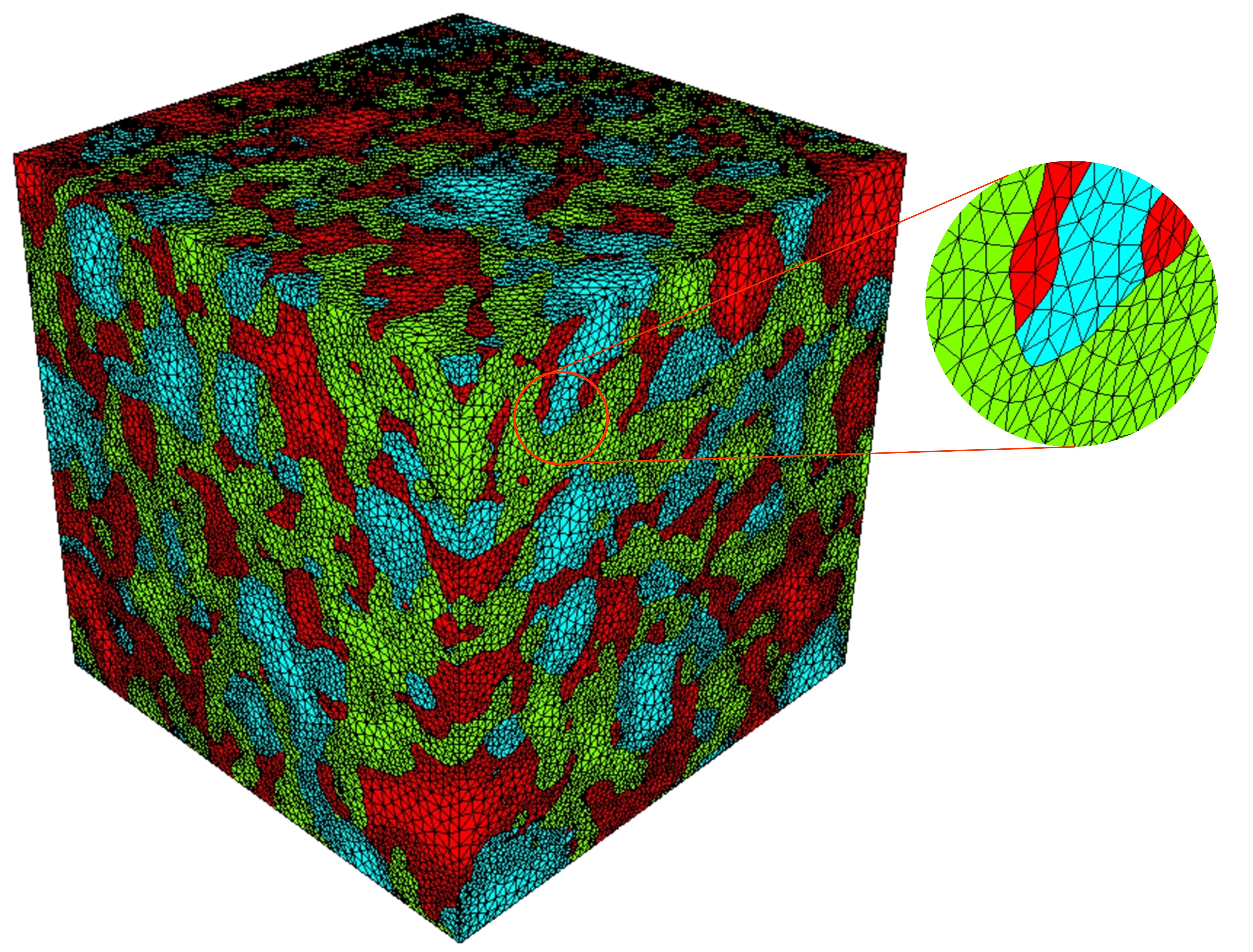}
		\text{(d) Image-based meshing}
	\end{minipage}  
	\begin{minipage}[t]{0.32\textwidth}  
		\centering  
		\includegraphics[width=0.95\textwidth]{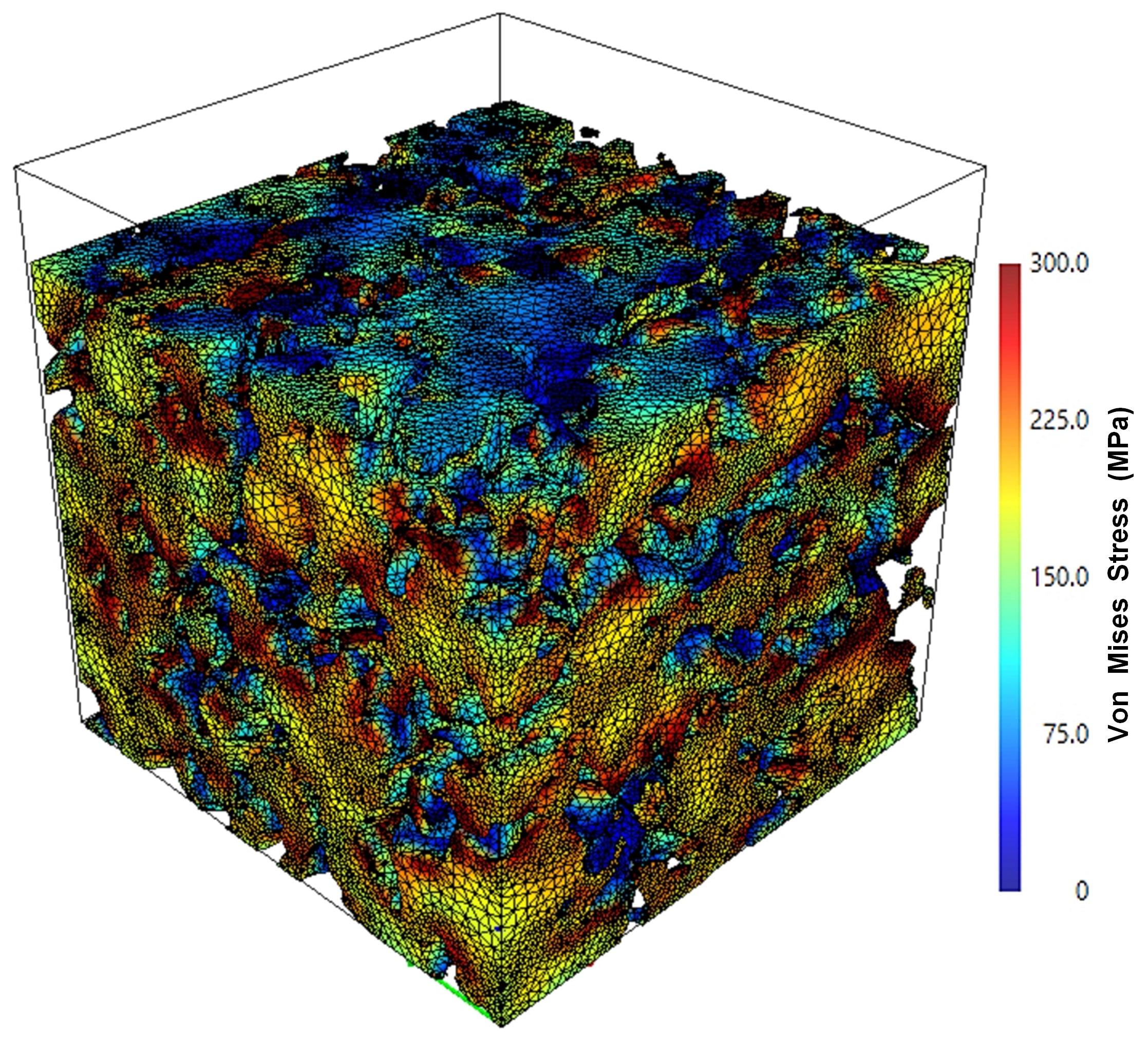} 
		\text{(e) Stress field in solid phases}
	\end{minipage}  
	\begin{minipage}[t]{0.32\textwidth}  
		\centering  
		\includegraphics[width=0.95\textwidth]{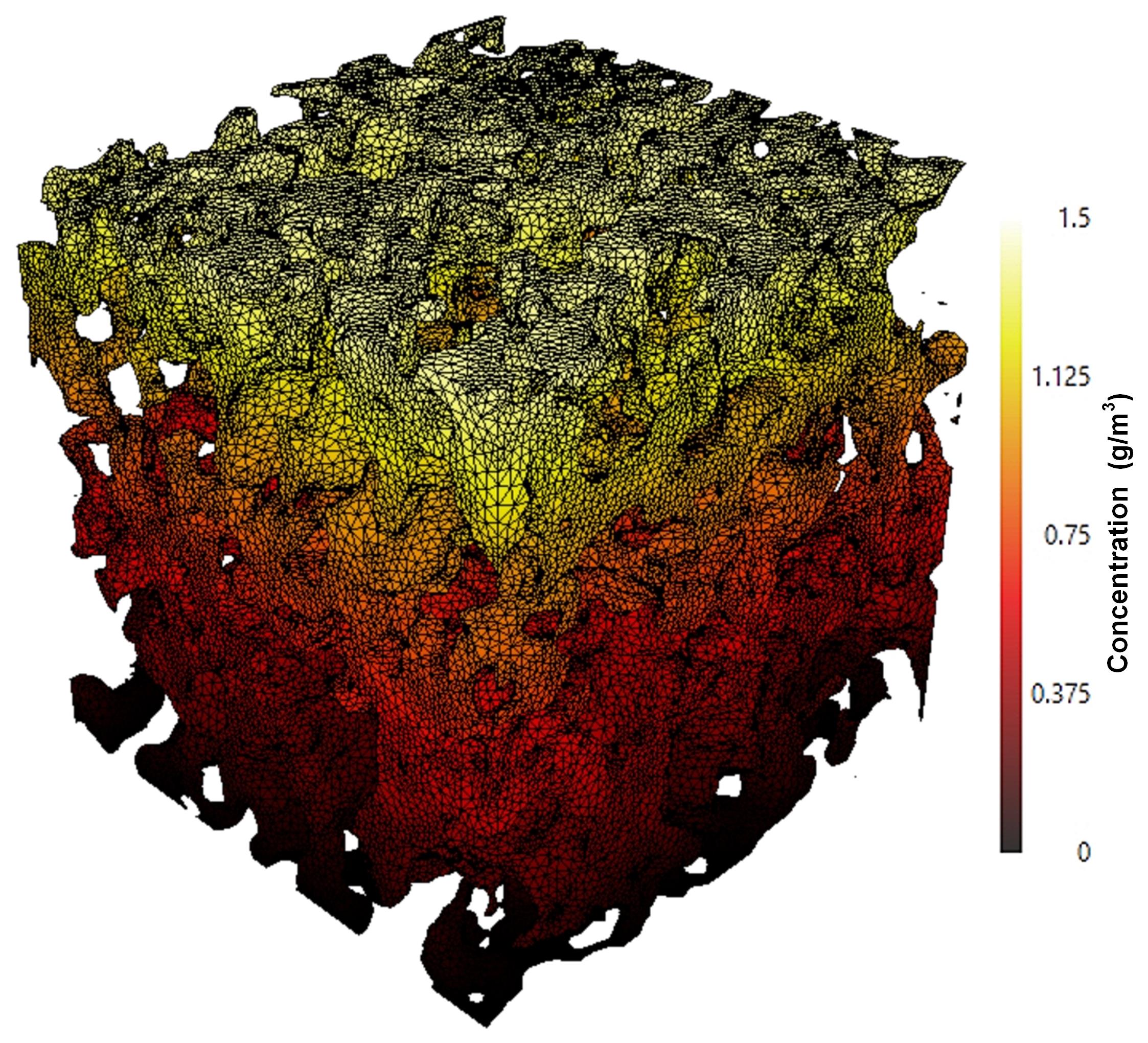}  
		\text{(f) Concentration field in pore space}
	\end{minipage}
	\caption{Image-based poro/micro-mechanical modelling on the reference (top raw) and the reconstructed (bottom raw) 3D microstructures of the porous SOFC anode: (a) and (d) Mesh generation on multiphase microstructures: red regions denote YSZ, cyan regions denote Ni, and green regions represent pore space; (b) and (e) Von Mises stress fields in the solid phases under axial compression; (c) and (f) Concentration fields of molecular diffusion within pore space.}
	\label{Fig:Ex2_numerical_simulation}
\end{figure}

\vspace{-2pt}
The mechanical and diffusional properties of porous SOFC anodes are essential for their functionality and longevity in electrochemical systems. Mechanical properties, such as stiffness, maintain structural integrity under operational stresses, while effective diffusivity governs the efficiency of gas transport through the porous network. These properties are largely determined by the microstructure, including phase distribution, porosity, and pore connectivity. To evaluate these features, image-based meshing of both reference and reconstructed 3D microstructures is conducted for poro/micro-mechanical modelling. As depicted in Figure \ref{Fig:Ex2_numerical_simulation}a and d, unstructured tetrahedral meshes are generated for 3D digital microstructures, and these volumetric meshes precisely capture the complex interfaces and shared nodes between phases within this heterogeneous material. 

\begin{figure}[H]\footnotesize
	\begin{minipage}[t]{0.5\textwidth}
		\centering  
		\includegraphics[width=0.8\textwidth]{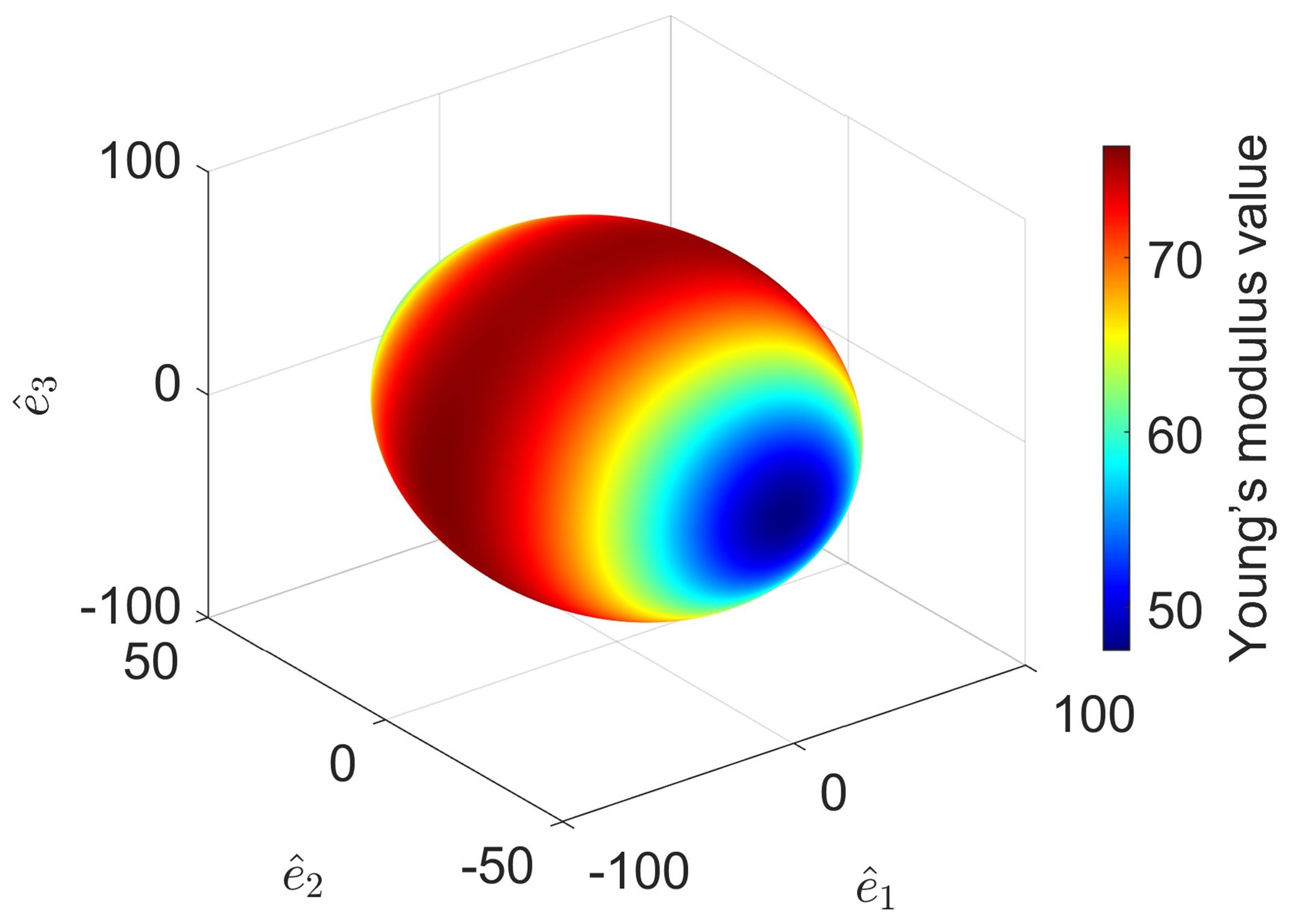}
		\text{(a) Volumetric representation of stiffness tensor (reference)}
	\end{minipage}  
    \smallskip\smallskip 
	\begin{minipage}[t]{0.5\textwidth}  
		\centering  
		\includegraphics[width=0.8\textwidth]{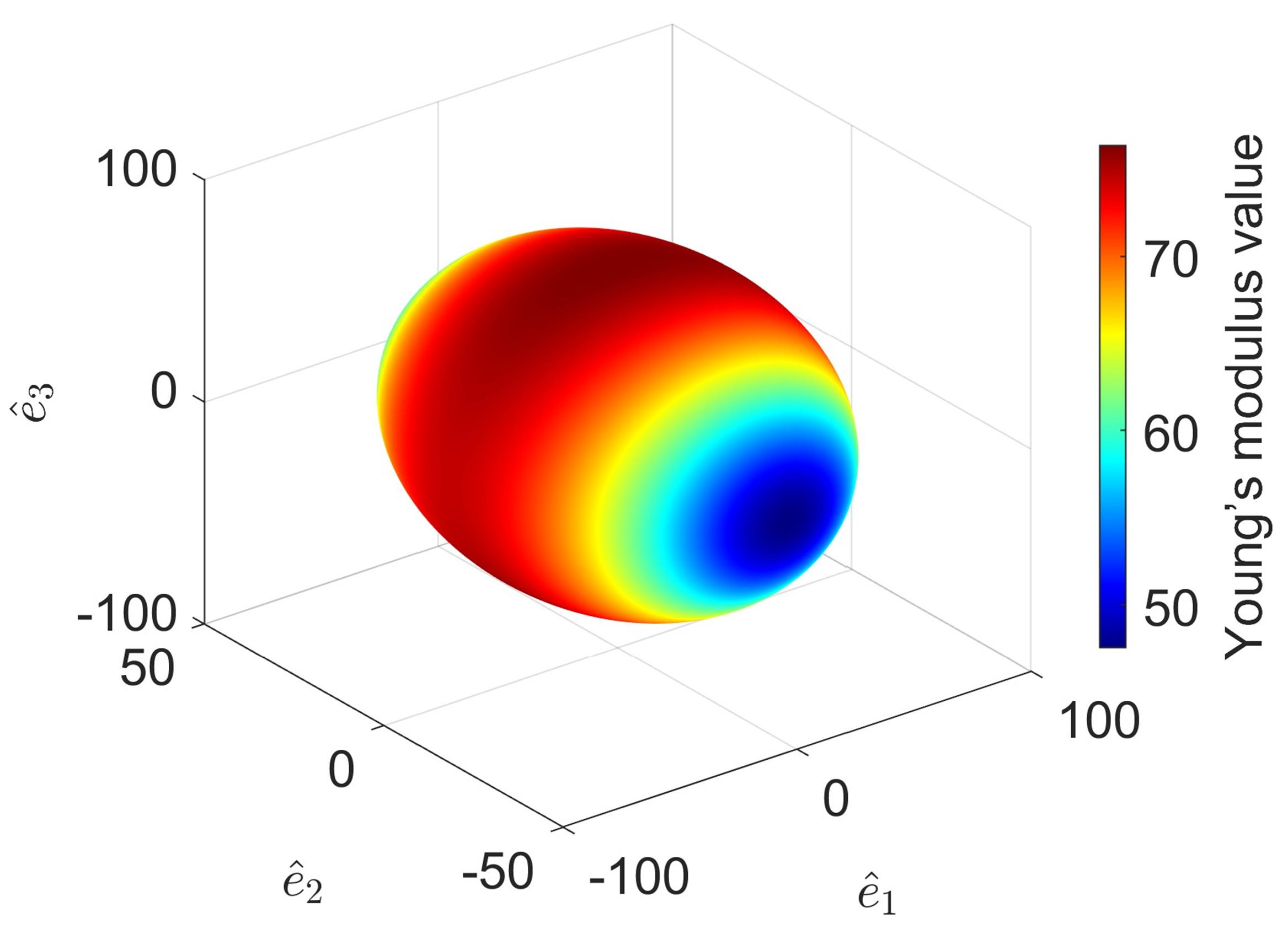}
		\text{(b) Volumetric representation of stiffness tensor (reconstruction)}
	\end{minipage}  
    \smallskip\smallskip  
    \begin{minipage}[t]{0.33\textwidth} 
		\centering  
		\includegraphics[width=0.98\textwidth]{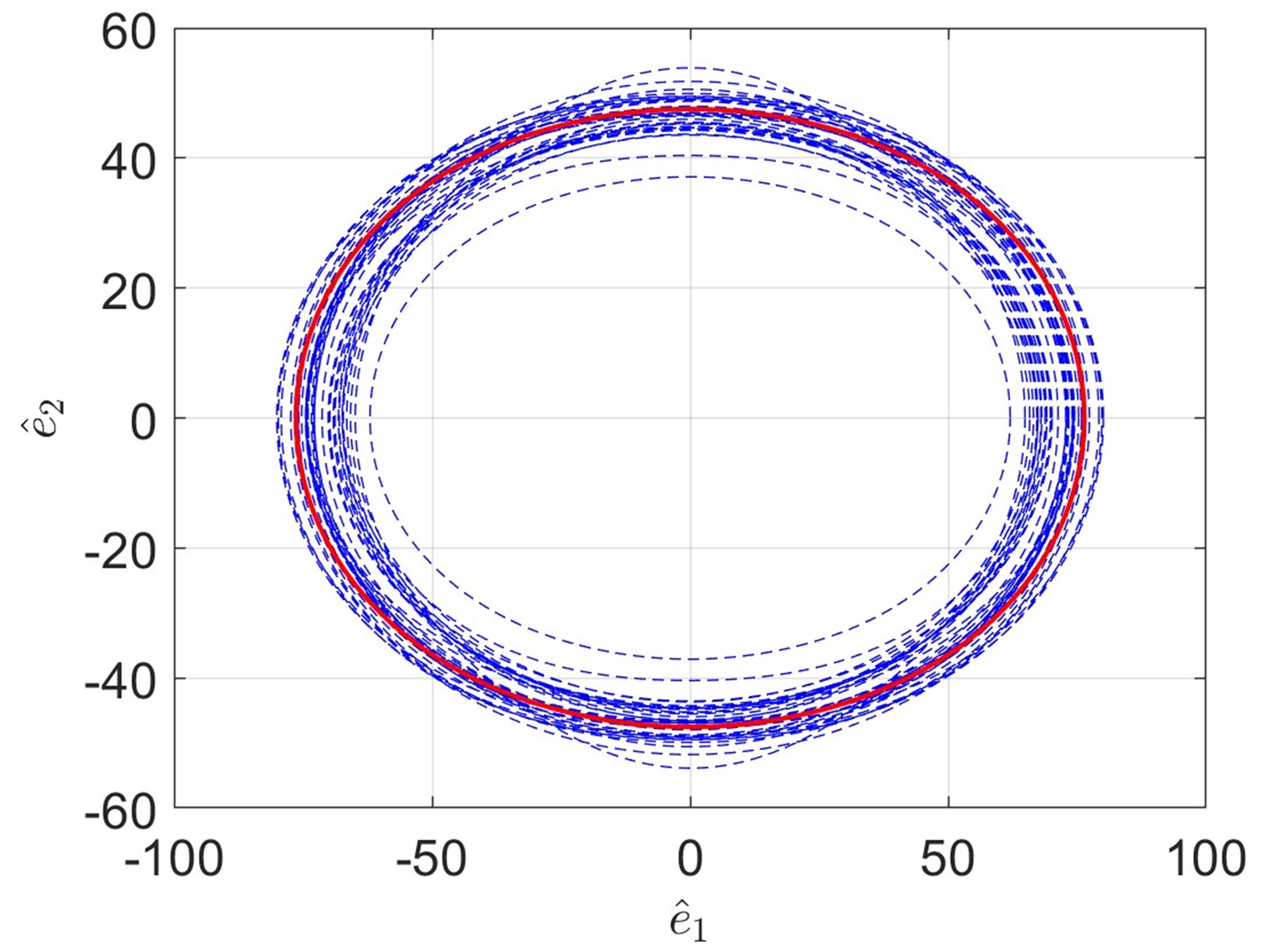}
		\text{(c) Stiffness tensor on $xy$-plane}
	\end{minipage}  
	\begin{minipage}[t]{0.33\textwidth}  
		\centering  
		\includegraphics[width=0.98\textwidth]{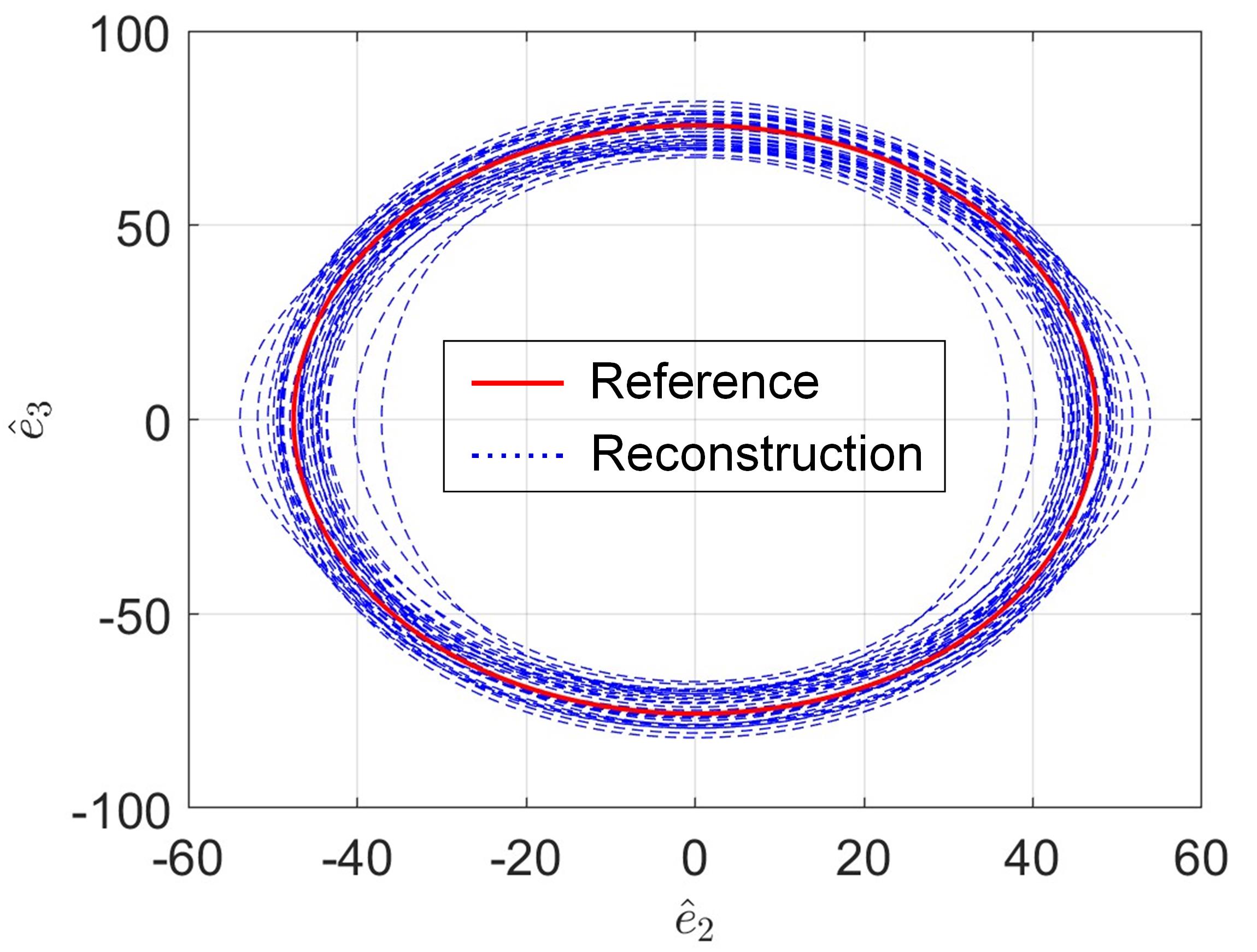} 
		\text{(d) Stiffness tensor on $yz$-plane}
	\end{minipage}  
	\begin{minipage}[t]{0.33\textwidth}  
		\centering  
		\includegraphics[width=0.98\textwidth]{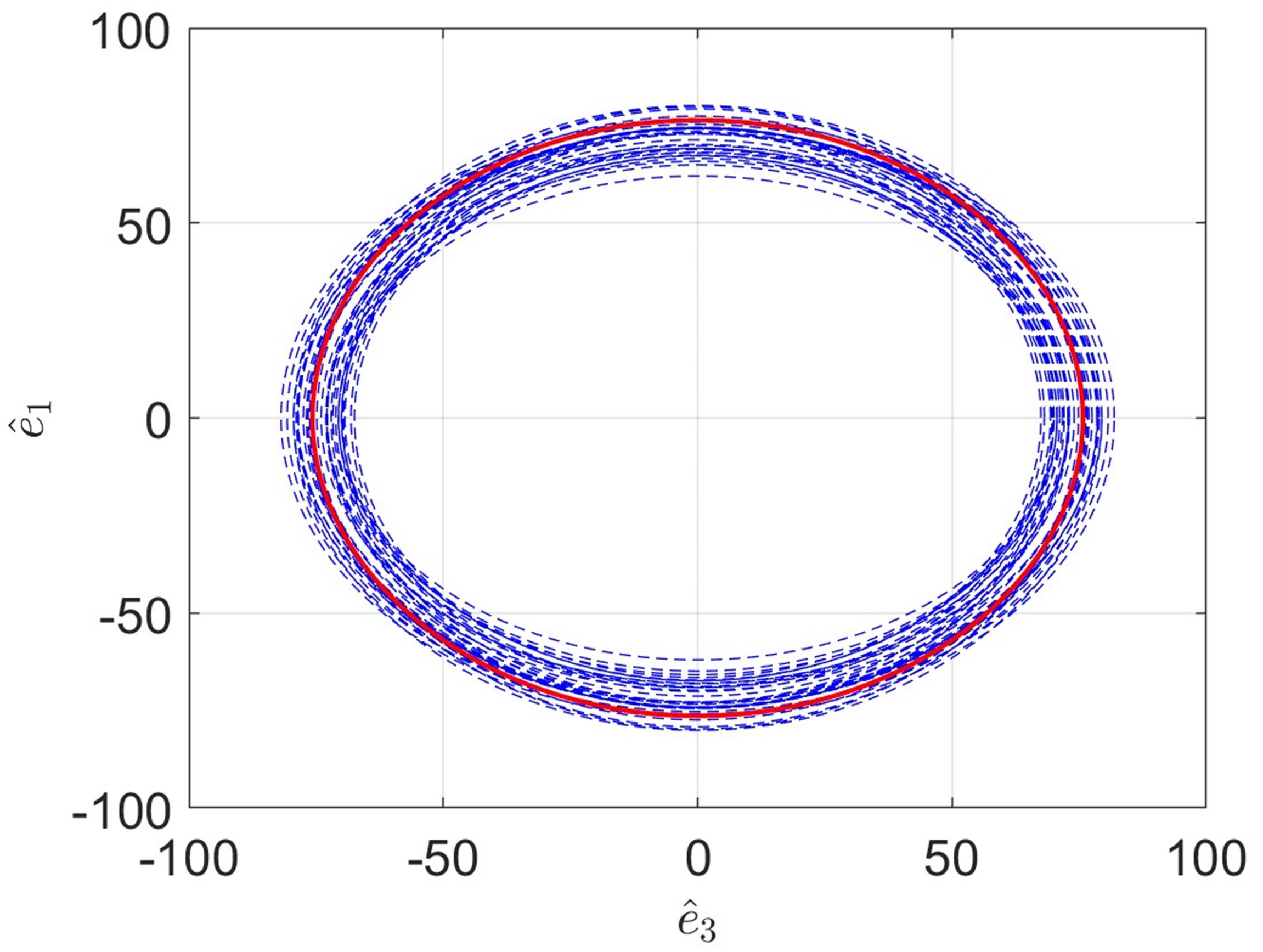}  
		\text{(e) Stiffness tensor on $zx$-plane}
	\end{minipage}
	\caption{Comparison of stiffness tensors between the reference and reconstructed 3D microstructures of the porous SOFC anode.}
	\label{Fig:Ex2_stiffness_tensor}
\end{figure}

FEM is applied to the RVEs of the porous SOFC anode to derive stress-strain relationships, enabling the extraction of the stiffness tensor. Figures \ref{Fig:Ex2_numerical_simulation}b and e display the stress fields in the reference and reconstructed 3D microstructures under axial compression, providing insights into the mechanical behaviour of the material. Volumetric stiffness tensors, as shown in Figures \ref{Fig:Ex2_stiffness_tensor}a and b, are computed for both the reference and a representative 3D reconstructed microstructure, revealing similar elasticity surfaces in terms of shape and size, thereby demonstrating consistency between the two microstructures. Additionally, stiffness tensors from thirty reconstructed samples are computed and projected onto three principal planes, as depicted in Figures \ref{Fig:Ex2_stiffness_tensor}c-e. The results reveal minor fluctuations around the reference tensor, with the mean closely matching the reference stiffness tensor. This consistency and the slight variations confirm the statistical equivalence of the reconstructed microstructures to the reference in terms of elasticity.

To compute the effective diffusivity tensor, FEM simulations are conducted on the image-based meshes of the porous SOFC anode. These simulations solve steady-state diffusion equations under prescribed concentration gradients along specific directions within the pore network. Figures \ref{Fig:Ex2_numerical_simulation}c and f illustrate the concentration fields for the reference and reconstructed 3D microstructures, highlighting the pathways of diffusive transport.
The effective diffusivity tensor \citep{ricketts2024stochastic} links the diffusion flux to concentration gradient via Fick's law, expressed as:
\begin{equation}
\begin{bmatrix}
J_{x} \\
J_{y} \\
J_{z}
\end{bmatrix}
=
\begin{bmatrix}
\lambda_{11} & \lambda_{12} & \lambda_{13} \\
\lambda_{21} & \lambda_{22} & \lambda_{23} \\
\lambda_{31} & \lambda_{32} & \lambda_{33}
\end{bmatrix}
\begin{bmatrix}
\smallskip
\frac{ \partial C}{ \partial x } \\\smallskip
\frac{ \partial C}{ \partial y } \\
\frac{ \partial C}{ \partial z }
\end{bmatrix}
\label{Eq:Fick's_law}
\end{equation}
where $J_x$, $J_y$ and $J_z$ denote the diffusion flux components along $x$-, $y$- and $z$- directions, respectively; $C$ represents the concentration; and $\lambda_{ij}$ is the element of the effective diffusivity tensor.

\begin{figure}[h]\footnotesize
	\begin{minipage}[t]{0.5\textwidth}
		\centering  
		\includegraphics[width=0.7\textwidth]{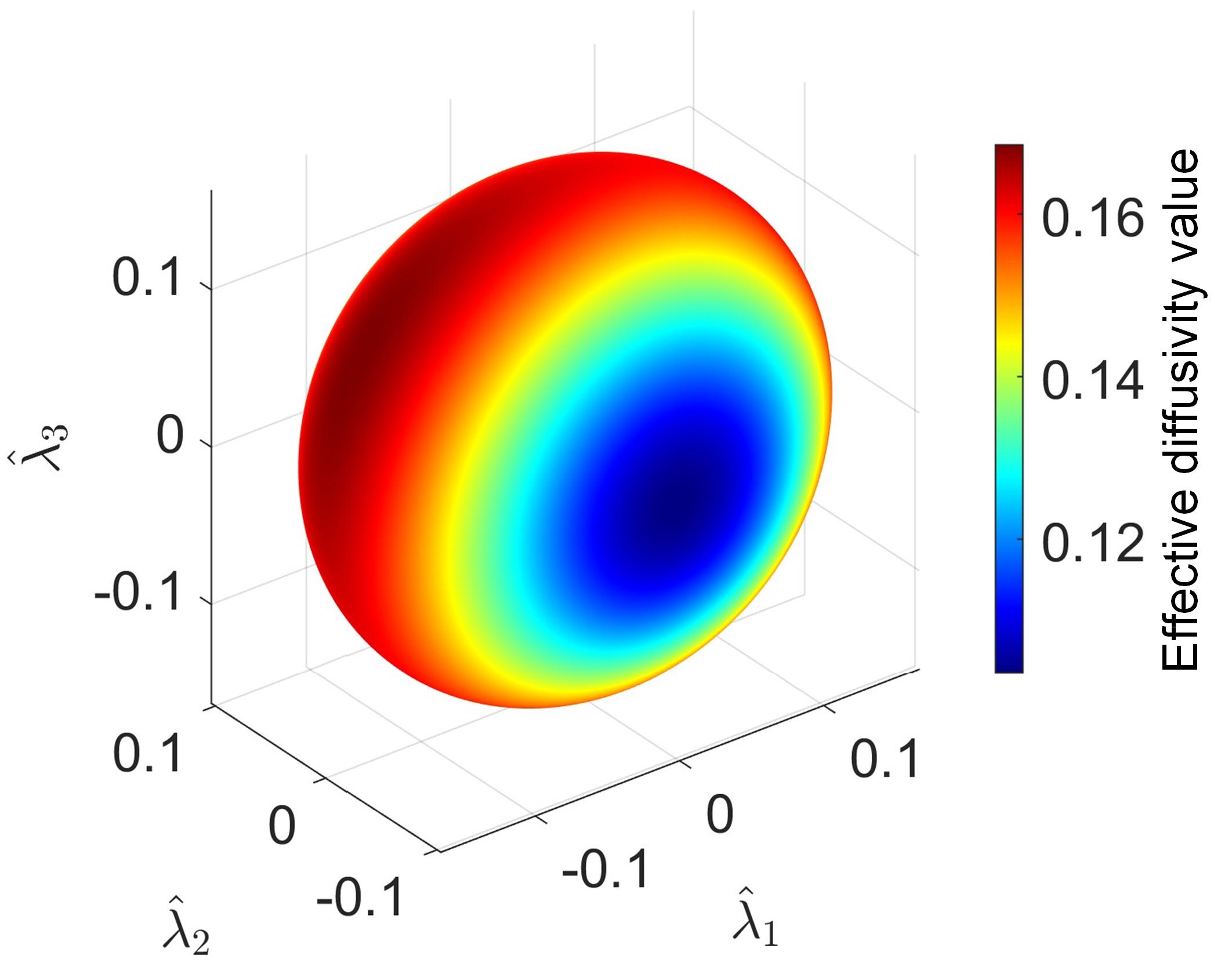}
		\text{(a) Volumetric representation of diffusivity tensor (reference)}
	\end{minipage}  
    \smallskip\smallskip 
	\begin{minipage}[t]{0.5\textwidth}  
		\centering  
		\includegraphics[width=0.7\textwidth]{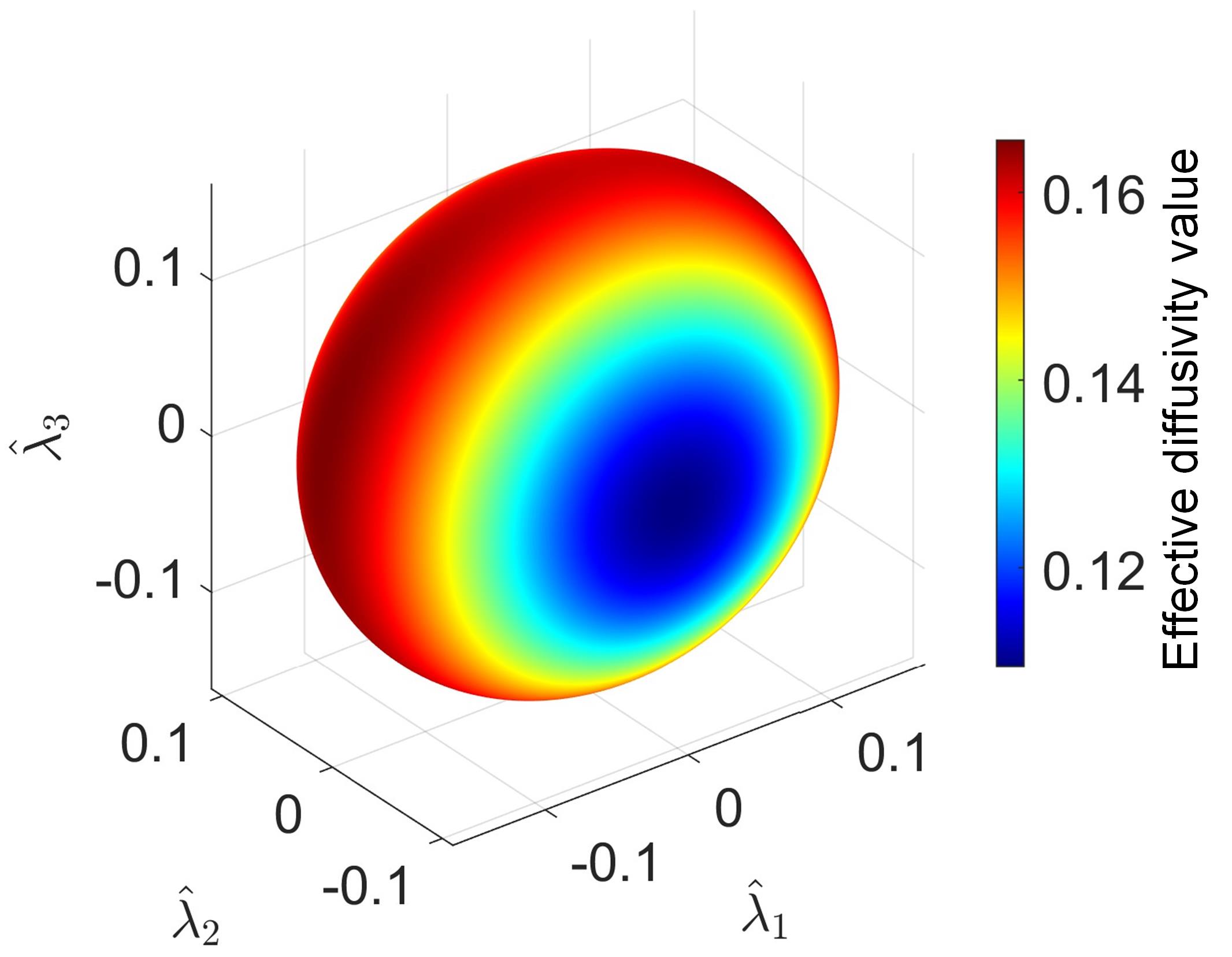}
		\text{(b) Volumetric representation of diffusivity tensor (reconstruction)}
	\end{minipage}  
    \smallskip\smallskip  
    \begin{minipage}[t]{0.33\textwidth} 
		\centering  
		\includegraphics[width=0.98\textwidth]{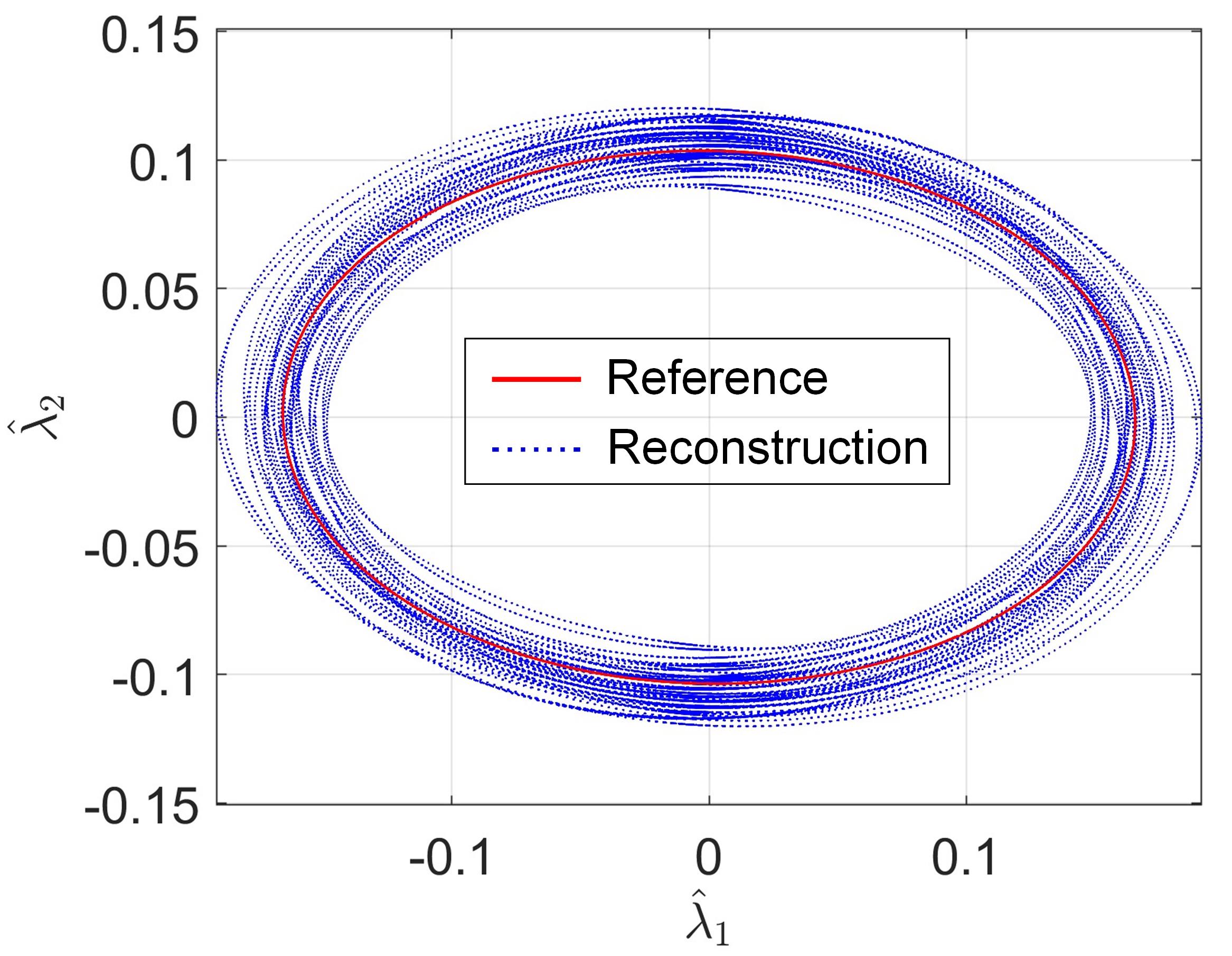}
		\text{(c) Effective diffusivity tensor on $xy$-plane}
	\end{minipage}  
	\begin{minipage}[t]{0.33\textwidth}  
		\centering  
		\includegraphics[width=0.98\textwidth]{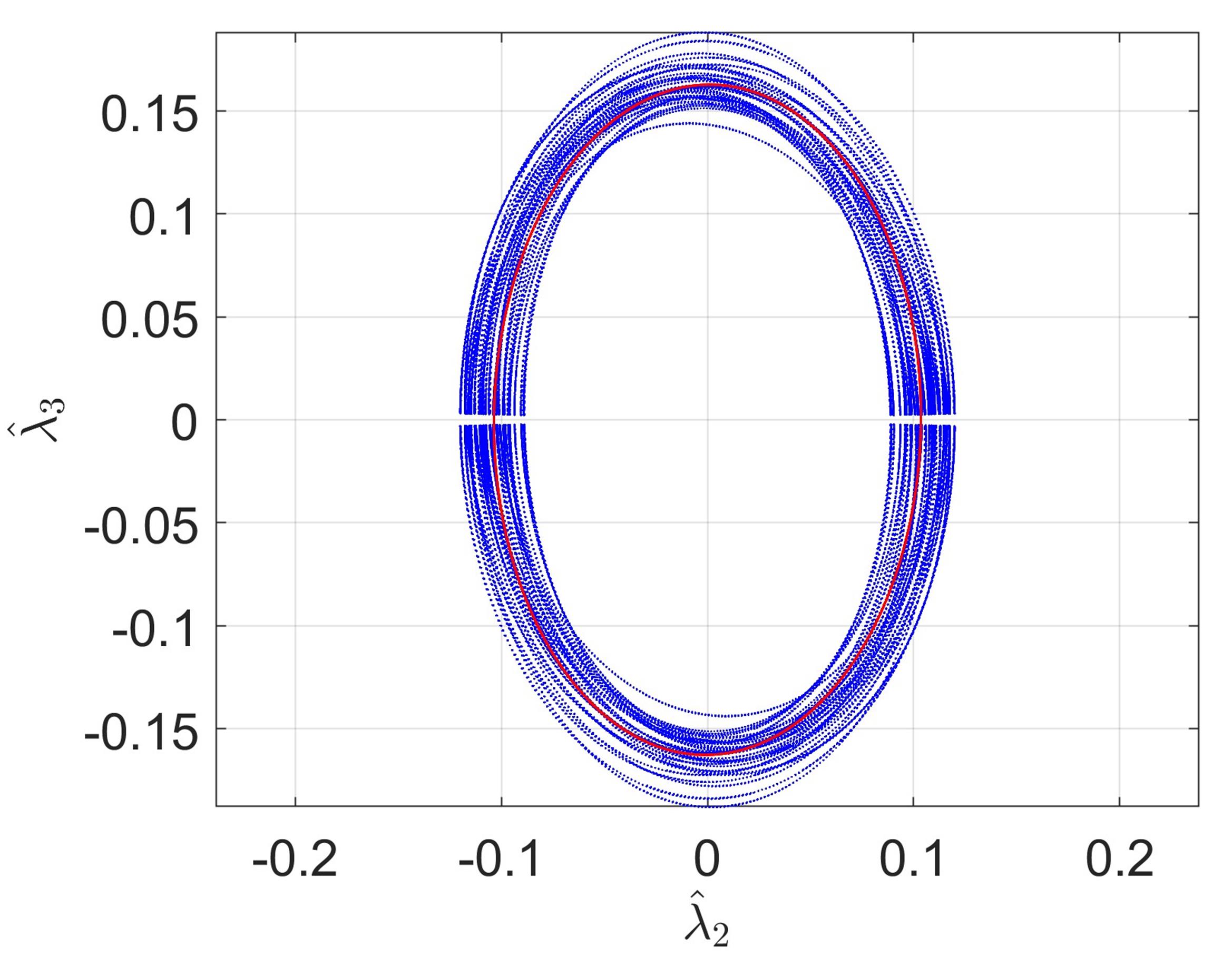} 
		\text{(d) Effective diffusivity tensor on $yz$-plane}
	\end{minipage}  
	\begin{minipage}[t]{0.33\textwidth}  
		\centering  
		\includegraphics[width=0.98\textwidth]{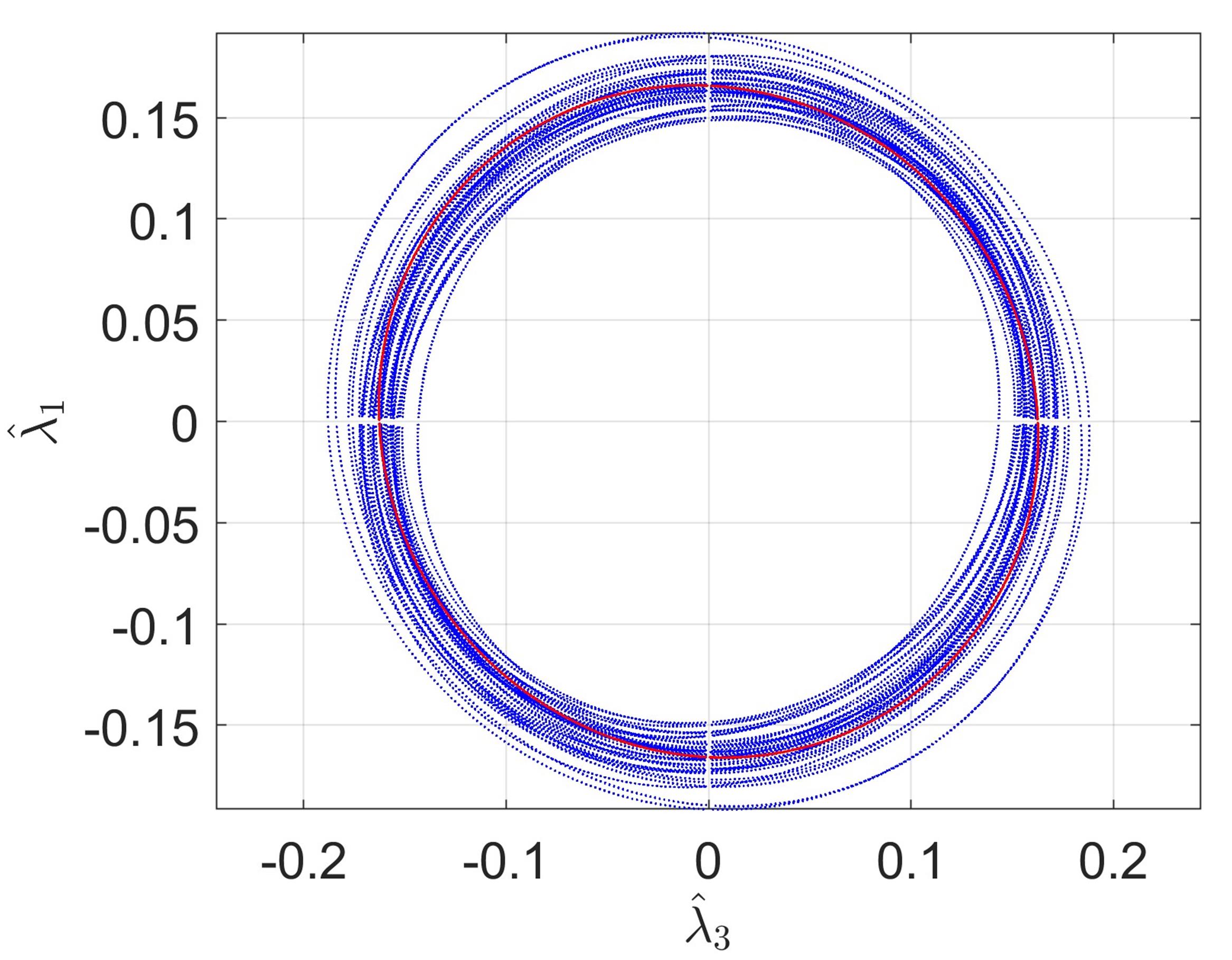}  
		\text{(e) Effective diffusivity tensor on $zx$-plane}
	\end{minipage}
	\caption{Comparison of effective diffusivity tensors between the reference and reconstructed 3D microstructures of the porous SOFC anode.}
	\label{Fig:Ex2_diffusivity_tensor}
\end{figure}

Figures \ref{Fig:Ex2_diffusivity_tensor}a and b depict volumetric representations of the effective diffusivity tensors for both the reference microstructure and a representative reconstructed sample. The similarity in shape and size between the two diffusivity surfaces demonstrates strong consistency between the reference and reconstructed microstructures. Additionally, volumetric diffusivity tensors for thirty reconstructed microstructures are computed and projected onto three principal planes, as shown in Figures \ref{Fig:Ex2_diffusivity_tensor}c-e. The results indicate that the diffusivity tensors of the reconstructed samples exhibit minor fluctuations around the reference tensor, with the mean closely matching the reference value. These findings confirm the statistical equivalence of the reference and reconstructed microstructures in terms of effective diffusivity, supported by the observed consistency and minor variations.

\subsection{Example 3: Mechanical and thermal-conduction properties}
\label{Subsec5.3}
\begin{figure}[H]\footnotesize
	\begin{minipage}[t]{0.36\textwidth}
		\centering  
		\includegraphics[width=1.0\textwidth]{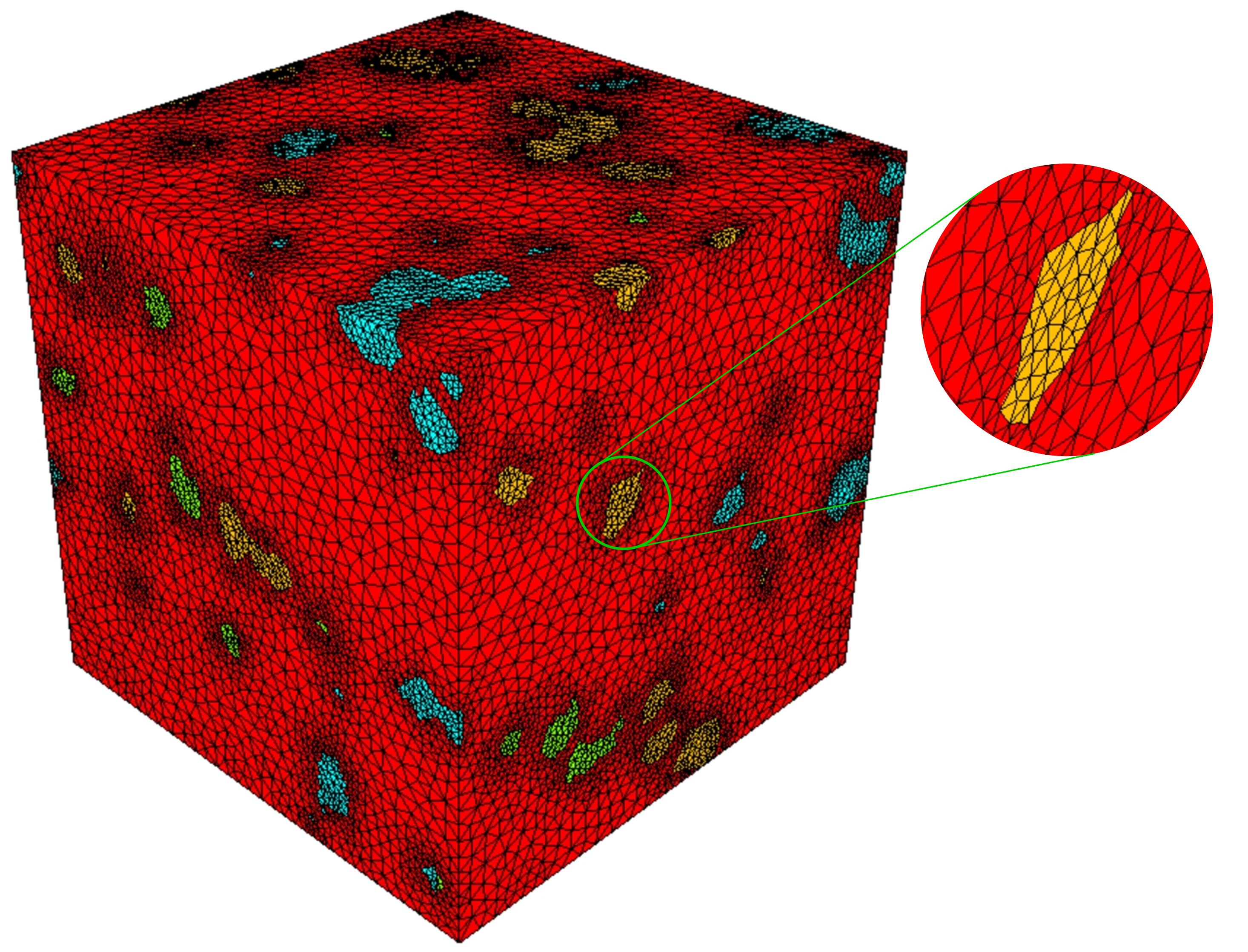}
		\text{(a) Image-based meshing}
	\end{minipage}  
    \smallskip
	\begin{minipage}[t]{0.32\textwidth}  
		\centering  
		\includegraphics[width=0.95\textwidth]{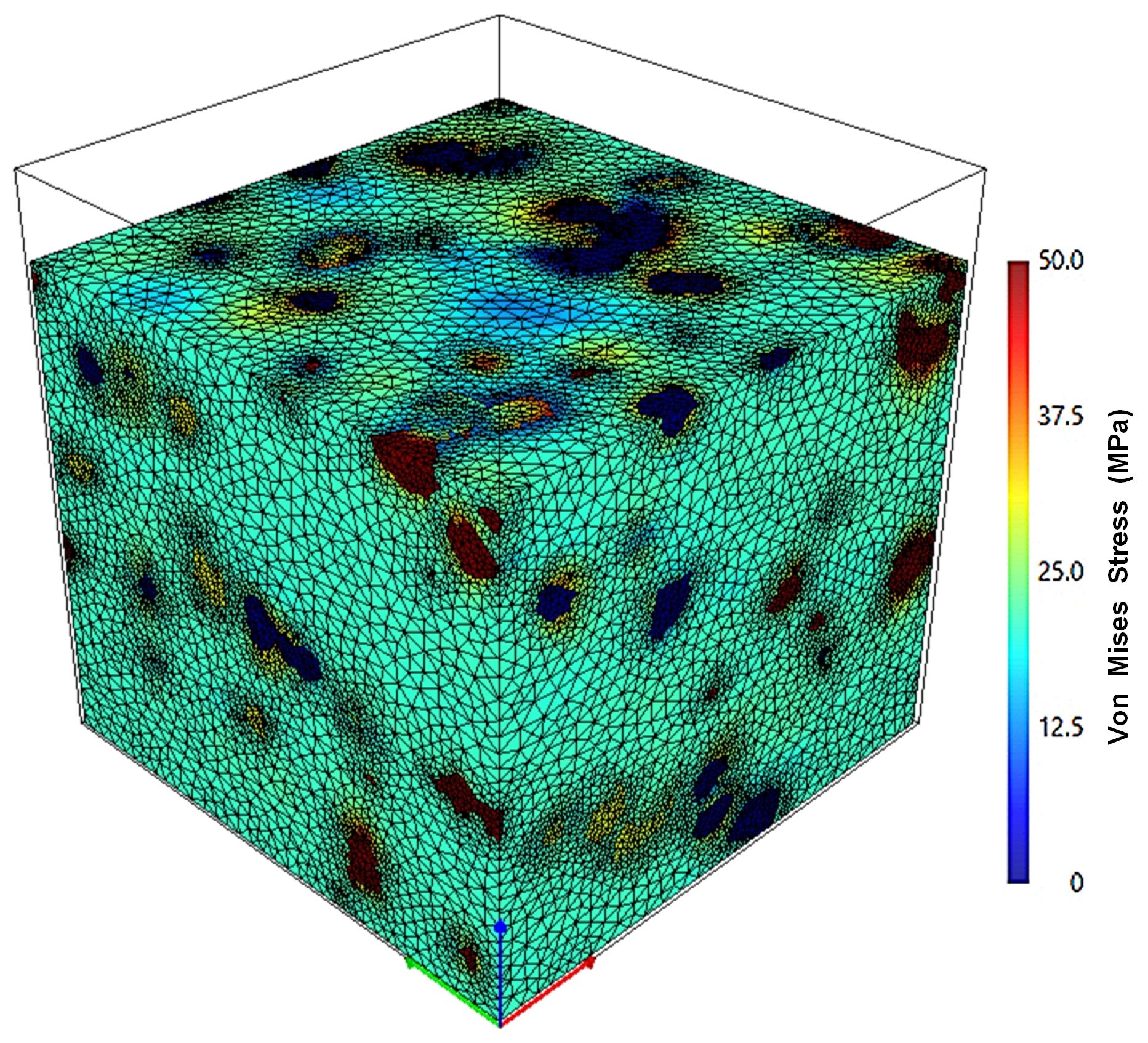}
		\text{(b) Stress field}
	\end{minipage}  
    \smallskip
	\begin{minipage}[t]{0.32\textwidth}  
		\centering  
		\includegraphics[width=0.95\textwidth]{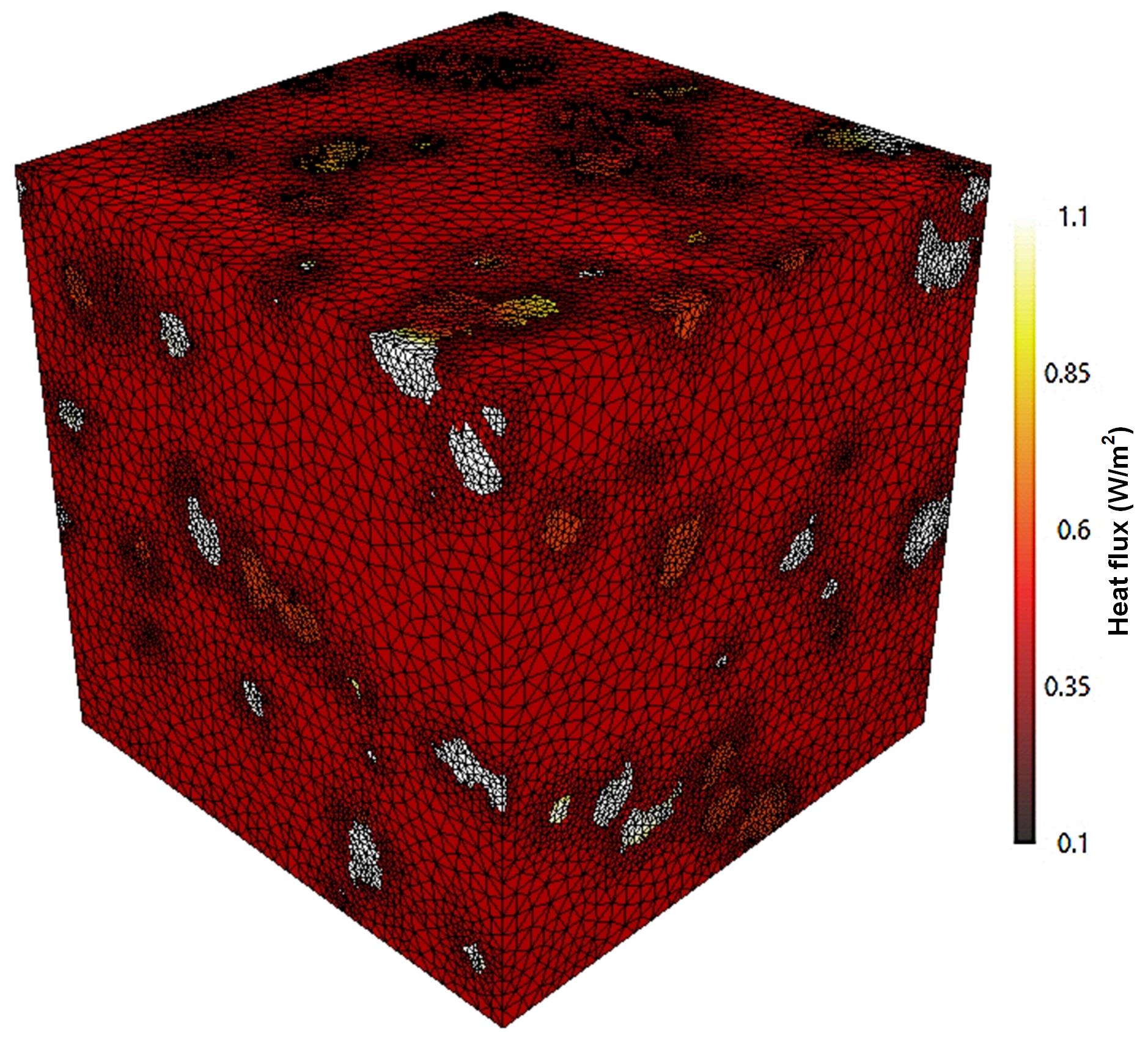}
		\text{(c) Heat flux field}
	\end{minipage}  
    \smallskip
	\begin{minipage}[t]{0.36\textwidth} 
		\centering  
		\includegraphics[width=1.0\textwidth]{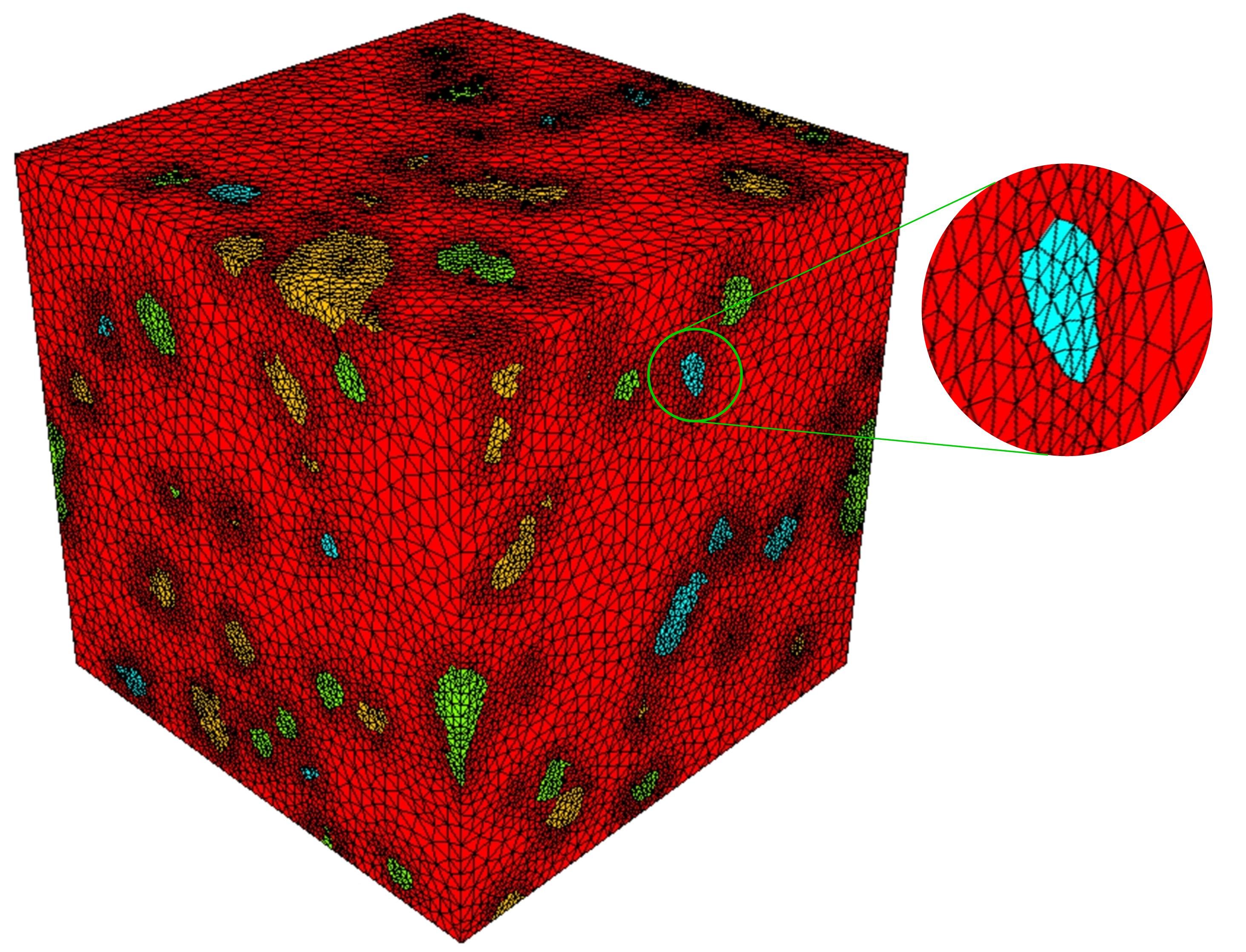}
		\text{(d) Image-based meshing}
	\end{minipage}  
	\begin{minipage}[t]{0.32\textwidth}  
		\centering  
		\includegraphics[width=0.95\textwidth]{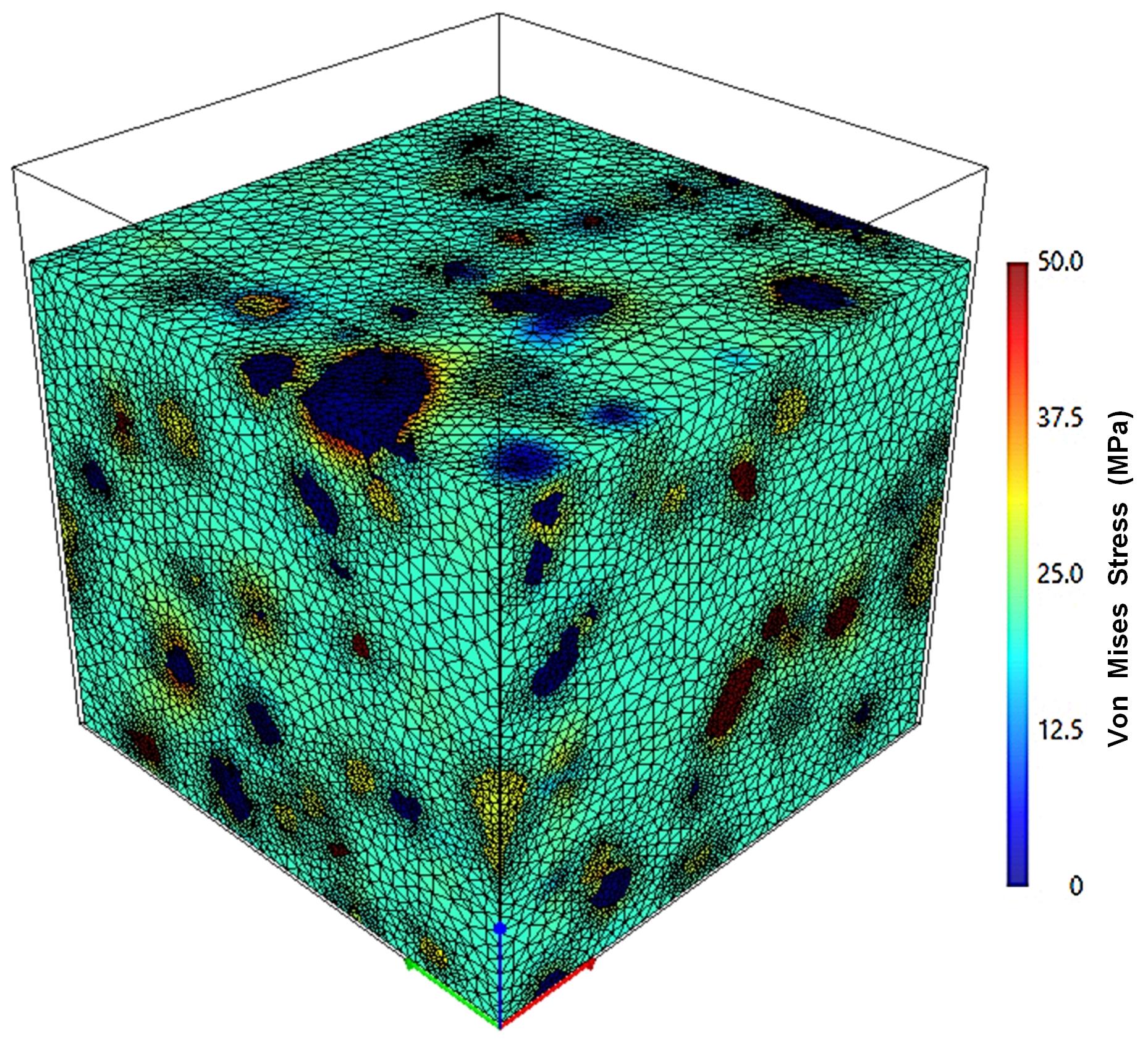} 
		\text{(e) Stress field}
	\end{minipage}  
	\begin{minipage}[t]{0.32\textwidth}  
		\centering  
		\includegraphics[width=0.95\textwidth]{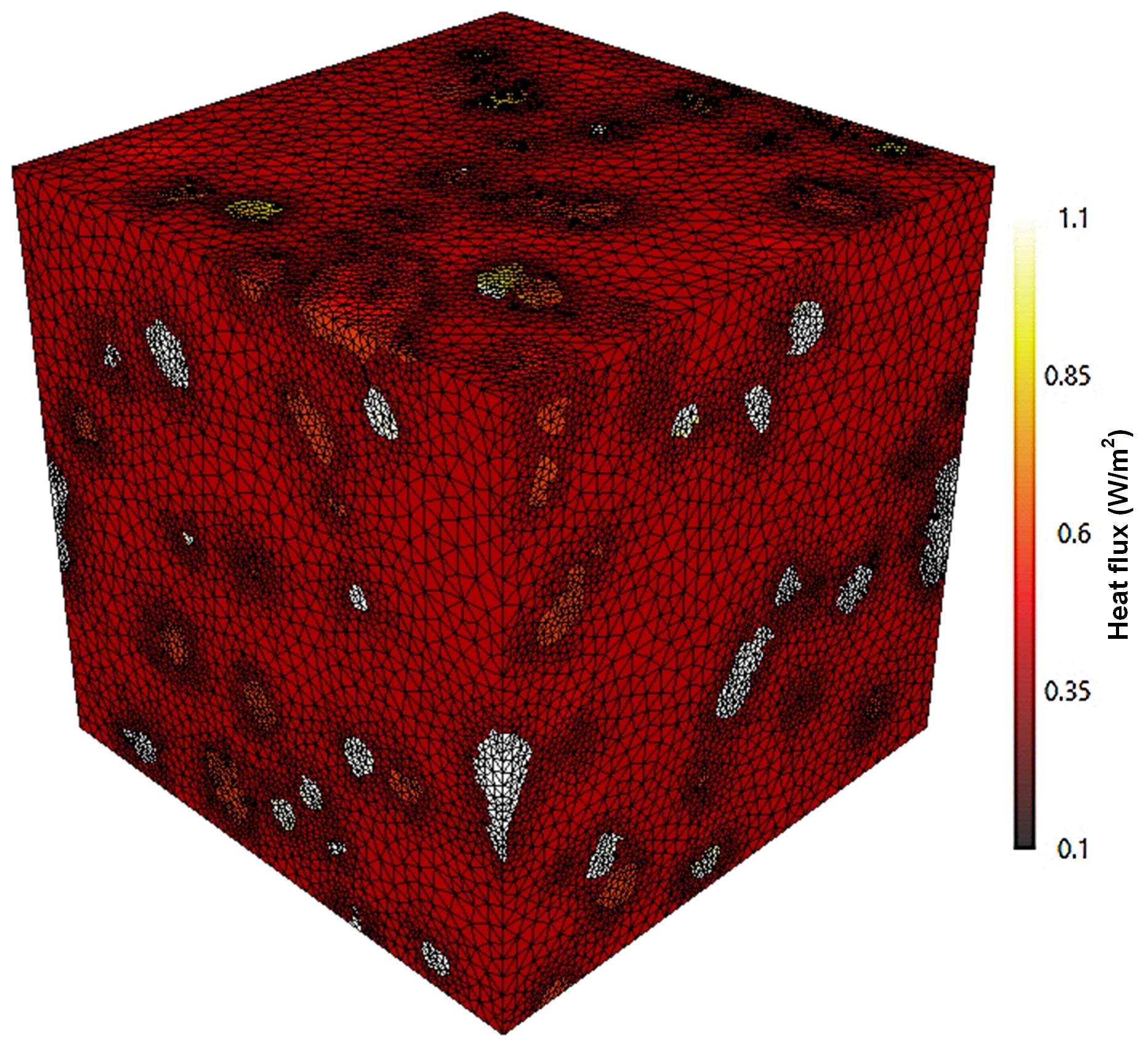}  
		\text{(f) Heat flux field}
	\end{minipage}
	\caption{(a) and (d) Mesh generation on multiphase microstructures, where red regions correspond to calcium silicate hydrate, cyan regions represent anhydrous grains, green regions indicate portlandite (calcium hydroxide) crystals, and yellow regions denote water-filled pores; (b) and (e) Von Mises stress fields under axial compression; (c) and (f) Heat flux distribution at steady state.}
	\label{Fig:Ex3_numerical_simulation}
\end{figure}

\vspace{-2pt}
The mechanical and thermal-conduction properties of composite cement paste play a pivotal role in its structural integrity and thermal performance. Mechanical properties, such as stiffness and strength, enable the material to withstand external loads, while effective thermal conductivity determines its heat transfer capability and impacts its susceptibility to thermal cracking. These properties are governed by the microstructure, including phase composition, porosity, and the connectivity of the solid matrix and pore network. To evaluate these properties, image-based meshing is applied to both reference and reconstructed 3D microstructures, facilitating micro-mechanical and thermal simulations. As depicted in Figure \ref{Fig:Ex3_numerical_simulation}a and d, the unstructured tetrahedral meshes are designed to precisely capture the intricate interfaces and interactions between phases, accurately representing the complex microstructures of this heterogeneous material.

\begin{figure}[H]\footnotesize
	\begin{minipage}[t]{0.5\textwidth}
		\centering  
		\includegraphics[width=0.8\textwidth]{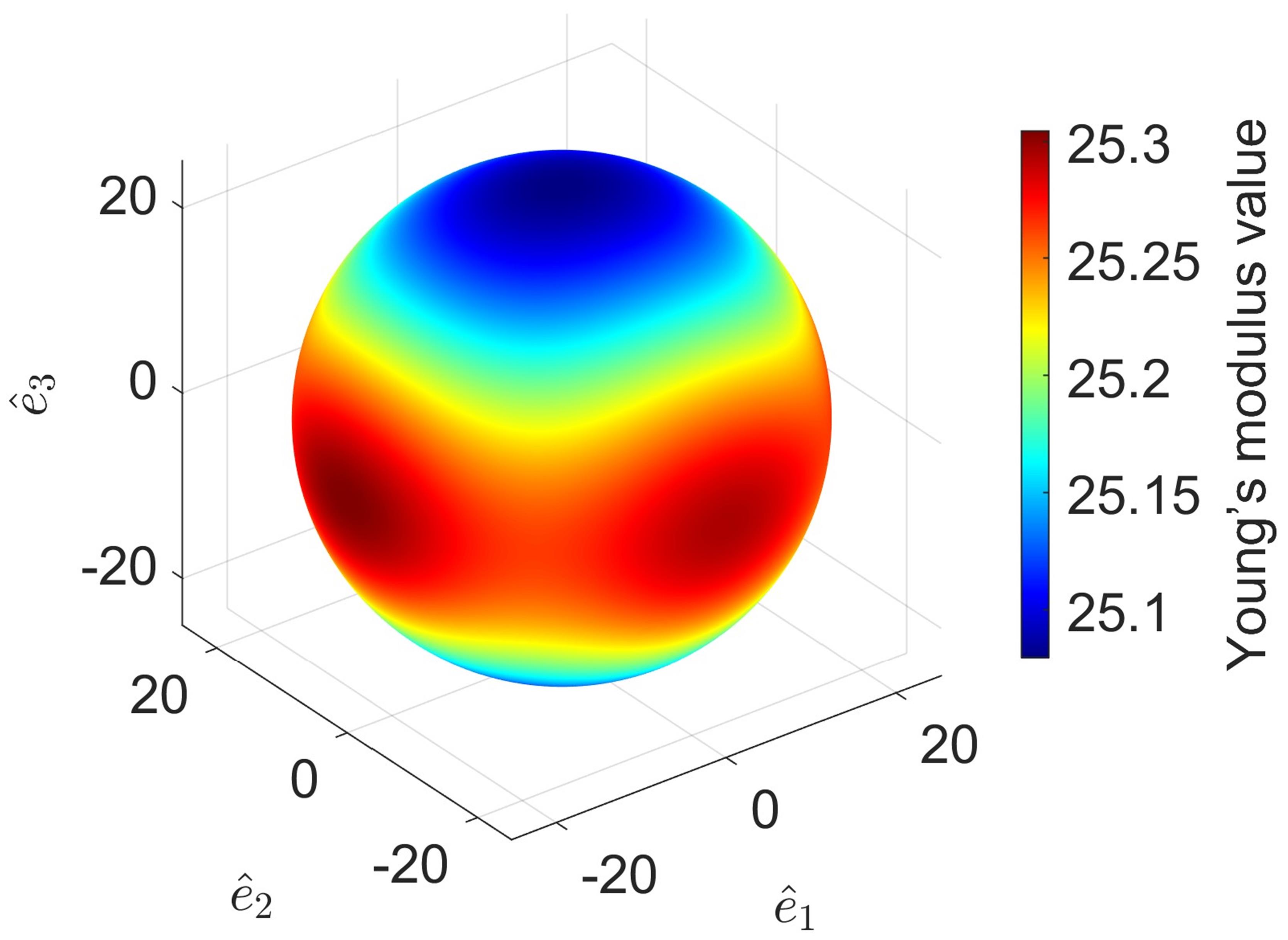}
		\text{(a) Volumetric representation of stiffness tensor (reference)}
	\end{minipage}  
    \smallskip\smallskip 
	\begin{minipage}[t]{0.5\textwidth}  
		\centering  
		\includegraphics[width=0.8\textwidth]{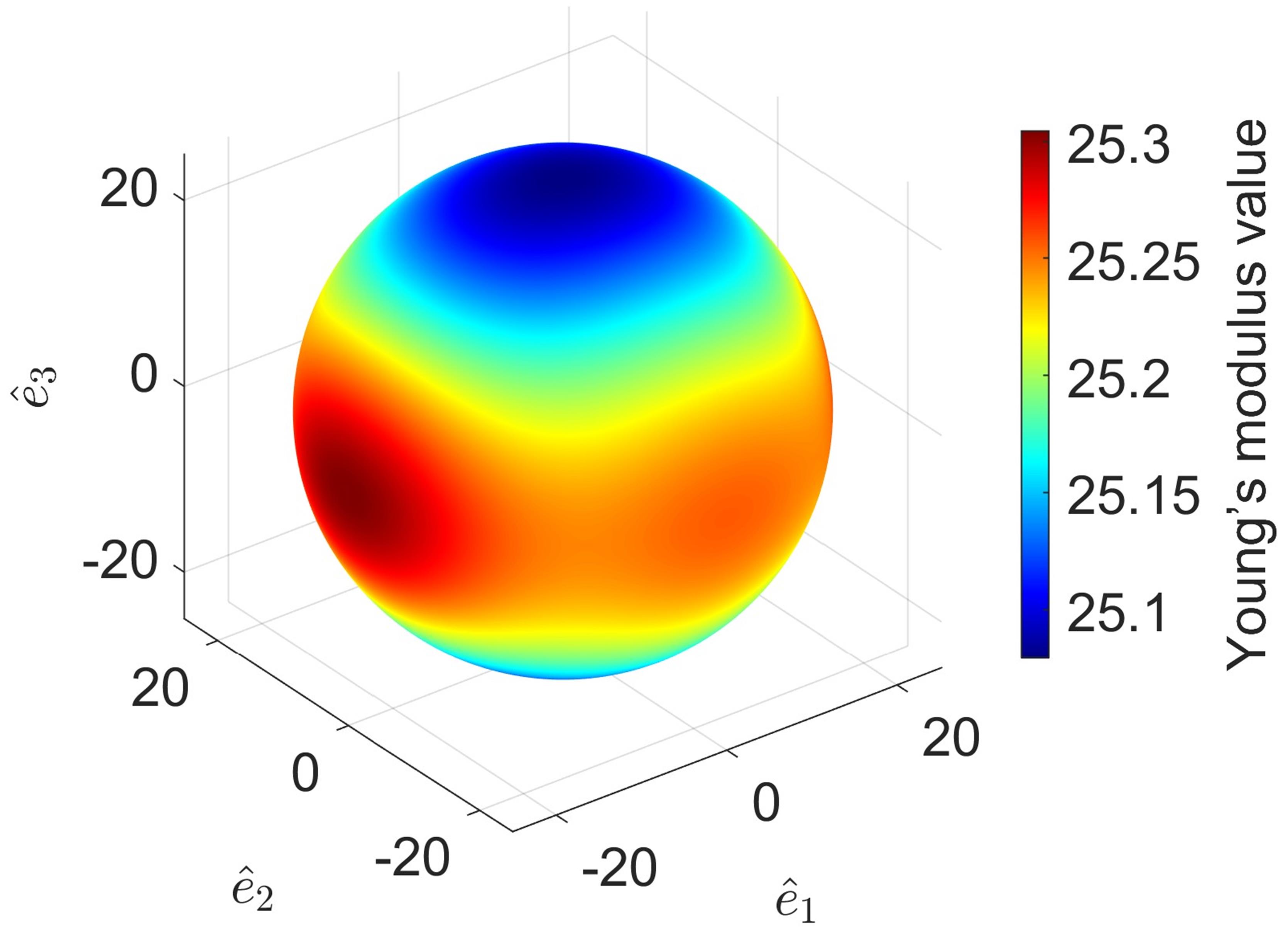}
		\text{(b) Volumetric representation of stiffness tensor (reconstruction)}
	\end{minipage}  
    \smallskip\smallskip  
    \begin{minipage}[t]{0.33\textwidth} 
		\centering  
		\includegraphics[width=0.98\textwidth]{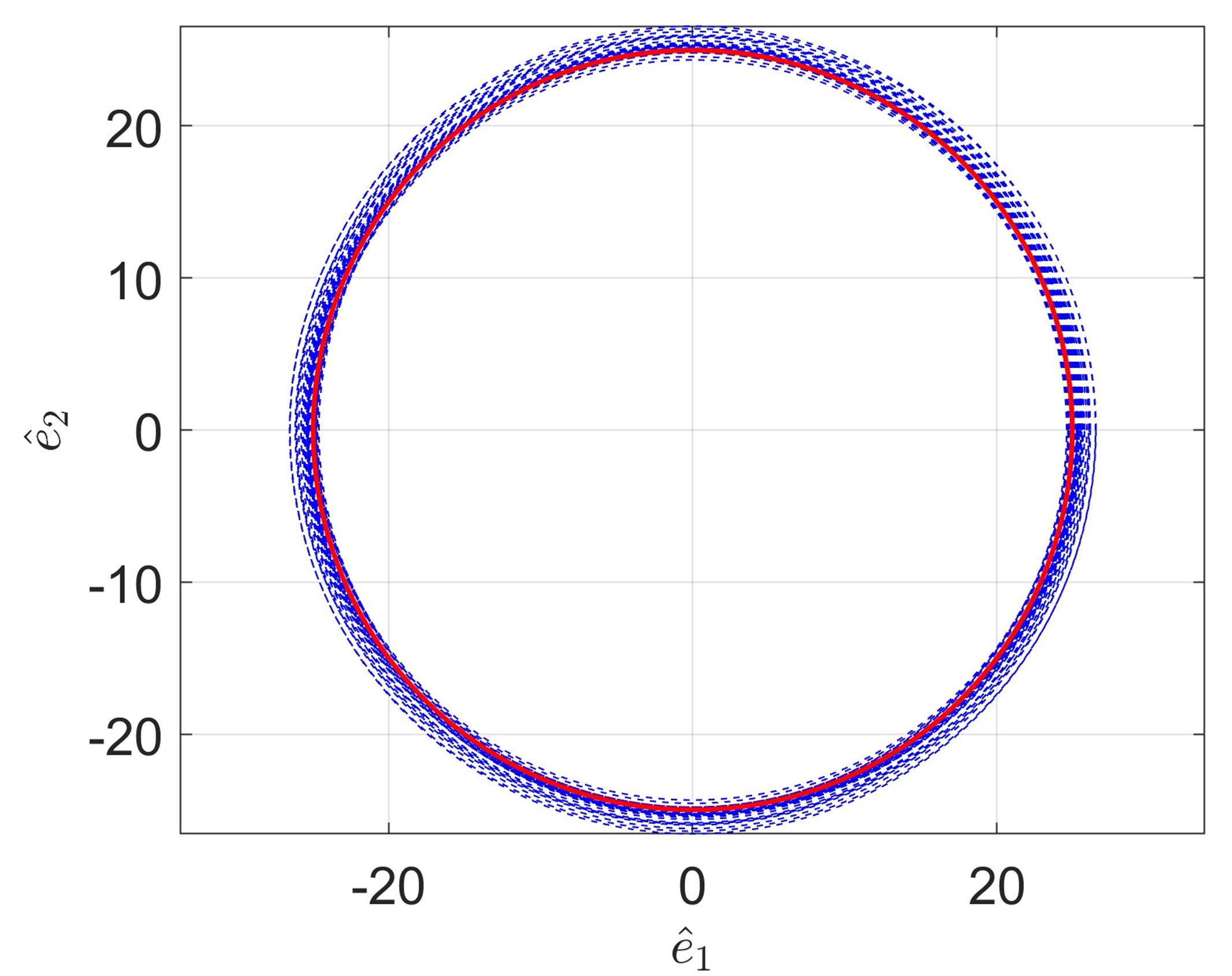}
		\text{(c) Stiffness tensor on $xy$-plane}
	\end{minipage}  
	\begin{minipage}[t]{0.33\textwidth}  
		\centering  
		\includegraphics[width=0.98\textwidth]{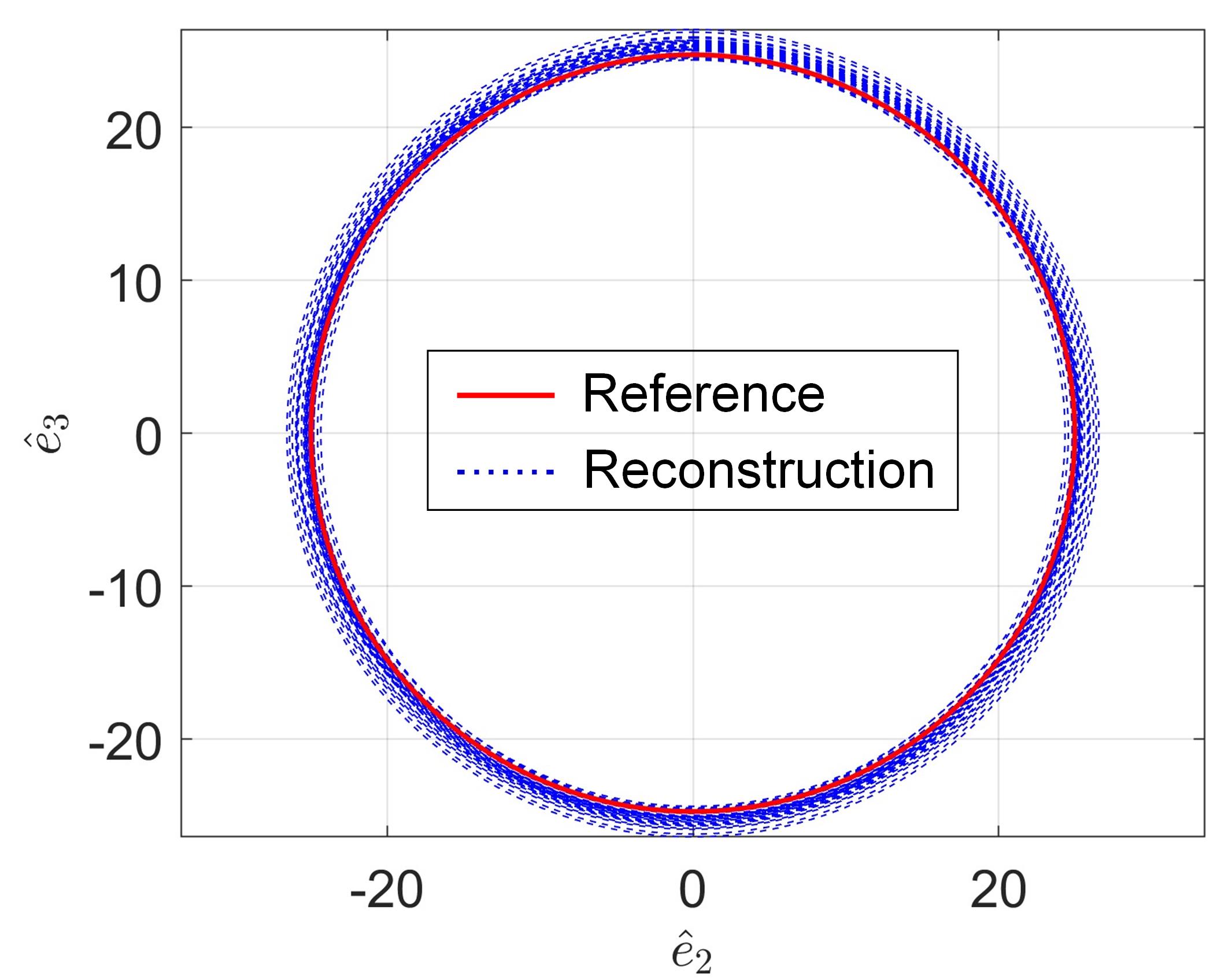} 
		\text{(d) Stiffness tensor on $yz$-plane}
	\end{minipage}  
	\begin{minipage}[t]{0.33\textwidth}  
		\centering  
		\includegraphics[width=0.98\textwidth]{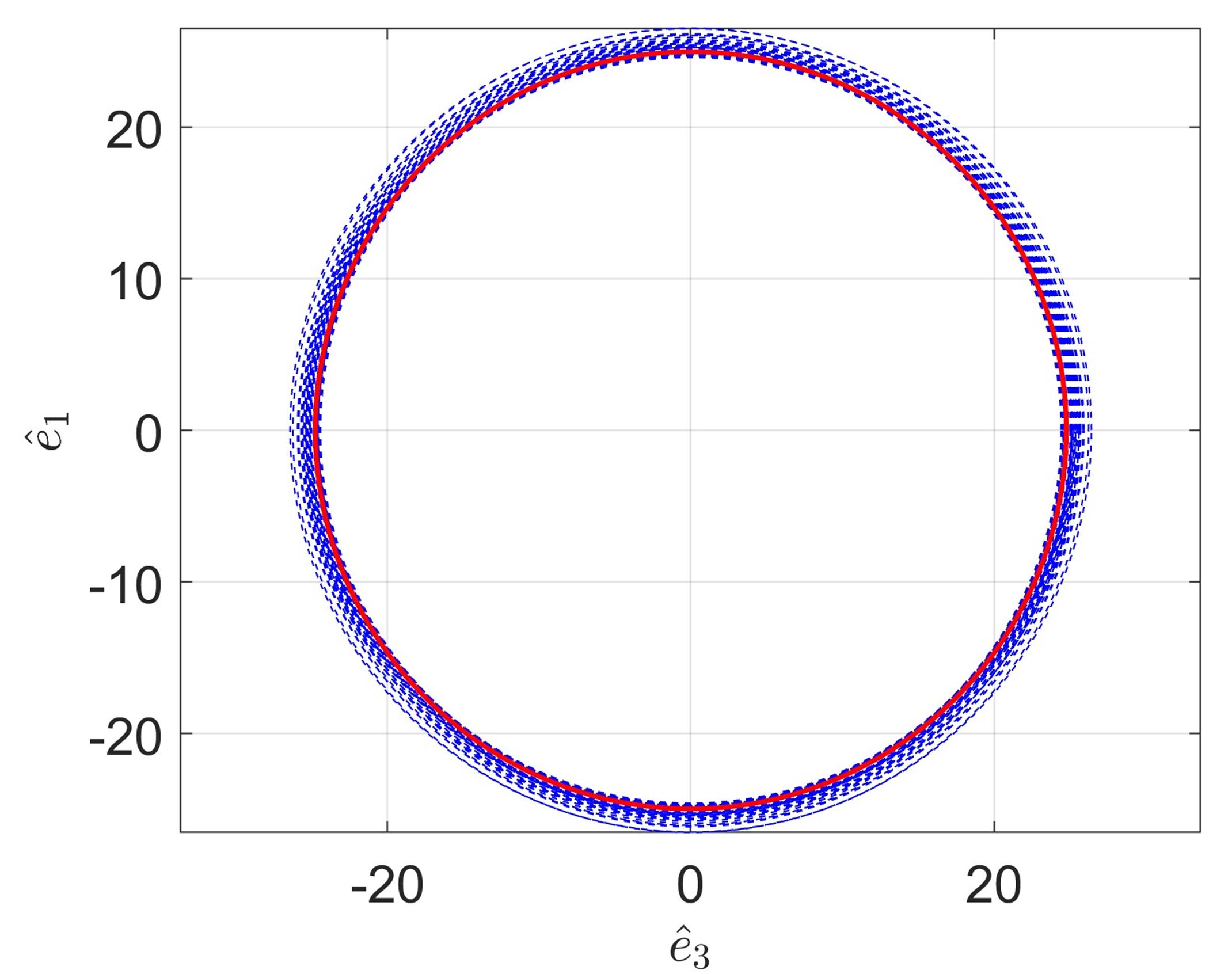}  
		\text{(e) Stiffness tensor on $zx$-plane}
	\end{minipage}
	\caption{Comparison of stiffness tensors between the reference and reconstructed 3D microstructures of the composite cement past.}
	\label{Fig:Ex3_stiffness_tensor}
\end{figure}

FEM is applied to the RVEs of composite cement paste to extract stress-strain relationships, enabling the derivation of the stiffness tensor. Figures \ref{Fig:Ex3_numerical_simulation}b and e illustrate the Von Mises stress fields under axial compression for the reference and reconstructed 3D microstructures, offering insights into their mechanical behaviour. Volumetric stiffness tensors are calculated for both the reference and a representative reconstructed 3D microstructure, as illustrated in Figures \ref{Fig:Ex3_stiffness_tensor}a and b. The resulting elasticity surfaces exhibit comparable shape and size, underscoring the alignment between the reference and reconstructed microstructures. Furthermore, stiffness tensors of 30 reconstructed microstructure samples are analysed and projected onto three principal planes, as shown in Figures \ref{Fig:Ex3_stiffness_tensor}c-e. The projections indicate minor deviations from the reference tensor, with the average values closely aligning with the reference stiffness tensor. These findings highlight the reconstructed microstructures' ability to statistically replicate the elastic properties of the reference, supported by their consistency and minor variations.

To compute the effective thermal conductivity tensors, FEM simulations are performed on image-based meshes of the composite cement paste. These simulations solve steady-state heat conduction equations under imposed temperature gradients in specific directions within the heterogeneous microstructures. Figures \ref{Fig:Ex3_numerical_simulation}c and f depict the heat flux fields for the reference and reconstructed thirty microstructures, illustrating the pathways of heat transfer through the solid phases and water-filled pores.
The effective thermal conductivity tensor relates the heat flux to the temperature gradient through Fourier's law, expressed as:
\begin{equation}
\begin{bmatrix}
H_{x} \\
H_{y} \\
H_{z}
\end{bmatrix}
=
\begin{bmatrix}
\chi_{11} & \chi_{12} & \chi_{13} \\
\chi_{21} & \chi_{22} & \chi_{23} \\
\chi_{31} & \chi_{32} & \chi_{33}
\end{bmatrix}
\begin{bmatrix}
\smallskip
\frac{ \partial T}{ \partial x } \\\smallskip
\frac{ \partial T}{ \partial y } \\
\frac{ \partial T}{ \partial z }
\end{bmatrix}
\label{Eq:Fick's_law}
\end{equation}
where $H_x$, $H_y$ and $H_z$ represent the heat flux components along $x$-, $y$- and $z$- directions, respectively; $T$ represents the temperature; and $\chi_{ij}$ is the element of the effective thermal conductivity tensor.

\begin{figure}[h]\footnotesize
	\begin{minipage}[t]{0.5\textwidth}
		\centering  
		\includegraphics[width=0.8\textwidth]{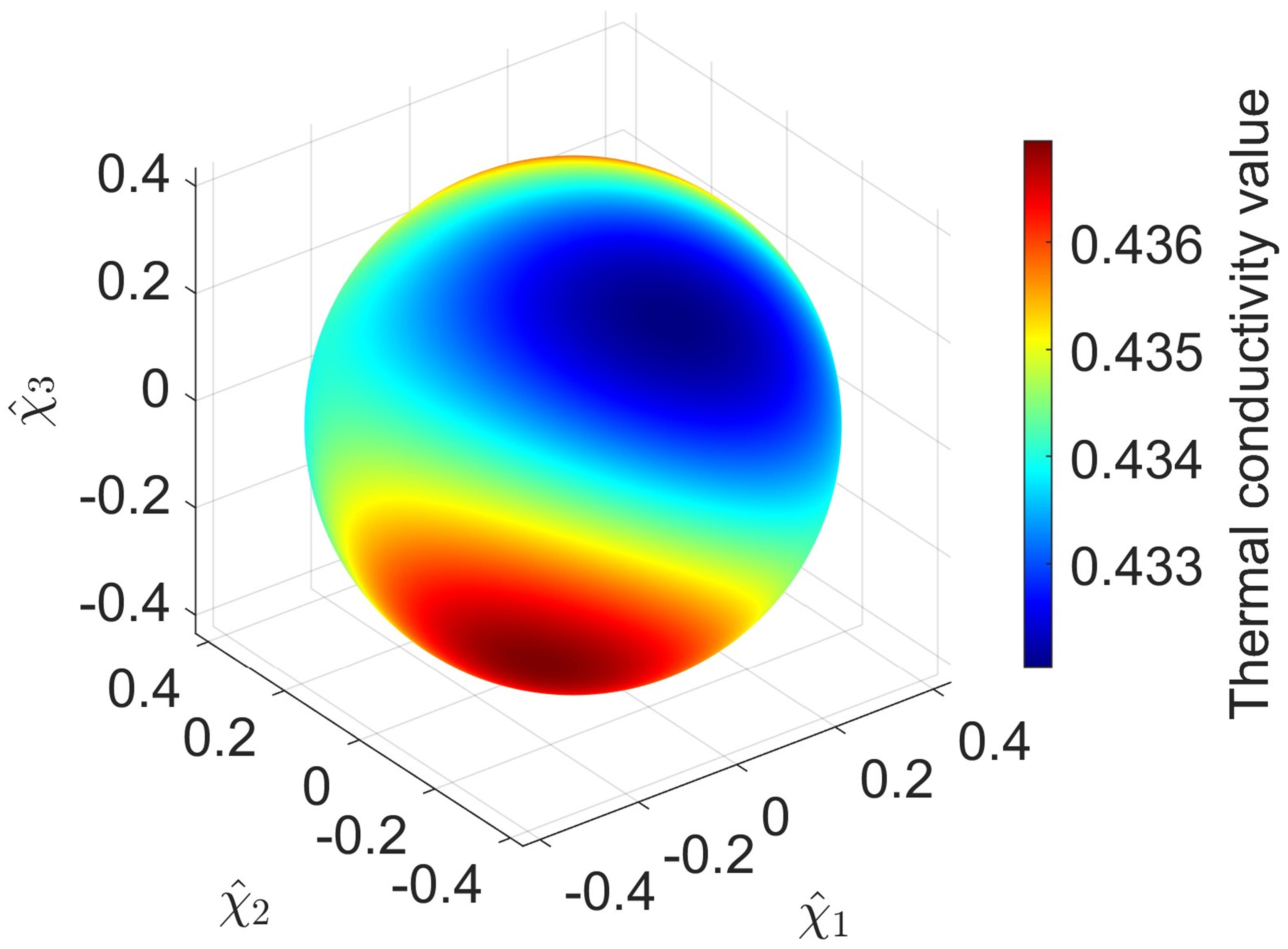}
		\text{(a) Volumetric representation of conductivity tensor (reference)}
	\end{minipage}  
    \smallskip\smallskip 
	\begin{minipage}[t]{0.5\textwidth}  
		\centering  
		\includegraphics[width=0.8\textwidth]{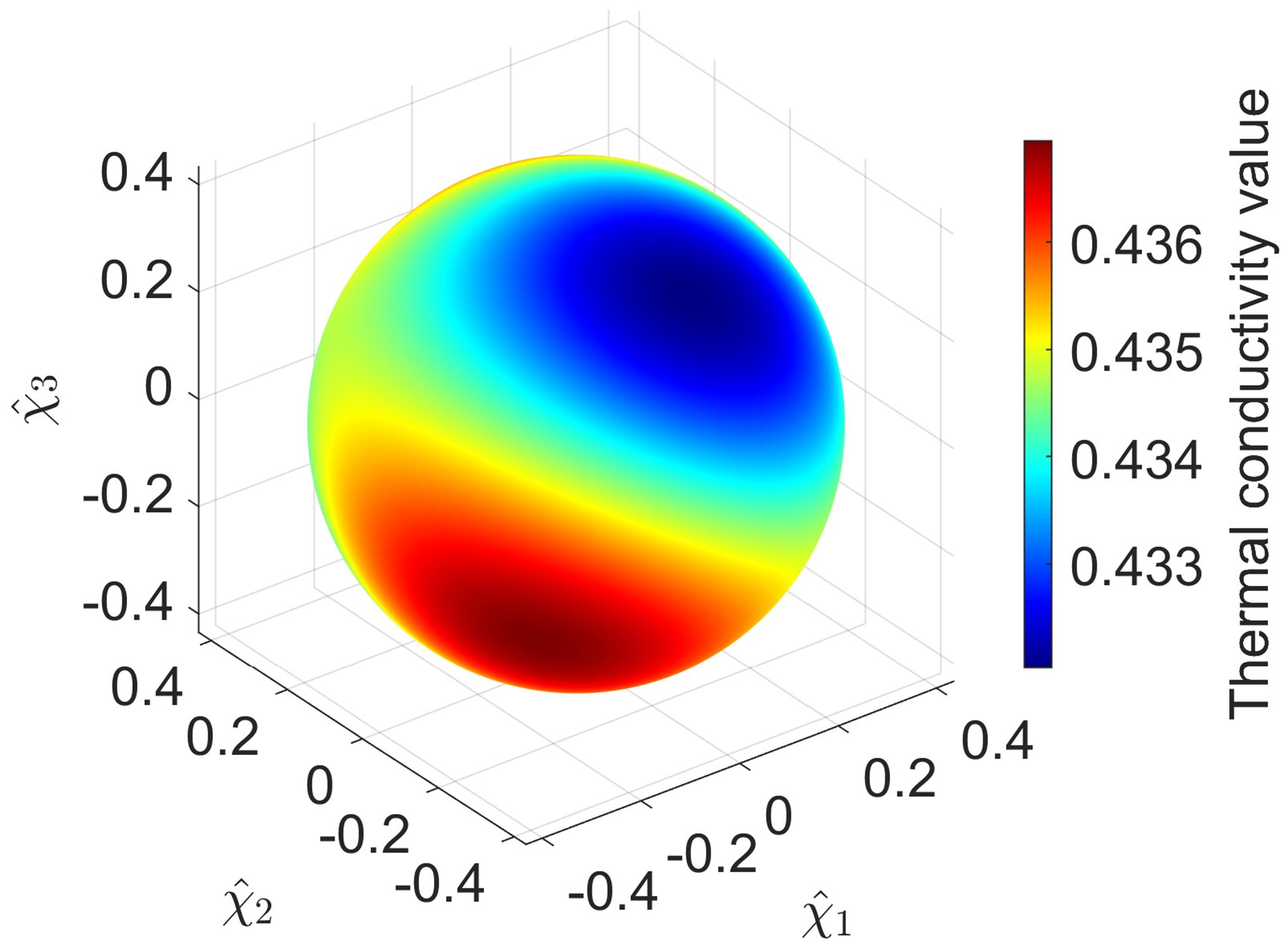}
		\text{(b) Volumetric representation of conductivity tensor (reconstruction)}
	\end{minipage}  
    \smallskip\smallskip  
    \begin{minipage}[t]{0.33\textwidth} 
		\centering  
		\includegraphics[width=0.98\textwidth]{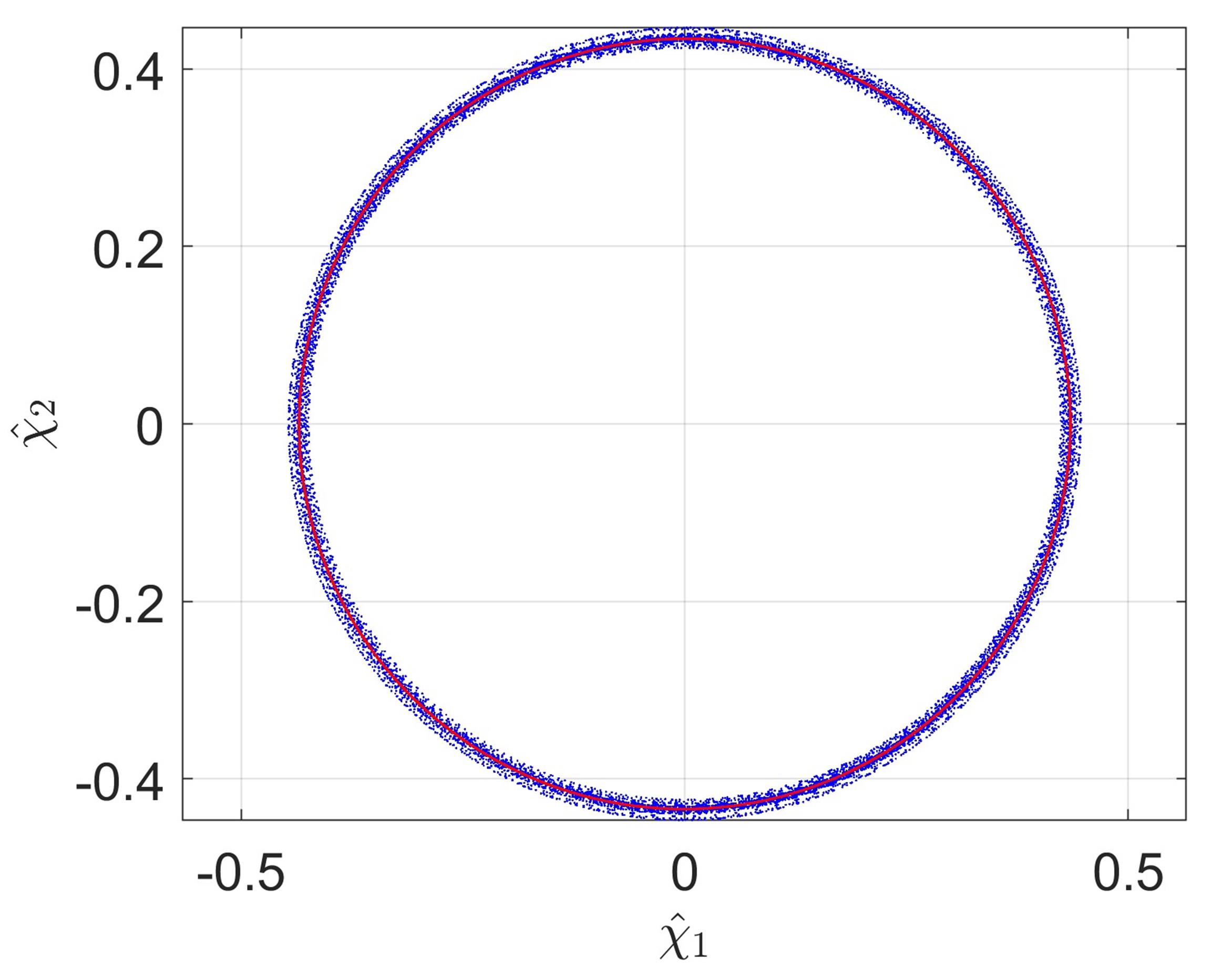}
		\text{(c) Thermal conductivity tensor on $xy$-plane}
	\end{minipage}  
	\begin{minipage}[t]{0.33\textwidth}  
		\centering  
		\includegraphics[width=0.98\textwidth]{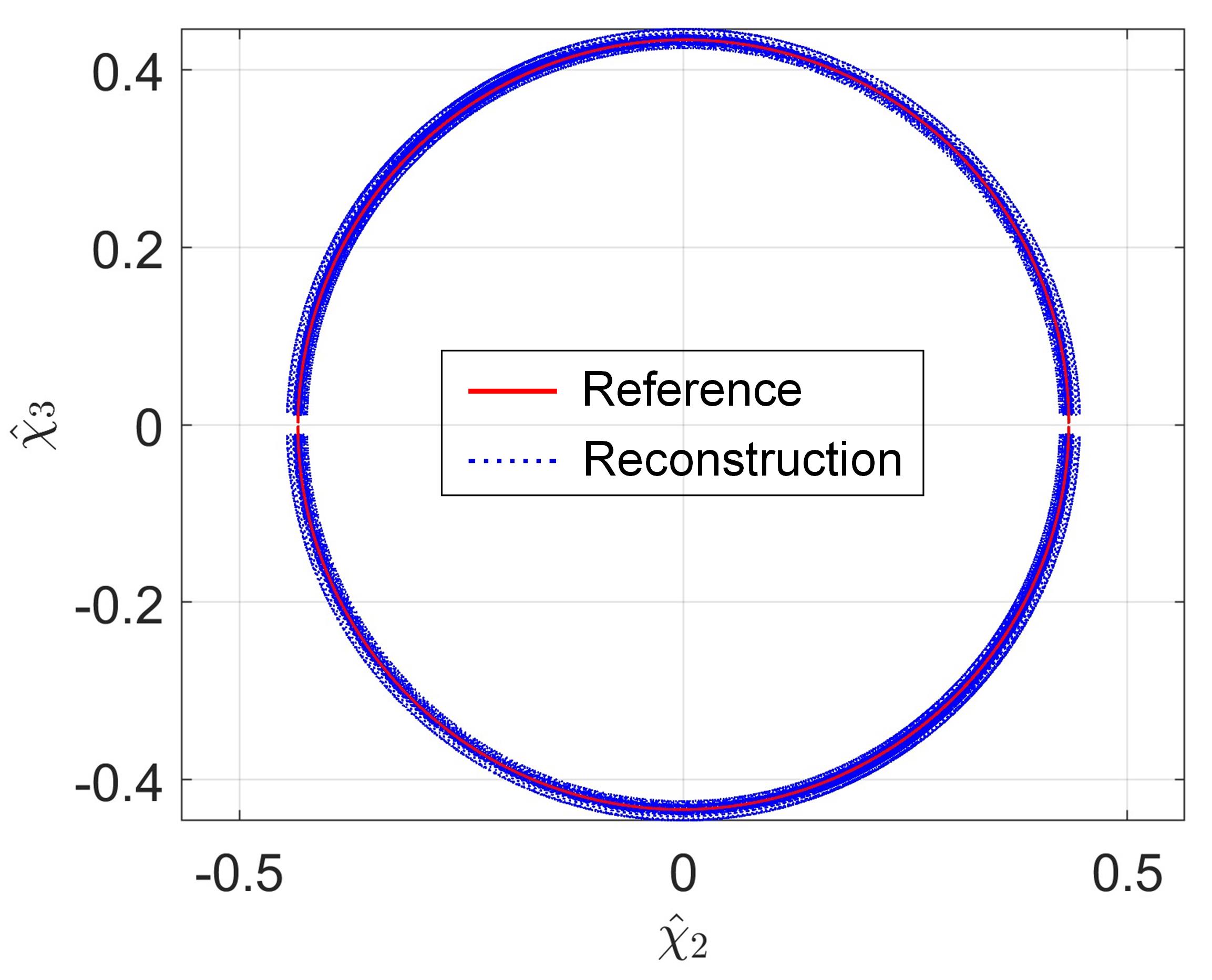} 
		\text{(d) Thermal conductivity tensor on $yz$-plane}
	\end{minipage}  
	\begin{minipage}[t]{0.33\textwidth}  
		\centering  
		\includegraphics[width=0.98\textwidth]{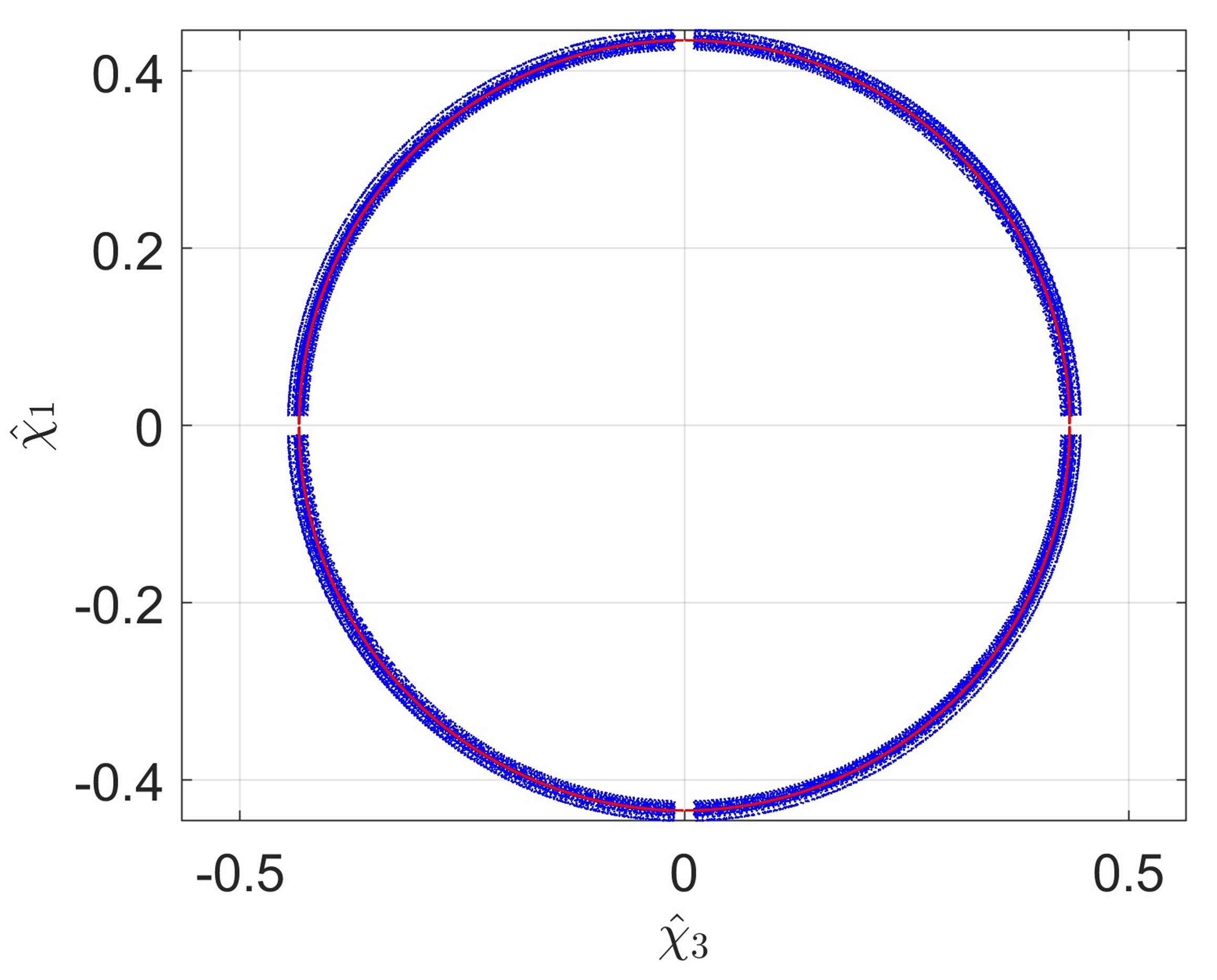}  
		\text{(e) Thermal conductivity tensor on $zx$-plane}
	\end{minipage}
	\caption{Comparison of effective thermal conductivity tensors between the reference and reconstructed 3D microstructures of the composite cement past.}
	\label{Fig:Ex3_conductivity_tensor}
\end{figure}

Figures \ref{Fig:Ex3_conductivity_tensor}a and b illustrate the 3D volumetric distributions of the effective thermal conductivity tensors for both the reference and a selected reconstructed microstructure. The comparable shape and size of the conductivity surfaces for both structures highlight a high degree of alignment between the reference and the reconstructed 3D microstructures. Furthermore, the thermal conductivity tensors for thirty reconstructed microstructures are calculated and projected onto three principal planes, as shown in Figures \ref{Fig:Ex3_conductivity_tensor}c-e. The analysis reveals that the tensors from the reconstructed samples fluctuate slightly around the reference tensor, with their average value closely approximating the reference. These results underscore the statistical equivalence between the reference and reconstructed microstructures, validating the accuracy of the stochastic microstructure reconstruction process in preserving effective thermal conductivity.

\section{Discussion and conclusions}
\label{Section6:Discussion_and_conclusions}
\vspace{-2pt}
\subsection{Discussion}
\label{Subsec6.1:Discussion}
\vspace{-2pt}
The proposed SENN-based framework for stochastic microstructure reconstruction is a versatile tool for advancing microstructure-property analyses in materials science, energy storage, and other related fields.
By reducing reliance on costly volumetric microscopy techniques, this framework is particularly valuable for studying opaque multiphase composites, where acquiring sufficient 3D digital microstructures is challenging. Its ability to synthesise statistically equivalent 3D microstructures from limited 2D exemplars enables large-scale statistical analyses, providing a cost-efficient pathway to better understand and optimise the performance and functionality of multiphase composites.
The key features of this innovative SENN-based framework include: 
\begin{itemize}[topsep=0pt]
\setlength{\itemsep}{0pt}
\setlength{\parsep}{0pt}
\setlength{\parskip}{0pt}
    \item \textbf{Multiphase reconstruction}: Unlike most reconstruction methods tailored to bi-phase microstructures, this framework specialises in reconstructing multiphase microstructures with complex morphologies. It demonstrates a strong capability to accurately capture the random distribution of phases and the intricate interfaces between them;
    \item \textbf{Accuracy and efficiency}: By integrating SENN with Gibbs sampling, the framework achieves precise statistical equivalence between reconstructed and reference microstructures while maintaining computational efficiency. Its hierarchical characterisation and multi-level reconstruction approach further ensure seamless capture and preservation of local, regional, and global features across various length scales;
    \item \textbf{Small data requirement}: In contrast to traditional machine learning-based approaches that require extensive 3D datasets, this SENN-based framework can reliably synthesise 3D microstructures using only a limited number of 2D exemplars. This feature is particularly advantageous in cases where obtaining volumetric datasets is prohibitively expensive or technically infeasible;
    \item \textbf{Model interpretability}: SENN explicitly encodes morphological statistics, providing clear insights into the model's predictions and fostering trust in the reconstruction process. This interpretability also enhances the understanding of the relationships between 2D morphological features and the 3D counterparts.   
\end{itemize}

However, certain limitations of the SENN-based framework should be noted. First, the framework relies on the quality and diversity of the input 2D exemplars; inadequate representation of critical features in 2D data may introduce biases in the reconstructed 3D microstructures. Second, the Markov random field assumption used for statistical characterisation may be less effective for multiphase microstructures with extremely strong heterogeneity. Third, while Gibbs sampling is effective for synthesising microstructures, it can become computationally intensive for extremely large datasets or less effective for highly complex morphologies. 
Future work could address these challenges by exploring more scalable synthesis techniques and enhancing the framework's adaptability to multiphase composites with evolving morphologies. Furthermore, integrating the framework with advanced optimisation algorithms could enhance its predictive capability, enabling direct linkage between desired macroscopic properties and reconstructed microstructures. Such developments would further expand the framework’s applicability and utility in real-world material design and analysis.

\subsection{Conclusions}
\label{Section6.2:Conclusions}
\vspace{-2pt}
This study presents a novel machine learning-based framework for stochastic microstructure reconstruction, addressing the critical challenge of generating 3D digital microstructures of random multiphase composites from limited 2D exemplars (cross-sectional images). Leveraging the innovative Statistics-Encoded Neural Network (SENN), the framework encodes and extrapolates 2D morphological statistics to synthesise statistically equivalent 3D microstructures, significantly reducing reliance on costly volumetric microscopy techniques. By integrating hierarchical characterisation and multi-level reconstruction, the SENN-based framework captures microstructural features at local, regional, and global scales with high fidelity, achieving statistical equivalence to reference microstructures. The reconstructed 3D microstructures exhibit strong morphological similarity to reference datasets, as confirmed by quantitative descriptors such as volume fraction, surface area density, triple-phase boundary density, and geometrical tortuosity. Furthermore, quantitative analysis of effective macroscopic properties, including stiffness, permeability, diffusivity, and thermal conductivity tensors, further affirmed the statistical equivalence and practical applicability of the reconstructed 3D microstructures for image-based poro/micro-mechanical modelling.

In conclusion, this work makes several key contributions to image-based poro/micro-mechanical modelling and MPR analysis. First, the SENN-based framework provides a cost-efficient solution for generating 3D multiphase microstructures, mitigating the need for extensive volumetric microscopy imaging. Second, its integration of hierarchical characterisation and multi-level reconstruction approaches ensures the accurate representation and capture of microstructural features across multiple relevant length scales, making the framework particularly effective for complex, multiphase composites. Finally, by enabling large-scale statistical analysis of MPRs, the SENN-based framework opens new avenues for materials design and optimisation across diverse applications, including materials science, structural engineering, and electrochemical devices.

\appendix
\renewcommand\thetable{\Alph{section}\arabic{table}} 
\section{Parameter settings for training statistics-encoded neural networks}
\label{appendix}
\setcounter{table}{0}
To train SENN models effectively for approximating 2D morphological statistics, data events extracted from 2D exemplars (cross-sectional images) are randomly divided into three subsets: 70\% for training, 20\% for validation, and 10\% for testing. This data-splitting strategy reduces the risk of overfitting and ensures robust model performance. The theoretical framework for SENN-based characterisation is outlined in Section \ref{Section2:Statistical_microstructure_characterization}, while the parameter settings for training SENN models in the three case studies are provided in Table \ref{Tab:SENN_parameters}.

\begin{table}[h]
    \fontsize{8}{10}\selectfont
    \renewcommand\arraystretch{1.02}
	\caption{Parameter settings for multi-level SENN-based characterisation of multiphase microstructures}
	\begin{tabular}{|c|c|c|cccccc|}
		\hline
		\multirow{3}{*}{\thead{\textbf{Multiphase} \\\textbf{composite}}}    & \multirow{3}{*}{\textbf{Level}}    & \multirow{3}{*}{\textbf{Size of data templates}} & \multicolumn{6}{c|}{\textbf{Statistics-encoded neural network}}                                                                                                                                                                                                                                        \\ \cline{4-9} 
		&                           &                              & \multicolumn{1}{c|}{\multirow{2}{*}{\thead{Learning\\rate}}}                  & \multicolumn{1}{c|}{\multirow{2}{*}{\thead{Epoch\\number}}}        & \multicolumn{4}{c|}{Number of neurons in each hidden layer}                                                                                                                    \\ \cline{6-9} 
		&                           &                              & \multicolumn{1}{c|}{}                      & \multicolumn{1}{c|}{}                              & \multicolumn{1}{c|}{Layer 1}                  & \multicolumn{1}{c|}{Layer 2}                 & \multicolumn{1}{c|}{Layer 3}                 & Layer 4                 \\ \hline
		\multirow{12}{*}{\thead{Silver-based \\electrode}}        & \multirow{4}{*}{Low}  & Data collection on three planes & \multicolumn{1}{c|}{\multirow{4}{*}{0.00001}}  & \multicolumn{1}{c|}{\multirow{4}{*}{200}}          & \multicolumn{1}{c|}{\multirow{4}{*}{128}}  & \multicolumn{1}{c|}{\multirow{4}{*}{64}}  & \multicolumn{1}{c|}{\multirow{4}{*}{16}}  & \multirow{4}{*}{--}  \\ \cline{3-3}
		&                           & $xy$-plane: $R_0$=10, $R_1$=8, $R_2$=12   & \multicolumn{1}{c|}{}                          & \multicolumn{1}{c|}{}                              & \multicolumn{1}{c|}{}                      & \multicolumn{1}{c|}{}                     & \multicolumn{1}{c|}{}                     &                      \\ \cline{3-3}
		&                           & $yz$-plane: $R_0$=16, $R_1$=15, $R_2$=15  & \multicolumn{1}{c|}{}                          & \multicolumn{1}{c|}{}                              & \multicolumn{1}{c|}{}                      & \multicolumn{1}{c|}{}                     & \multicolumn{1}{c|}{}                     &                         \\ \cline{3-3}
		&                           & $zx$-plane: $R_0$=12, $R_1$=10, $R_2$=9  & \multicolumn{1}{c|}{}                          & \multicolumn{1}{c|}{}                              & \multicolumn{1}{c|}{}                      & \multicolumn{1}{c|}{}                     & \multicolumn{1}{c|}{}                     &                      \\ \cline{2-9} 
		& \multirow{4}{*}{Middle} & Data collection on three planes  & \multicolumn{1}{c|}{\multirow{4}{*}{0.00001}} & \multicolumn{1}{c|}{\multirow{4}{*}{200}}        & \multicolumn{1}{c|}{\multirow{4}{*}{128}} & \multicolumn{1}{c|}{\multirow{4}{*}{64}} & \multicolumn{1}{c|}{\multirow{4}{*}{32}} & \multirow{4}{*}{--} \\ \cline{3-3}
		&                           & $xy$-plane: $R_0$=12, $R_1$=15, $R_2$=13  & \multicolumn{1}{c|}{}                          & \multicolumn{1}{c|}{}                              & \multicolumn{1}{c|}{}                      & \multicolumn{1}{c|}{}                     & \multicolumn{1}{c|}{}                     &                      \\ \cline{3-3}
		&                           & $yz$-plane: $R_0$=10, $R_1$=12, $R_2$=10  & \multicolumn{1}{c|}{}                          & \multicolumn{1}{c|}{}                              & \multicolumn{1}{c|}{}                      & \multicolumn{1}{c|}{}                     & \multicolumn{1}{c|}{}                     &                      \\ \cline{3-3}
		&                           & $zx$-plane: $R_0$=10, $R_1$=10, $R_2$=8  & \multicolumn{1}{c|}{}                          & \multicolumn{1}{c|}{}                              & \multicolumn{1}{c|}{}                      & \multicolumn{1}{c|}{}                     & \multicolumn{1}{c|}{}                     &                        \\ \cline{2-9} 
		& \multirow{4}{*}{High} & Data collection on three planes  & \multicolumn{1}{c|}{\multirow{4}{*}{0.00001}} & \multicolumn{1}{c|}{\multirow{4}{*}{250}}      & \multicolumn{1}{c|}{\multirow{4}{*}{128}}   & \multicolumn{1}{c|}{\multirow{4}{*}{64}}  & \multicolumn{1}{c|}{\multirow{4}{*}{32}}  & \multirow{4}{*}{--} \\ \cline{3-3}
		&                           & $xy$-plane: $R_0$=6, $R_1$=16, $R_2$=22  & \multicolumn{1}{c|}{}                          & \multicolumn{1}{c|}{}                              & \multicolumn{1}{c|}{}                      & \multicolumn{1}{c|}{}                     & \multicolumn{1}{c|}{}                     &                      \\ \cline{3-3}
		&                           & $yz$-plane: $R_0$=8, $R_1$=22, $R_2$=21  & \multicolumn{1}{c|}{}                          & \multicolumn{1}{c|}{}                              & \multicolumn{1}{c|}{}                      & \multicolumn{1}{c|}{}                     & \multicolumn{1}{c|}{}                     &                      \\ \cline{3-3}
		&                           & $zx$-plane: $R_0$=8, $R_1$=16, $R_2$=14  & \multicolumn{1}{c|}{}                          & \multicolumn{1}{c|}{}                              & \multicolumn{1}{c|}{}                      & \multicolumn{1}{c|}{}                     & \multicolumn{1}{c|}{}                     &                        \\ \hline
		\multirow{14}{*}{\thead{Porous\\ SOFC\\ anode}} & \multirow{4}{*}{Low}  & Data collection on three planes   & \multicolumn{1}{c|}{\multirow{4}{*}{0.00001}}  & \multicolumn{1}{c|}{\multirow{4}{*}{400}}          & \multicolumn{1}{c|}{\multirow{4}{*}{128}}  & \multicolumn{1}{c|}{\multirow{4}{*}{64}}  & \multicolumn{1}{c|}{\multirow{4}{*}{16}}  & \multirow{4}{*}{--}  \\ \cline{3-3}
		&                           & $xy$-plane: $R_0$=20, $R_1$=18, $R_2$=22  & \multicolumn{1}{c|}{}                          & \multicolumn{1}{c|}{}                              & \multicolumn{1}{c|}{}                      & \multicolumn{1}{c|}{}                     & \multicolumn{1}{c|}{}                     &                      \\ \cline{3-3}
		&                           & $yz$-plane: $R_0$=16, $R_1$=22, $R_2$=22  & \multicolumn{1}{c|}{}                          & \multicolumn{1}{c|}{}                              & \multicolumn{1}{c|}{}                      & \multicolumn{1}{c|}{}                     & \multicolumn{1}{c|}{}                     &                      \\ \cline{3-3}
		&                           & $zx$-plane: $R_0$=18, $R_1$=16, $R_2$=12  & \multicolumn{1}{c|}{}                          & \multicolumn{1}{c|}{}                              & \multicolumn{1}{c|}{}                      & \multicolumn{1}{c|}{}                     & \multicolumn{1}{c|}{}                     &                      \\ \cline{2-9} 
		& \multirow{5}{*}{Middle}  & Data collection on nine planes  & \multicolumn{1}{c|}{\multirow{5}{*}{0.00001}}  & \multicolumn{1}{c|}{\multirow{5}{*}{200}}          & \multicolumn{1}{c|}{\multirow{5}{*}{128}}  & \multicolumn{1}{c|}{\multirow{5}{*}{64}}  & \multicolumn{1}{c|}{\multirow{5}{*}{32}}  & \multirow{5}{*}{--}  \\ \cline{3-3}
		&                           & $xy$-plane: $R_0$=25, $R_1$=14, $R_2$=23   & \multicolumn{1}{c|}{}                          & \multicolumn{1}{c|}{}                              & \multicolumn{1}{c|}{}                      & \multicolumn{1}{c|}{}                     & \multicolumn{1}{c|}{}                     &                      \\ \cline{3-3}
		&                           & $yz$-plane: $R_0$=14, $R_1$=22, $R_2$=20  & \multicolumn{1}{c|}{}                          & \multicolumn{1}{c|}{}                              & \multicolumn{1}{c|}{}                      & \multicolumn{1}{c|}{}                     & \multicolumn{1}{c|}{}                     &                      \\ \cline{3-3}
		&                           & $zx$-plane: $R_0$=20, $R_1$=10, $R_2$=19 & \multicolumn{1}{c|}{}                          & \multicolumn{1}{c|}{}                              & \multicolumn{1}{c|}{}                      & \multicolumn{1}{c|}{}                     & \multicolumn{1}{c|}{}                     &                      \\ \cline{3-3}
		&                           & Other planes: $R_0$=15, $R_1$=18, $R_2$=20  & \multicolumn{1}{c|}{}                          & \multicolumn{1}{c|}{}                              & \multicolumn{1}{c|}{}                      & \multicolumn{1}{c|}{}                     & \multicolumn{1}{c|}{}                     &                      \\ \cline{2-9} 
		& \multirow{5}{*}{High} & Data collection on nine planes   & \multicolumn{1}{c|}{\multirow{5}{*}{0.00001}} & \multicolumn{1}{c|}{\multirow{5}{*}{250}}         & \multicolumn{1}{c|}{\multirow{5}{*}{128}}  & \multicolumn{1}{c|}{\multirow{5}{*}{64}} & \multicolumn{1}{c|}{\multirow{5}{*}{32}} & \multirow{5}{*}{--} \\ \cline{3-3}
		&                           & $xy$-plane: $R_0$=23, $R_1$=11, $R_2$=21  & \multicolumn{1}{c|}{}                          & \multicolumn{1}{c|}{}                              & \multicolumn{1}{c|}{}                      & \multicolumn{1}{c|}{}                     & \multicolumn{1}{c|}{}                     &                      \\ \cline{3-3}
		&                           & $yz$-plane: $R_0$=10, $R_1$=20, $R_2$=18  & \multicolumn{1}{c|}{}                          & \multicolumn{1}{c|}{}                              & \multicolumn{1}{c|}{}                      & \multicolumn{1}{c|}{}                     & \multicolumn{1}{c|}{}                     &                      \\ \cline{3-3}
		&                           & $zx$-plane: $R_0$=18, $R_1$=8, $R_2$=16  & \multicolumn{1}{c|}{}                          & \multicolumn{1}{c|}{}                              & \multicolumn{1}{c|}{}                      & \multicolumn{1}{c|}{}                     & \multicolumn{1}{c|}{}                     &                      \\ \cline{3-3}
		&                           & Other planes: $R_0$=14, $R_1$=20, $R_2$=16  & \multicolumn{1}{c|}{}                          & \multicolumn{1}{c|}{}                              & \multicolumn{1}{c|}{}                      & \multicolumn{1}{c|}{}                     & \multicolumn{1}{c|}{}                     &                      \\ \hline
		\multirow{8}{*}{\thead{Composite\\ cement\\ past}}   & \multirow{4}{*}{Low}  & Data collection on three planes    & \multicolumn{1}{c|}{\multirow{4}{*}{0.00005}}  & \multicolumn{1}{c|}{\multirow{4}{*}{250}}          & \multicolumn{1}{c|}{\multirow{4}{*}{128}}  & \multicolumn{1}{c|}{\multirow{4}{*}{64}}  & \multicolumn{1}{c|}{\multirow{4}{*}{32}}  & \multirow{4}{*}{16}  
        \\ \cline{3-3}
		&                           & $xy$-plane: $R_0$=20, $R_1$=$R_2$=$R_3$=10  & \multicolumn{1}{c|}{}                          & \multicolumn{1}{c|}{}                              & \multicolumn{1}{c|}{}                      & \multicolumn{1}{c|}{}                     & \multicolumn{1}{c|}{}                     &   
        \\ \cline{3-3}
		&                           & $yz$-plane: $R_0$=20, $R_1$=$R_2$=$R_3$=10  & \multicolumn{1}{c|}{}                          & \multicolumn{1}{c|}{}                              & \multicolumn{1}{c|}{}                      & \multicolumn{1}{c|}{}                     & \multicolumn{1}{c|}{}                     &   
        \\ \cline{3-3}
		&                           & $zx$-plane: $R_0$=20, $R_1$=$R_2$=$R_3$=10  & \multicolumn{1}{c|}{}                          & \multicolumn{1}{c|}{}                              & \multicolumn{1}{c|}{}                      & \multicolumn{1}{c|}{}                     & \multicolumn{1}{c|}{}                     &   
        \\ \cline{2-9} 
		& \multirow{4}{*}{High}  & Data collection on three planes   & \multicolumn{1}{c|}{\multirow{4}{*}{0.00005}}  & \multicolumn{1}{c|}{\multirow{4}{*}{250}}          & \multicolumn{1}{c|}{\multirow{4}{*}{128}}  & \multicolumn{1}{c|}{\multirow{4}{*}{64}}  & \multicolumn{1}{c|}{\multirow{4}{*}{32}}  & \multirow{4}{*}{16}  \\ \cline{3-3}
		&                           & $xy$-plane: $R_0$=15, $R_1$=$R_2$=$R_3$=25 & \multicolumn{1}{c|}{}                          & \multicolumn{1}{c|}{}                              & \multicolumn{1}{c|}{}                      & \multicolumn{1}{c|}{}                     & \multicolumn{1}{c|}{}                     &   
        \\ \cline{3-3}
		&                           & $yz$-plane: $R_0$=15, $R_1$=$R_2$=$R_3$=25 & \multicolumn{1}{c|}{}                          & \multicolumn{1}{c|}{}                              & \multicolumn{1}{c|}{}                      & \multicolumn{1}{c|}{}                     & \multicolumn{1}{c|}{}                     &   
        \\ \cline{3-3}
		&                           & $zx$-plane: $R_0$=15, $R_1$=$R_2$=$R_3$=25 & \multicolumn{1}{c|}{}                          & \multicolumn{1}{c|}{}                              & \multicolumn{1}{c|}{}                      & \multicolumn{1}{c|}{}                     & \multicolumn{1}{c|}{}                     &   
        \\ \hline
	\end{tabular}
	\label{Tab:SENN_parameters}
\end{table}

\section*{Acknowledgements}
\noindent 
Wei Tan acknowledges the financial support from the EPSRC New Investigator Award [grant number: EP/V049259/1]. This work was selected by the ERC and funded by UK Research and Innovation (UKRI) under the UK government’s Horizon Europe funding guarantee [grant number EP/Y037103/1].

\section*{Data availability}
\noindent 
Data will be made available on request.

\section*{Declarations}
\noindent 
The authors declare that they have no conflict of interest in this paper.

\bibliographystyle{unsrt} 
\bibliography{bibliography}

\end{document}